\documentclass[a4]{JHEP3} 
\usepackage{feynmp} 
\usepackage{amsmath}
\usepackage{graphicx}
\usepackage{pstricks}
\usepackage{graphpap}
\usepackage{cite}

\psset{linewidth=2pt}
\psset{unit=1pt}

\newcommand{\epemto}{e^+e^-\to}
\newcommand{\m}{-}
\newcommand{\p}{+}

\def\as{\alpha_{\mbox{\tiny S}}}
\newcommand{\nnb}{\nonumber}
\title{Implementing the ME+PS merging algorithm}
\author{A.\ Sch{\"a}licke\\
        Institut f{\"u}r Theoretische Physik,\\
        TU Dresden\\
        D-01062 Dresden, Germany\\
        E-mail: dreas@theory.phy.tu-dresden.de}
\author{F.\ Krauss\\
        Institut f{\"u}r Theoretische Physik,\\
        TU Dresden\\
        D-01062 Dresden, Germany\\
        E-mail: krauss@theory.phy.tu-dresden.de}

\abstract{
\noindent 
The method to merge matrix elements for multi particle production and
parton showers in $e^+e^-$ annihilations and hadronic collisions and
its implementation into the new event generator SHERPA is described in
detail. Examples highlighting different aspects of it are thoroughly
discussed, some results for various cases are presented. In addition, 
a way to extend this method to general electroweak interactions is
presented.} 


\newcommand{\graphtwojet}{
\begin{fmfgraph*}(100,100) 
  \fmfbottom{i0,i1} 
  \fmftop{o0,o1} 
  \fmf{plain,tension=2.}{i0,v0} 
  \fmfv{label=0}{i0} 
  \fmf{plain,tension=2.,label=$Q$}{v0,v1} 
  \fmf{phantom,tension=2.}{v1,i1} 
  \fmfv{label=1}{i1} 
  \fmf{gluon}{v1,o1} 
  \fmf{gluon}{v0,o0} 
  \fmffreeze
  \fmf{plain}{v1,ii1} 
  \fmf{fermion}{ii1,i1} 
\end{fmfgraph*} 
\psccurve[linestyle=dashed]
(-70,56)(-30,56)(-10,36)(-30,16)(-70,16)(-90,36)
}

\newcommand{\graphthreejet}{
\begin{fmfgraph*}(100,100) 
  \fmfbottom{i0,i1} 
  \fmftop{o0,oa,o1,o2}
  \fmf{plain,tension=2}{i0,v0}   
  \fmfv{label=0}{i0} 
  \fmf{plain,tension=2.,label=$Q$}{v0,v1} 
  \fmfv{label=1}{i1} 
  \fmf{plain,tension=2}{v1,i1}   
  \fmf{phantom}{v1,o2}
  \fmf{gluon}{v0,o0} 
  \fmffreeze
  \fmf{plain}{i0,ii0,v0} 
  \fmf{plain}{v1,ii1} 
  \fmf{fermion}{ii1,i1} 
  \fmf{gluon}{v1,v2} 
  \fmf{plain}{v2,o1} 
  \fmf{fermion}{v2,o2} 
  \fmfv{label=$Q_1$,label.angle=180}{v2}
\end{fmfgraph*} 
\psccurve[linestyle=dashed]
(-70,56)(-30,56)(-10,36)(-30,16)(-70,16)(-90,36)
}

\begin{document}
\begin{fmffile}{agraphs}
\section{Introduction}

\noindent
The analysis of multi-particle final states becomes increasingly
important in the search for the production and decay of new, heavy
particles. Therefore, in order to guide such analyses, their
simulation in Monte Carlo event generators should be as correct as
possible. There are two complementary approaches to model the
production of multi-particle final states: First, employing
fixed-order perturbation theory, exact matrix elements at tree-level
or beyond describe particle production in specific processes through
Feynman diagrams, taking into account all quantum interferences at the
corresponding level of accuracy. Alternatively, the parton shower
approach organises the emission of secondary partons in such a way
that all leading collinear or soft logarithms of the form 
$\alpha_S^n\log^{2n}$ are resummed. The former way of modelling
particle production has the benefit of being well-defined and exact up
to given fixed-order accuracy for large angle or high energy emission
of partons, whereas the second approach correctly treats the soft
and collinear regions of phase space. Of course, a combination of both
approaches allows for a better description of particle production over
the full available phase space. A way of merging multi-particle
matrix elements at tree-level with the subsequent parton shower
consistently at leading logarithmic accuracy and taking into account
important parts of the next-to leading logarithms was formulated first
for the process $\epemto$hadrons in \cite{Catani:2001cc}. The
principles of its application to hadronic processes have been
discussed in \cite{Krauss:2002up}. In both cases, the phase space of
particle production is divided into two disjoint regimes, one of jet
production covered by the corresponding matrix elements, and one of
jet evolution modelled through the parton shower. The separation in
both cases, $\epemto$hadrons and hadronic collisions, is achieved
through a $k_\perp$ jet measure \cite{Catani:1991hj,Catani:1992zp,
Catani:1993hr}. The implementation of this algorithm is discussed in
detail in this publication; it forms the cornerstone of the new event
generator SHERPA \cite{Gleisberg:2003xi}. The approach been extended
\cite{Lonnblad:2002sy} to cover also the dipole shower formulation of
multiple parton emission \cite{Lonnblad:1992tz}. A somewhat related
approach has been taken in \cite{Mangano:2001xp}. The algorithm has
been tested in a variety of versions, ranging from its prime example
$\epemto$hadrons \cite{Kuhn:2000dk} over the production of gauge
bosons in hadronic collisions at the Tevatron \cite{Mrenna:2003if,
Krauss:2004bs} or the LHC \cite{wz@LHC} to $W$-pair production 
at the Tevatron \cite{wpair@Tevatron}.  

\noindent
To some extent, however, all existing algorithms so far assume that
there is a signal process (like, e.g., $\epemto q\bar q$ or 
$q\bar q\to l\bar\nu_l$) with one specific order in the electroweak
coupling constant and that all additional jets are emitted through
strong interactions. This implies that all matrix elements have the
same order in the electroweak coupling constant and that they form 
a hierarchy of extra orders in $\alpha_s$, related to extra jets.
Despite its apparent success there is one question that remains to be
answered. This is the question of how to deal with situations where
both electroweak and strong amplitudes contribute significantly to the
same final state. For example, at LEP~II both pure QCD amplitudes and
$W$ boson pair production amplitudes contribute to the total cross
section of 4-jet production processes. Depending on the specific
kinematical situation, their relative amount may vary; however, they
exhibit different properties. This is exemplified by their colour
flows, being responsible for the kinematical domain in which hadrons
are formed. It is clear that a consistent merging procedure for such
processes is highly desirable. Such a merging algorithm has to take
proper care of relevant coupling constants, and it has to reliably
predict the corresponding colour structure.

\noindent
In the next section, Sec.\ \ref{Algo_Sec}, the algorithm is discussed
for both $e^+e^-$ and hadronic processes. Certain aspects presented
here have not been covered before. They include the treatment of jet
production beyond the availability of corresponding matrix elements
and some ways of using variable jet resolution scales for different
jet multiplicities. In addition, some first steps into the direction
of treating matrix elements, where electroweak and strong interactions
compete with each other are reported. The presentation proceeds with
examples highlighting the ideas underlying the algorithm, cf.\ Sec.\
\ref{Example_Sec}. Finally, a large amount of results indicating its
quality are presented in Sec.\ \ref{Results_Sec}. Details on the
specific implementation of the algorithm into SHERPA are given in the
appendix, Sec.\ \ref{Det_Sec}.

\section{The algorithm}\label{Algo_Sec}

\noindent
In this section the merging algorithm together with its extensions, as
implemented in SHERPA, will be discussed. It 
can be divided into three parts. First of all, a sample of matrix
elements has to be defined, from which processes are selected for the
generation of individual events. The four-momenta of the particles are
distributed according to the corresponding matrix element. Then,
having fixed the number, flavours and four-momenta of the particles, a
pseudo parton shower history is constructed through backwards
clustering of the particles according to the $k_\perp$ algorithm. Here, 
care has to be taken in situations, where one particular clustering
allows for different colour flows. The nodal values of this clustering
serve as input for the construction of a weight applied on the matrix
element. This weight resums higher order effects at leading
logarithmic accuracy and consists of Sudakov form factors for coloured
lines (e.g., quarks or gluons) and of ratios of the strong coupling
constant. If the event is accepted, parton showers are attached to the
outgoing particles. For this, again the nodes of the clustering serve
as input. Inside the parton shower, those emissions are vetoed that
lead to additional unwanted jets. At this point there is some
subtlety, since it is obviously impossible to calculate
matrix elements for an infinite number of additional jets. Hence,
there is some maximal number $n_{\rm max}$ of jets covered by the
matrix elements, higher jet multiplicities must be accounted for by
the parton shower. This leads to a somewhat modified treatment of the
parton shower for those events with $n_{\rm max}$ jets stemming from
the matrix elements. 

\noindent
Apart from this special treatment for configurations with the
largest number of jets produced through matrix elements, there are
cases where a similar treatment is necessary for configurations
with the minimal number of jets. Examples include both electroweak
such as $\epemto W^{+}W^{-} \to jets$ and the QCD production of jets
in hadronic collisions. Using the $k_\perp$ measure to separate
matrix elements with the minimal number of jets $n_{\rm min}$ from 
higher jet multiplicities with $n>n_{\rm min}$ also restricts the
phase space for the minimal number of jets. Since the separation of
different jet multiplicities is slightly washed out by the parton
shower and hadronisation, the lowest jet multiplicity samples
experience a loss of events at the phase space boundary, which is not
compensated for by smaller jet multiplicities. To deal with this
problem one may try to use generation cuts that are much tighter than
the analysis cuts - an option that is clearly not very
efficient. Alternatively, a lower jet definition cut may be used for
the lowest jet multiplicity. This idea leads to an extension 
of the algorithm, which enables a merging of processes with different
jet multiplicities and different separation cuts. In fact, this
algorithm  is closely related to the highest multiplicity treatment.

\noindent
In the following, the algorithm and some of its refinements are discussed
in greater detail, dividing the procedure into three steps, namely
matrix element generation, parton clustering and Sudakov weight
construction and into, finally, running the parton showers. 

\subsection{Combination of matrix elements}
\begin{enumerate}
\item Composition of the process samples:\\
  In each run, processes with a fixed identical number of electroweak
  couplings may be combined, which differ only in the number of extra
  jets produced through QCD. All strongly interacting particles are
  subject to phase space cuts according to the $k_\perp$ algorithm
  \cite{Catani:1991hj,Catani:1992zp,Catani:1993hr}. This is necessary
  in order to allow a reweighting of the matrix element with Sudakov
  form factors. Phase space cuts on other particles are not needed
  unless for the sake of avoiding potential infrared divergencies. As
  an example consider the case of a (massless) lepton pair, which have
  to be cut through, e.g.\ a cone algorithm or by demanding some
  minimal invariant masses. 

  \noindent 
  In the original algorithm, however, it was implicitly assumed that
  any gauge boson of the electroweak interactions is connected to
  maximally one strongly interacting line only; in other words, it was
  implicitly assumed that any photon, $W$ or $Z$ boson would couple to
  one quark line only. The present proposal aims at widening the scope
  in such a way that competition between strong and electroweak
  interactions is possible.
\item Selection of a particular process:\\
  For all the processes contributing in a single run, labelled with
  $i$, total cross sections are evaluated at tree-level through
  \begin{align}
    \sigma_i^{(0)} = 
    \int {\rm d}\Omega |{\cal M}(\mu_R,\,\mu_F) |^2\,,
  \end{align}
  where ${\rm d}\Omega$ denotes the integral over the available phase
  space. For convenience, here any eventual integration over the
  Bjorken-$x$ of incoming partons is subsumed in ${\rm d}\Omega$. The 
  matrix element ${\cal M}$ eventually is extended by parton
  distribution functions; it is evaluated at $\mu_R = \mu_F =
  Q_{\rm cut}$, the cut parameter of the $k_\perp$ algorithm\footnote{    
    A comment is in order here: Often in hadronic collisions, it
    proves useful to use the $k_\perp$ algorithm with a parameter $D$,
    which can be identified as a pseudo cone-size. In such a case, the
    $Q_{\rm cut}$ value is rescaled by $D$.}. 
  
  \noindent
  The probability ${\cal P}_i$ for a process $i$ to be selected for
  event generation is then given by
  \begin{align}
    {\cal P}_i^{(0)} &= \frac{\sigma_i^{(0)}}{\sum_j\sigma_j^{(0)}}\,.
  \end{align}
  Having selected the process, four-momenta for the incoming and
  outgoing particles are chosen according to its matrix element. 
\item Highest multiplicity treatment:\\
  For those processes that have the highest multiplicity of jets in
  the matrix element, i.e.\ where $n=n_{\rm max}$, already during
  integration the scale $Q_s$ of the softest jet produced through QCD
  is determined according to the $k_\perp$ algorithm. Then, the
  parton distribution functions in the matrix element are taken at the
  factorisation scale $\mu_F = Q_s$, whereas the renormalisation scale
  used in the evaluation of the coupling constant remains at 
  $\mu_R = Q_{\rm cut}$. Of course, this will later on affect the 
  Sudakov weights and the parton showering as well; at that point,
  however, it implicitly takes into account the possibility of having
  softer extra jets emitted in the initial state parton shower. 
\item Multi-cut treatment:\\
  There is some condition that the multi-cut treatment does not lead
  to ambiguities, namely that the jet definition becomes tighter with
  increasing jet multiplicity. In other words, for each jet
  multiplicity $n$, a jet separation cut $Q_{\rm  cut}^{(n)}$ is
  defined such that $Q_{\rm  cut}^{(n-1)}\le Q_{\rm  cut}^{(n)}$. 
  For the calculation of the corresponding a priori cross sections 
  $\sigma^{(n)}$ the factorisation scales are also set dynamically to 
  \begin{align}\label{eq:Qmin_def}
    \mu_F = Q_{\rm min} =
    \mbox{\rm min}\left\{Q_s,Q_{\rm cut}^{(n+1)}\right\}\,,
  \end{align}
  where $Q_s$ is the scale of the softest jet in the process. Again,
  the renormalisation scale is fixed at $\mu_R = Q_{\rm cut}^{(n)}$.
\end{enumerate}

\subsection{Pseudo parton shower history}
\begin{enumerate}
\item Clustering of particles:\\
  In the original version of the merging procedure, only QCD
  clusterings have been considered. There, for each allowed pair of
  partons a relative transverse momentum has been defined.
  According to the $k_\perp$ algorithm, its square reads
  \begin{align}\label{yij}
  Q_{ij}^2 &= 
  2\,\mbox{\rm min}\{E_i^2, E_j^2\}(1-\cos\theta_{ij})
  \end{align}
  for a pair of partons in $e^+e^-$ collisions. In hadronic
  collisions it is given by
  \begin{align}\label{y_out} 
  Q_{ij}^2 &= 
  2\,\mbox{\rm min}\{p_{\perp,i}^2, p_{\perp,j}^2\} \, 
  \left[\cosh^2(\eta_i-\eta_j)-\cos^2(\phi_i-\phi_j)\right]
  \end{align}
  for the clustering of two final state hadrons, and by
  \begin{align}\label{y_in} 
  Q_{i}^2 &= 
  p_{\perp,i}^2
  \end{align}
  when a final state parton is to be clustered with an initial state
  particle. In the original algorithm, the pair with the lowest
  $k_\perp$ has been clustered. In order to prohibit ``illegal''
  clusterings, such as, for instance, the clustering of two quarks
  instead of a quark-anti-quark pair, only those pairs have been
  considered that correspond to a Feynman diagram contributing to the
  process in question. 

  \noindent
  Going beyond this, some new problems may manifest themselves, which
  are related to the possibility of having competing colour flows. A
  good example for this is the possible competition of clustering a
  quark-anti-quark pair into either a gluon or a $Z$-boson. To resolve
  this ambiguity, there are, in principle, two options: One would be
  to globally select a specific colour configuration according to the
  relative weight of different colour-ordered amplitudes. This is
  clearly the preferred choice. The other one, that will be pursued
  as a proposal here, is to try to decide locally which colour
  configuration to chose. 

  \noindent
  For this, relative weights are constructed for each possible
  clustering, which take into account the coupling and pole structure
  of the underlying Feynman diagram(s). In each case, the contribution
  ${\cal W}_{ij}$ to the weight for a specific clustering of two
  particles $ij\to k$ can be written as 
  \begin{align}
    {\cal W}_{ij;k}(q) &= \frac{[g_{ij}(\mu_R)g_{kl}(\mu_R)]^2}
    {(q^2-m_k^2)^2+m_k^2\Gamma_k^2}\,,   
  \end{align}
  where $g_{ij}(\mu_R)$ is the coupling constant at the vertex
  and $g_{kl}(\mu_R)$ denotes the coupling constant of the resulting
  propagator in the potential next clustering $kl\to m$. In addition,
  the propagator term contains the mass and the (fixed) width of
  particle $k$. Clearly, $m_k$ and $\Gamma_k$ may be zero, for
  instance for gluons or massless quarks. In the equation above,
  $q^2$ is the square of the four-momentum $q$ of the pair $ij$.  

  \noindent
  Now, for each allowed clustering of pairs $ij$, all potential 
  ${\cal W}_{ij}(q)$ with different coupling structure are added,
  and the pair with the largest total weight
  \begin{align}
    {\cal W}_{ij}^{\rm tot}(q) &= \sum\limits_k{\cal W}_{ij;k}(q)
  \end{align}
  is selected. The emerging propagator $k$ is then chosen according to
  the relative probability 
  \begin{align}
    {\cal P}_k &= 
    \frac{{\cal W}_{ij;k}(q)}{{\cal W}_{ij}^{\rm tot}(q)}\,.
  \end{align}

\item Core $2\to 2$ process:\\
  In the original as well as in the extended version of the merging
  algorithm, proposed here, this clustering procedure is repeated recursively, until
  a core $2\to 2$ process is recovered. It defines the initial colour
  flow in the large $N_c$ limit necessary for the fragmentation. In
  addition, through this choice of an initial colour flow, the hard process scales
  $Q_h$ for the partons in this process are defined. The following
  cases must be considered: 
  \begin{itemize}
    \item Two particles with and two particles without colour quantum
      number, for instance $e^+e^-\to q\bar q$, $q\bar q\to e^+e^-$,
      or $gg\to H\to \tau\tau$ in an effective model for the $ggH$
      coupling. Then, for the two coloured objects, $Q_h^2 = \hat s$. 
    \item Three particles with colour quantum numbers and one without,
      for instance in $q\bar q\to Wg$. Then, for the incoming
      particles $Q_h^2 = \hat s$, and for the outgoing ones $Q_h$ is
      given by their transverse momentum. 
    \item Four particles with colour. In this case, often different
      colour structures are competing. The selection is then made
      according to relative contributions which can usually be
      connected with the $\hat s$, $\hat t$ or $\hat u$ channel
      exchange of colour. The hard scale for all four particles is
      then chosen according to this selection. Hence, usually, the
      minimum of $\hat u$ and $\hat t$ is the relevant scale. 
  \end{itemize}
\item Construction of the Sudakov weight: \\
  Having fixed the parton shower sequence and the hard scales $Q_h$,
  the Sudakov weight can be calculated. To a large extent, the
  construction prescription  for this is the same for the original
  approach as well as for its proposed extension. In both cases, the
  Sudakov weights consists of ratios of the strong coupling constant
  taken at the varying nodal scales and at the fixed renormalisation
  scale and of Sudakov form factors, which in next-to-leading
  logarithmic (NLL) approximation are given by 
  \begin{align}
    \Delta(Q,Q_0) = \exp\left[-\int\limits_{Q_0}^Q
      {\rm d}q\Gamma(Q,q)\right]\,.
  \end{align}
  Here, $\Gamma(Q,q)$ is the integrated splitting function for the
  particle in question. For convenience, it incorporates a factor
  $\alpha_S(q)/q$. Integrated splitting functions for different
  splittings are listed in the appendix. In terms of these
  constituents the Sudakov weight is constructed as a product of
  \begin{itemize}
  \item
    factors $\alpha_S(Q)/\alpha_S(\mu_R)$ for each node with
    $Q=k_\perp$ which involves a strong coupling constant;
  \item 
    factors $\Delta(Q_1,Q_{\rm cut})/\Delta(Q_2,Q_{\rm cut})$ for each
    internal line (propagator) that carries colour quantum numbers,
    where the arguments are given by the nodal $k_\perp$ scales
    $Q_1$ and $Q_2$;
  \item 
    and of factors $\Delta(Q_1,Q_{\rm cut})$ for colour-charged
    outgoing lines emerging at a node with $Q_1=k_\perp$.
  \end{itemize}

  \noindent
  At this point it should be noted that in such cases where coloured
  particles are produced through $s$-channel electroweak interactions
  the nodal scale value of the vertex should be the invariant mass 
  $\hat s$  of the particles rather than their transverse momentum,
  which again is beyond the scope or the original algorithm.

\item Highest multiplicity treatment:\\
  In case, a hard process with the maximal number of jets accommodated
  by the matrix elements has been chosen, the parton shower must be
  able to produce higher jet configurations. Of course, these
  additional jets may in principle emerge at transverse momenta larger
  than the jet definition cut $Q_{\rm cut}$. On the other hand, it is
  clear that they should be softer than the softest jet produced by
  the matrix element in order to ensure that the matrix element is
  used to cover the hard regions of phase space. Since the Sudakov
  form factors forming the weight attached to the matrix elements can
  be identified as a no-radiation probability between two scales, the
  soft scales of the Sudakov form factors need to be modified. Because
  in this situation radiation from the parton shower must be softer
  than $Q_s$, the scale of the softest jet in the matrix element,
  rather than $Q_{\rm cut}$, this modification amounts to a
  replacement $Q_{\rm cut} \to Q_s$ in all Sudakov form factors,
  i.e.\ for both internal and external lines.
\item Multi-cut treatment:\\
  Similarly to the highest multiplicity treatment discussed above, in
  the multi-cut treatment the soft scales of the Sudakov weight are
  also set dynamically, i.e.\ in dependence on the actual kinematical
  configuration. The scale $Q_{\rm cut}$ in the Sudakov form factors 
  for an $n$ jet process is replaced by $Q_{\rm min}$ defined according to
  Eq.\ \eqref{eq:Qmin_def}. In case the process is purely
  electroweak, like $\epemto ZZ \to q\bar{q} q\bar{q}$, this
  prescription translates into completely switching off all Sudakov
  form factors if $Q_s \le Q_{\rm cut}^{(n+1)}$.  
\end{enumerate}

\subsection{Starting the parton shower}
\begin{enumerate}
\item Vetoing emissions:\\
  According to the paradigm of the merging procedure, inside the
  parton shower\footnote{
    It should be noted here that SHERPA employs a parton shower
    ordered by virtualities, supplemented by an explicit veto on
    rising opening angles in branching processes. This is an apparent
    mismatch to the transverse momenta taken as scales so far. Thus,
    in the following it should be understood that all scales $Q$
    emerging from the parton shower denote the transverse momentum
    that can be approximated from the splitting kinematics formulated
    in terms of $t$, the respective virtual mass.}
  all emissions leading to extra unwanted jets are vetoed. This is
  implemented in the following way: 
  \begin{itemize}
    \item The probability for no branching resolvable at a scale
      $t_0$, usually the infrared cut-off of the parton shower,
      between two scales $t_1$ and $t_2$ is given by the ratio
      \begin{align}
        {\cal P}_{\rm no}(t_1,t_2) = 
        \frac{\Delta(t_1,t_0)}{\Delta(t_2,t_0)}
      \end{align}
      of Sudakov form factors. Equating this with a random number
      allows to solve this for $t_2$, the scale of the next trial
      emission\footnote{It should be noted here that in usual parton
        shower programs the Sudakov form factors rely on the integral
        over splitting functions rather than on integrated splitting
        functions. Therefore, usually a splitting variable $z$ is
        selected with a second random number. In SHERPAs parton shower
        module APACIC, only then transverse momenta can be constructed
        from $t$ and $z$. This, however, is primarily a technical
        issue.}.
    \item Having at hand the transverse momentum related to this
      trial emission, it can be compared with the jet resolution  
      of the $k_\perp$ algorithm. If this particular emission would
      give rise to an unwanted jet, the next trial emission is 
      constructed with its upper scale $t_1$ equal to the actual scale
      $t_2$ of the vetoed emission.
  \end{itemize}
  For a single parton line starting at some scale $Q$ the matrix
  element correction weight reads
  \begin{align}
    W_{\rm ME} = \Delta(Q,Q_{\rm cut})\,.
  \end{align}
  The combined weight of all possible rejections due to vetoed
  emissions in the parton shower for the same parton, starting at
  scale $Q$ at NLL accuracy reads 
  \begin{align}
    W_{\rm PS} = 
    \left[1+
      \int\limits_{Q_{\rm cut}}^Q{\rm d}q\Gamma(Q,q) +
      \int\limits_{Q_{\rm cut}}^Q{\rm d}q\Gamma(Q,q) 
        \int\limits_{Q_{\rm cut}}^q{\rm d}q'\Gamma(Q,q')+
      \dots\right] = \Delta^{-1}(Q,Q_{\rm cut})\,.
  \end{align}
  Combining both thus formally leads to a cancellation of large logarithms
  of the form $\log Q/Q_{\rm cut}$ at NLL precision. However, there
  are remaining dependencies on $Q_{\rm cut}$, some of which are due 
  to the fact that the actual implementation of the parton shower is at a
  different level of logarithmic accuracy. 

  \noindent
  For the highest multiplicities or for the multi-cut treatment,
  the veto of course is performed w.r.t.\ $Q_s$ or $Q_{\rm min}$,
  respectively. In case the process is purely electroweak no shower
  veto is applied at all as long as $Q_{\rm min}\le Q_{\rm cut}$.  
\item Starting scales:\\
  The reasoning above immediately implies which starting scales are to
  be chosen for the parton shower evolution of each parton. In each
  case it should be the scale where the parton was first produced, in
  accordance with how the Sudakov weights are constructed and how the
  vetoing applied in the parton shower cancels the dependence on
  the jet resolution scale. 

  \noindent
  There is one last minor point to be discussed, namely the scale, at
  which the parton density functions are evaluated in the backward
  evolution of initial state showers. Remember that there, in order to
  recover the correct parton distribution functions at each step of
  the space-like evolution, the ratio of Sudakov form factors
  describing the no-branching probability between $t_1$ and $t_2$ are
  supplemented with corresponding factors, namely,
  \begin{align}
    {\cal P}_{\rm no}(t_1,t_2) = 
    \frac{f(x,\,t_2)}{f(x,\,t_1)}\cdot
    \frac{\Delta(t_1,t_0)}{\Delta(t_2,t_0)}\,.
  \end{align}
  If this expression is to describe the first emission through the
  parton shower along an incoming parton line, the hard scale $t_1$ in
  the parton distribution function is replaced by either $Q_{\rm cut}$
  (or $Q_s$ or $Q_{\rm min}$, if the process in question has the
  maximal number of jets in the matrix element, or if the multi-cut
  treatment is active).
\end{enumerate}

\section{Examples}\label{Example_Sec}

\noindent
In this section, the algorithms discussed above are illustrated
through some examples, namely
\begin{enumerate}
\item $\epemto \mbox{jets}$,
\item $p \bar{p} \to W + \mbox{jets}$ ,
\item $p \bar{p} \to \mbox{jets}$ , and
\item $\epemto d \bar{d} u \bar{u} (g)$.
\end{enumerate}

\subsection{Example I -- $\epemto \mbox{jets}$}

\begin{figure}[h!]
\begin{center}
\begin{pspicture}(230,110)
\put(0,10){
\begin{fmfgraph*}(100,100) 
  \fmfbottom{i0,i1} 
  \fmftop{o2,o1,o0} 
  \fmf{plain,l.side=left,label=$e^\m$}{i0,v0} 
  \fmf{fermion,l.side=left,label=$e^\p$}{v0,i1} 
  \fmf{boson,tension=2.}{v0,v1} 
  \fmf{plain}{v1,v3} 
  \fmf{fermion}{v3,o0} 
  \fmf{plain}{v2,v1} 
  \fmf{plain}{v2,o2} 
  \fmffreeze
  \fmf{gluon}{o1,v3} 
  \fmfblob{1mm}{v1,v3}
  \fmfv{label=$Q$,label.angle=-150}{v1}
  \fmfv{label=$Q_1$,label.angle=-70}{v3}
\end{fmfgraph*}}
\put(130,10){
\begin{fmfgraph*}(100,100) 
  \fmfbottom{i0,i1} 
  \fmftop{o2,o1,o0} 
  \fmf{plain,l.side=left,label=$e^\m$}{i0,v0} 
  \fmf{fermion,l.side=left,label=$e^\p$}{v0,i1} 
  \fmf{boson,tension=2.}{v0,v1} 
  \fmf{plain}{v1,v3} 
  \fmf{fermion}{v3,o0} 
  \fmf{plain}{v2,v1} 
  \fmf{plain}{v2,o2} 
  \fmffreeze
  \fmf{gluon}{v2,o1} 
  \fmfblob{1mm}{v1,v2}
  \fmfv{label=$Q$,label.angle=-30}{v1}
  \fmfv{label=$Q_1$,label.angle=-90}{v2}
\end{fmfgraph*}}
\put(50,5){(a)}
\put(180,5){(b)}
\end{pspicture}
\caption{\label{3j_topos} The diagrams contributing to $\epemto
  3\mbox{jets}$. }
\end{center}
\end{figure}
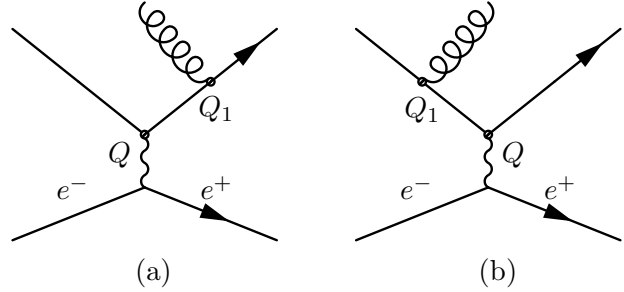

\noindent
As a first example for the original version of the algorithm, consider
the process $\epemto \mbox{jets}$ at LEP~I. Choosing a jet resolution of
$Q_{\rm cut}=5.77$ GeV ($y_{\rm cut}=0.004$ in the Durham scheme),
the a-priori cross sections $\sigma^{(0)}_i$ and the resulting
effective cross sections $\sigma_i=\sigma^{(0)}_i \cdot W_{\rm ME}$ 
for a specific choice of $\alpha_S=0.127$ are given by
\begin{equation}
\begin{array}{l@{\,=\,}r@{\extracolsep{15mm}}
              l@{\extracolsep{1mm}\,=\,}r@{\extracolsep{15mm}}
              l@{\extracolsep{0mm}\,=\,}r}
\sigma_2^{(0)}        & 40.46 \,\mbox{\rm nb}      &
\sigma_2^{\rm nmax=5} & 18.80 \,\mbox{\rm nb}      & ({\cal R}_2 & 38.8\%)\nnb\\
\sigma_3^{(0)}        & 43.38 \,\mbox{\rm nb}      &
\sigma_3^{\rm nmax=5} & 21.10 \,\mbox{\rm nb}      & ({\cal R}_3 & 43.5\%)\nnb\\
\sigma_4^{(0)}        & 14.05 \,\mbox{\rm nb}      &
\sigma_4^{\rm nmax=5} & 6.90  \,\mbox{\rm nb}      & ({\cal R}_4 & 14.2\%)\nnb\\
\sigma_5^{(0)}        & 2.80  \,\mbox{\rm nb}      &
\sigma_5^{\rm nmax=5} & 1.69  \,\mbox{\rm nb}      & ({\cal R}_5 & 3.5\%)\nnb\\
\multicolumn{2}{c}{}&
\sigma_{\rm tot}^{\rm nmax=5} & 
\multicolumn{3}{l}{
\sum\limits_{i=2}^5\sigma_i^{\rm nmax=5} = 48.49 \,\mbox{\rm nb}}
\end{array}
\end{equation}
Assume now that at some point a three-jet event is chosen with a 
$q\bar qg$ final state. The diagrams contributing to this process are
depicted in Fig.\ \ref{3j_topos}. 
There are two allowed clusterings, namely $qg$ and $\bar qg$. If the
former leads to a smaller $k_\perp$, i.e.\ if $Q(qg)<Q(\bar qg)$ this
clustering is selected, of course with scale $Q_1=Q(qg)$, and the core
$2\to 2$ process is readily recovered. Its associated hard scale is
$Q=\sqrt{s}$. The Sudakov weight is constructed, leading to 
\begin{align}
  {\cal W}_{\rm ME} &= \Delta_{\bar q}(Q,Q_{\rm cut})\,
  \frac{\Delta_q(Q,Q_{\rm cut})}{\Delta_q(Q_1,Q_{\rm cut})}\,
  \frac{\alpha_S(Q_1)}{\alpha_S(Q_{\rm cut})}\,
  \Delta_q(Q_1,Q_{\rm cut})\Delta_g(Q_1,Q_{\rm cut})\nonumber\\
  &=\Delta_{\bar q}(Q,Q_{\rm cut})\,\Delta_q(Q,Q_{\rm cut})\,
  \frac{\alpha_S(Q_1)}{\alpha_S(Q_{\rm cut})}\,
  \Delta_g(Q_1,Q_{\rm cut})\,,
\end{align}
where the first factor in the first line corresponds to the anti-quark
line, the second factor is for the internal quark line, the ratio of
the strong coupling constants applies for the vertex, and the two last
factors correspond to the two outgoing lines. Emissions in the parton
shower for all three lines are vetoed if their transverse momentum is
larger than $Q_{\rm cut}$; the start scales for the parton shower
evolution are $t$ for both the quark and the anti-quark line, 
and $t_1$ for the gluon.

\noindent
If, in contrast in the simulation the matrix elements are restricted
by $n_{\rm max}=3$, the highest multiplicity treatment would apply to
the three-jet configuration. Consequently, the cross sections and
rates change according to  
\begin{equation}
\begin{array}{l@{\,=\,}r@{\extracolsep{15mm}}
              l@{\extracolsep{1mm}\,=\,}r@{\extracolsep{15mm}}
              l@{\extracolsep{0mm}\,=\,}r}
 \sigma_2^{(0)}        & 40.46 \,\mbox{\rm nb}      &
 \sigma_2^{\rm nmax=3} & 18.80 \,\mbox{\rm nb}      & ({\cal R}_2 & 36.3\%)\\
 \sigma_3^{(0)}        & 43.38 \,\mbox{\rm nb}      &
 \sigma_3^{\rm nmax=3} & 32.94 \,\mbox{\rm nb}      & ({\cal R}_3 & 63.7\%)\\
\multicolumn{2}{c}{}&
    \sigma_{\rm tot}^{\rm nmax=3} &
\multicolumn{3}{l}{
    \sum\limits_{i=2}^3\sigma_i^{\rm nmax=3} = 51.74 \,\mbox{\rm nb}\,,}
  \end{array}
\end{equation}
and the weight for the three-jet configuration would be given by
\begin{align}
  \tilde{\cal W}_{\rm ME} &= \Delta_{\bar q}(Q,Q_1)\,
  \frac{\Delta_q(Q,Q_1)}{\Delta_q(Q_1,Q_1)}\,
  \frac{\alpha_S(Q_1)}{\alpha_S(Q_1)}\,
  \Delta_q(Q_1,Q_1)\Delta_g(Q_1,Q_1)\nonumber\\
  &=\Delta_{\bar q}(Q,Q_1)\,\Delta_q(Q,Q_1)\,.
\end{align}
Emissions in the parton shower for all three lines would then be
vetoed if their transverse momentum was larger than $Q_1$; the start
scales for the parton shower evolution again are $t$ for both the
quark and the anti-quark line, and $t_1$ for the gluon.

\subsection{Example II -- $p \bar{p} \to W + \mbox{jets}$}

\noindent
The next example that will be considered is a case where both initial
and final state emissions may occur. Hence, the reconstruction of the
pseudo parton shower history and the evaluation of the corresponding
weight is more involved.

\noindent
Again, the starting point will be the
calculation of cross sections. For $Q_{\rm cut}=20$ GeV, $\alpha_S=0.118$,
and by using the CTEQ6L parton distribution functions, they read
\begin{equation}
\begin{array}{l@{\,=\,}r@{\extracolsep{15mm}}
              l@{\extracolsep{1mm}\,=\,}r@{\extracolsep{15mm}}
              l@{\extracolsep{0mm}\,\,}r}
    \sigma_0^{(0)} & 953.03 \mbox{\rm pb}&
    \sigma_0^{\rm nmax=3} & 825.77 \mbox{\rm pb}\\
    \sigma_1^{(0)} & 155.76 \mbox{\rm pb} &
    \sigma_1^{\rm nmax=3} & 108.35 \mbox{\rm pb}\\
    \sigma_2^{(0)} & 36.75 \mbox{\rm pb}& 
    \sigma_2^{\rm nmax=3} & 20.10 \mbox{\rm pb}\\
    \sigma_3^{(0)} & 7.22 \mbox{\rm pb}& 
    \sigma_3^{\rm nmax=3} & 3.32 \mbox{\rm pb}\\
\multicolumn{2}{c}{}&
    \sigma_{\rm tot}^{\rm nmax=3} &
\multicolumn{3}{l}{
    \sum\limits_{i=0}^3\sigma_i^{\rm nmax=3} = 957.54  \mbox{\rm pb}\,.}
  \end{array}
\end{equation}

\noindent
In the following, the construction of the weights for different
multiplicities and the starting conditions for the subsequent parton
shower will be briefly discussed.
\begin{enumerate}
\item $n=0$:\\
  Starting with the lowest multiplicity of jets produced in the matrix
  element, $n=0$, the leading order contributions to $W^\m$ production
  are recovered. They are of the Drell--Yan type, i.e.\ processes of
  the form  
  \[
  q \,\bar{q}' \to e \bar{\nu}_e\;.
  \]
  Obviously, this is already $2\to 2$ process, therefore clustering
  does not take place. Due to the absence of any strong interaction,
  the rejection weight is merely given by two quark Sudakov form factors:
  \begin{align}
    {\cal W_{\rm ME}} = 
    \Delta_{q}(Q,Q_{\rm cut})\,\Delta_{\bar{q}'}(Q,Q_{\rm cut})\,,
  \end{align}
  where the hard scale $Q$ is fixed by the invariant mass of the fermion
  pair, $Q^2 = M_{e \bar{\nu}_e}^2$.

  \noindent
  The parton shower for both the quark and the anti-quark in the
  initial state starts with scale $t$, for the first emission. However,
  the parton distribution weight is taken at $\mu_F=Q_{\rm cut}$,
  i.e.\ it is given by ${\cal W}_{\rm PDF}=f(x,q)/f(x,\mu_F)$ rather
  than by ${\cal W}_{\rm PDF}=f(x,q)/f(x,Q)$. Also, the jet veto
  inside the parton shower is performed w.r.t.\ $Q_{\rm cut}$. 
\item $n=1$:\\
  For $n=1$ jets, different cluster configurations are possible, two of which
  are exhibited in Fig.\ \ref{fig_w1jet}.
  \begin{figure}
    \begin{center}
      \includegraphics[width=230pt]{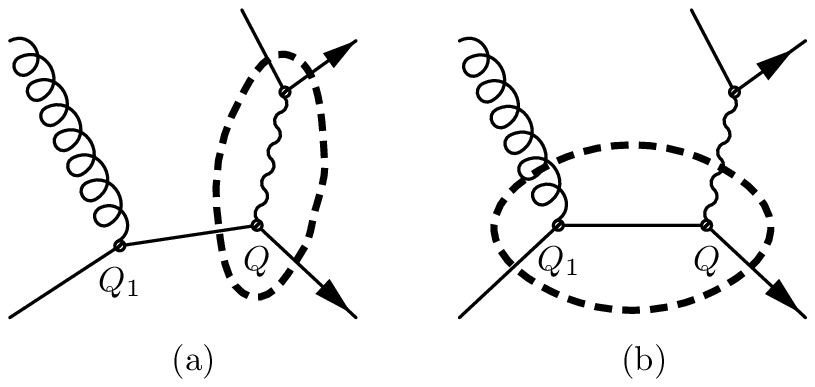}
      \caption{\label{fig_w1jet}Two possible cluster configurations of
        $W+1$ jet events. The dashed line highlights the hard $2 \to
        2$ process.} 
    \end{center}
  \end{figure}
  The hard $2 \to 2$ process either is again of the Drell--Yan type 
  (Fig.\ \ref{fig_w1jet}a) or, for example, of the type 
  $q\bar{q}' \to gW$ (Fig.\ \ref{fig_w1jet}b). The respective weights
  in both cases read:
  \begin{align}
    {\cal W}_{\rm ME}^{(a)} &= {\cal W}_{\rm ME}^{(b)} = 
    \Delta_{q}(Q,Q_{\rm cut})\,\Delta_{\bar{q}'}(Q,Q_{\rm cut})\,
    \Delta_{g}(Q_1,Q_{\rm cut})\frac{\alpha_s(Q_1)}{\alpha_s(Q_{\rm cut})}\;.
  \end{align}
  where $Q$ is the scale of the core $2\to 2$ process and the nodal
  value $Q_1$ is given by the transverse momentum of the extra jet. 
  For the first configuration, $Q^2=p_W^2=M_{e\nu}^2$, and the gluon
  jet tends to be soft, i.e.\ $Q_1$ preferentially is close to 
  $Q_{\rm cut}$. The second configuration differs from the first only
  by the result of the clustering and in the scale of the core
  process, now given by 
  \begin{align}
    Q^2 = p_{\perp,g}^2 + M_{e \bar{\nu}_e}^2 = M_{\perp,W}^2\,.
  \end{align}
  The transverse momentum of the gluon jet $p_{\perp,g}^2$
  now is of the order of the $W$-boson mass. In the first case, $Q_1= p_{\perp,g}$
  emerges as a part of the clustering procedure, whereas in the second
  case, it is read off directly from the core process. It is
  important, however, that the scale in both cases is defined in the
  same way in order to guarantee a smooth transition between the
  regime where clustering (a) and the regime where clustering (b) is
  chosen. 

  \noindent
  In both cases considered here, the parton shower for both the quark
  and the anti-quark in the initial state again starts with the
  respective scale $t$, and the parton distribution weights are
  treated in the same manner as before. The parton shower for the
  final state jet in contrast starts at $t_1$, all emissions in the
  three parton showers are vetoed if their transverse momentum exceeds
  $Q_{\rm cut}$. 
\item $n=2$:\\
  \noindent
  Many processes contribute to the production of two extra jets, some
  illustrative examples are displayed in Fig.\ \ref{fig_w2jet}.
  \begin{figure}
    \begin{center}
      \includegraphics[width=230pt]{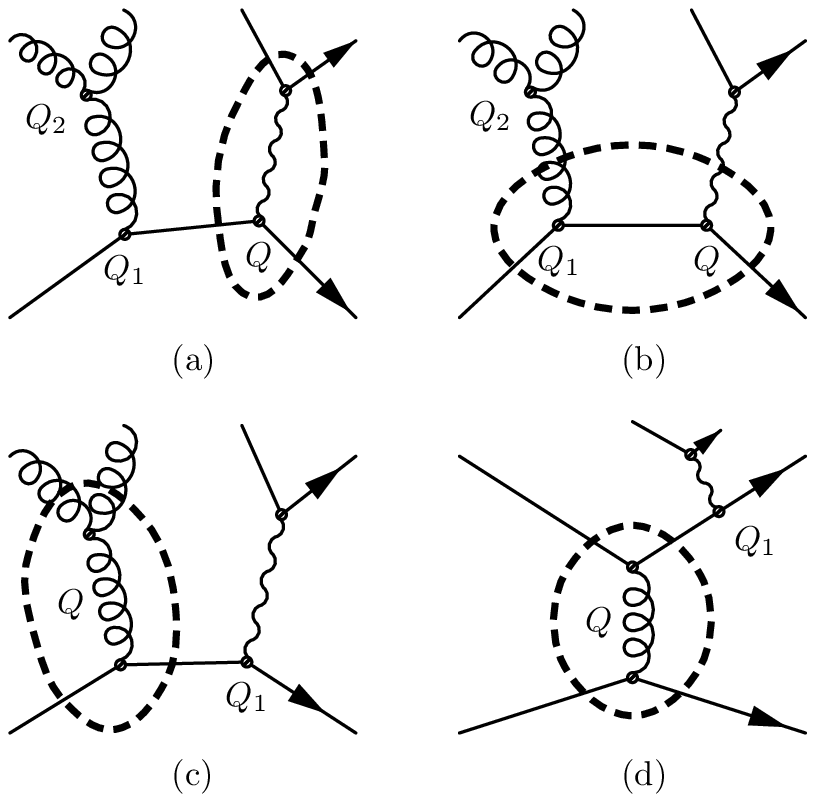}
      \caption{\label{fig_w2jet}Four possible cluster configurations
        of a W+2 jet event. The dashed line highlights the hard $2 \to
        2$ process, being either of Drell--Yan type (a), a vector
        boson production (b) or a pure QCD process (c,d).} 
    \end{center}
  \end{figure}
  The cases a) and b) displayed there are very similar to the example
  with one extra jet only. The corresponding weights read:
  \begin{align}
    {\cal W}_{\rm ME}^{(a)} &= {\cal W}_{\rm ME}^{(b)} = 
    \Delta_{q}(Q,Q_{\rm cut})\,\Delta_{\bar{q}'}(Q,Q_{\rm cut})\,
    \Delta_{g}(Q_1,Q_{\rm cut})\,\Delta_{g}(Q_2,Q_{\rm cut}) 
    \frac{\alpha_s(Q_1)}{\alpha_s(Q_{\rm cut})}\,
    \frac{\alpha_s(Q_2)}{\alpha_s(Q_{\rm cut})}\;.
  \end{align}
  The nodal value $Q_2$ is given by the $k_\perp$-algorithm, again it
  is the transverse momentum of the gluon. The scales $Q_1$ and $Q$
  are chosen in full analogy to the one-jet case discussed above.

  \noindent
  In contrast a new situation arises when a pure QCD process has been
  chosen as the ``core'' $2 \to 2$ process, see for instance Fig.\
  \ref{fig_w2jet}c). Since the ``core'' process is not resolved, there
  is only one scale available, $Q^2=p_\perp^2$, the transverse
  momentum of the outgoing jets. The correction weight in this case
  thus reads: 
  \begin{align}
    {\cal W}_{\rm ME}^{(c)} &= 
    \Delta_{q}(Q,Q_{\rm cut})\,
    \frac{\Delta_{q}(Q,Q_{\rm cut})}{\Delta_{q}(Q_1,Q_{\rm cut})}\,
    \Delta_{\bar{q}'}(Q_1,Q_{\rm cut})\,[\Delta_{g}(Q,Q_{\rm cut})]^2 
    \left[\frac{\alpha_s(Q)}{\alpha_s(Q_{\rm cut})}\right]^2\,.
  \end{align}
  In this case, the Sudakov form factors in the denominator
  corresponding to the internal quark line and its external
  continuation cancel only, if both quarks have the same mass, which
  is not necessarily the case\footnote{This example shows that the
  prescription implicitly deals with flavour changing currents as
  well.}.

  \noindent
  In contrast to the case exhibited in diagram \ref{fig_w2jet}c),
  where the boson was clustered with an initial state parton, Fig.\
  \ref{fig_w2jet}d) pictures an example configuration, where the boson
  is clustered with a final state parton. In this case, the
  corresponding correction weight is given by  
  \begin{align}\label{Eq_Wd_w2jet} 
    {\cal W}_{\rm ME}^{(d)} &= 
    [\Delta_{q}(Q,Q_{\rm cut})]^2\,
    \frac{\Delta_{q'}(Q,Q_{\rm cut})}{\Delta_{q'}(Q_1,Q_{\rm cut})}\,
    \Delta_{q''}(Q_1,Q_{\rm cut})\Delta_{\bar{q}'}(Q,Q_{\rm cut}) 
    \left[\frac{\alpha_s(Q)}{\alpha_s(Q_{\rm cut})}\right]^2\,.
  \end{align}

  \noindent
  The starting conditions for the parton showers for the first two
  cases, Fig.\ \ref{fig_w2jet}c) and \ref{fig_w2jet}b), are very
  similar to the $n=1$ case: The initial state partons 
  start their evolution at $t$, the two extra jets start their
  evolution at $t_1$ and $t_2$, respectively, and all are subject to a
  jet veto inside the parton shower with transverse momentum 
  $Q_{\rm cut}$. For the last two cases, the situation changes. There,
  the electroweak boson does not play any significant role for the
  parton shower; all four parton showers start at their common QCD
  core process scale, $t$. Of course, again, emissions with transverse
  momentum larger than $Q_{\rm cut}$ from any of the four shower seeds
  are vetoed. It should be noted here that there is a potential
  mismatch of logarithms of correction weight and veto weight in the
  quark line that changes its flavour. This happens if the two quarks
  adjacent to the electroweak boson have different mass; mass effects,
  however, usually can be safely neglected as long as no top quarks
  are present.  
\end{enumerate}

\noindent
The extension to higher multiplicities is straightforward. However,
assume again for illustrative reasons that $n_{\rm max}=2$,
leading to the application of the highest multiplicity treatment for
the two-jet configuration. Then, during cross section evaluation the
factorisation scale is set dynamically to $\mu_F = Q_s$, i.e.\ to the
nodal value of the softest emission. This leads to the following cross
sections

\begin{equation}
 \begin{array}{l@{\,=\,}r@{\extracolsep{15mm}}
               l@{\extracolsep{1mm}\,=\,}r@{\extracolsep{15mm}}
               l@{\extracolsep{0mm}\,\,}r}

    \sigma_0^{(0)} & 953.03 \mbox{\rm pb}&
    \sigma_0^{\rm nmax=2} & 825.77 \mbox{\rm pb}\\
    \sigma_1^{(0)} & 155.76 \mbox{\rm pb} &
    \sigma_1^{\rm nmax=2} & 108.35 \mbox{\rm pb}\\
    \sigma_2^{(0)} & 34.29 \mbox{\rm pb}& 
    \sigma_2^{\rm nmax=2} & 22.01 \mbox{\rm pb}\\
\multicolumn{2}{c}{}&
    \sigma_{\rm tot}^{\rm nmax=2} &
\multicolumn{3}{l}{
    \sum\limits_{i=0}^2\sigma_i^{\rm nmax=2} = 956.13 \mbox{\rm pb}\,.}
  \end{array}
\end{equation}
Assuming that $Q_2<Q_1$, the correction weight for the diagram a) in
Fig.\ \ref{fig_w2jet} would read
\begin{align}
  \tilde{\cal W}_{\rm ME}^{(a)} &= 
  \Delta_{q}(Q,Q_2)\,\Delta_{\bar{q}'}(Q,Q_2)\,
  \Delta_{g}(Q_1,Q_2)\,\Delta_{g}(Q_2,Q_2) 
  \frac{\alpha_s(Q_1)}{\alpha_s(Q_2)}\,
  \frac{\alpha_s(Q_2)}{\alpha_s(Q_2)}\nonumber\\
  &=\Delta_{q}(Q,Q_2)\,\Delta_{\bar{q}'}(Q,Q_2)\,
  \Delta_{g}(Q_1,Q_2)\,
  \frac{\alpha_s(Q_1)}{\alpha_s(Q_2)}\;.
\end{align}
The parton showers for the four legs would start at $t$ for the two
quark lines, and at $t_1$ and $t_2$ for the two gluon lines,
respectively. Vetos would be applied for emissions with a $k_\perp$
larger than $Q_2$, which implies that there would be no jet veto in
the parton shower evolution of the second gluon line. Of course, the
scales in the parton distribution weights of the first initial state
radiation inside the shower would also be adjusted. 

\noindent
The situation is even more extreme when considering diagram d). There,
the softest QCD radiation actually is at the core process. Thus, no
correction weight whatsoever would be applied, cf.\ Eq.\
\ref{Eq_Wd_w2jet}, if the masses of $q'$ and $q''$ are identical, 
\begin{align}
  \tilde{\cal W}_{\rm ME}^{(d)} &= 
  [\Delta_{q}(Q,Q)]^2\,
  \frac{\Delta_{q'}(Q,Q)}{\Delta_{q'}(Q_1,Q)}\,
  \Delta_{q''}(Q_1,Q)\Delta_{\bar{q}'}(Q,Q) 
  \left[\frac{\alpha_s(Q)}{\alpha_s(Q)}\right]^2 =
  \frac{\Delta_{q''}(Q_1,Q)}{\Delta_{q'}(Q_1,Q)}\,.
\end{align}
All parton showers for all four legs would start at $t$, and the veto
would be applied for emissions larger than $Q$, but this phase space
region is kinematically excluded anyway. 

\subsection{Example III -- $p \bar{p} \to \mbox{jets}$}

\noindent
In this example the operation of the multi-cut treatment is
illustrated through the case of $p \bar{p} \to \le 3\mbox{jets}$. The
two-jet sample here is generated with a jet resolution cut of 
$Q_{\rm cut}^{(2)}=20$~GeV, and the three-jet sample is produced with
$Q_{\rm cut}^{(3)}=30$~GeV. 
The corresponding a-priori cross sections read

\begin{equation}
\begin{array}{l@{\,=\,}r@{\extracolsep{15mm}}
              l@{\extracolsep{1mm}\,=\,}r@{\extracolsep{15mm}}
              ll}
    \sigma_2^{(0)}        & 30.423  \mbox{\rm mb} &      
    \sigma_2^{\rm nmax=3} & 13.903  \mbox{\rm mb} & \\
    \sigma_3^{(0)}        &  0.133  \mbox{\rm mb} &     
    \sigma_3^{\rm nmax=3} &  0.092  \mbox{\rm mb} & \\
\multicolumn{2}{c}{}&
\sigma_{\rm tot}^{\rm nmax=3} & 
\multicolumn{3}{l}{
\sum\limits_{i=2}^3\sigma_i^{\rm nmax=3} = 13.995 \,\mbox{\rm mb}}\,.
  \end{array}
\end{equation}
In their calculation, the factorisation scale of the two-jet events
has consistently been set to $Q_{\rm min}$ defined as
\begin{align}
  Q_{\rm min} = {\rm min}\left\{p_\perp,Q_{\rm cut}^{(3)}\right\}\,,
\end{align}
where $p_\perp$ is the transverse momentum of the outgoing jets.
In contrast, in the evaluation of the cross section of the three-jet
events, the factorisation scale has consistently been set to $Q_s$,
the scale of the softest jet.

\noindent
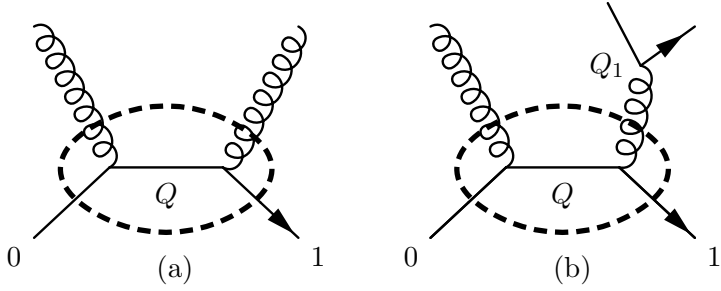
\begin{figure}
\begin{center}
\begin{pspicture}(250,130)
%
\put(0,10){\graphtwojet}
\put(150,10){\graphthreejet}
\put(50,5){(a)}
\put(200,5){(b)}
\end{pspicture}
\end{center}
\caption{\label{fig:jets} Examples for the production two and
  three jets. }
\end{figure}
In Fig.\ \ref{fig:jets}, exemplary diagrams for the two processes,
the production of two and three jets, are depicted. For a typical two
jet event, such as the one in diagram a), the weight reads
\begin{align}
  {\cal W}_{\rm ME}^{(a)} &= 
  [\Delta_{q}(Q,Q_{\rm min})]^2\,[\Delta_{g}(Q,Q_{\rm min})]^2\,
  \left[\frac{\alpha_s(Q_{\rm min})}
    {\alpha_s(Q_{\rm cut}^{(2)})}\right]^2\,.
\end{align}
The shower for all four legs starts at $t$, with a veto on emissions
harder than $Q_{\rm min}$, and again the parton distribution weight in
the first emission of each of the initial state shower evolutions is
adjusted.

\noindent
For the three-jet event depicted in diagram b), the weight reads
\begin{align}
  {\cal W}_{\rm ME}^{(b)} &= 
  [\Delta_{q}(Q,Q_s)]^2\,\Delta_{g}(Q,Q_s)
  \frac{\Delta_{g}(Q,Q_s)}{\Delta_{g}(Q_1,Q_s)}
  [\Delta_{q}(Q_1,Q_s)]^2
  \left[\frac{\alpha_s(Q)}
    {\alpha_s(Q_{\rm cut}^{(3)})}\right]^2\,
  \frac{\alpha_s(Q_1)}{\alpha_s(Q_{\rm cut}^{(3)})}\,.
\end{align}
The parton shower for the two incoming quarks, for the outgoing gluon
and for the harder of the two outgoing quarks starts at $t$, the
parton shower of the softer of the two quarks emerging from the gluon
line starts at $t_1$. The veto is performed w.r.t.\ the scale $Q_1$. 

\subsection{Example IV -- $\epemto d \bar{d} u \bar{u} (g)$}

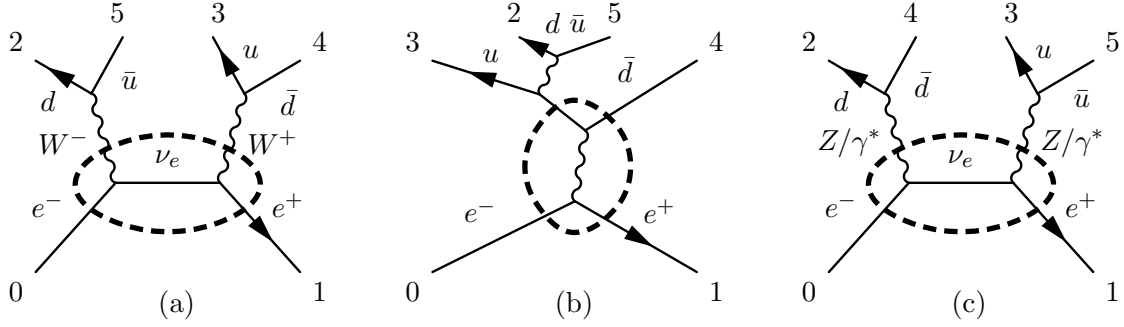
\begin{figure}
\begin{center}
\begin{pspicture}(360,110)
%
\put(-20,10){
\begin{fmfgraph*}(100,100) 
  \fmfbottom{i0,i1} 
  \fmftop{o3,o2,o1,o0} 
  \fmf{plain,label=$e^\m$}{i0,v0} 
  \fmfv{label=0}{i0} 
  \fmf{plain,label=$\nu_e$}{v1,v0} 
  \fmfv{label=1}{i1} 
  \fmf{fermion,l.side=left,label=$e^\p$}{v1,i1} 
  \fmf{boson,l.side=right,label=$W^\p$}{v1,v2} 
  \fmfv{label=4}{o0} 
  \fmf{plain,label=$\bar d$}{v2,o0} 
  \fmfv{label=3}{o1} 
  \fmf{fermion,label=$u$}{v2,o1} 
  \fmf{boson,label=$W^\m$}{v0,v3} 
  \fmfv{label=5}{o2} 
  \fmf{plain,label=$\bar u$}{v3,o2} 
  \fmfv{label=2}{o3} 
  \fmf{fermion,label=$d$}{v3,o3} 
\end{fmfgraph*} 
\psccurve[linestyle=dashed]
(-70,29)(-30,29)(-15,43)(-30,58)(-70,58)(-85,43)
}
\put(130,10){
\begin{fmfgraph*}(100,100) 
  \fmfbottom{i0,i1} 
  \fmftop{o3,o2,o1,o0} 
  \fmf{plain,label=$e^\m$}{i0,v0} 
  \fmfv{label=0}{i0} 
  \fmfv{label=1}{i1} 
  \fmf{fermion,l.side=left,label=$e^\p$}{v0,i1} 
  \fmf{boson,tension=2.}{v0,v1} 
  \fmfv{label=4}{o0} 
  \fmf{plain,label=$\bar d$}{v1,o0} 
  \fmf{plain,tension=2.}{v2,v1} 
  \fmf{photon}{v2,v3} 
  \fmfv{label=5}{o1} 
  \fmf{plain,label=$\bar u$}{v3,o1} 
  \fmfv{label=2}{o2} 
  \fmf{fermion,label=$d$}{v3,o2} 
  \fmfv{label=3}{o3} 
  \fmf{fermion,label=$u$}{v2,o3} 
\end{fmfgraph*} 
\psccurve[linestyle=dashed]
(-65,50)(-60,65)(-47,75)(-30,65)(-25,50)(-30,35)(-47,25)(-60,35)
}
\put(280,10){
\begin{fmfgraph*}(100,100) 
  \fmfbottom{i0,i1} 
  \fmftop{o3,o2,o1,o0} 
  \fmf{plain,label=$e^\m$}{i0,v0} 
  \fmfv{label=0}{i0} 
  \fmf{plain,label=$\nu_e$}{v1,v0} 
  \fmfv{label=1}{i1} 
  \fmf{fermion,l.side=left,label=$e^\p$}{v1,i1} 
  \fmf{boson,l.side=right,label=$Z/\gamma^{*}$}{v1,v2} 
  \fmfv{label=5}{o0} 
  \fmf{plain,label=$\bar u$}{v2,o0} 
  \fmfv{label=3}{o1} 
  \fmf{fermion,label=$u$}{v2,o1} 
  \fmf{boson,label=$Z/\gamma^{*}$}{v0,v3} 
  \fmfv{label=4}{o2} 
  \fmf{plain,label=$\bar d$}{v3,o2} 
  \fmfv{label=2}{o3} 
  \fmf{fermion,label=$d$}{v3,o3} 
\end{fmfgraph*} 
\psccurve[linestyle=dashed]
(-70,29)(-30,29)(-15,43)(-30,58)(-70,58)(-85,43)
}
\put(30,5){(a)}
\put(180,5){(b)}
\put(330,5){(c)}
\end{pspicture}
\end{center}
\label{fig:eeww1}
\caption{Possible cluster configurations in
  $\epemto d \bar{d} u \bar{u} (g)$. The dashed line indicates the
  core $2 \to 2$ process.}
\end{figure}

\noindent
In the process $\epemto d \bar{d} u \bar{u}$, there are basically
three classes of subprocesses that can emerge as the core $2\to 2$
process, namely
\begin{itemize}
\item $e^{+}e^{-}\to W^{+}\,W^{-}$,
\item $e^{+}e^{-}\to Z^{0}/\gamma\,Z^{0}/\gamma$, and
\item $e^{+}e^{-}\to d\bar{d}$ or  $e^{+}e^{-}\to u\bar{u}$\,, 
\end{itemize}
all of which are depicted in Fig.\ \ref{fig:eeww1}.
The first two are electroweak processes, with $W$ pair production
usually largely dominating, whereas the latter can either lead to a
QCD or to an electroweak topology. Interferences between QCD and
electroweak diagrams are negligible, therefore it is convenient to
consider both contributions as independent. 

\noindent
In the following, the focus will be mainly on the electroweak
contributions. There exist 4 different possibilities for the first
clustering, listed in Tab.\ \ref{tab:eeww1}. 
\begin{table}[h]
\begin{center}
\begin{tabular}{r@{\&}ll}
\hline
 $i$ & $j$ &  Probability \\[4mm]
\hline
 $2$ & $4$ &  $P_{24}=p_{24}^Z + p_{24}^\gamma $ \\[1mm]
 \multicolumn{2}{c}{}
           &  $\displaystyle p_{24}^Z = 
  \left[\frac{\alpha_{\rm QED}(\mu)}{\sin^2\theta_w}\right]^2 \,
   \frac{(g_{1,d}^2 + g_{2,d}^2)(g_{1,e}^2 + g_{2,e}^2)}{(q^2-M_Z^2)^2+M_Z^2 \Gamma_Z^2} $   \\[4mm]
 \multicolumn{2}{c}{}
           &  $\displaystyle p_{24}^\gamma = [ \alpha_{\rm QED}(\mu)]^2 \,
               \frac{Q_{d}^2 Q_{e}^2}{(q^2)^2}$ \\[4mm]
 $2$ & $5$ & $P_{25}=p_{25}^{W}$ \\[1mm]
 \multicolumn{2}{c}{}
           &  $\displaystyle p_{25}^{W} =  \left[\frac{\alpha_{\rm QED}(\mu)}
  {2\sin^2\theta_w}\right]^2 \,
   \frac{(M^{\rm CKM}_{ud})^2}{(q^2-M_W^2)^2+M_W^2 \Gamma_W^2}$ \\[4mm]
 $3$ & $4$ & $P_{34}=p_{34}^{W}$ \\[1mm]
 \multicolumn{2}{c}{}
           &  $\displaystyle p_{34}^{W} =  \left[\frac{\alpha_{\rm QED}(\mu)}
  {2\sin^2\theta_w}\right]^2 \,
   \frac{(M^{\rm CKM}_{ud})^2}{(q^2-M_W^2)^2+M_W^2 \Gamma_W^2}$ \\[4mm]
 $3$ & $5$ &  $P_{35}=p_{35}^Z + p_{35}^\gamma$ \\[1mm]
 \multicolumn{2}{c}{}
           &  $\displaystyle p_{35}^Z =  
   \left[\frac{\alpha_{\rm QED}(\mu)} {\sin^2\theta_w}\right]^2 \,
   \frac{(g_{1,u}^2 + g_{2,u}^2)(g_{1,e}^2 + g_{2,e}^2)}{(q^2-M_Z^2)^2+M_Z^2 \Gamma_Z^2}$   \\[4mm]
 \multicolumn{2}{c}{}
           &  $\displaystyle p_{35}^\gamma = [\alpha_{\rm QED}(\mu)]^2 \,
               \frac{Q_{u}^2Q_{e}^2}{(q^2)^2}$ \\[4mm]
\hline
\end{tabular}
\end{center}
\label{tab:eeww1} 
\caption{All possibilities for the (electroweak)
  first clustering of $\epemto d \bar{d} u \bar{u}$. For brevity of
  the example only one Feynman diagram is taken into account for each
  possible propagator flavour.}
\end{table}
After, for instance, choosing $2$\&$5$ (the $d\bar{u}$-pair) to be
clustered first and to become a $W^{-}$ boson, a second clustering
leads to the $2\to2$ core process. Of course, the first step restricts
the possibilities for any subsequent clustering - in this example
three options remain. Their probabilities are listed Tab.\
\ref{tab:eeww2}. 

\begin{table}[h]
\begin{center}
\begin{tabular}{r@{\&}ll}
\hline
 $i$ & $j$ &  Probability \\[4mm]
\hline
 $2$ & $3$ & $P_{23}=p_{23}^{d}$ \\[1mm]
 \multicolumn{2}{c}{}
           &  $\displaystyle p_{23}^{d} =  \frac12 \left[\frac{\alpha_{\rm QED}(\mu)}
  {\sin^2\theta_w}\right]^2 \,
   \frac{(M^{\rm CKM}_{ud})^2}{(q^2)^2} (g_{1,u}^2 + g_{2,u}^2)$ \\[4mm]
 $2$ & $4$ & $P_{24}=p_{24}^{\bar{u}}$  \\[1mm]
 \multicolumn{2}{c}{}
           &  $\displaystyle p_{24}^{\bar{u}} = \frac12 \left[\frac{\alpha_{\rm QED}(\mu)}
  {\sin^2\theta_w}\right]^2  \,   \frac{(M^{\rm
    CKM}_{ud})^2}{(q^2)^2} (g_{1,d}^2 + g_{2,d}^2)$ \\[4mm]
 $3$ & $4$ & $P_{34}=p_{34}^{W}$ \\[1mm]
 \multicolumn{2}{c}{}
           &  $\displaystyle p_{34}^{W} = \left[\frac{\alpha_{\rm QED}(\mu)}
  {2\sin^2\theta_w}\right]^2 \,
   \frac{(M^{\rm CKM}_{ud})^4}{(q^2-M_W^2)^2+M_W^2 \Gamma_W^2}$ \\[4mm]
\hline
\end{tabular}
\end{center}
\caption{\label{tab:eeww2} All possibilities for the clustering of
  $\epemto W^{-} \bar{d} u$. For brevity of the example only one
  Feynman diagram is taken into account for each possible propagator
  flavour.} 
\end{table}

One possible outcome of the clustering procedure is a $W$ pair
production process as depicted in Fig.\ \ref{fig:eeww1}a. The
evaluation of the Sudakov weight in this case yields
\begin{align}
 {\cal W}_{\rm ME}^{(a)} &= 
 \Delta_d(Q_1,Q_{\rm cut}) \Delta_{\bar{u}}(Q_1,Q_{\rm cut})
 \Delta_u(Q_2,Q_{\rm cut}) \Delta_{\bar{d}}(Q_2,Q_{\rm cut})
\end{align}
in the $WW$ case; when the $q\bar{q}$ production process is chosen
instead, cf.\ Fig.\ \ref{fig:eeww1}b, the correction weight is given by
\begin{align}
  {\cal W}_{\rm ME}^{(b)} &= 
  \Delta_d(Q_1,Q_{\rm cut}) \Delta_{\bar{u}}(Q_1,Q_{\rm cut})
  \Delta_u(Q,Q_{\rm cut}) \Delta_{\bar{d}}(Q,Q_{\rm cut})
\end{align}
Both weights look very similar, and indeed for massless quarks this
holds true for all Sudakov weights that can be obtained. However,
while in the first example both scales $Q_1$ and $Q_2$ are of the
order $M_W$, the relevant scales in the second case are more likely to be
$Q_1\approx M_W$ and $Q\approx 2 M_W$, of course depending on the
exact kinematical configuration. Of course, these different
clusterings result in different starting conditions for the shower.

\section{Results}\label{Results_Sec} 

\noindent
The detailed presentation of examples in the previous section, Sec.\
\ref{Example_Sec} will be supplemented with results in this section.
To validate the consistency of the approach, clearly a careful
examination is mandatory, checking whether the exclusive samples
prepared through the re-weighted matrix elements and further evolved
through the parton shower combine into a consistent, inclusive
sample. Any larger discontinuity that becomes visible, in particular
on scales comparable to the merging scale $Q_{\rm cut}$, may serve as
an indication for a mismatch of leading logarithms. Obviously, a good
way of scrutinising the radiation pattern in the interaction of matrix
elements and parton shower is to investigate differential jet rates,
especially in a $k_\perp$ scheme. These rates are defined through the
jet resolution in the corresponding scheme, where an $n+1$-jet event
turns into an $n$-jet event. 

\subsection{Results for $\epemto \mbox{jets}$ at LEP~I}

\begin{figure}
  \begin{center}
    \begin{picture}(430,160)
      \put(-15,0){\includegraphics[width=6.cm]
        {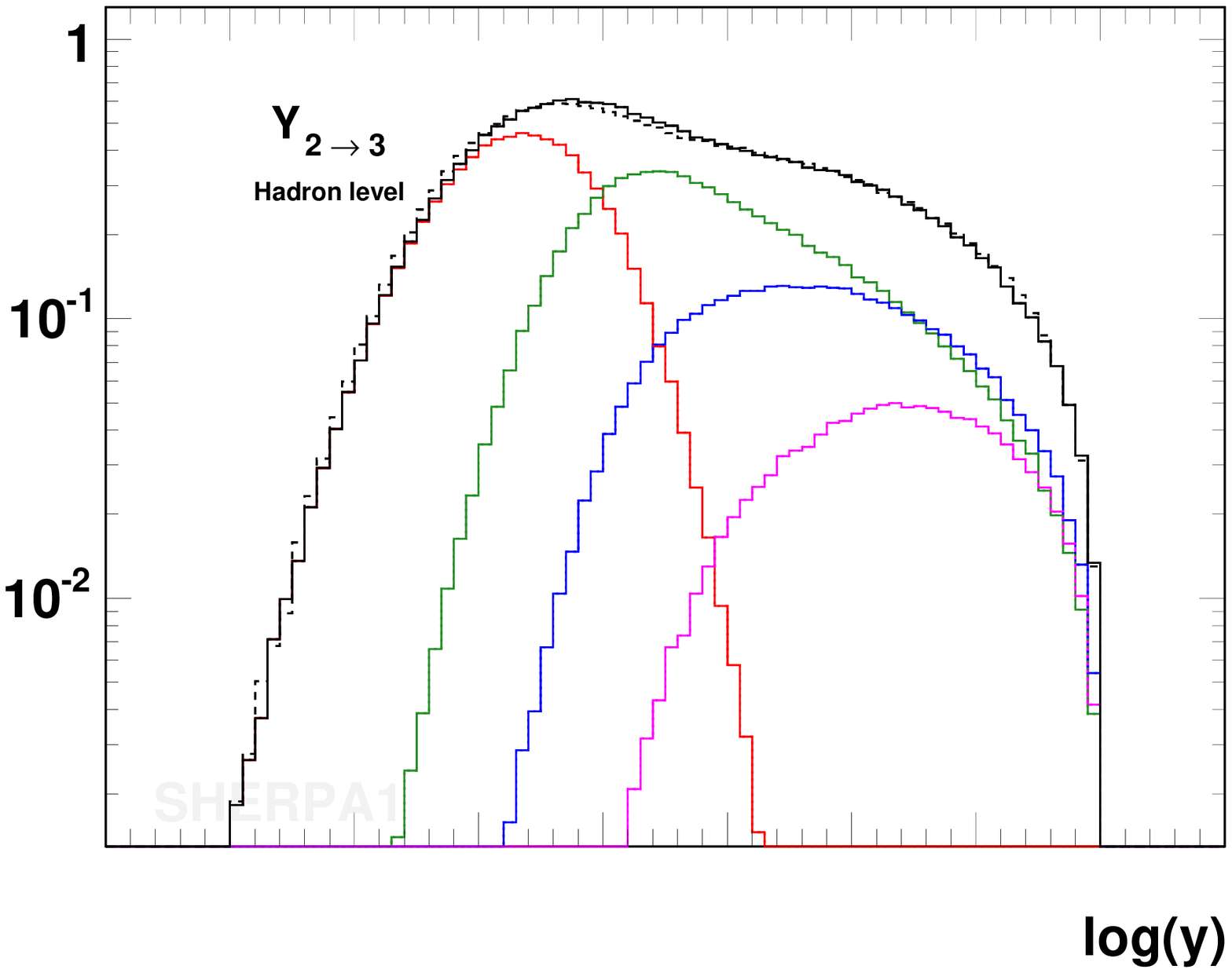}}
      \put(-15,0){\includegraphics[width=6.cm]
        {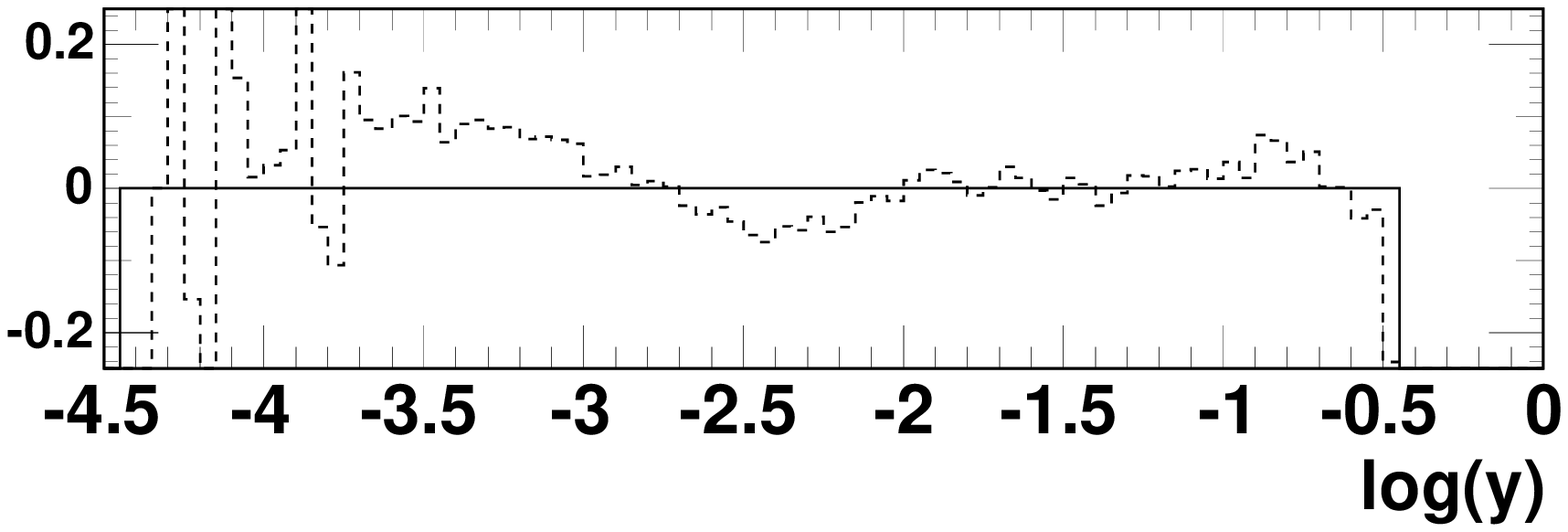}}
      \put(135,0){\includegraphics[width=6.cm]
        {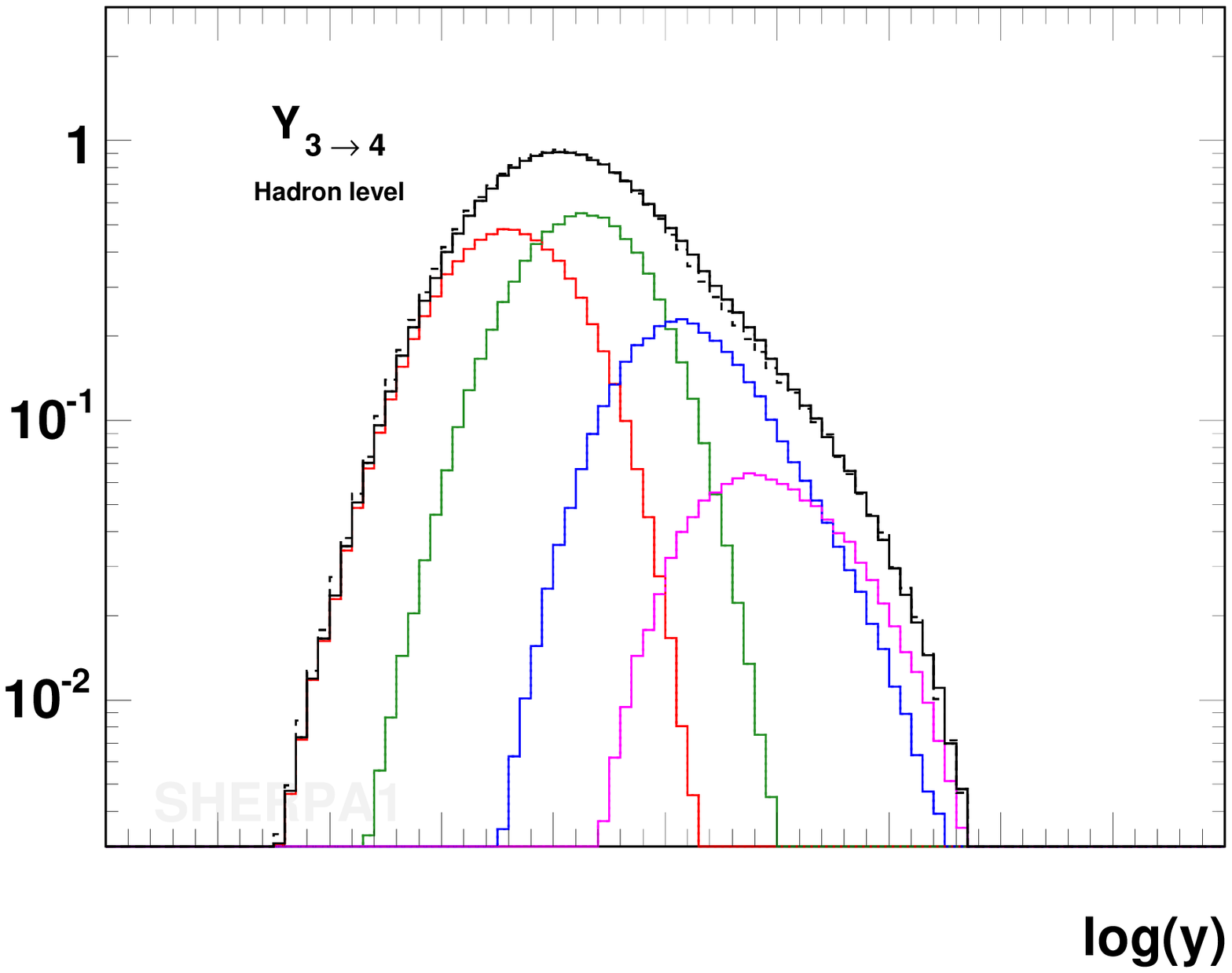}}
      \put(135,0){\includegraphics[width=6.cm]
        {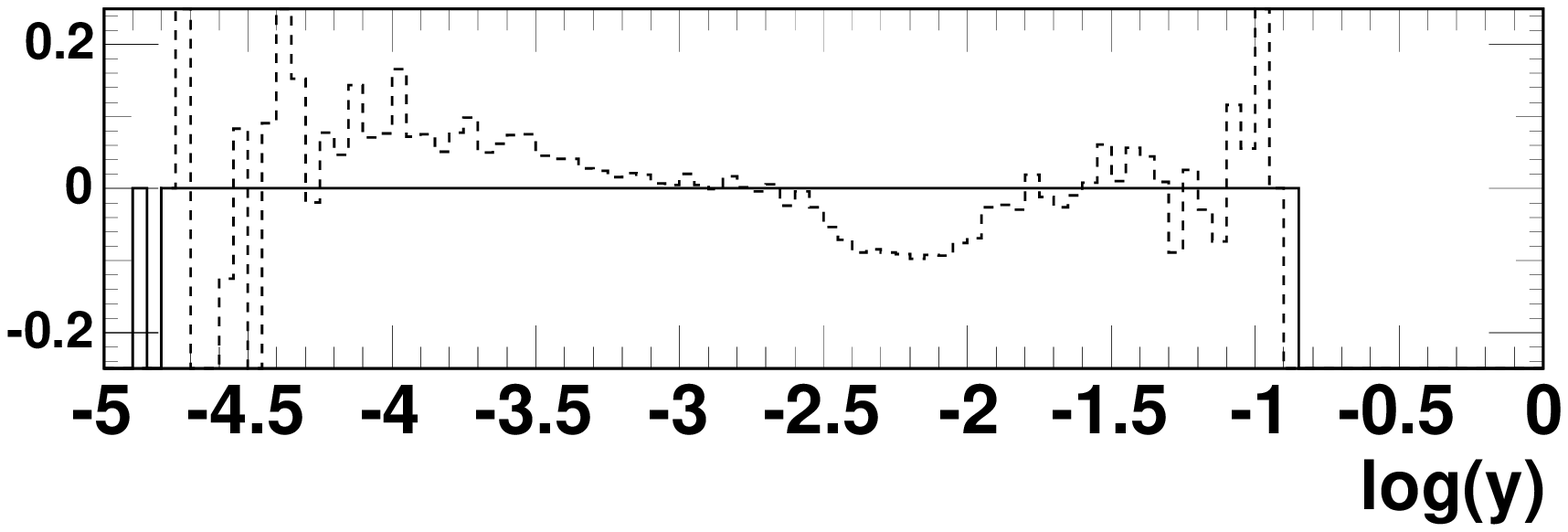}}
      \put(285,0){\includegraphics[width=6.cm]
        {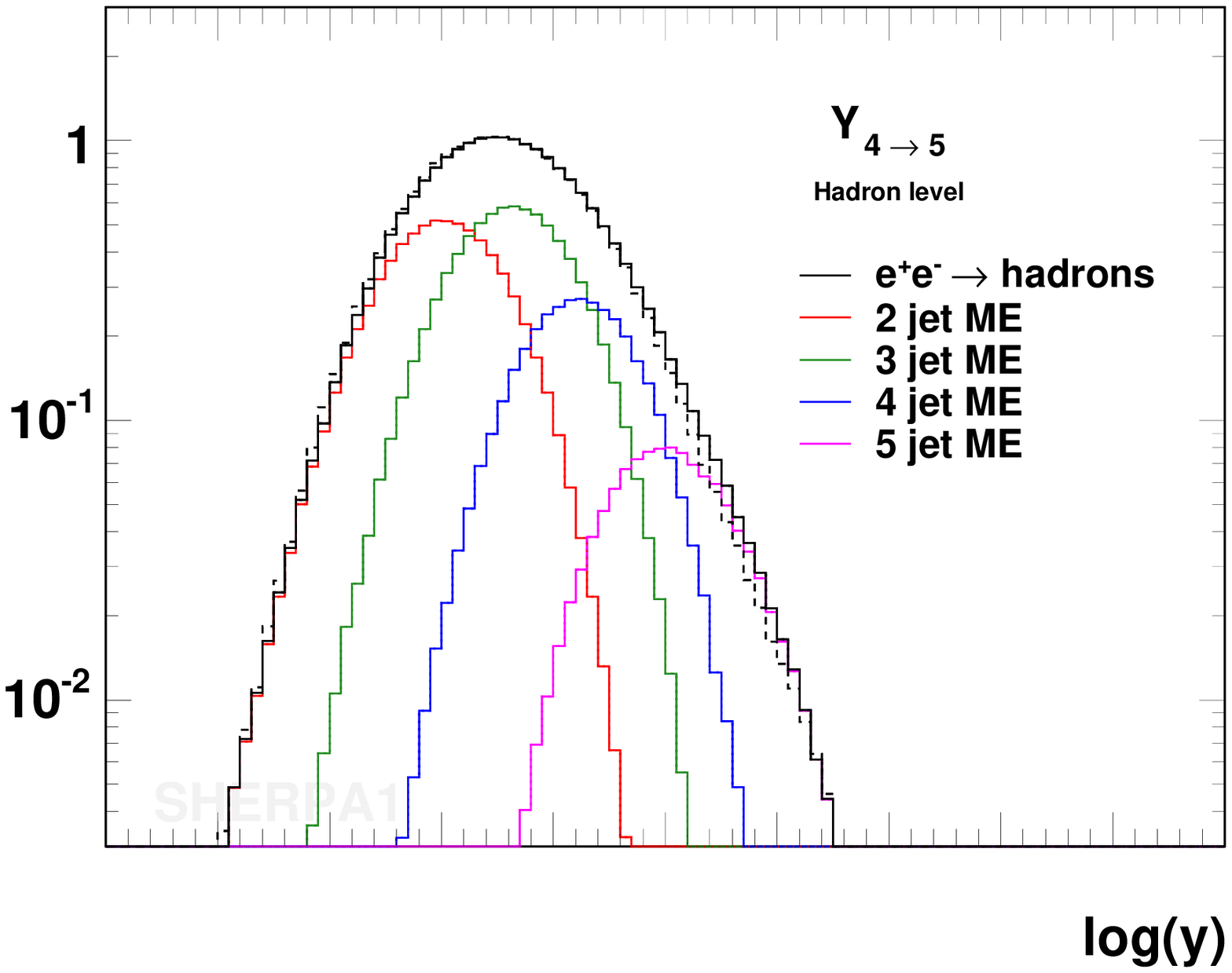}}
      \put(285,0){\includegraphics[width=6.cm]
        {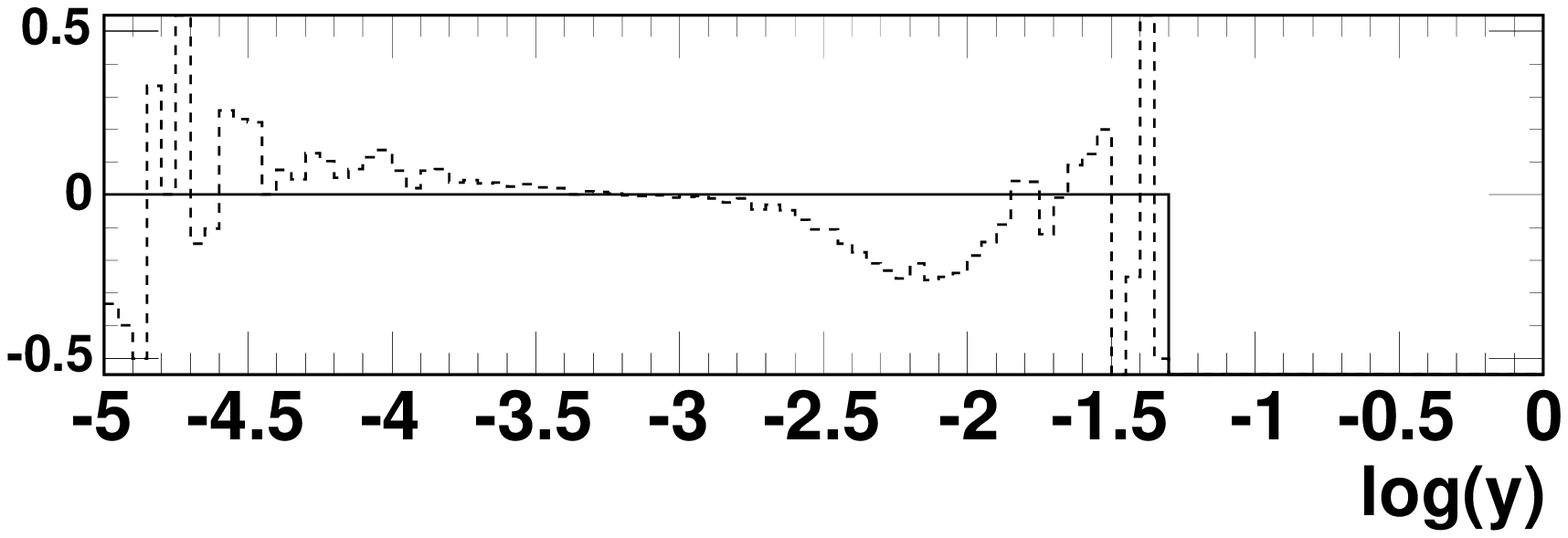}}
    \end{picture}
    \caption{Differential jet rates in the
      Durham scheme at LEP~I. Shown are the results obtained through
      the merging of matrix elements for up to five jets with the
      parton shower, with two different separation cuts.  The solid
      lines correspond to a cut at $y_{\rm cut}=10^{-2.5}$, and the
      dashed curve illustrates the result using $y_{\rm cut}=10^{-2}$. 
      In the former case coloured lines indicate the contributions
      from individual matrix elements: two jets (red), three jets
      (green), four jets (blue), and five jets (purple). 
      \label{fig:lep_diffdurham} 
    }
  \end{center}
\end{figure}

\noindent
To start with, differential jet rates in $\epemto \mbox{jets}$ are
compared. In $e^+e^-$ annihilations, it is often convenient to define 
a variable $y_{\rm cut}$ rather than $Q_{\rm cut}$; in the Durham 
scheme employed here, it is defined through
\begin{align}
y_{\rm cut} &= \frac{Q_{\rm cut}^2}{E_{\rm c.m.}^2}
\end{align}
implying that two particles $i$ and $j$ belong to different jets if
they are separated by a distance
\begin{align}
y_{ij} \ge 2\frac{\mbox{\rm min}\{E_i^2, E_j^2\}}{E_{\rm c.m.}^2}
(1-\cos\theta_{ij})\,.
\end{align}
In Fig.\ \ref{fig:lep_diffdurham}, results for differential jet rates
are shown ranging over four orders of magnitude in $y_{\rm cut}$.
The dependence on the actual value of $Q_{\rm cut}$ in the
generation of two different samples is barely visible. Also, the
distributions seem to be perfectly smooth around the generation
cut. Therefore, one may conclude that the merging in this case has been
accomplished with very high quality. 

\begin{figure}
  \begin{center}
    \begin{picture}(400,480)
      \put(-15,240){\includegraphics[width=8.0cm]
        {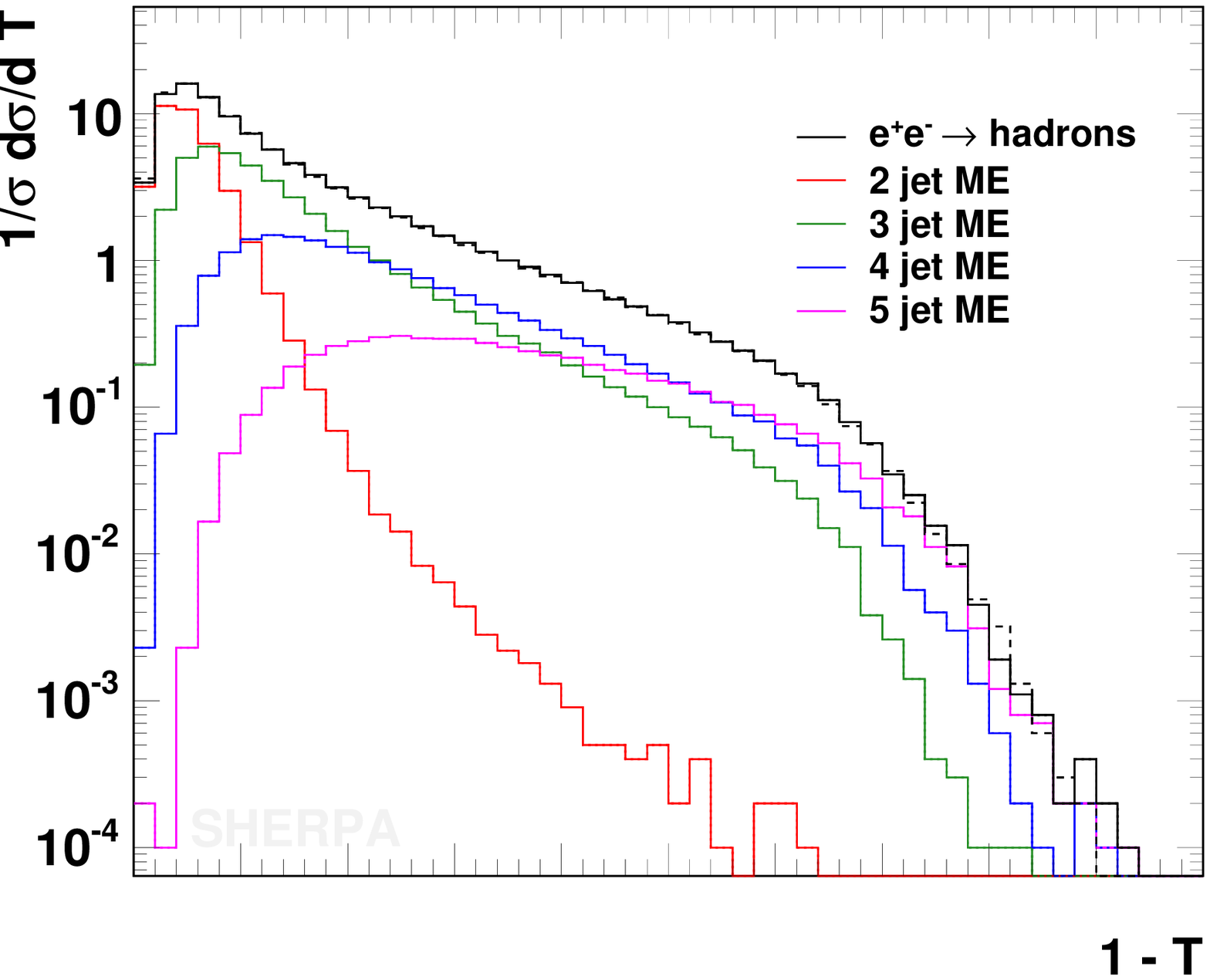}}
      \put(-15,240){\includegraphics[width=8.0cm]
        {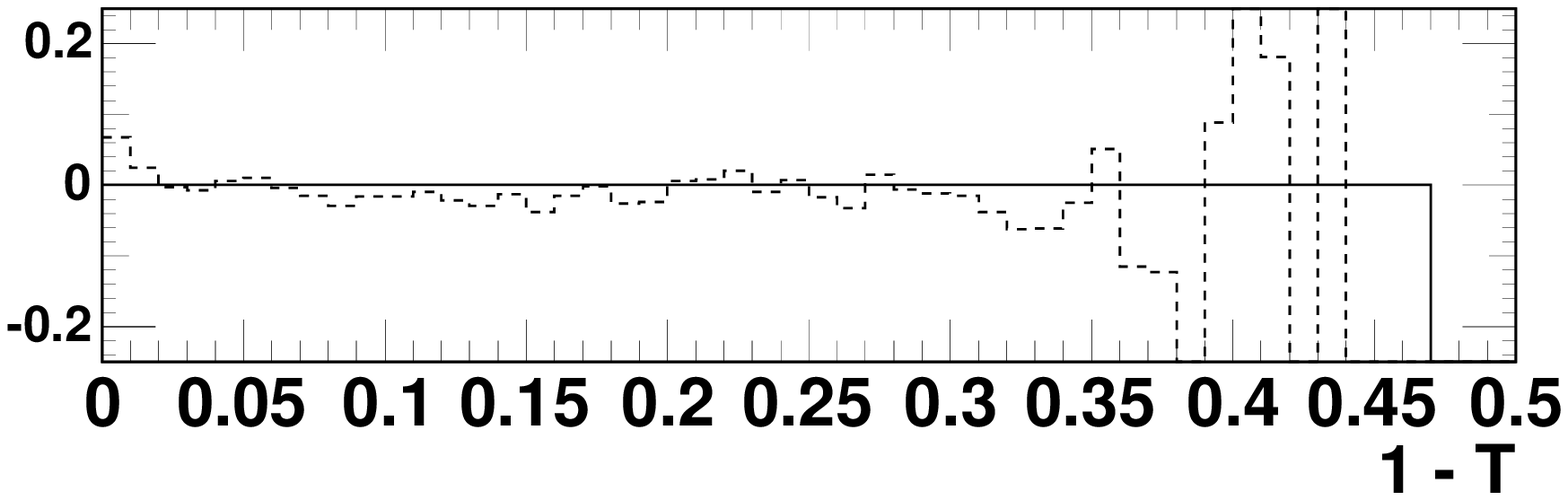}}
      \put(200,240){\includegraphics[width=8.0cm]
        {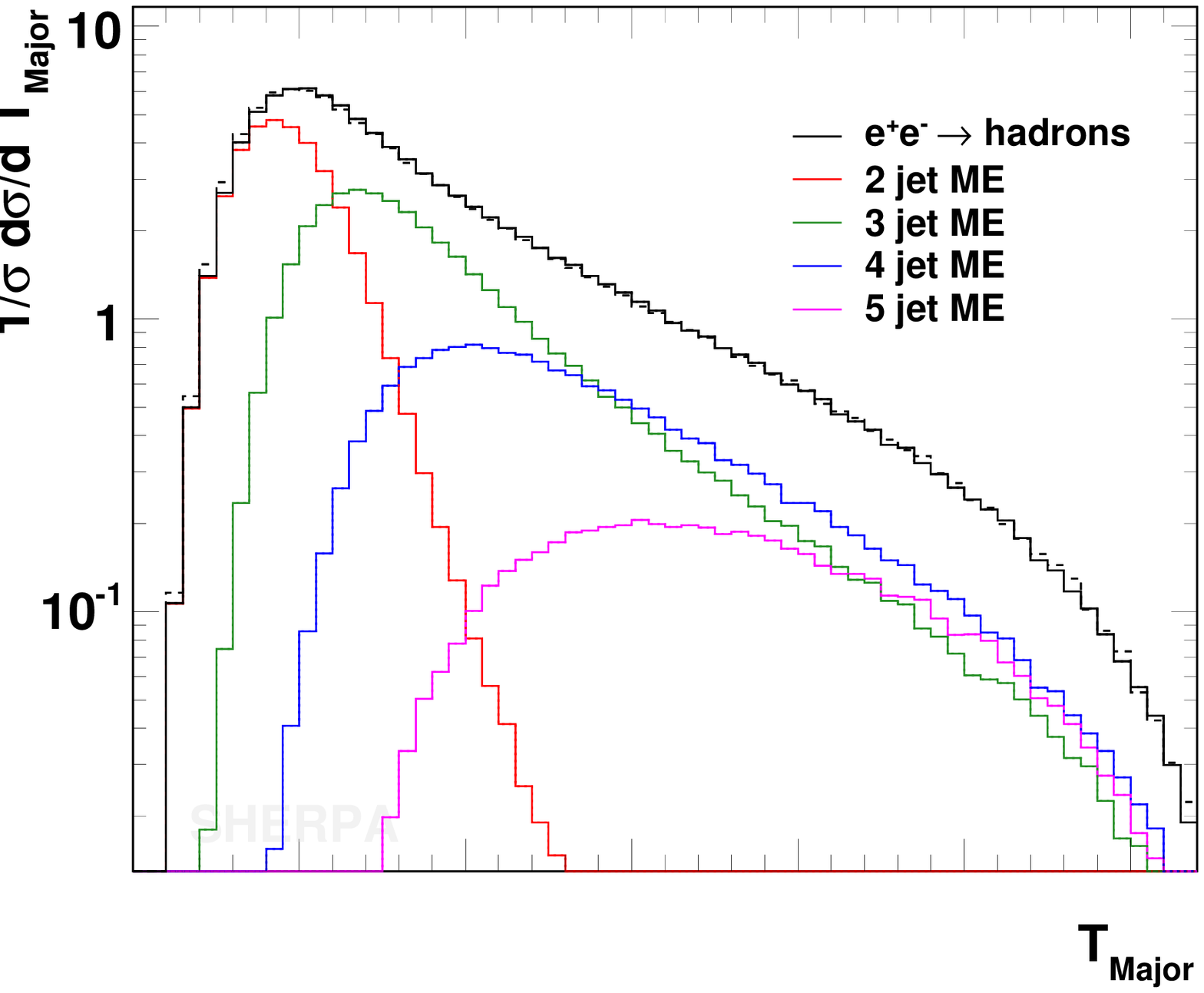}}
      \put(200,240){\includegraphics[width=8.0cm]
        {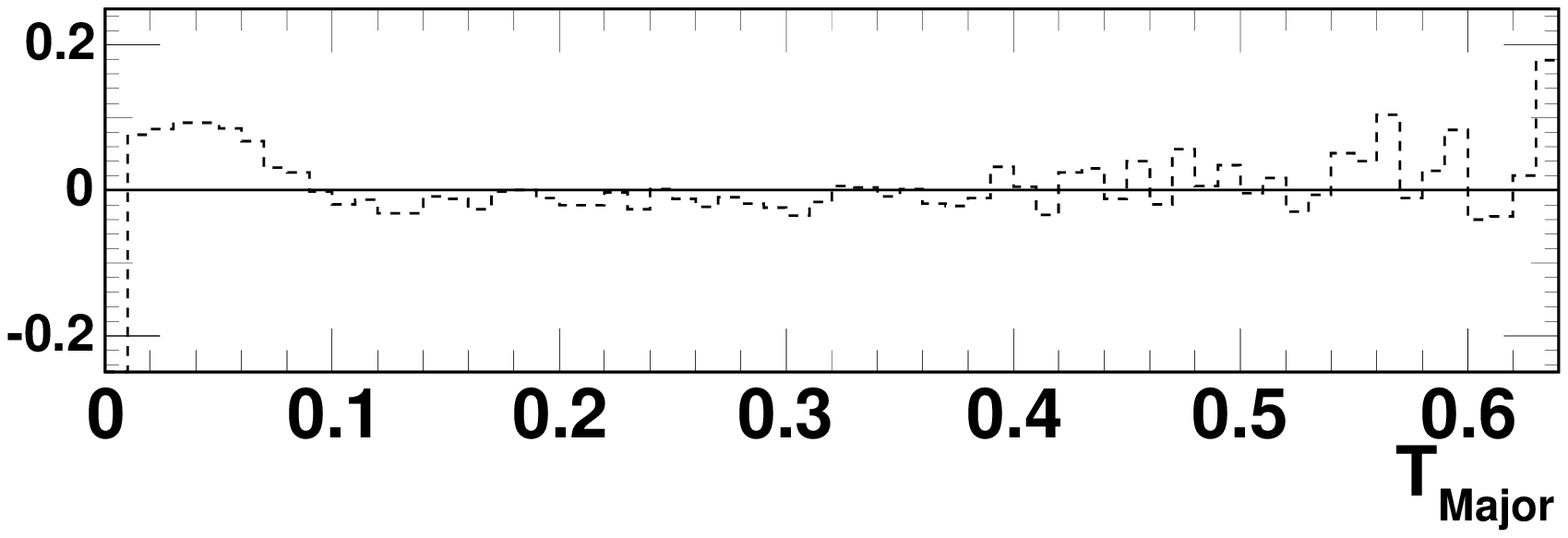}}
      \put(-15,0){\includegraphics[width=8.0cm]
        {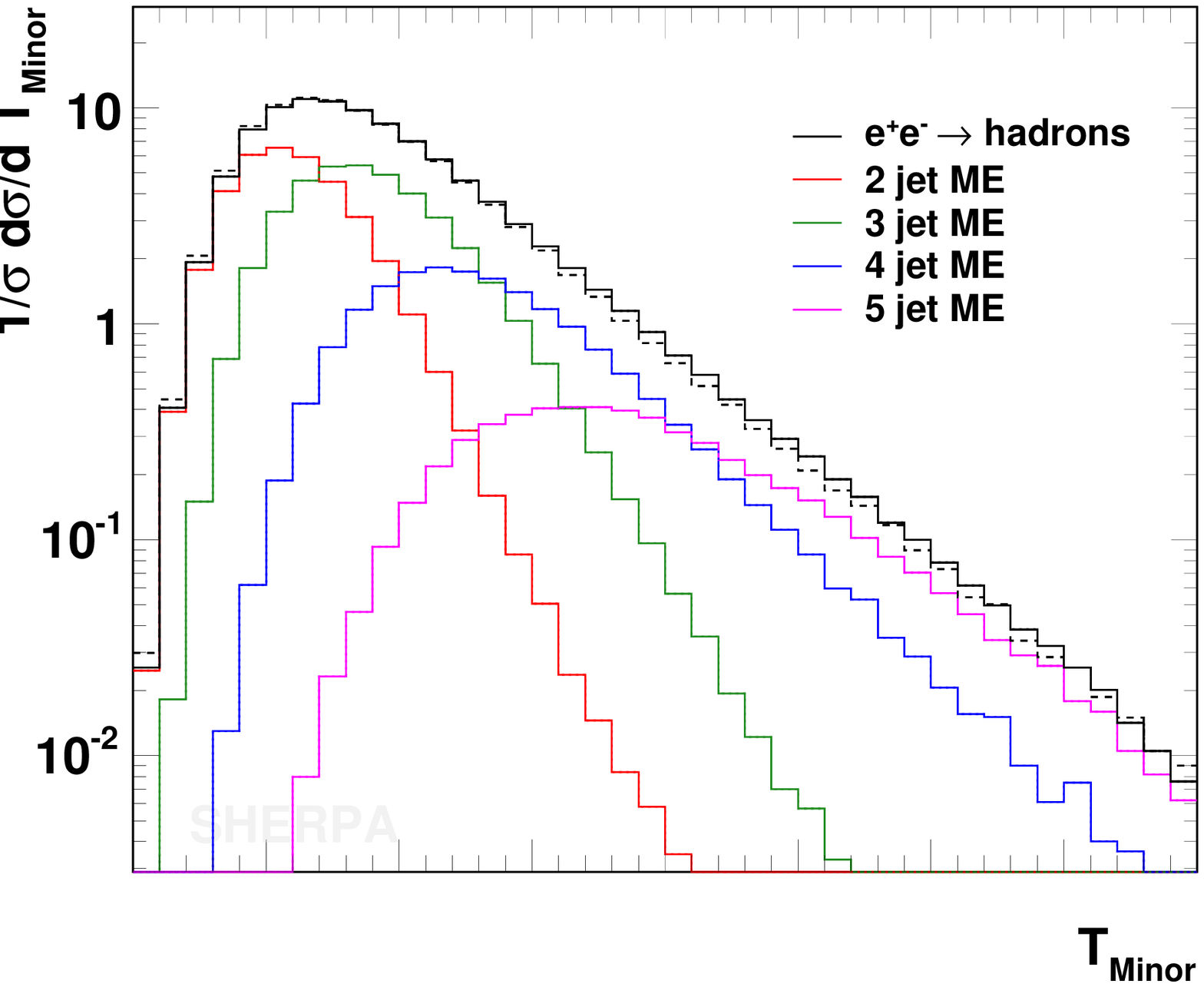}}
      \put(-15,0){\includegraphics[width=8.0cm]
        {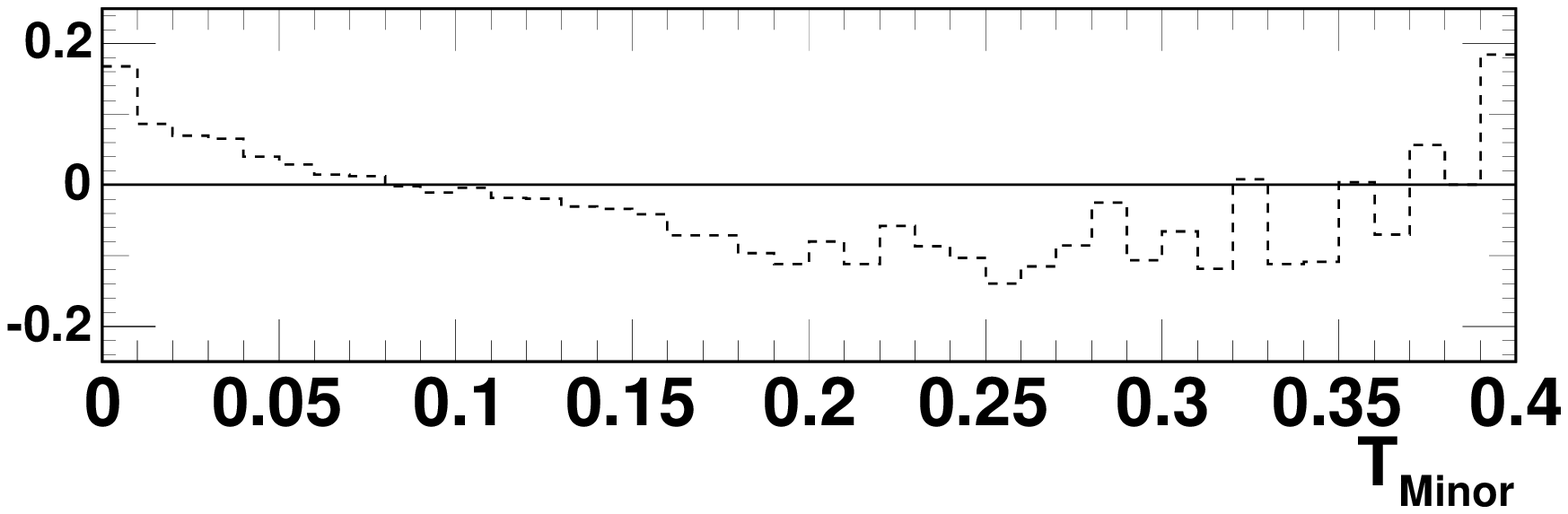}}
      \put(200,0){\includegraphics[width=8.0cm]
        {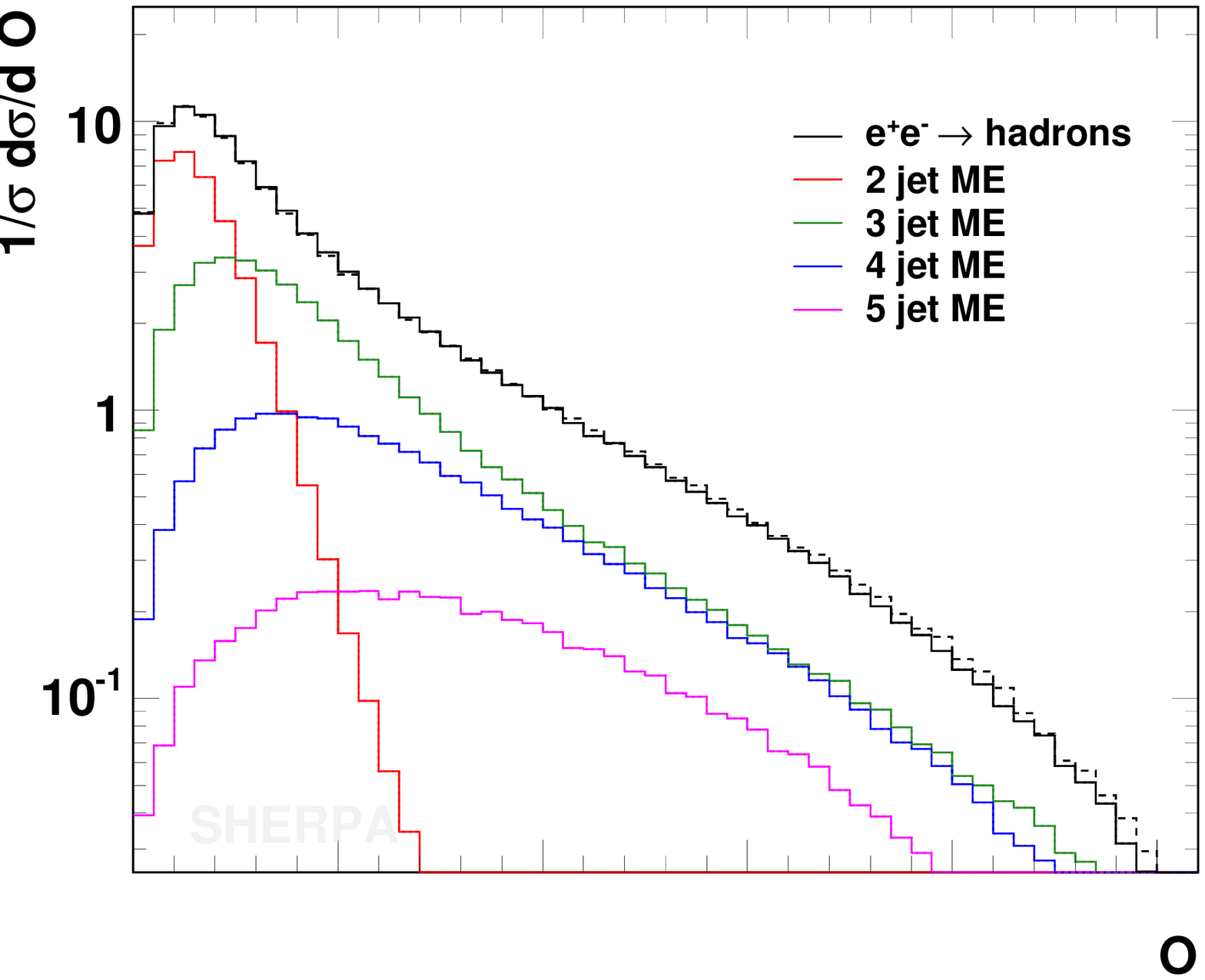}}
      \put(200,0){\includegraphics[width=8.0cm]
        {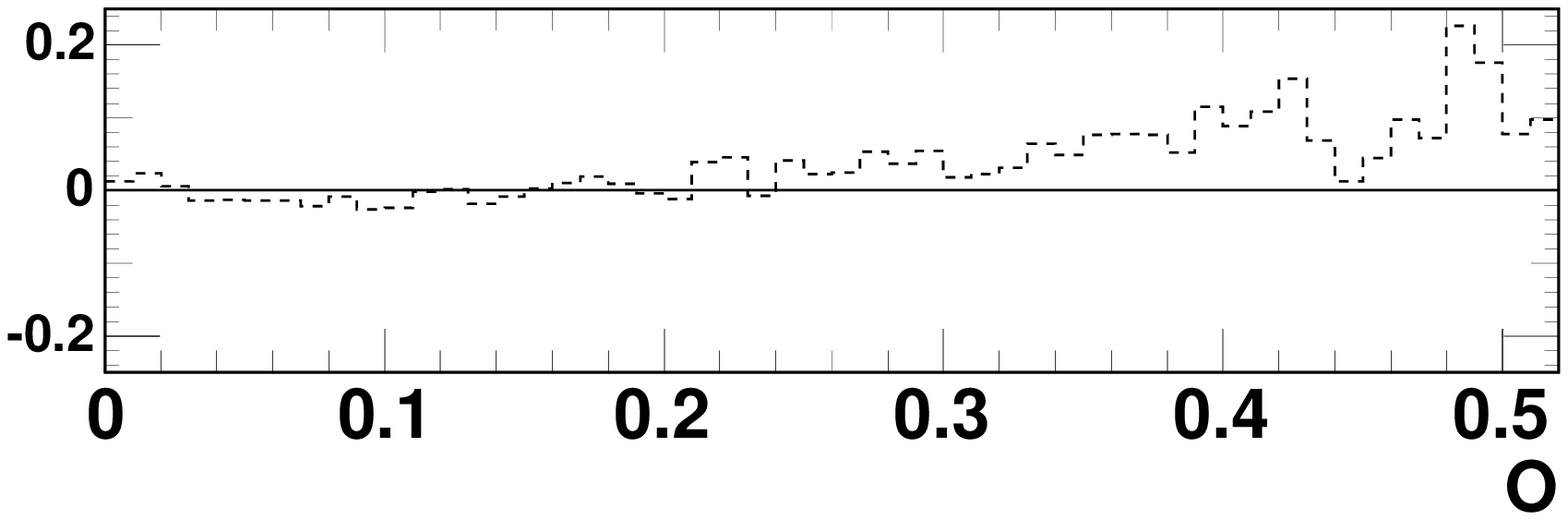}}
    \end{picture}
    \caption{\label{fig:lep_eventshapes} Thrust (top left), thrust-major (top
      right), thrust minor (bottom left), and oblateness (bottom right) at
      LEP~I. For definitions of these observables, cf.\ appendix
      \ref{app:observables1}. 
      The hadron level result of SHERPA is pictured for two different
      separation cuts $y=10^{-2.5}$ and $y=10^{-2}$. 
      Line styles and colours are the same as in Fig.\ \ref{fig:lep_diffdurham1}. }
  \end{center}
\end{figure}

\noindent
The samples generated by SHERPA also reproduce
event shape observables such as thrust, thrust-major, thrust-minor or
oblateness, cf.\ Fig.\ \ref{fig:lep_eventshapes}. Again, the
dependence on the generation cut is rather small, 
deviations are well below 20\%.  

\noindent
Going to more exclusive observables that are sensitive to the full
interference structure of matrix elements, various four-jet
correlations may be tested. Examples for such correlations are the
Bengtsson-\-Zerwas and the Nachtmann-\-Reiter angle, see the appendix
for their definition. In Fig.\ \ref{fig:lep_fourjet_angles}, data
taken at LEP~I \cite{Hendrikdipl} are compared with results of the
merged samples of SHERPA and with a ``shower''-only result. Of course,
the latter lacks the exact treatment of quantum interferences, which
is possible only through full matrix elements. Correspondingly, there
is a visible shape difference between data and the merged sample on
the one hand and the shower-only sample on the other hand. This
beautifully underlines the power of the merging approach.
\begin{figure}
  \begin{center}
    \begin{picture}(400,220)
      \put(-15,0){\includegraphics[width=8.0cm]
        {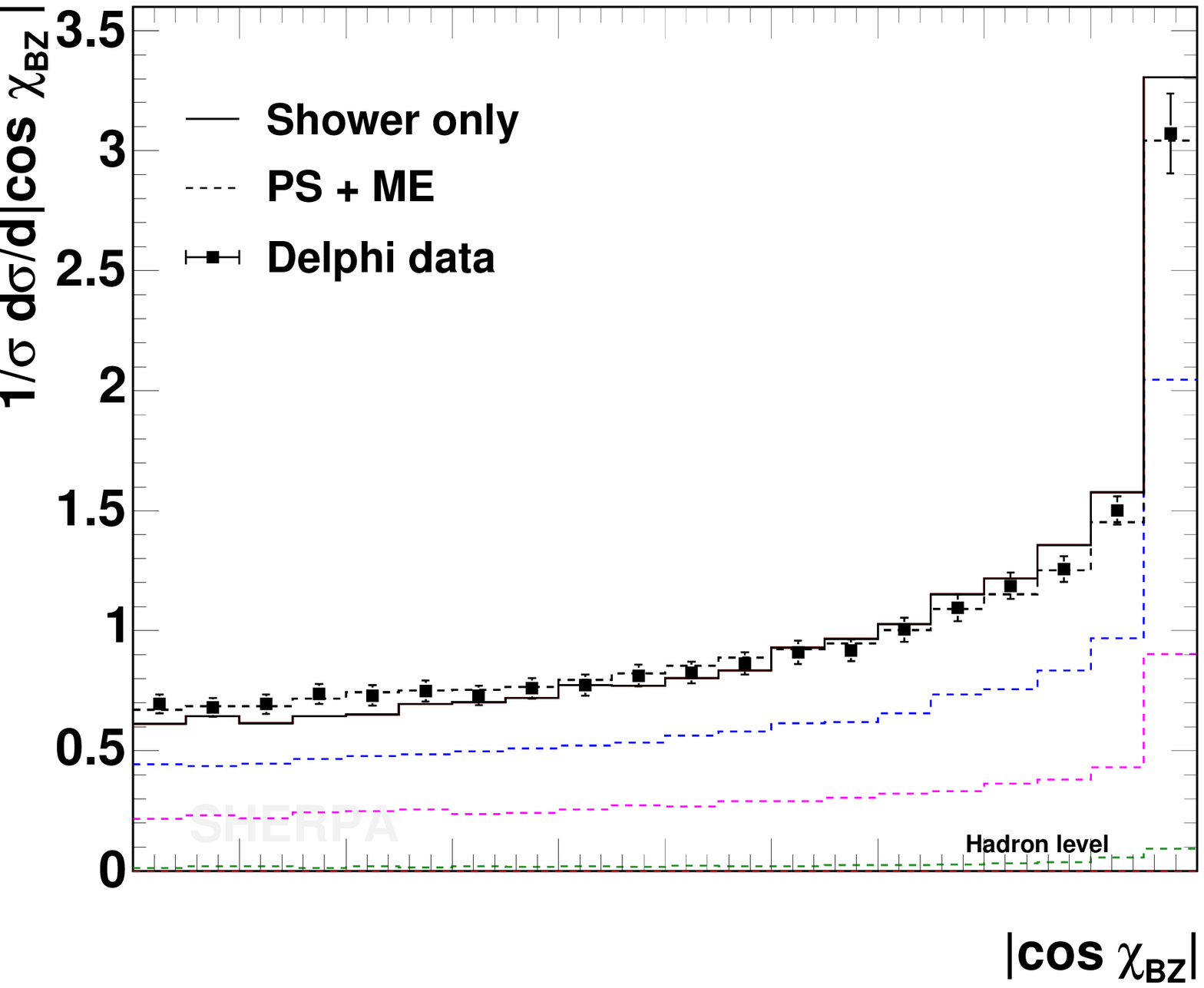}}
      \put(-15,0){\includegraphics[width=8.0cm]
        {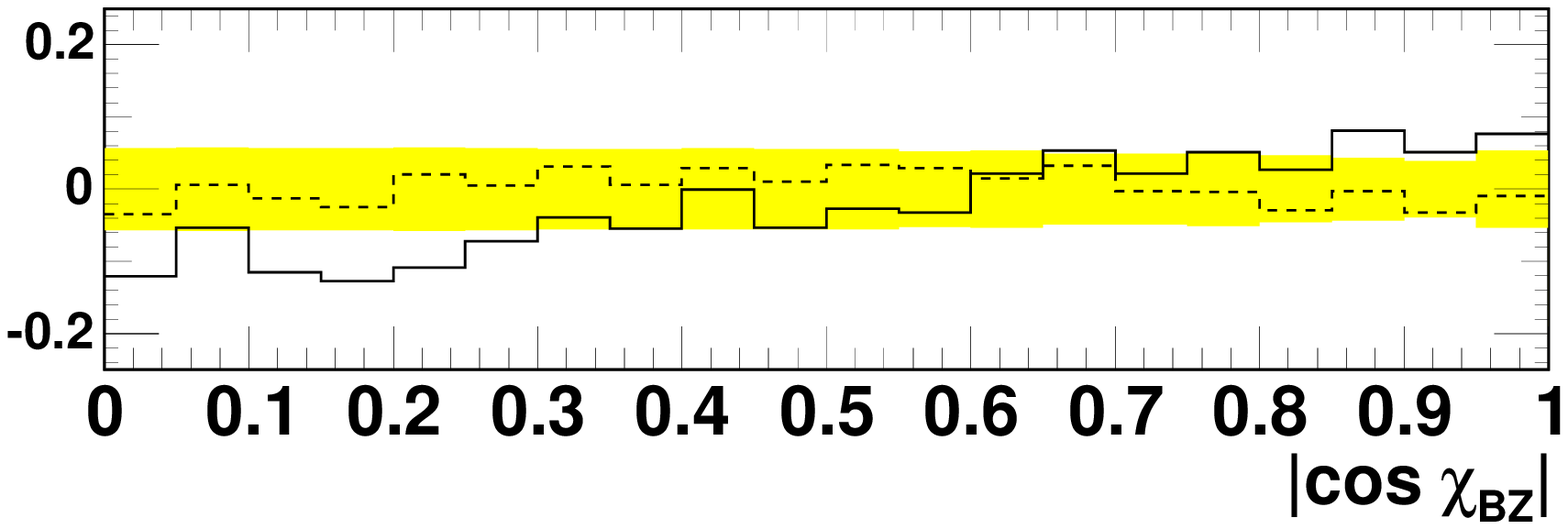}}
      \put(200,0){\includegraphics[width=8.0cm]
        {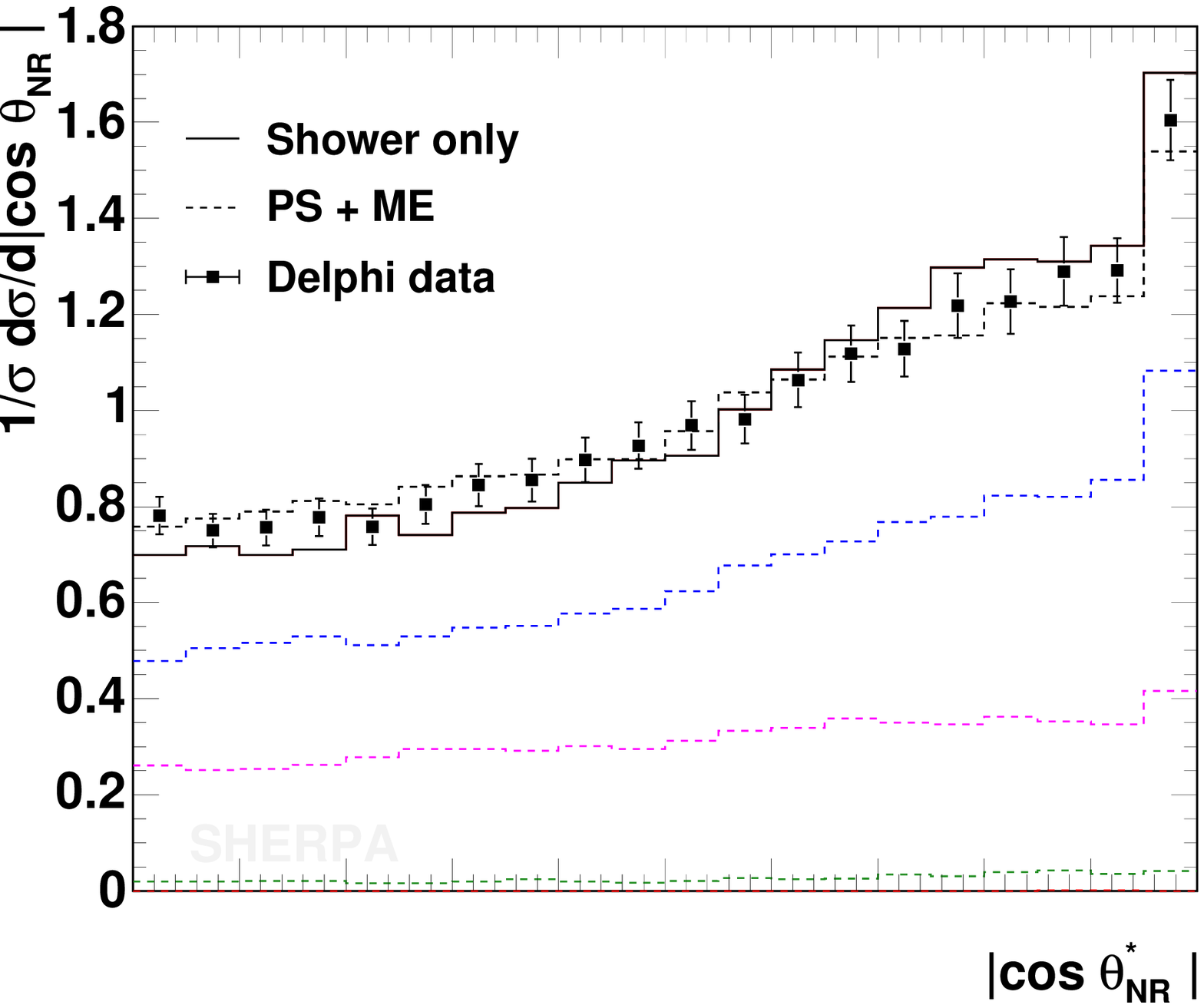}}
      \put(200,0){\includegraphics[width=8.0cm]
        {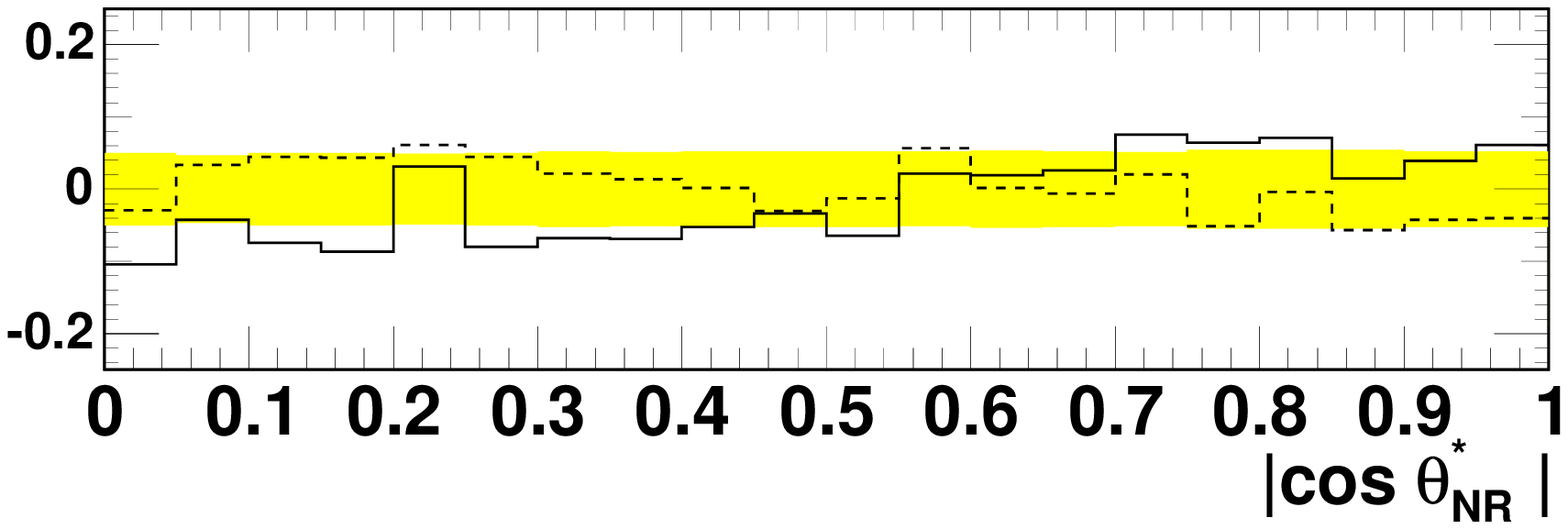}}
    \end{picture}
    \caption{\label{fig:lep_fourjet_angles} Four jet angle
      distributions. Shown are the Bengtsson-\-Zerwas angle (left) and
      the modified Nachtmann-Reiter angle (right). The data points are
      from a DELPHI measurement \cite{Hendrikdipl}.}
  \end{center}
\end{figure}

\subsection{Results for $p\bar{p}\to W+\mbox{jets}$ at Tevatron, Run I}

\begin{figure*}[h]
  \begin{center}
    \begin{pspicture}(430,430)
      \put(290,290){\includegraphics[width=5.3cm]
        {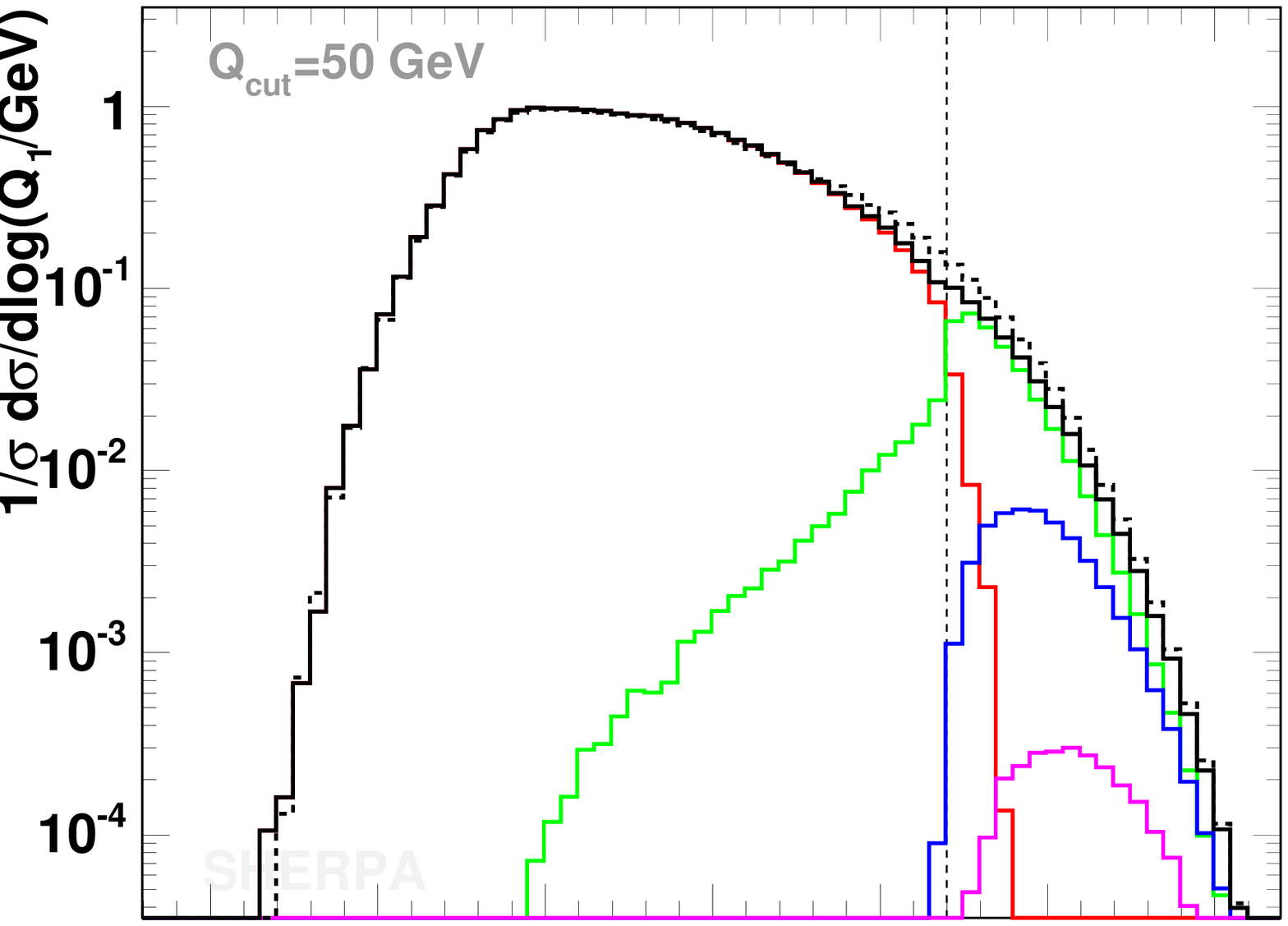}}
      \put(290,290){\includegraphics[width=5.3cm]
        {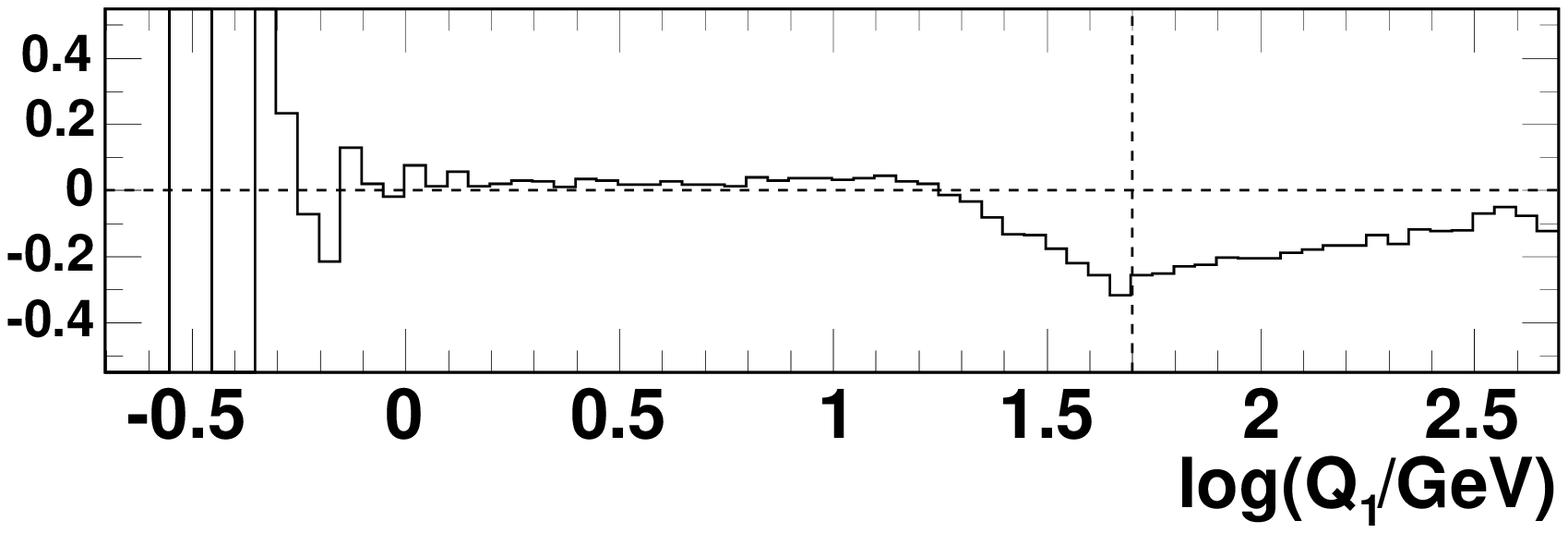}}
      \put(145,290){\includegraphics[width=5.3cm]
        {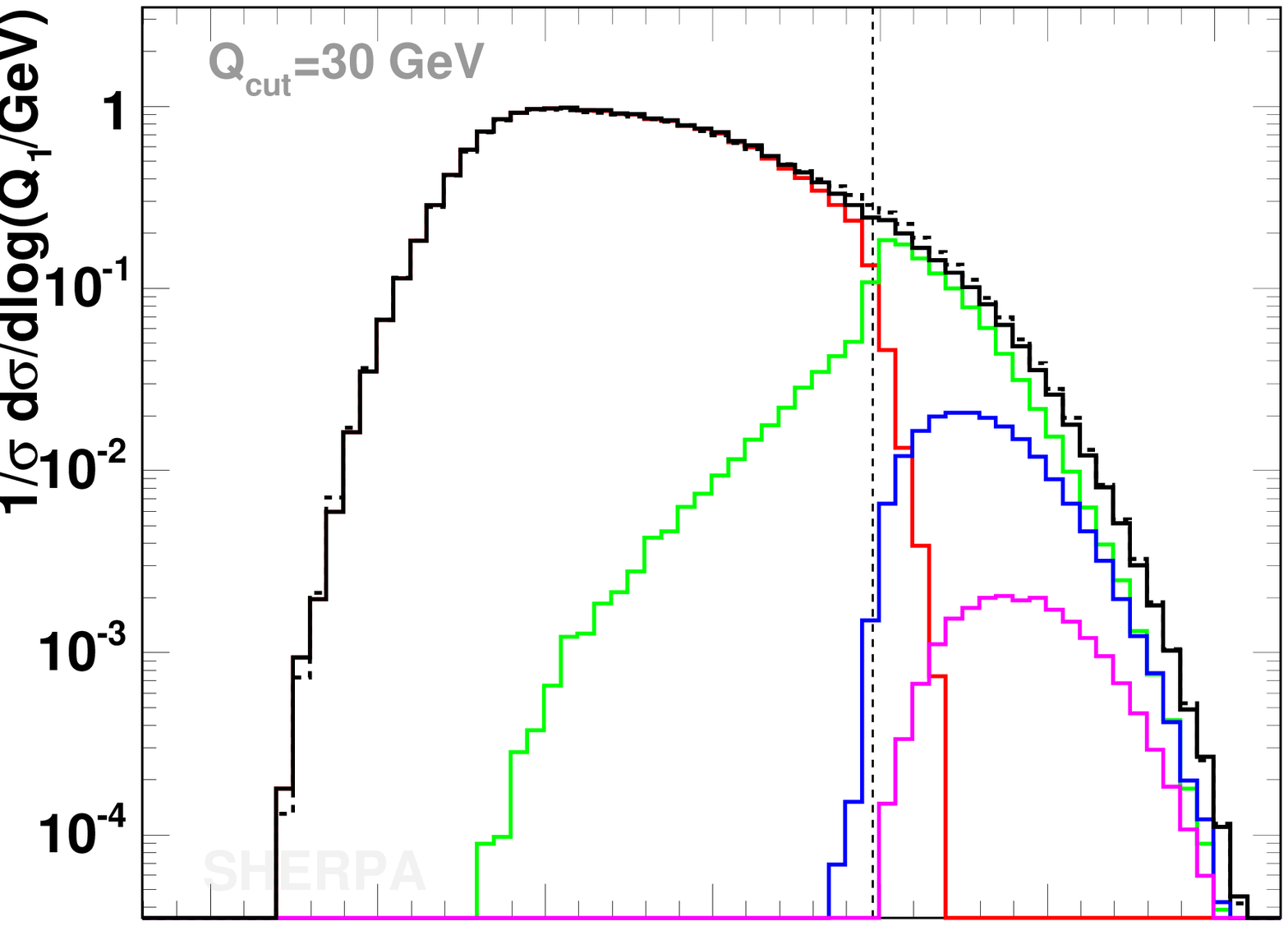}}
      \put(145,290){\includegraphics[width=5.3cm]
        {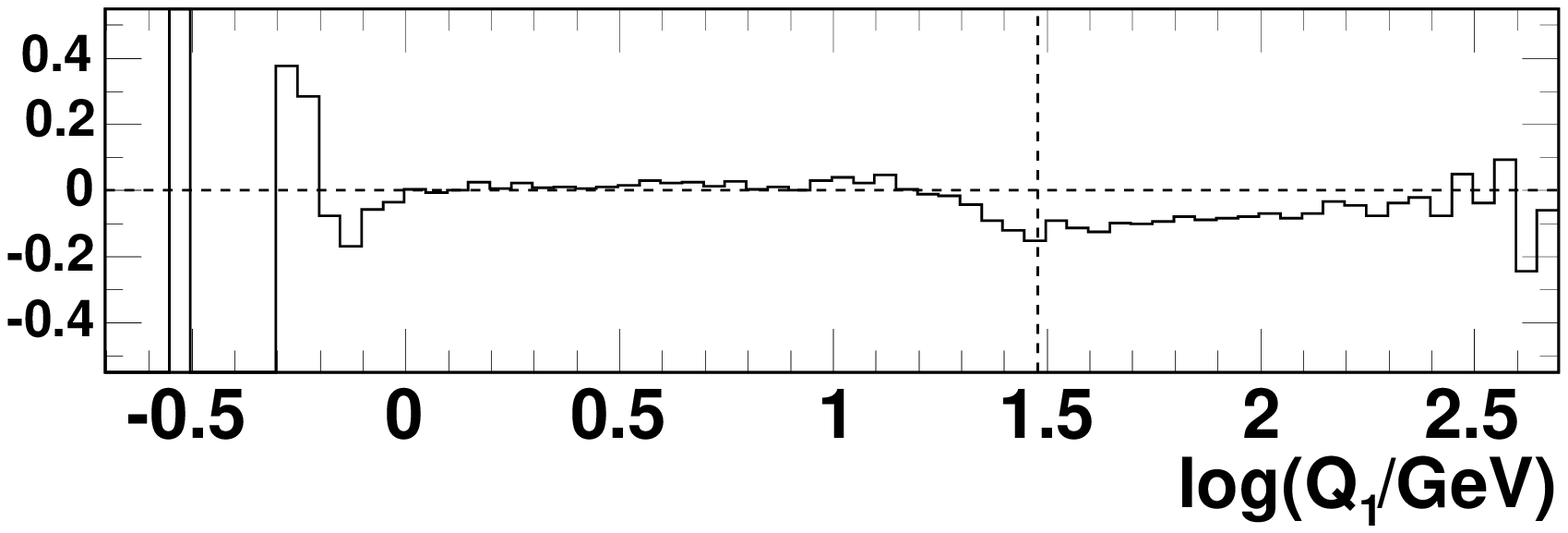}}
      \put(0,290){\includegraphics[width=5.3cm]
        {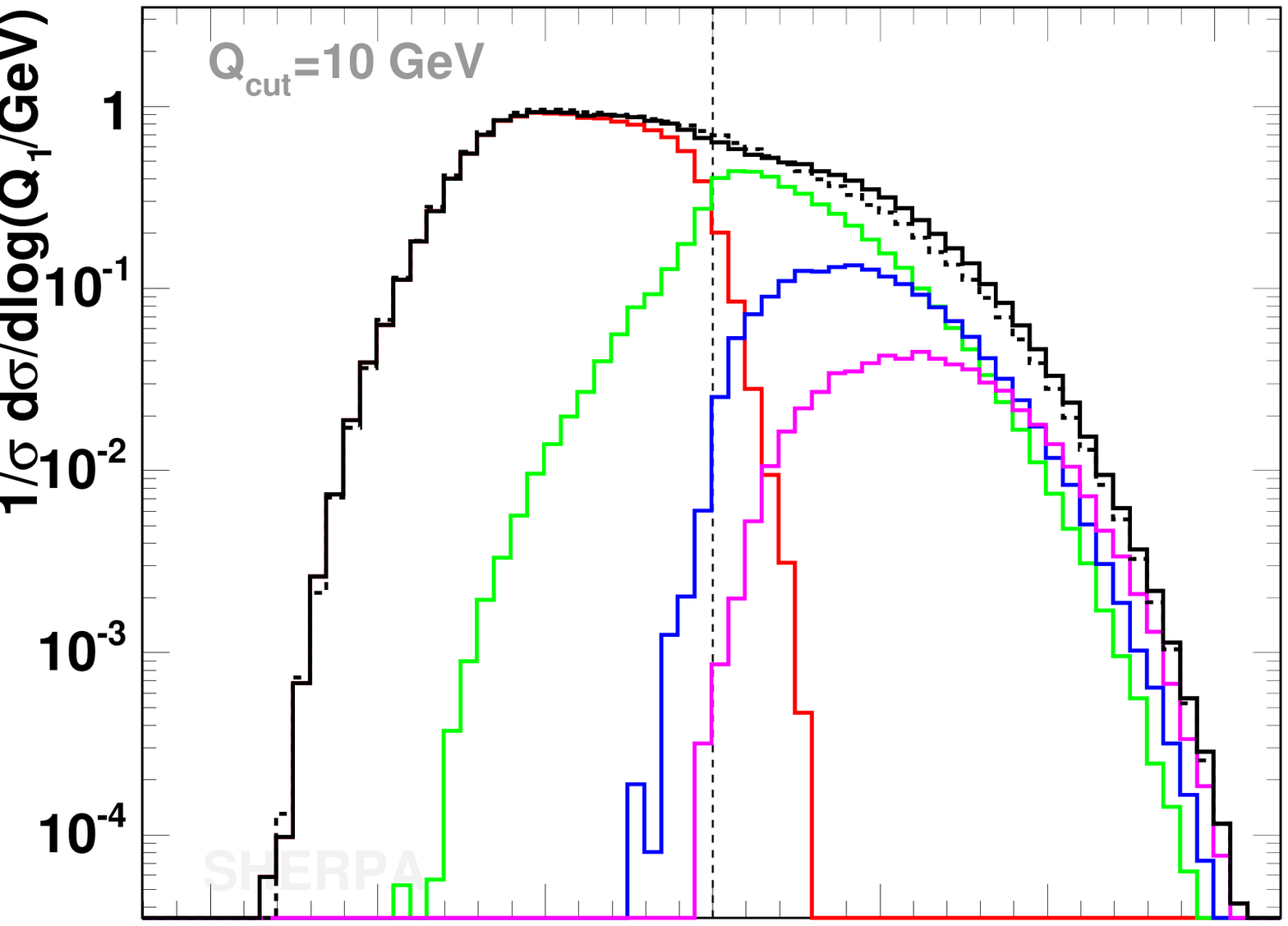}}
      \put(0,290){\includegraphics[width=5.3cm]
        {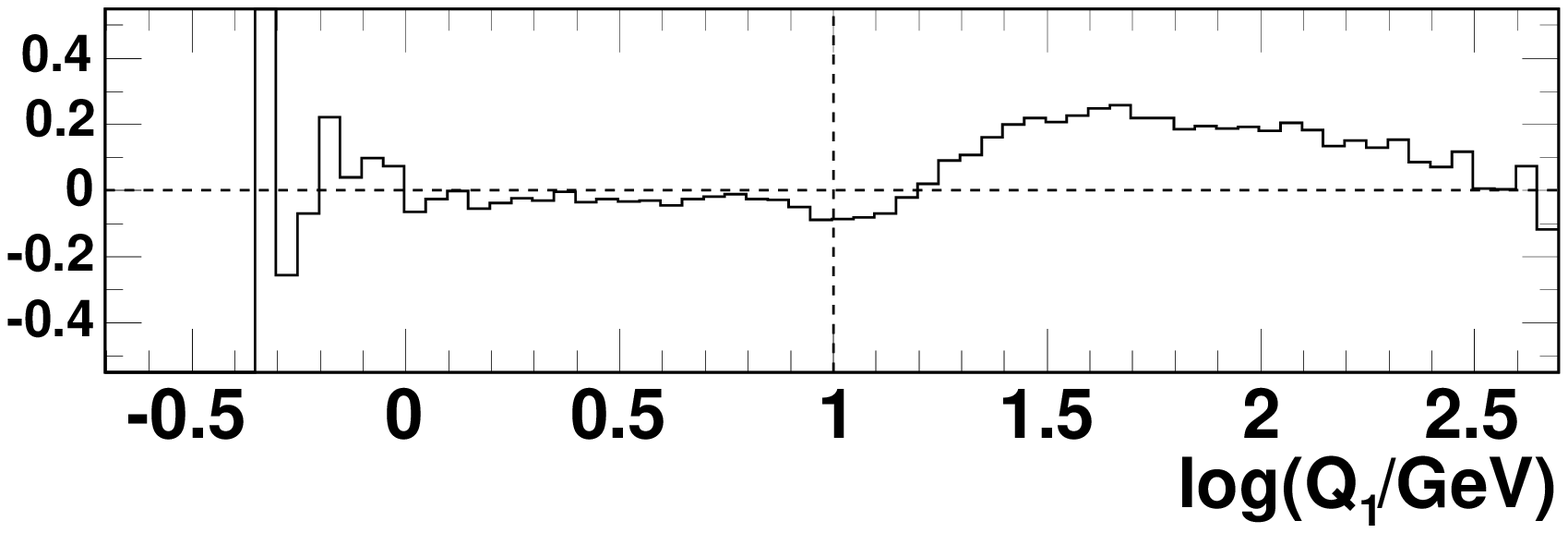}}
      \put(290,140){\includegraphics[width=5.3cm]
        {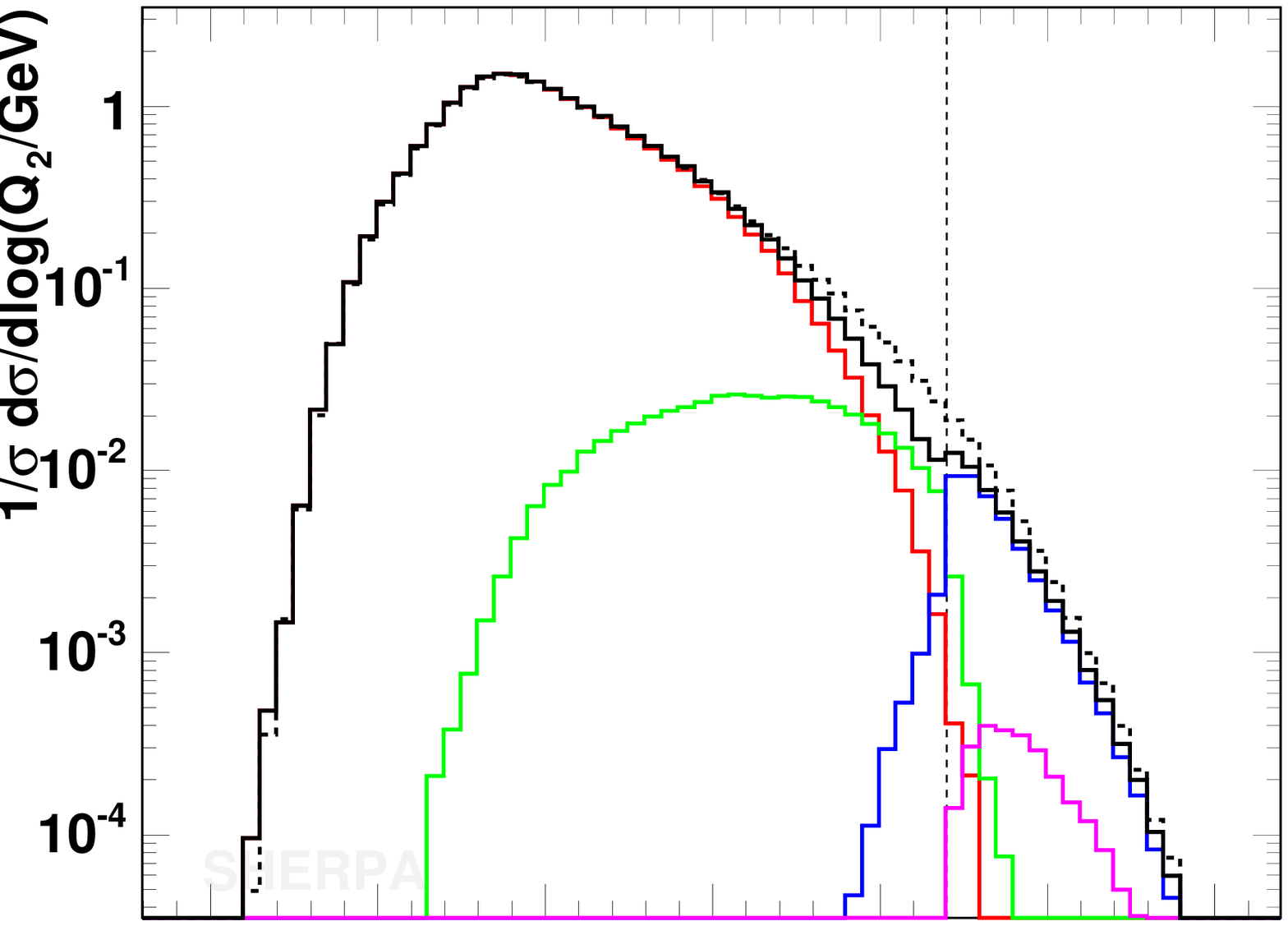}}
      \put(290,140){\includegraphics[width=5.3cm]
        {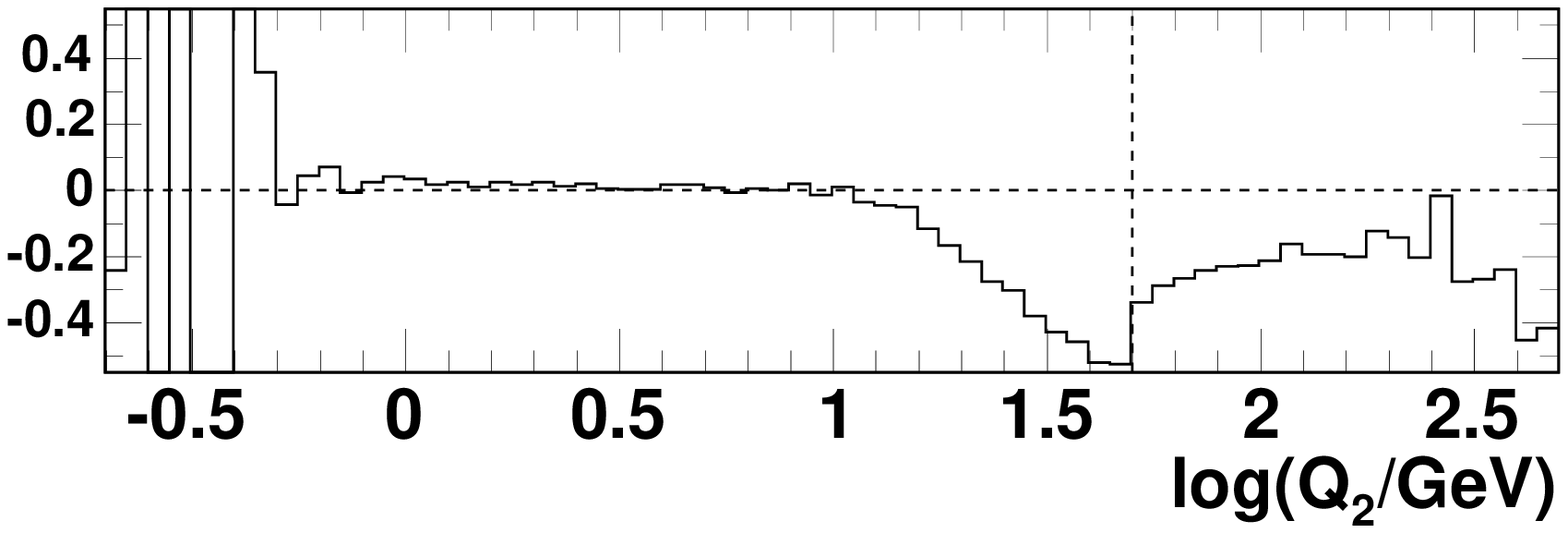}}
      \put(145,140){\includegraphics[width=5.3cm]
        {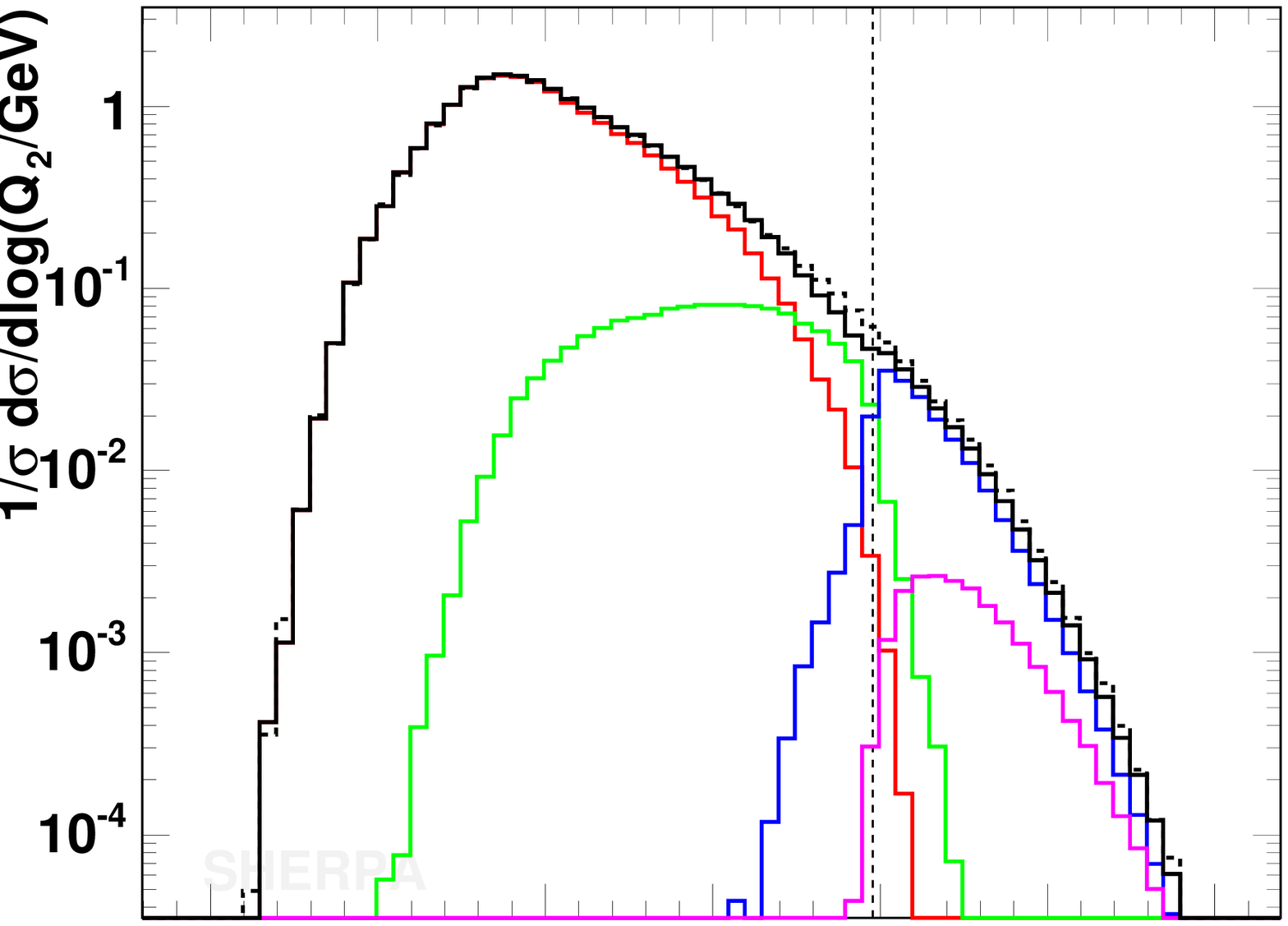}}
      \put(145,140){\includegraphics[width=5.3cm]
        {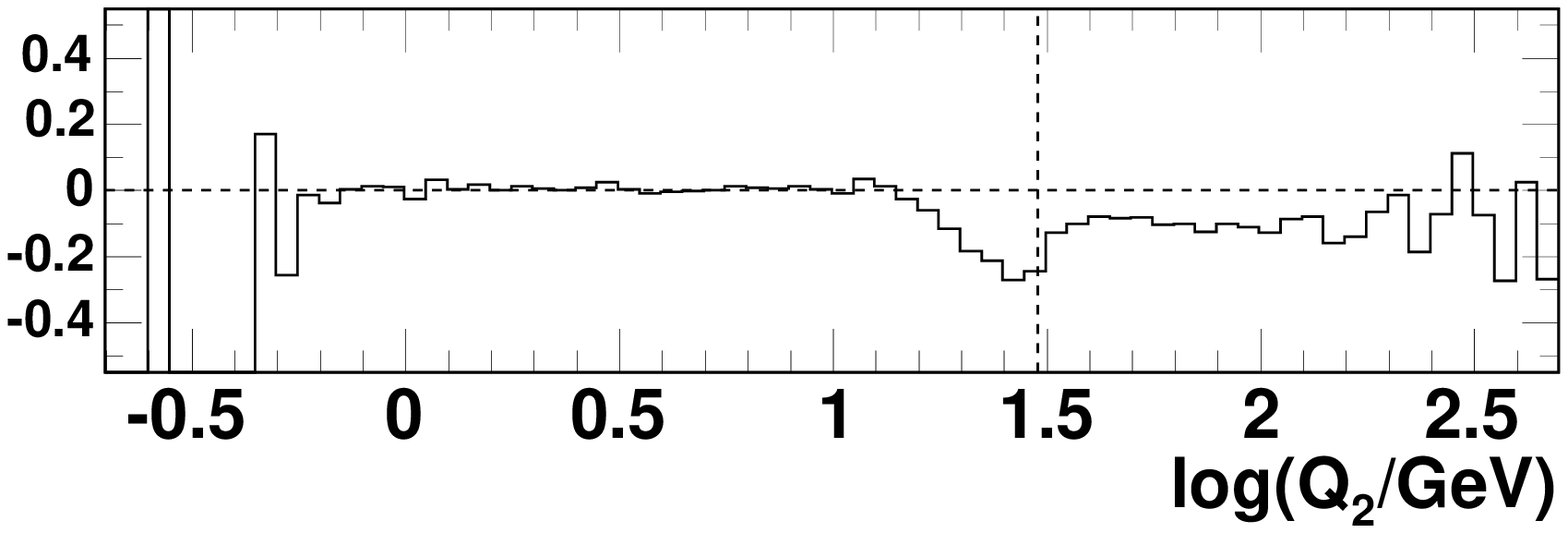}}
      \put(0,140){\includegraphics[width=5.3cm]
        {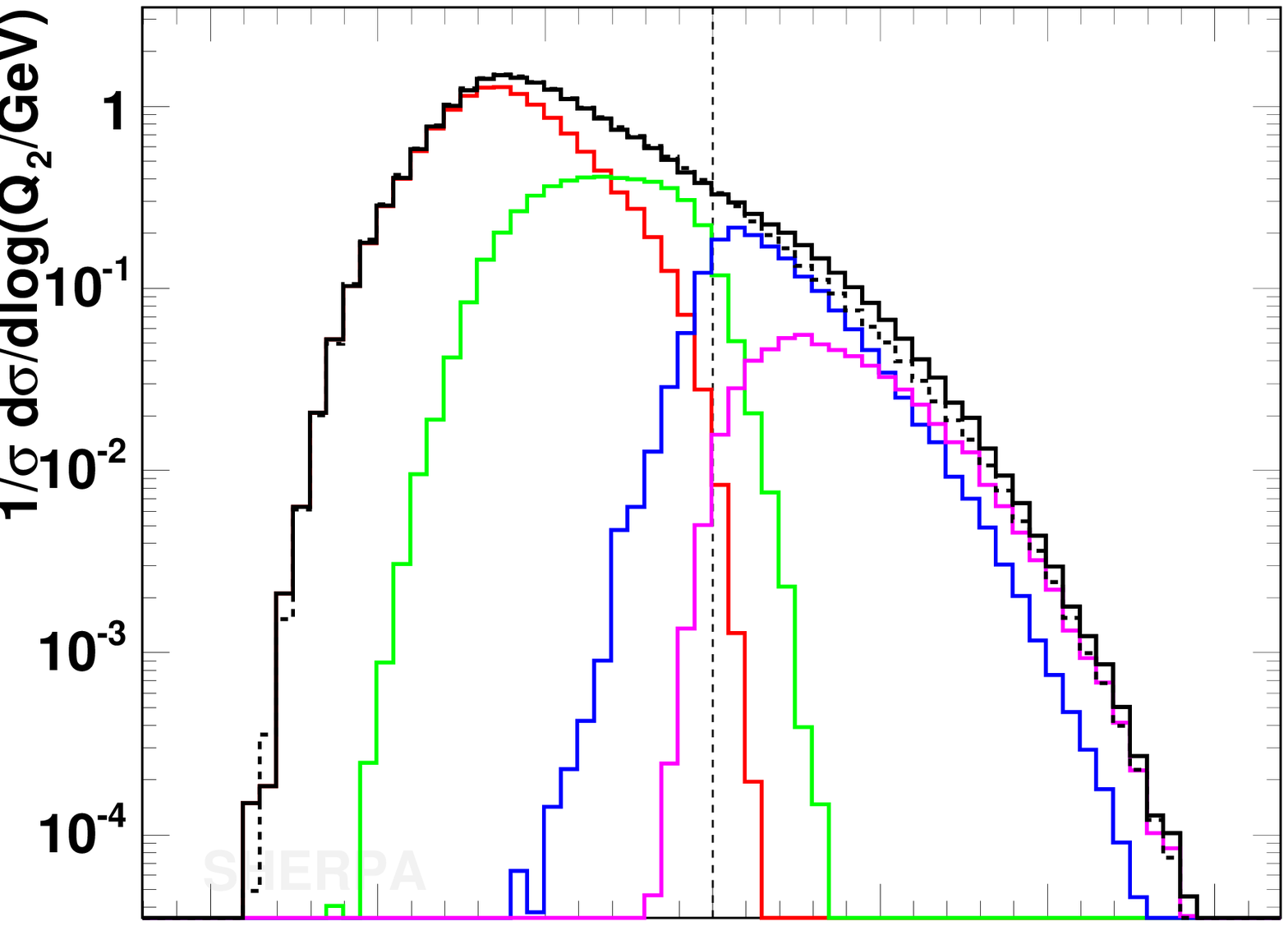}}
      \put(0,140){\includegraphics[width=5.3cm]
        {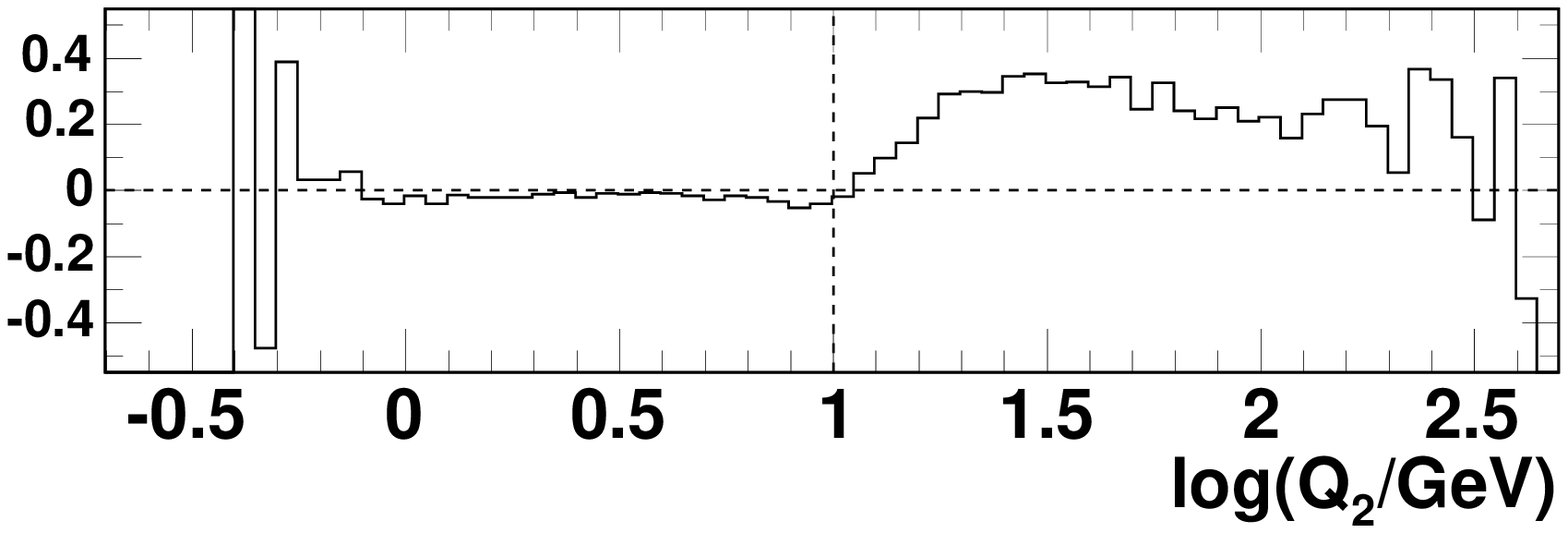}}
      \put(290,-10){\includegraphics[width=5.3cm]
        {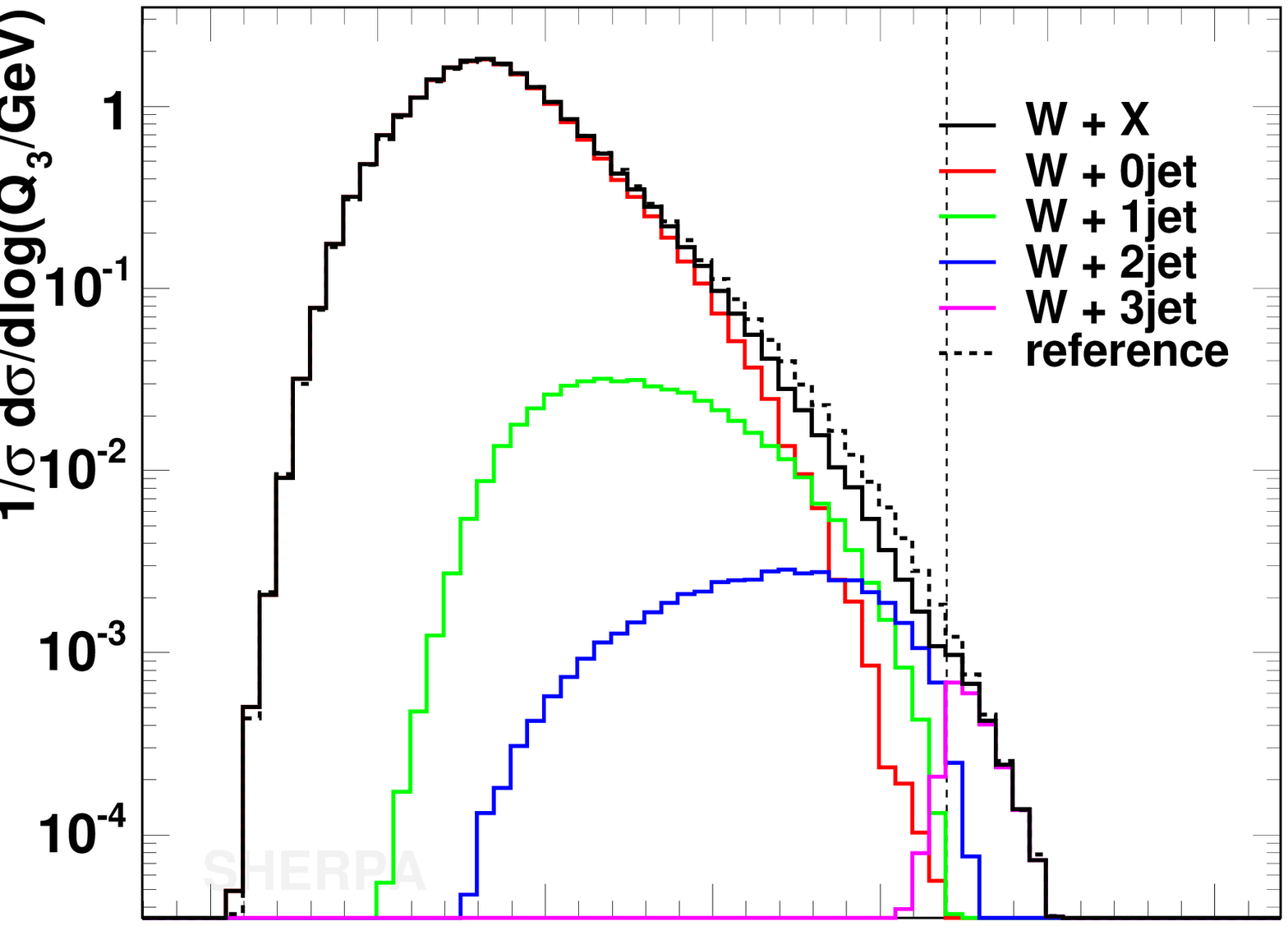}}
      \put(290,-10){\includegraphics[width=5.3cm]
        {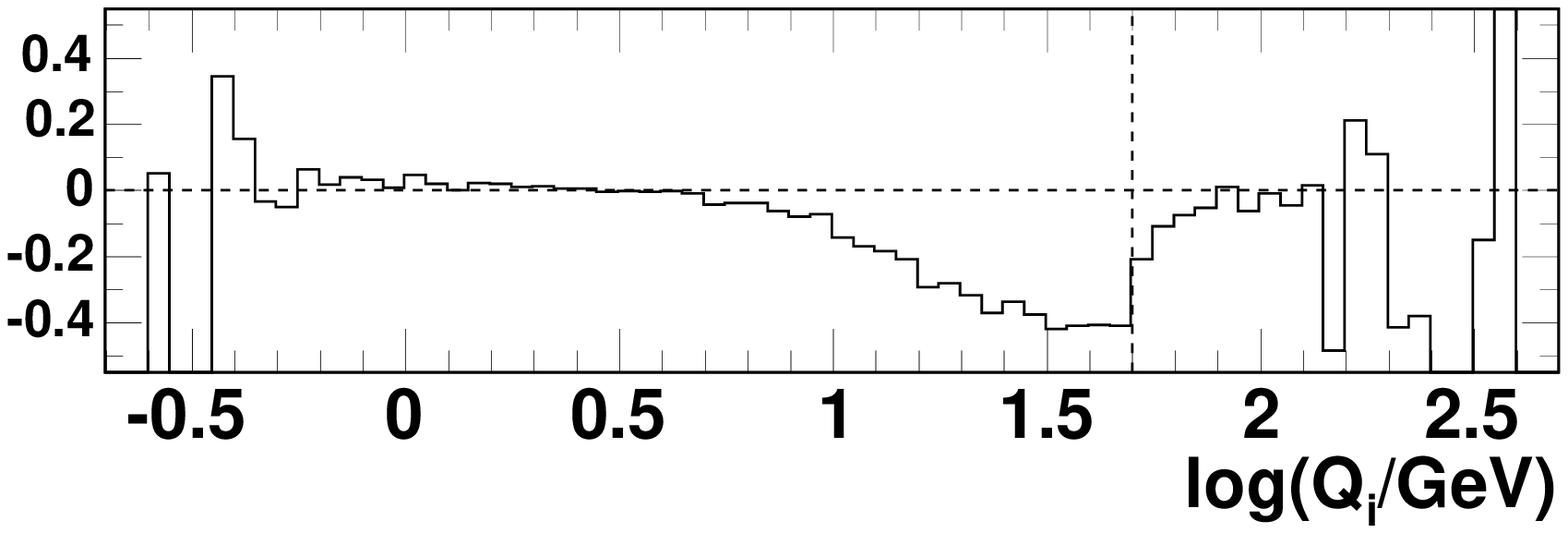}}
      \put(145,-10){\includegraphics[width=5.3cm]
        {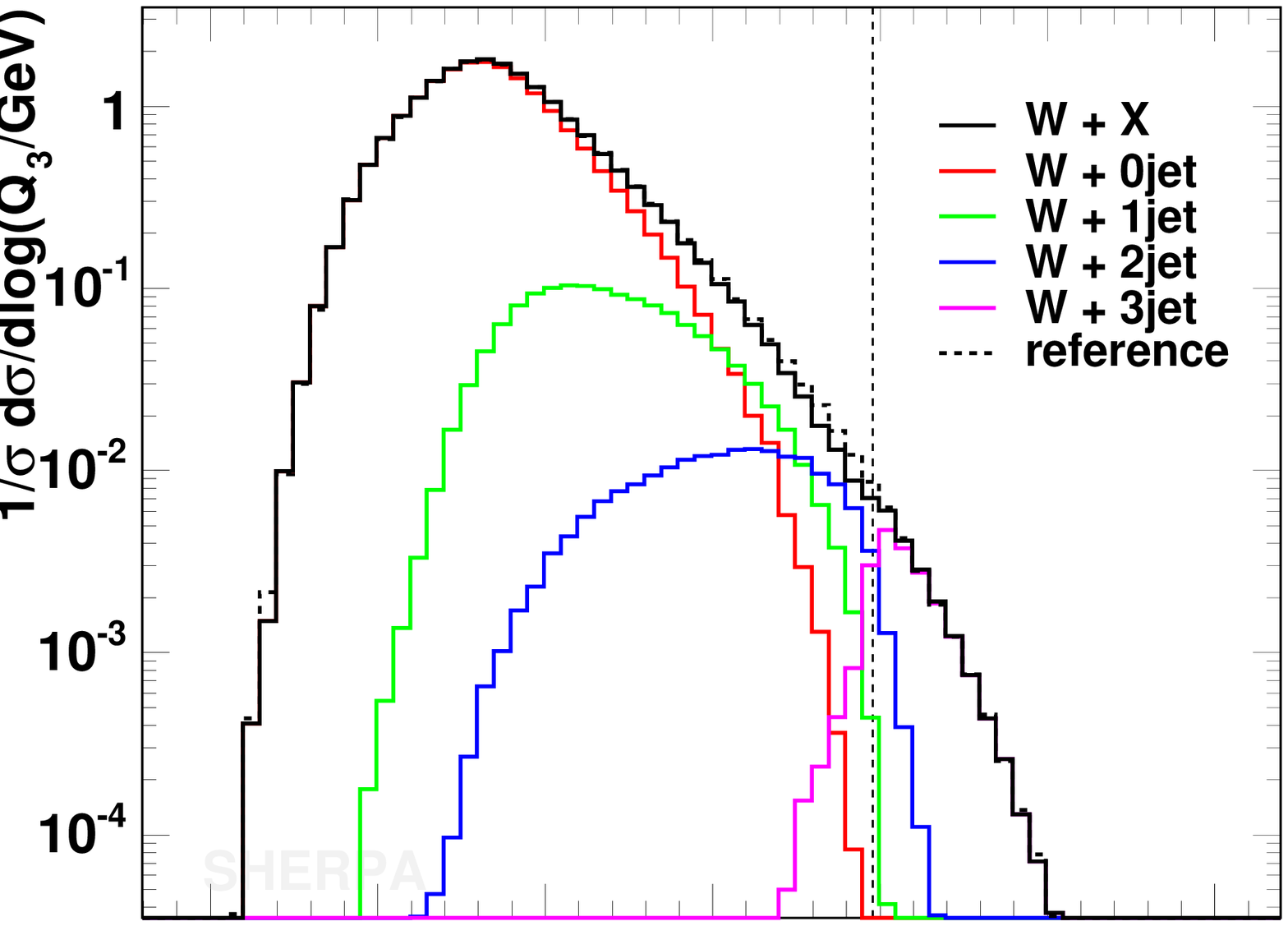}}
      \put(145,-10){\includegraphics[width=5.3cm]
        {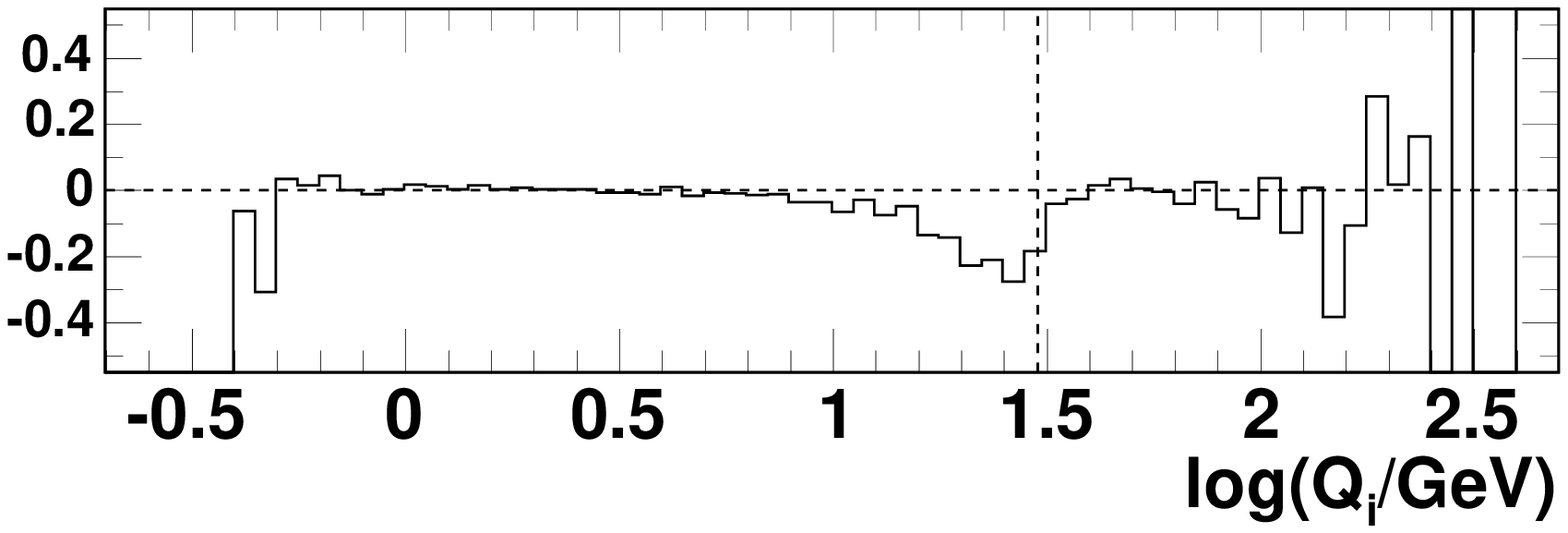}}
      \put(0,-10){\includegraphics[width=5.3cm]
        {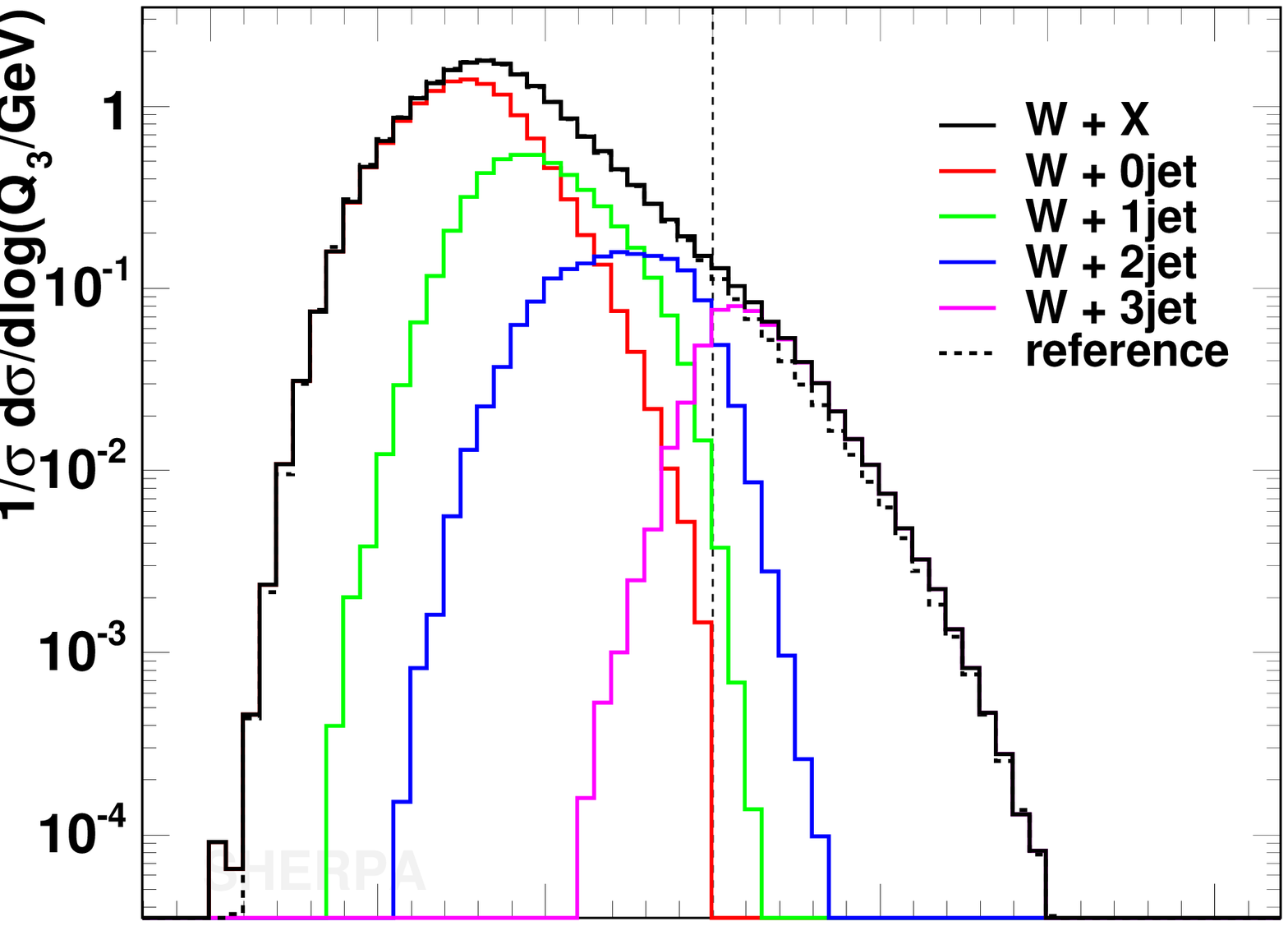}}
      \put(0,-10){\includegraphics[width=5.3cm]
        {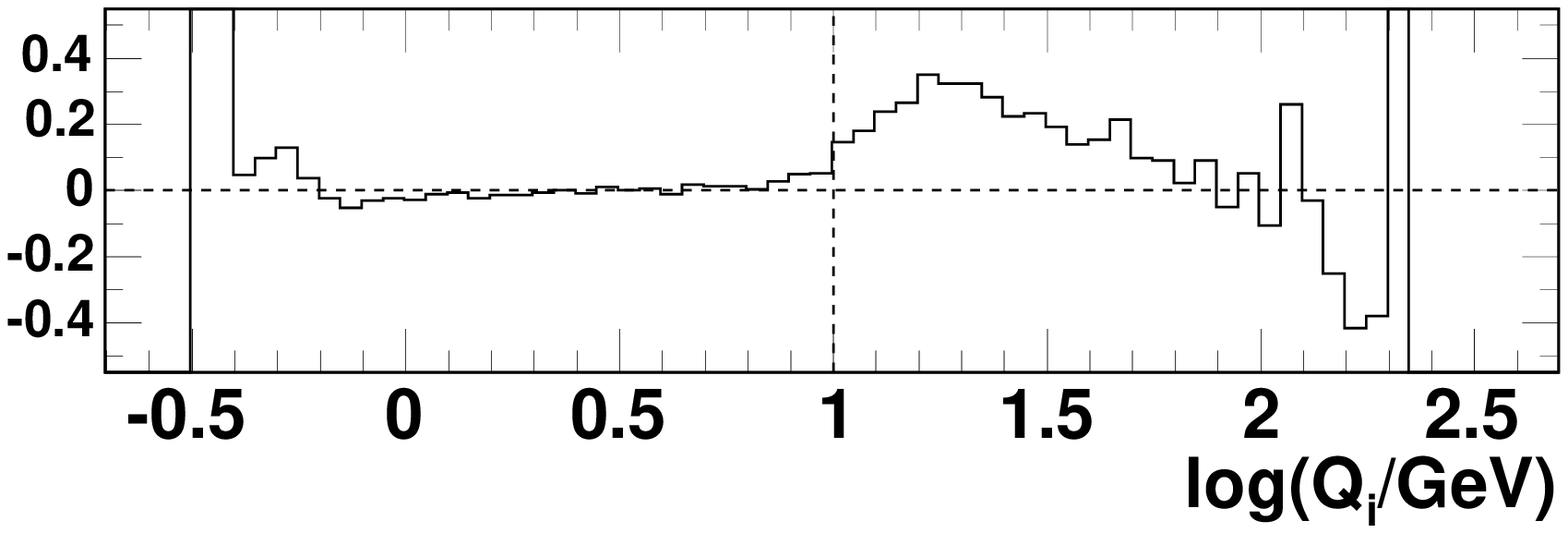}}
    \end{pspicture}
  \end{center}
  \caption{\label{ycut_diff_W_Tev}Differential jet rates for the  
    $1 \to 0$, $2 \to 1$ and $3 \to 2$ transition (top to bottom), 
    for $Q_{\rm cut}=10$~GeV,  $30$~GeV, and $50$~GeV (from left to
    right). In each plot, the results are compared with those for 
    $Q_{\rm cut}=20$ GeV. }
\end{figure*}

\noindent
The investigations for the case of $p\bar{p}\to W+\mbox{jets}$ start
with an analysis of differential jet rates in the $k_\perp$ algorithm
for this process at Tevatron, Run I. Results of SHERPA with different
jet resolution cuts during the generation of the respective sample are
exhibited in Fig.\ \ref{ycut_diff_W_Tev}. The results are not quite as
good as those obtained for the previous case of $e^+e^-$ annihilations
into jets, on the other hand, the example presented here is much more
complicated. This extra complication is due to a more intricate
radiation pattern with emissions in both the initial and the final
state. Still the relative differences are marginal, ranging up to
20\%. Only in the sample with the highest jet resolution cut
it becomes apparent that the parton shower is not able to
fill the phase space properly. This is the reason behind the visible
hole in the differential jet rates around the cut. 

\begin{figure*}[h]
  \begin{center}
    \begin{pspicture}(400,275)
      \put(250,130){\includegraphics[width=5.cm]
        {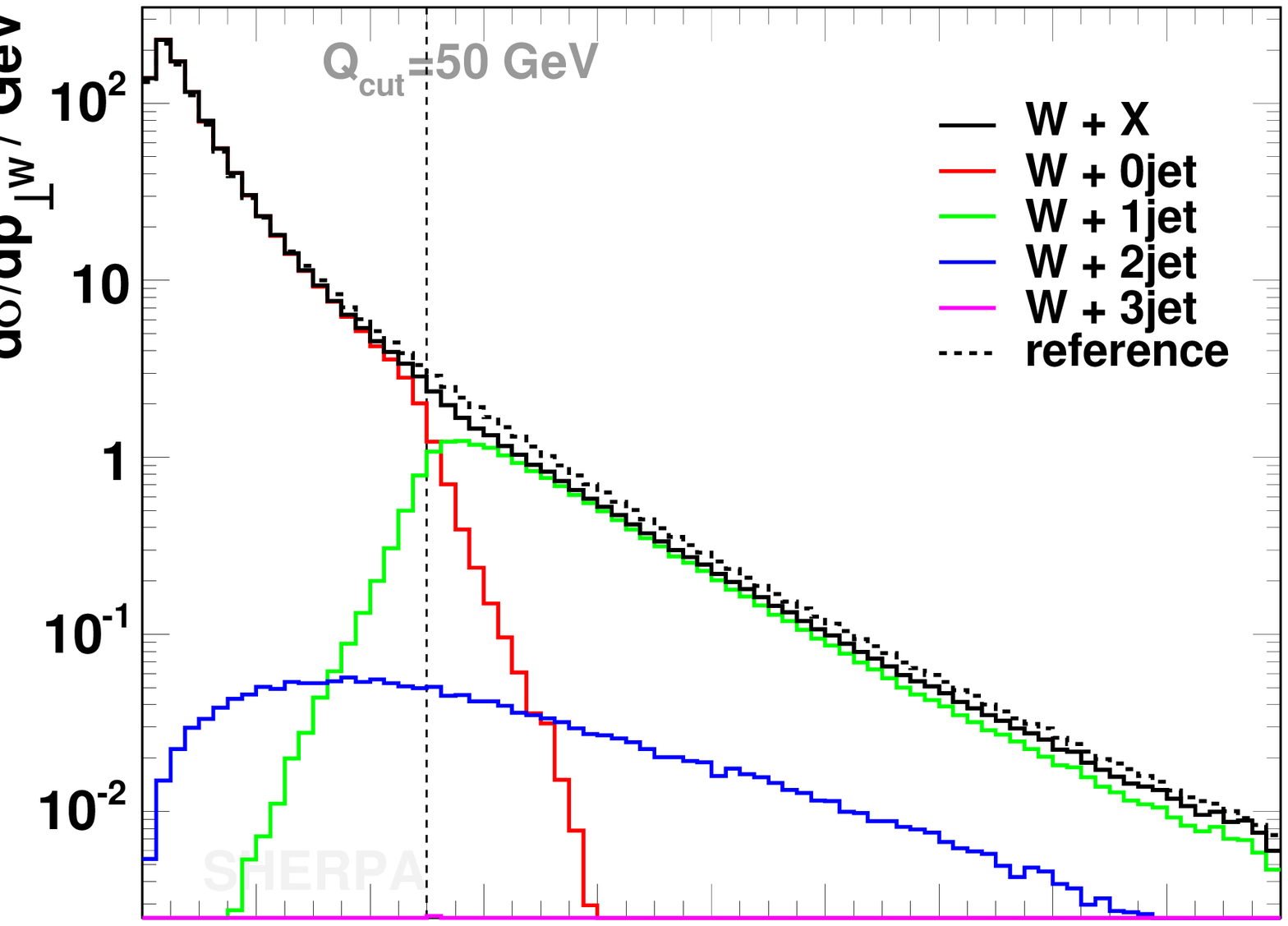}}
      \put(250,130){\includegraphics[width=5.cm]
        {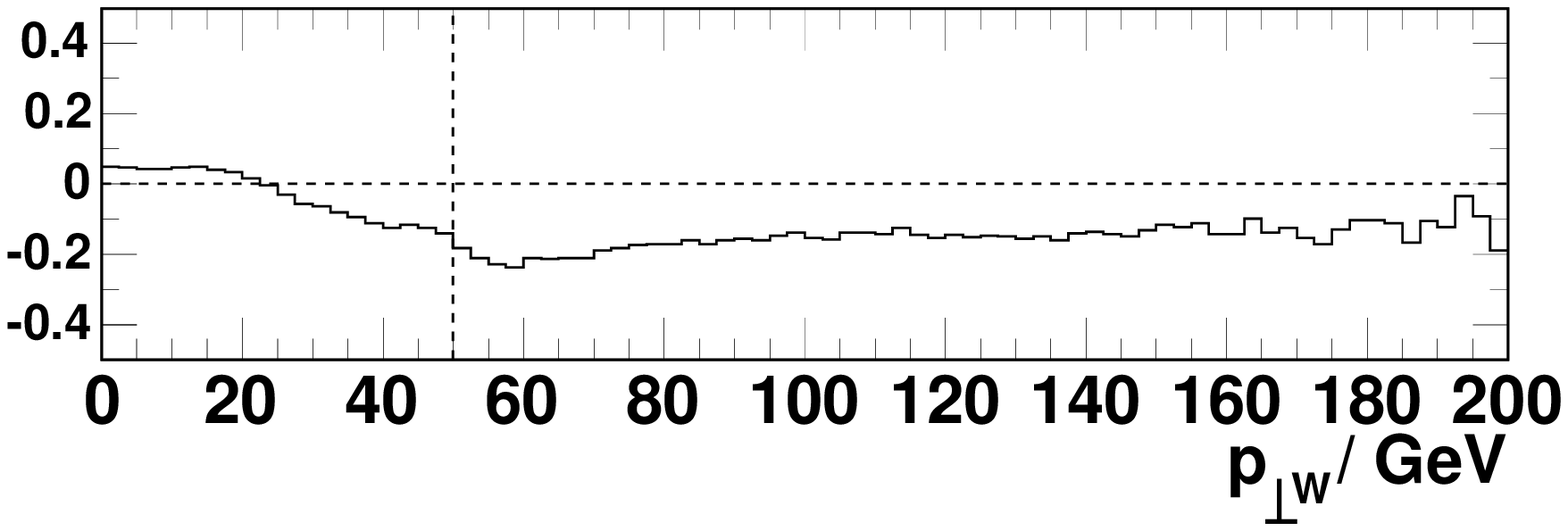}}
      \put(125,130){\includegraphics[width=5.cm]
        {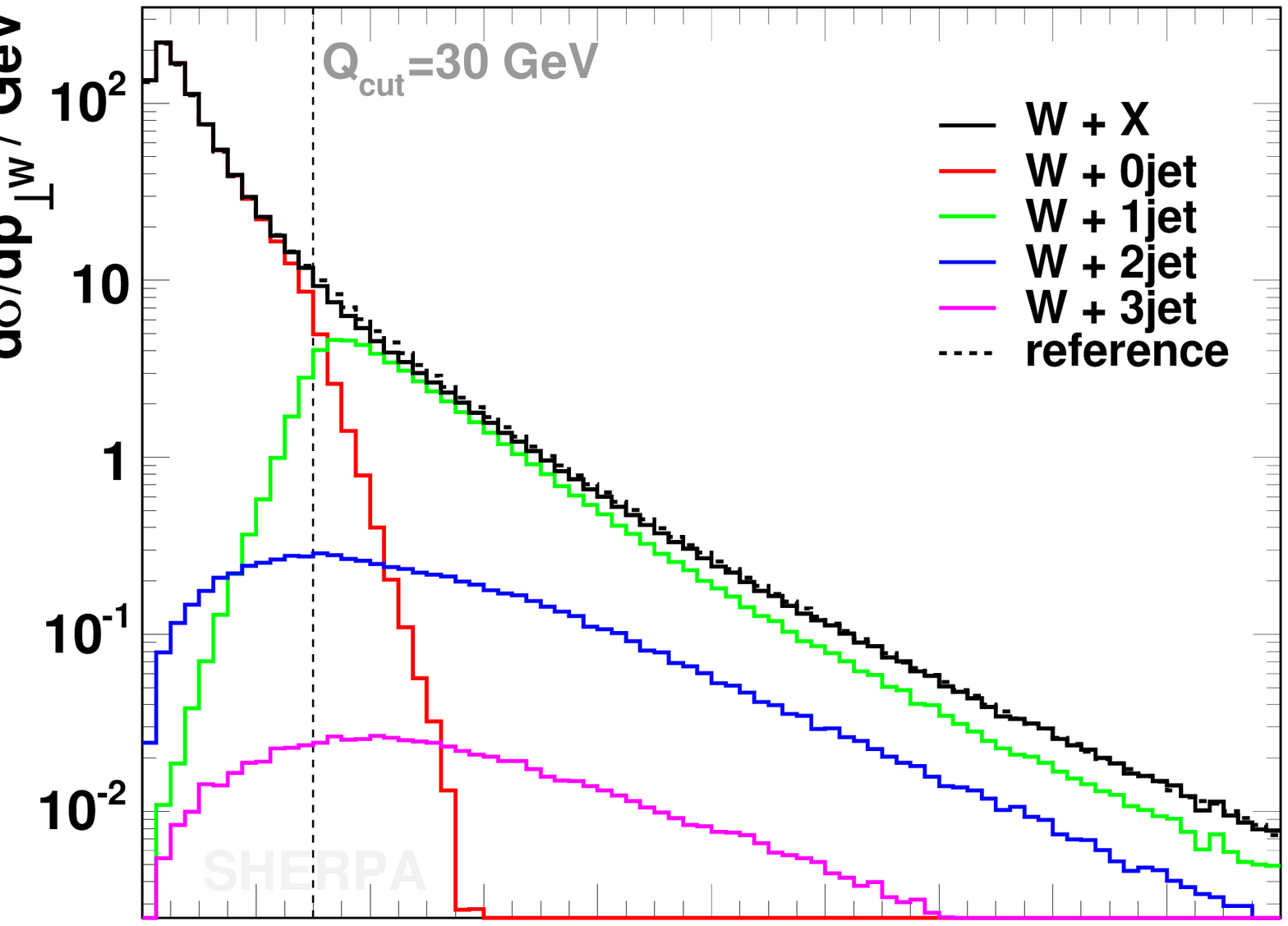}}
      \put(125,130){\includegraphics[width=5.cm]
        {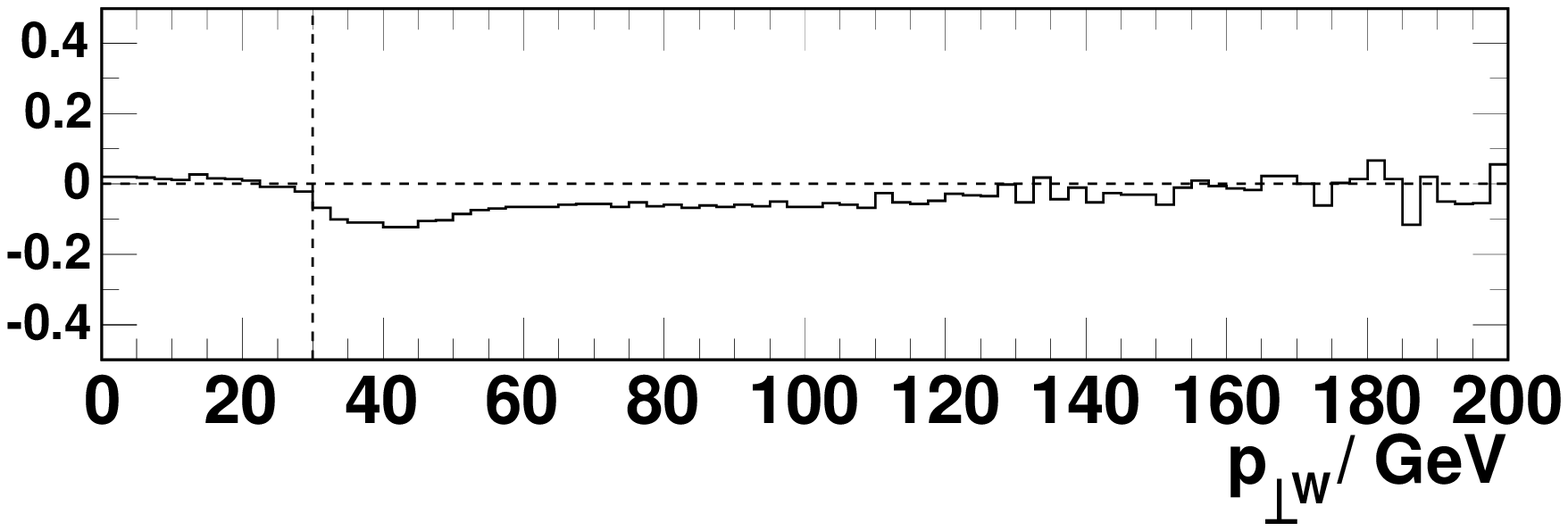}}
      \put(0,130){\includegraphics[width=5.cm]
        {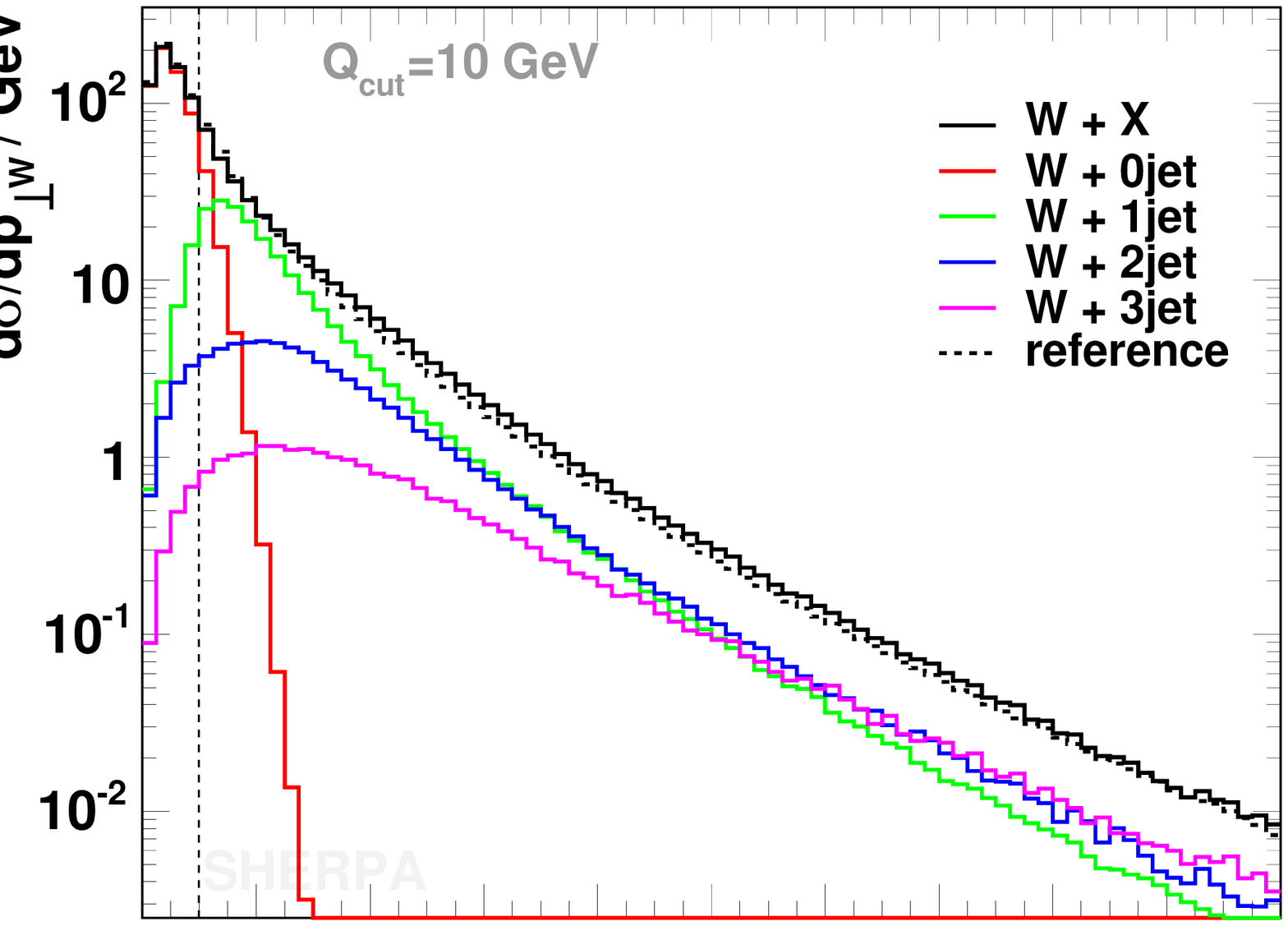}}
      \put(0,130){\includegraphics[width=5.cm]
        {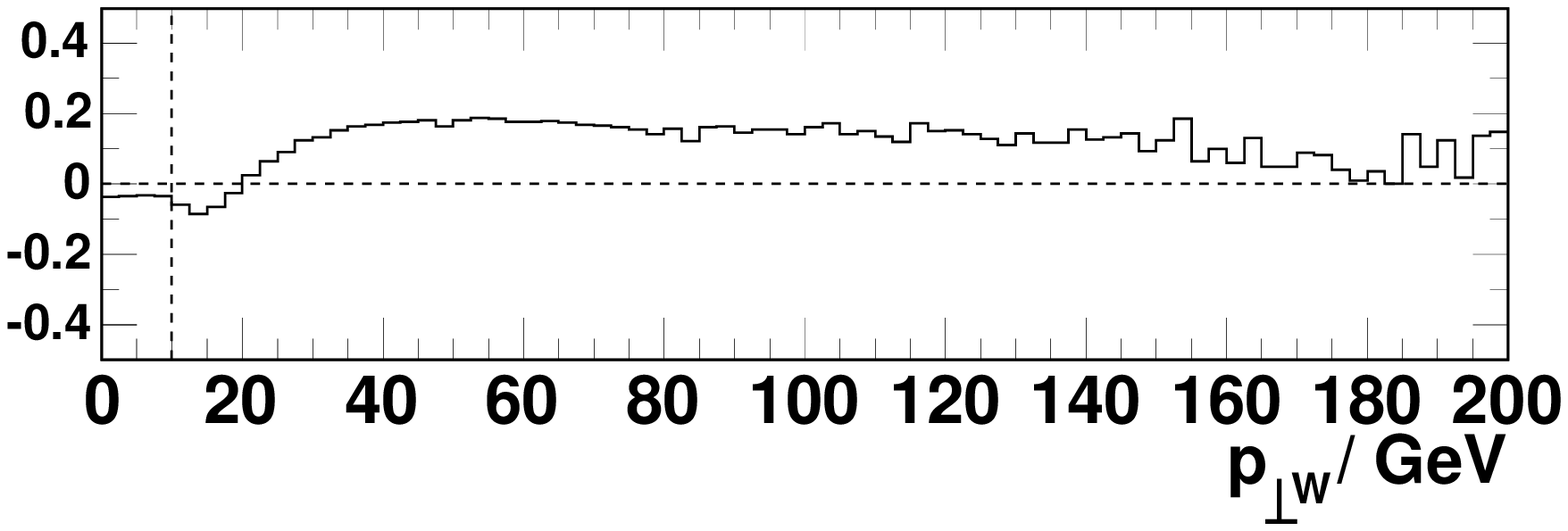}}
      \put(250,-10){\includegraphics[width=5.cm]
        {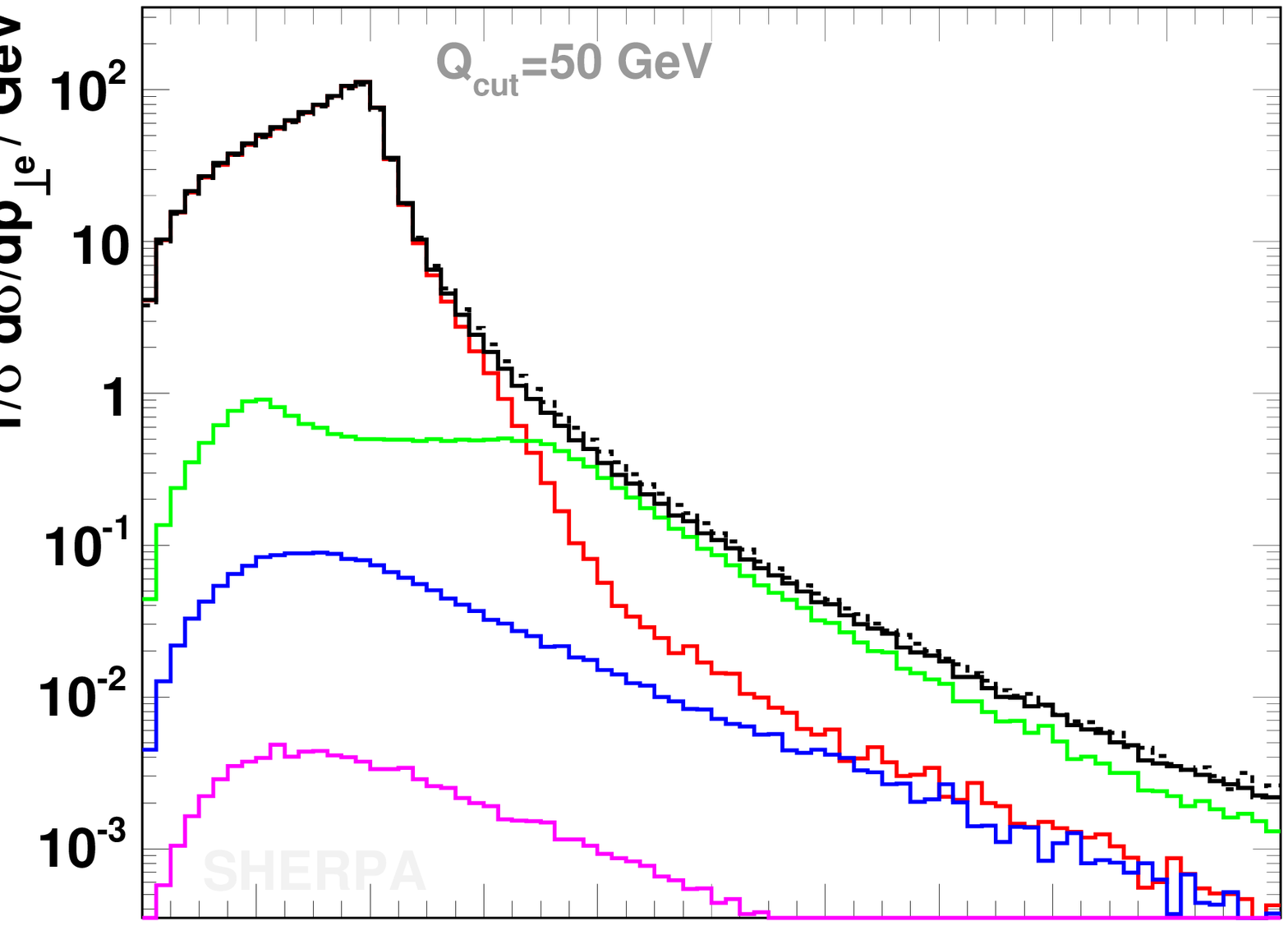}}
      \put(250,-10){\includegraphics[width=5.cm]
        {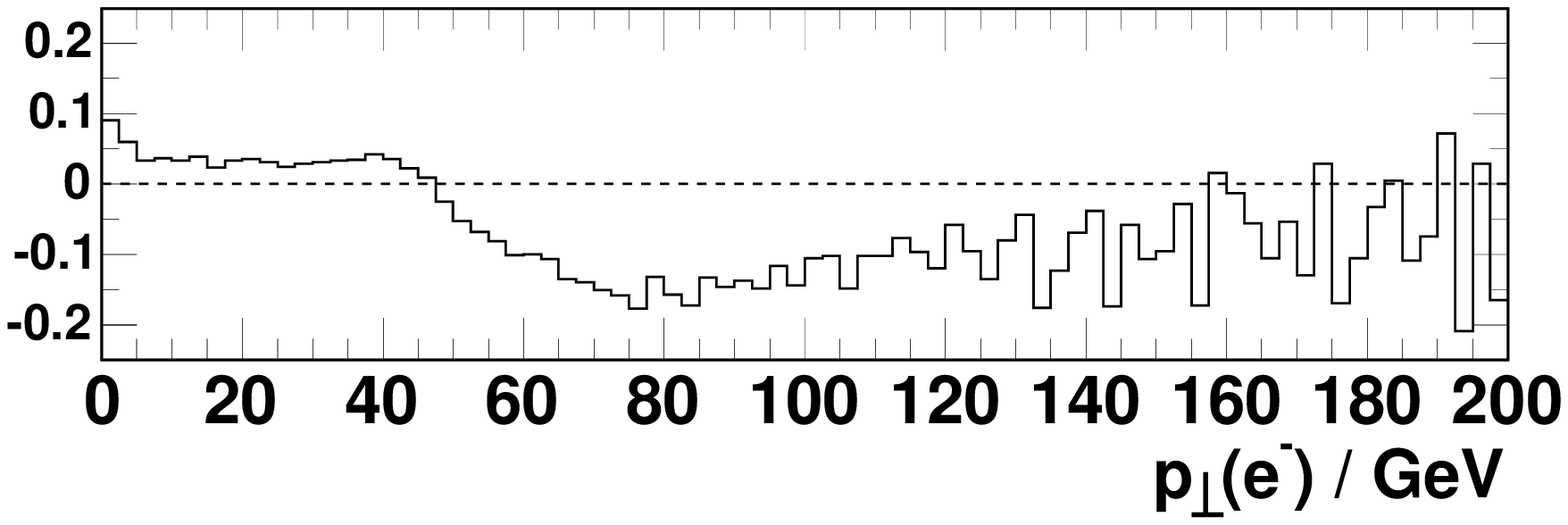}}
      \put(125,-10){\includegraphics[width=5.cm]
        {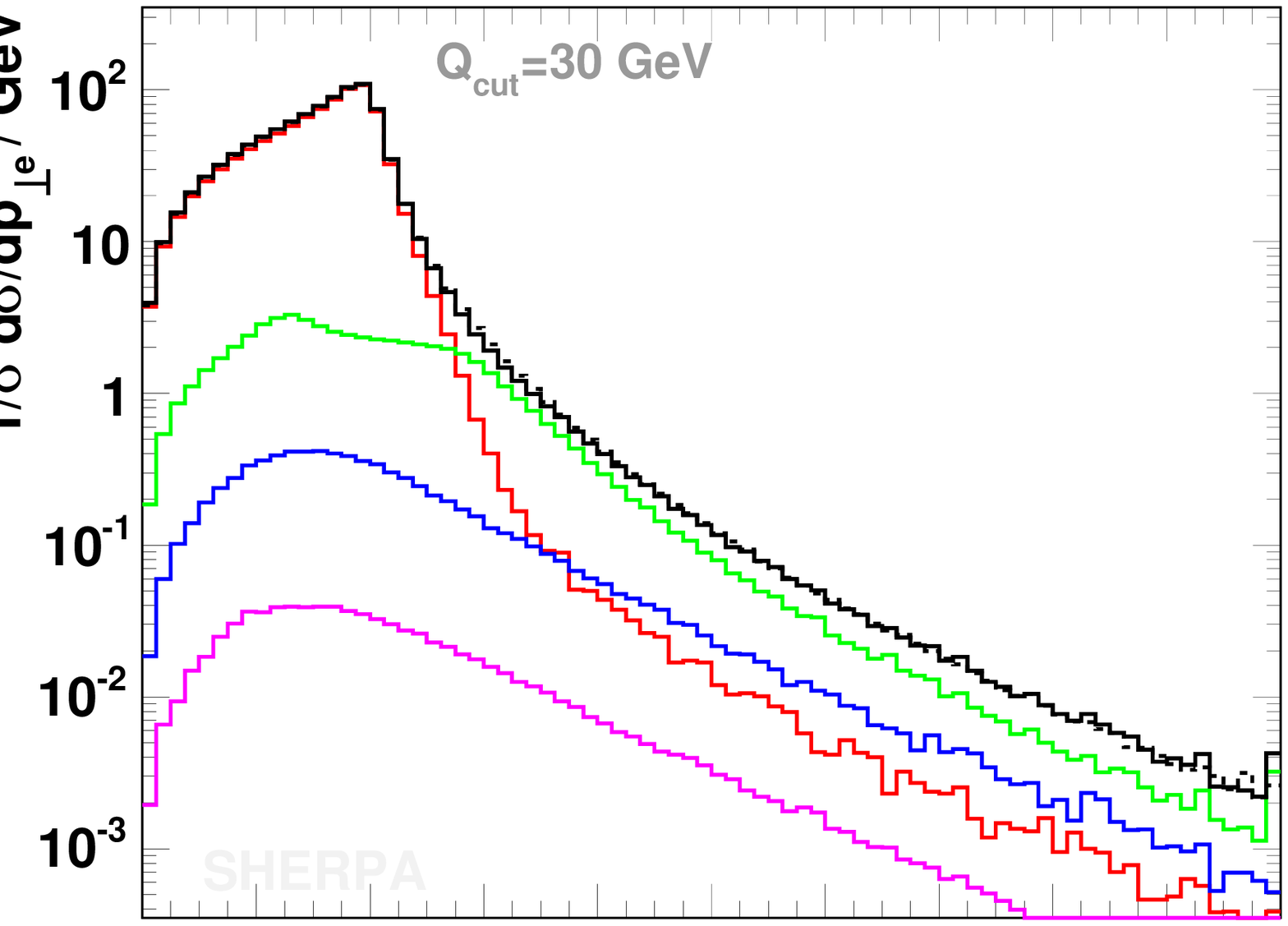}}
      \put(125,-10){\includegraphics[width=5.cm]
        {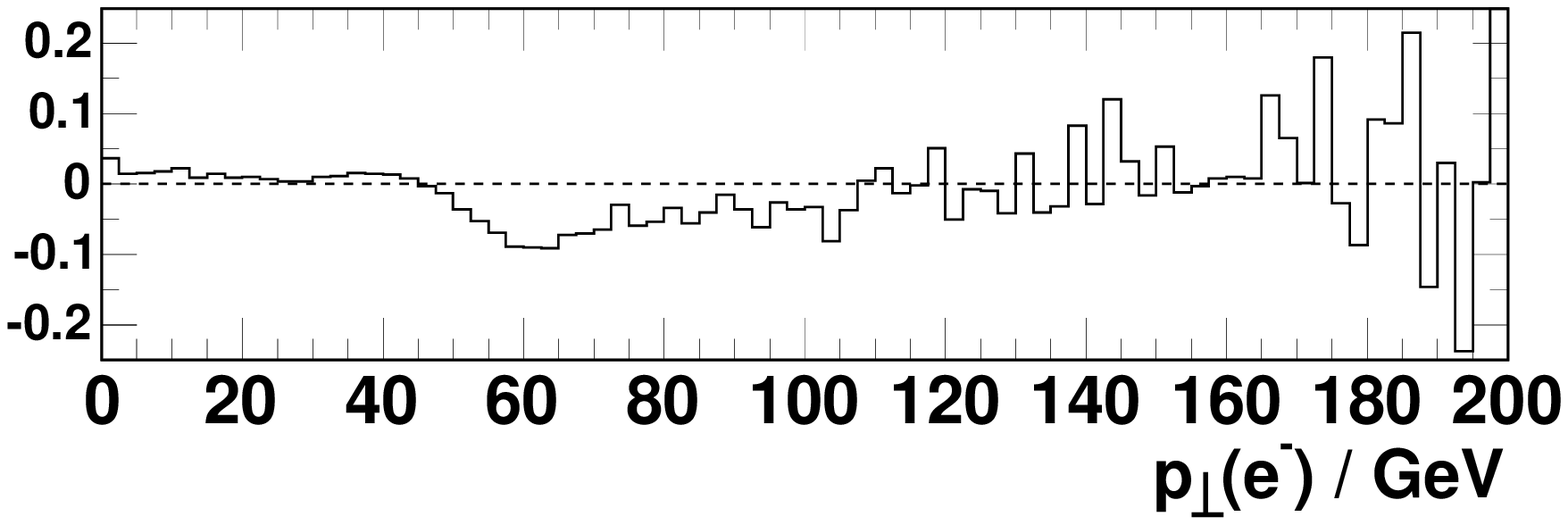}}
      \put(0,-10){\includegraphics[width=5.cm]
        {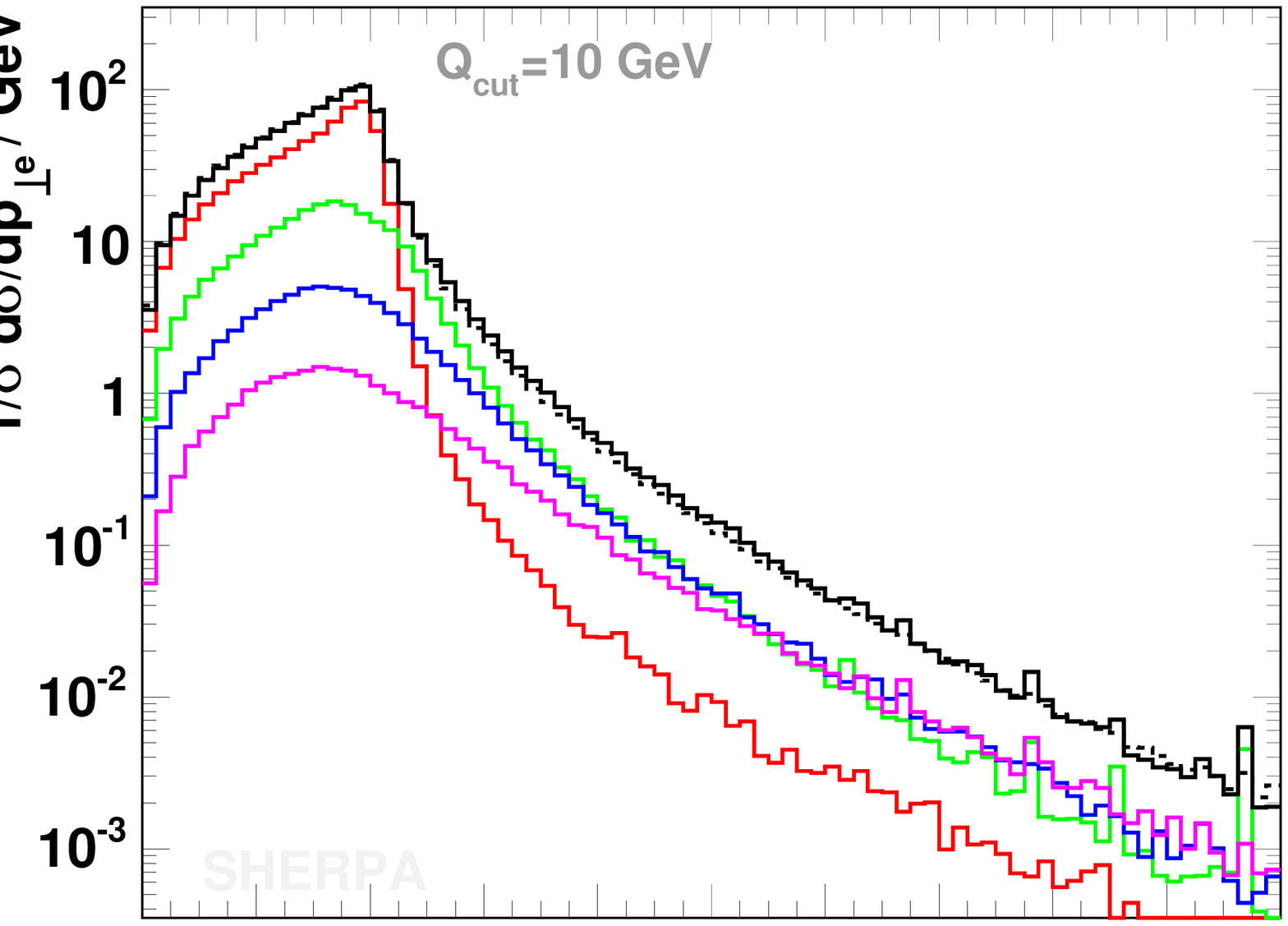}}
      \put(0,-10){\includegraphics[width=5.cm]
        {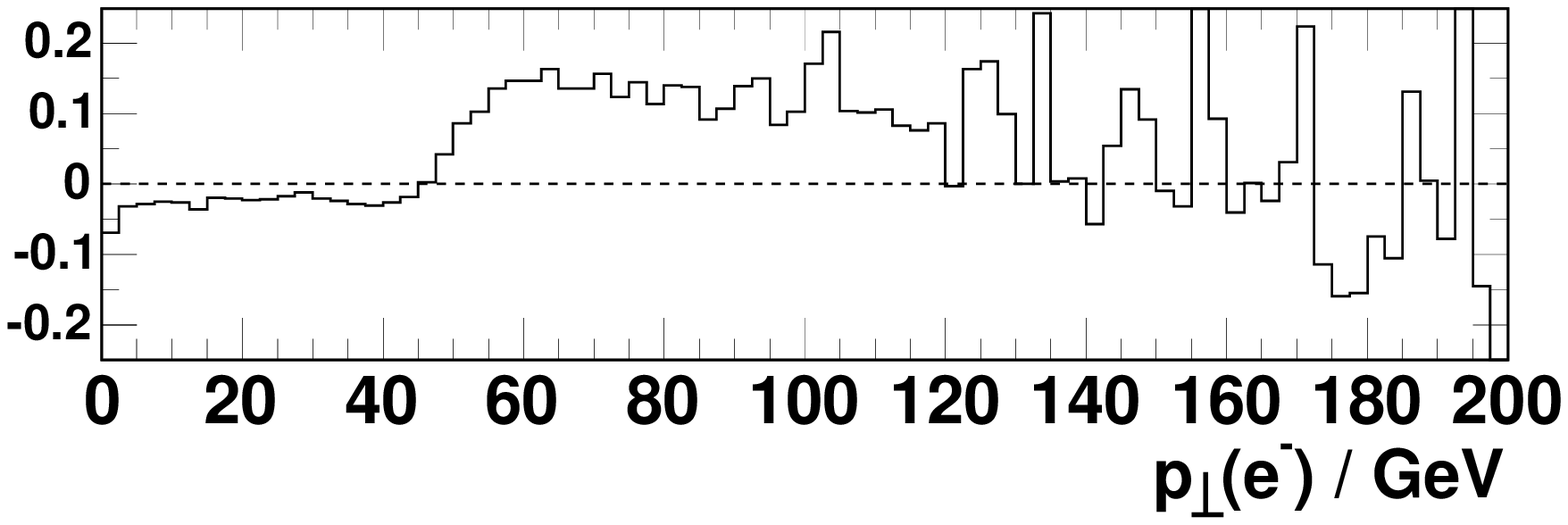}}
    \end{pspicture}
  \end{center}
  \caption{\label{ycut_pt_W_Tev}$p_\perp(W^-)$ and $p_\perp(e^-)$ for 
    $Q_{\rm cut}=10$ GeV, $30$ GeV and $50$ GeV in comparison with 
    $Q_{\rm cut}=20$ GeV. }
\end{figure*}

\begin{figure*}[h]
  \begin{center}
    \begin{pspicture}(400,150)
      \put(250,0){\includegraphics[width=5.cm]
        {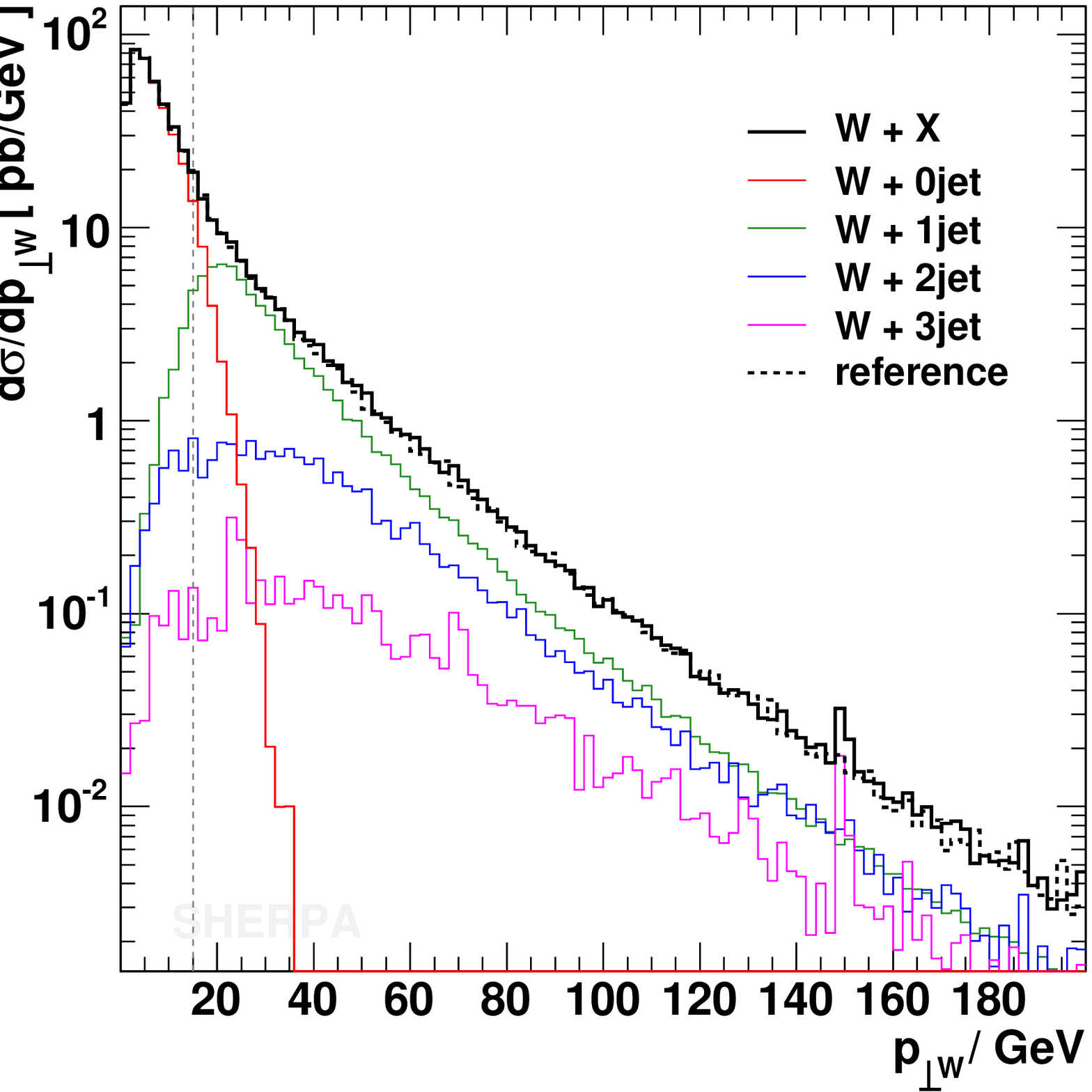}}
      \put(125,0){\includegraphics[width=5.cm]
        {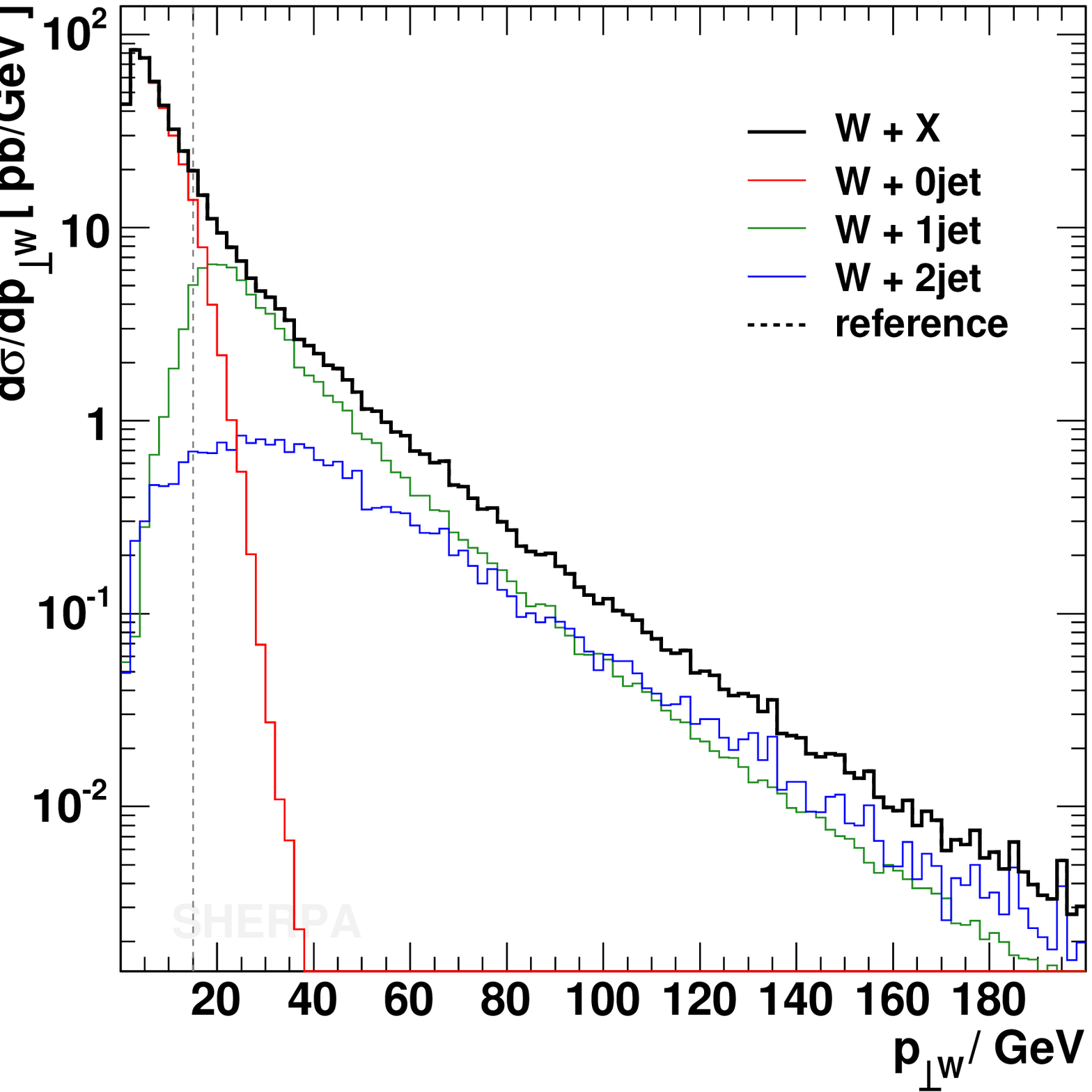}}
      \put(0,0){\includegraphics[width=5.cm]
        {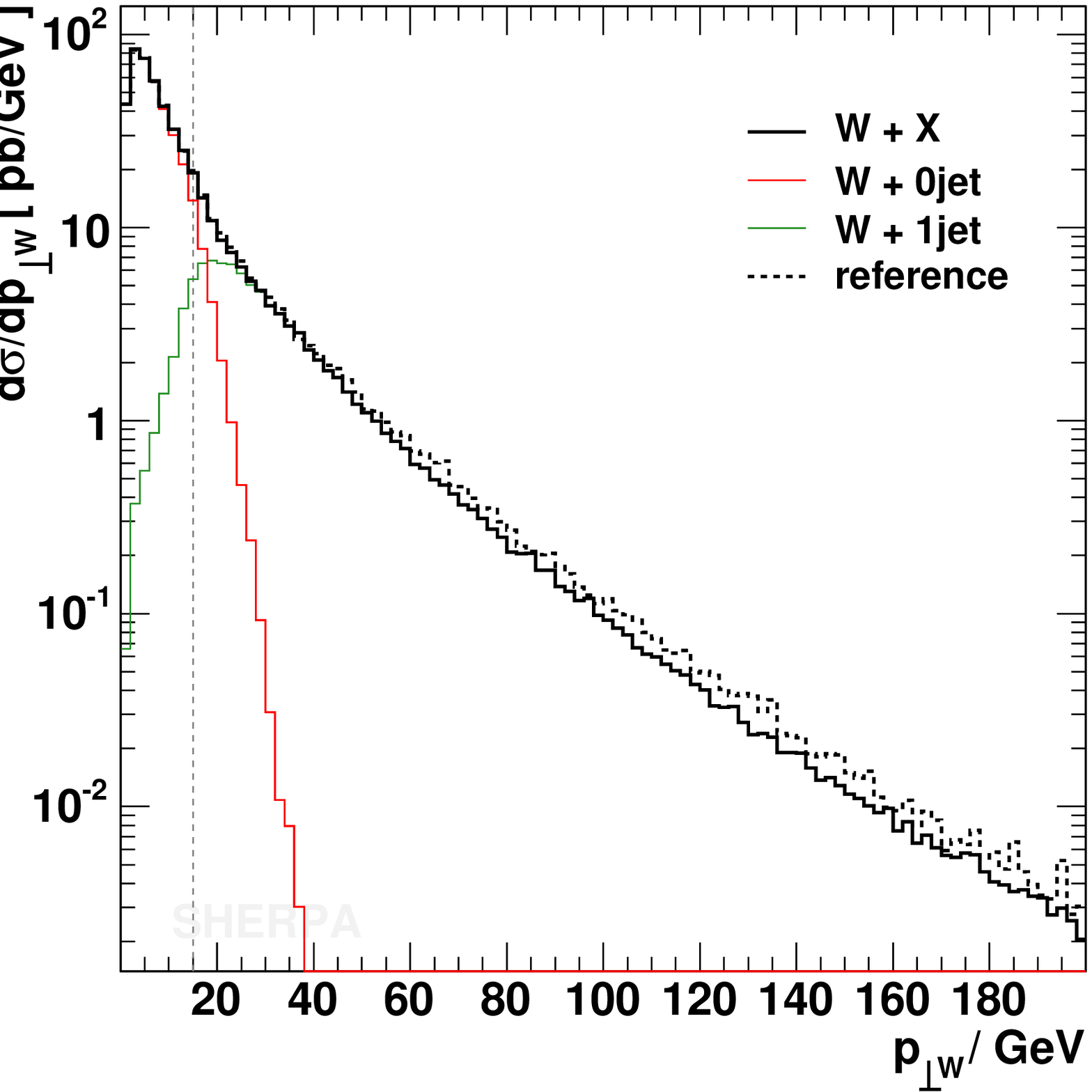}}
    \end{pspicture}
  \end{center}
  \caption{\label{nmax_pt_W_Tev}$p_\perp(W^-)$ for $Q_{\rm cut}=15$
    GeV and different maximal numbers of ME jets included. The dashed
    line corresponds to a maximal number of ME jets $n_{\rm max}=2$.} 
\end{figure*}

\noindent
In analogy to the event shapes above, the transverse momentum
distribution of the $W$ boson and of the electron produced in its
decay may be considered as inclusive observable. The dependence of
these observables from the jet resolution cut in the generation of the
samples and on the maximal number of jets covered by the matrix
elements is displayed in Figs.\ \ref{ycut_pt_W_Tev} and
\ref{nmax_pt_W_Tev}, respectively. In both cases it becomes apparent
that the dependences are negligible - results of different samples
produced under different conditions are in very good agreement with
each other. 

\begin{figure}
  \begin{center}
    \includegraphics[width=8.5cm]{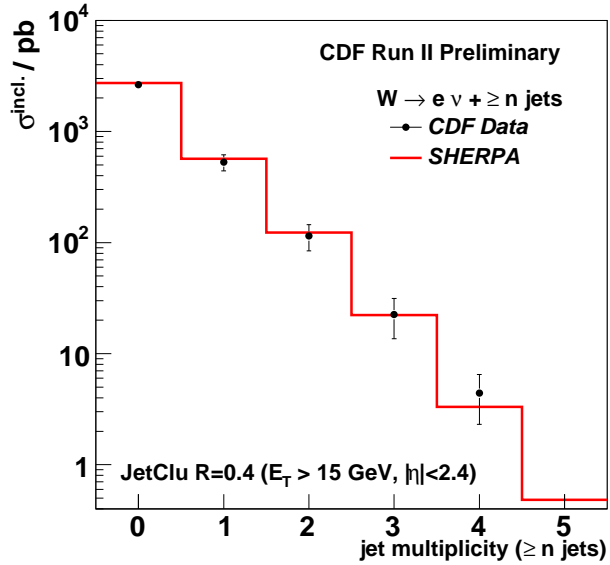}
    \caption{\label{fig:wjet_xsec1} Inclusive cross sections for the
      process $p \bar{p} \to W + n$ jets. The SHERPA prediction  is
      contrasted with the measurement by CDF \cite{Hesketh:2004qb}} 
  \end{center}
\end{figure}

\begin{figure}
  \begin{center}
    \includegraphics[width=8.5cm]{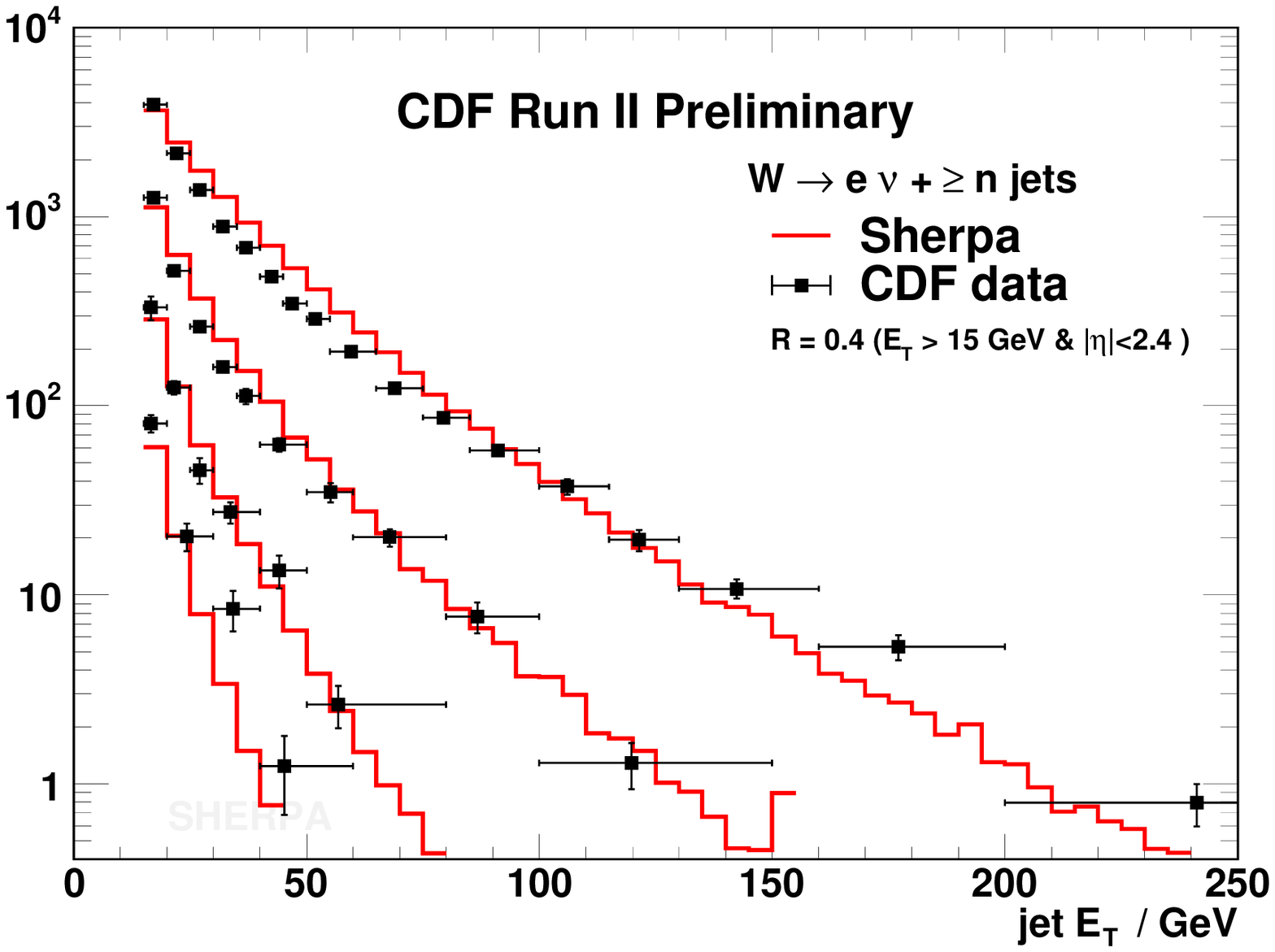}
    \caption{\label{fig:wjet_et1} Jet transverse energy
      distribution for the process $p \bar{p} \to W + n$ jets from a
      CDF \cite{Hesketh:2004qb} measurement. Shown are the highest
      $E_T$ jet distribution in inclusive $W+1jet$ events, the second
      highest $E_T$ jet distribution in inclusive $W+2jet$ events, the
      third highest $E_T$ jet distribution in inclusive $W+3jet$
      events, and the fourth highest $E_T$ jet distribution in
      inclusive $W+4jet$ events.
      The SHERPA  result includes matrix elements with up to 4 jets.} 
    \end{center}
  \end{figure}

\noindent
Finally, measured total cross sections for different jet
multiplicities in $W+{\rm jet}$ \cite{Hesketh:2004qb} are compared
with those obtained from SHERPA after reweighting the matrix elements
with up to 4 jets with the Sudakov weights and after applying a
constant $K$-factor of $1.44$ to all samples, that has been calculated
to match SHERPA with a NNLO prediction \cite{Hamberg:1990np,
Harlander:2002wh}. Taking into account the errors, the results are
in great agreement with each other, cf.\ Fig.\
\ref{fig:wjet_xsec1}. Correspondingly, the 
$p_\perp$ spectra of the jets are depicted in Fig.\ \ref{fig:wjet_et1}. There, 
the measurement of transverse energy distributions of jets for
different multiplicities \cite{Hesketh:2004qb} are compared with
results from SHERPA. Again, the results agree very well after applying
a global $K$-factor on the latter. In both cases, jets were defined
through a cone algorithm with cone size of $R=0.4$ and a transverse
energy of the jets of at least $E_T=15$ GeV. 

\subsection{Results for $p\bar{p}\to \mbox{jets}$ at Tevatron, Run I}

\begin{figure*}[h]
  \begin{center}
    \begin{pspicture}(300,250)
      \put(0,0){\includegraphics[width=10cm]
        {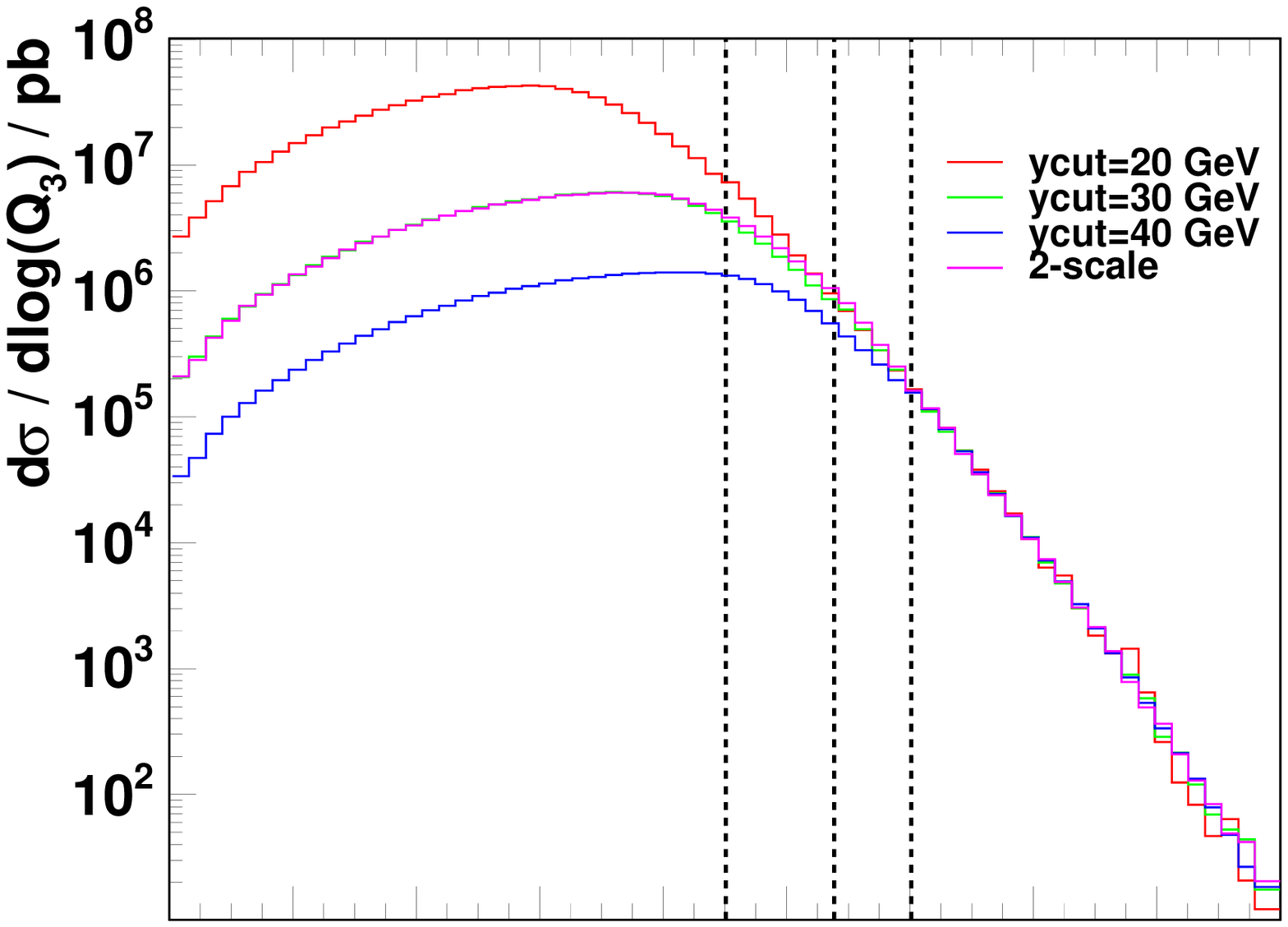}}
      \put(0,0){\includegraphics[width=10cm]
        {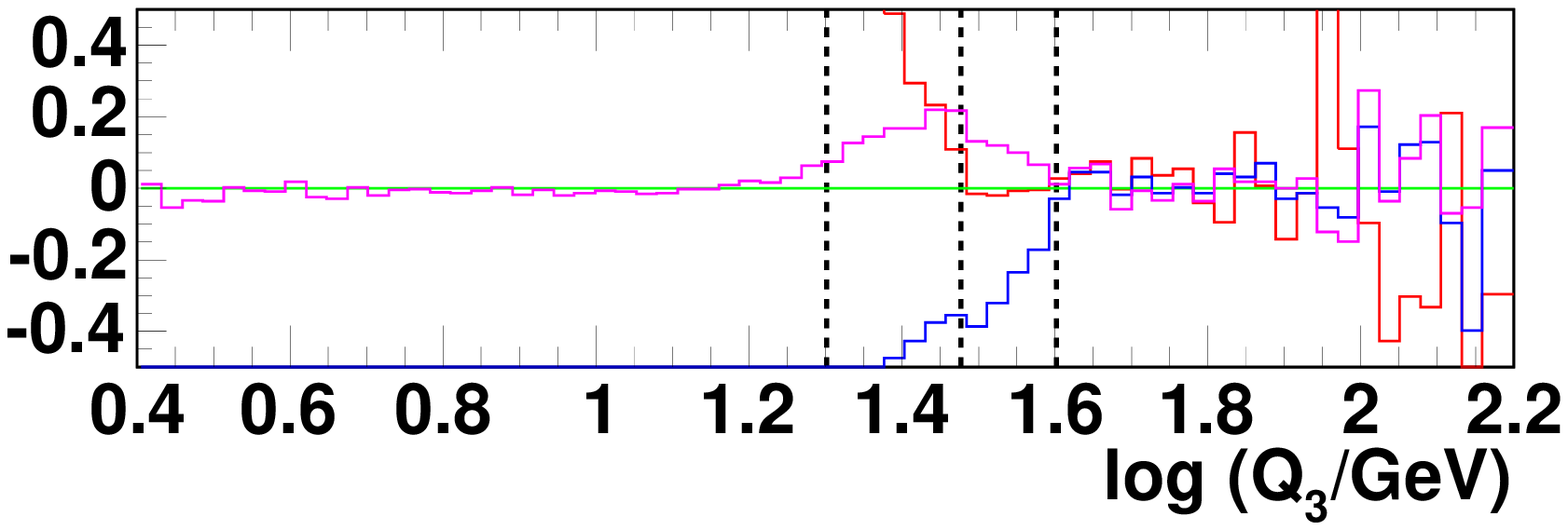}}
    \end{pspicture}
  \end{center}
  \caption{\label{fig:diff_jj_tev1_cut} Differential $3\to 2$ jet rate
    at the Tevatron, Run I. Different samples with different values of
    $Q_{\rm cut}$ are displayed in different colours, the respective
    $Q_{\rm cut}$ values are indicated with dashed vertical lines. 
    Clearly, above 40 GeV, all samples coincide, then successively,
    different samples die off. Apparently the sample with mixed cuts
    ($Q_{\rm cut}^{(2)}=30$~GeV and $Q_{\rm cut}^{(3)}=40$~GeV),
    depicted in purple, agress very well with the sample produced with
    the lower of the two cuts.}
\end{figure*}

\noindent
Before investigating in greater detail the consistency of the merging
prescription for jet production at hadron colliders, in particular at
the Tevatron, Run I, consider Fig.\ \ref{fig:diff_jj_tev1_cut}. There,
the differential $3\to 2$ jet rates for samples with $n_{\rm max}=3$ 
for $Q_{\rm cut}=20$, $30$, and $40$ GeV are compared with a the
result for a sample, where two different jet resolution cuts have been
applied for the different multiplicities, namely $Q_{\rm cut}^{(2)}=30$ 
GeV and $Q_{\rm cut}^{(2)}=40$ GeV. Obviously, for $Q\ge 40$ GeV the
four results are in fair agreement with each other, as expected. Below
$40$ GeV, the sample generated with $Q_{\rm cut}=40$ GeV starts to
undershoot the other three curves significantly, as expected. In
principle, there should be no contribution left at all, since there
are no matrix elements for any jet configurations populating this
regime. However, due to the parton shower, some of the jets produced
at higher $p_\perp$ values spread out, leading to some non-negligible
fraction of events migrating into that region. The same pattern
repeats itself at $Q$-values below $30$ GeV. This region is not filled
by the $Q_{\rm cut}=30$ GeV and the mixed sample any longer. This
implies that in order to describe jet observables at jet resolutions
above, say, $30$ GeV, a $Q_{\rm cut}\le 30$ GeV should be applied. Due
to the steep descend of cross sections this may not be very efficient,
rendering a multi-scale treatment the method of choice. 
\begin{figure*}[h]
  \begin{center}
    \begin{pspicture}(440,150)
      \put(290,0){\includegraphics[width=5.5cm]
        {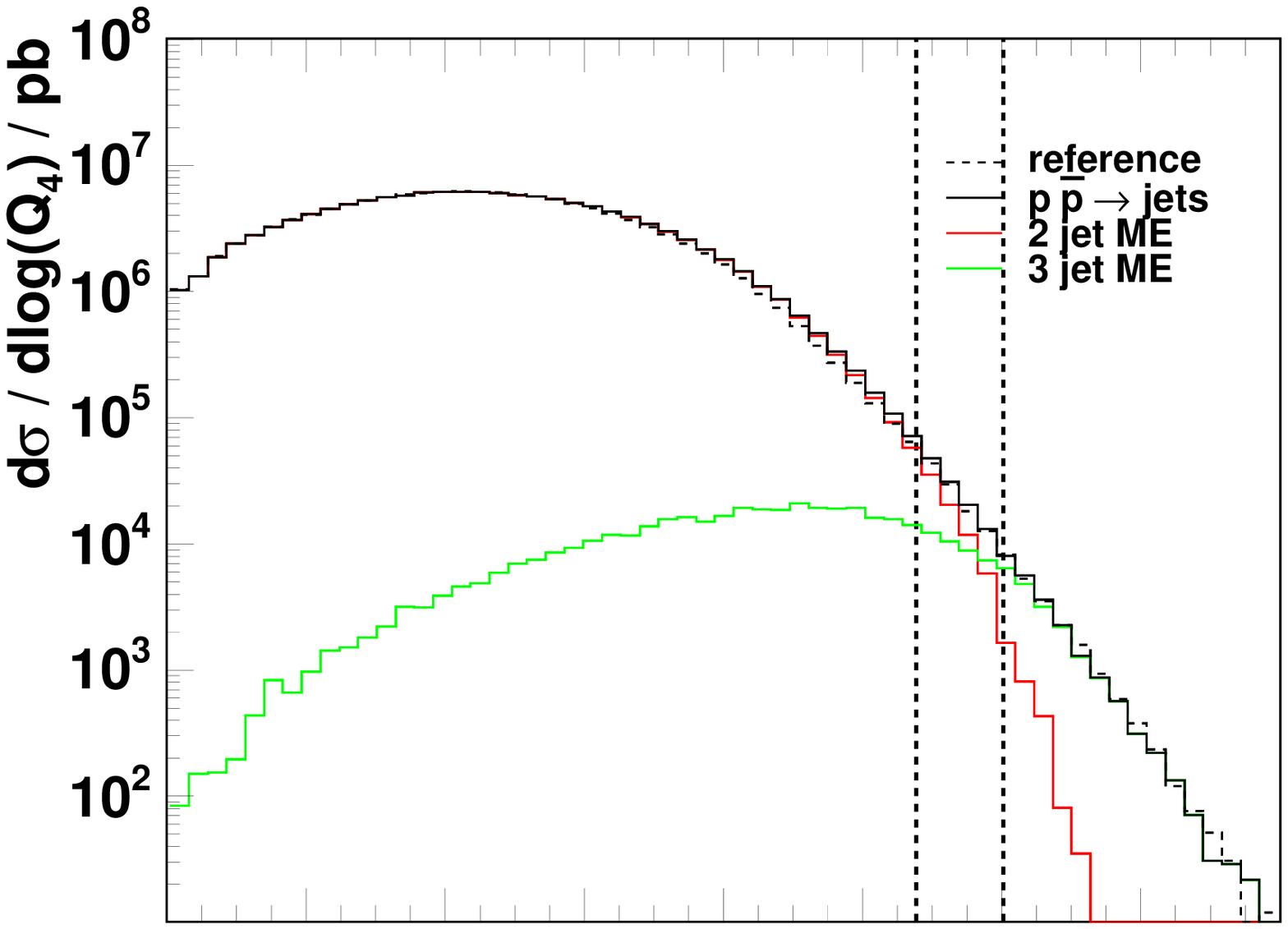}}
      \put(290,0){\includegraphics[width=5.5cm]
        {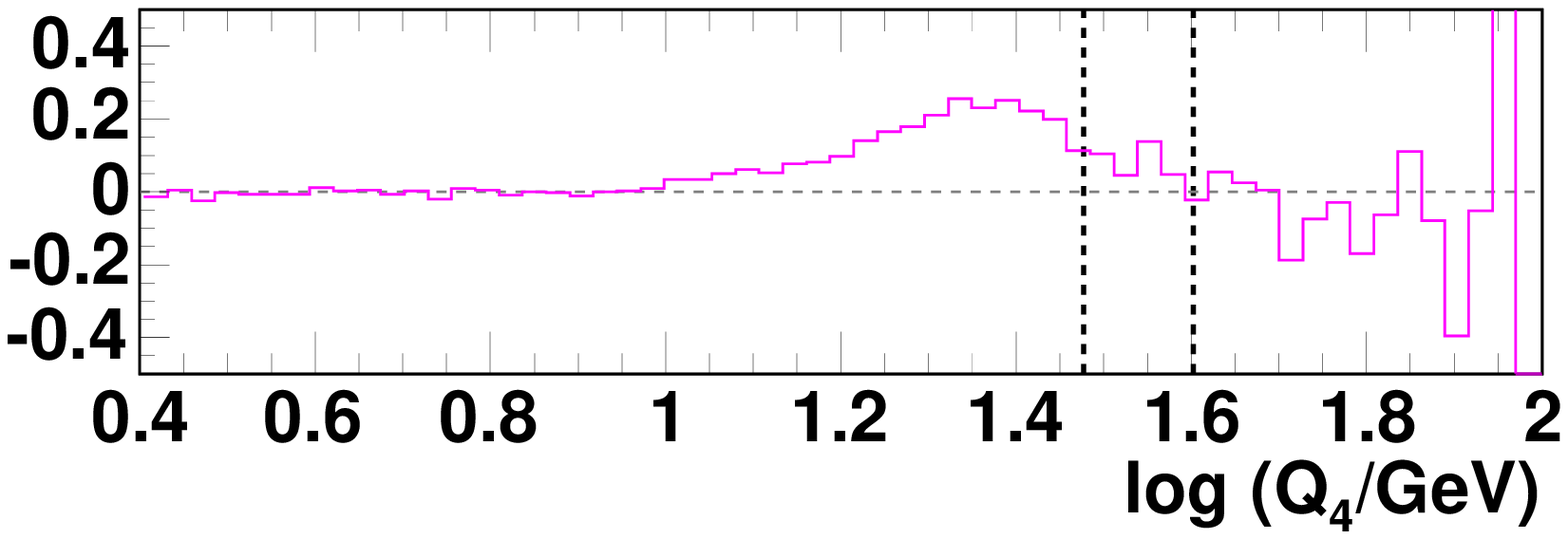}}
      \put(145,0){\includegraphics[width=5.5cm]
        {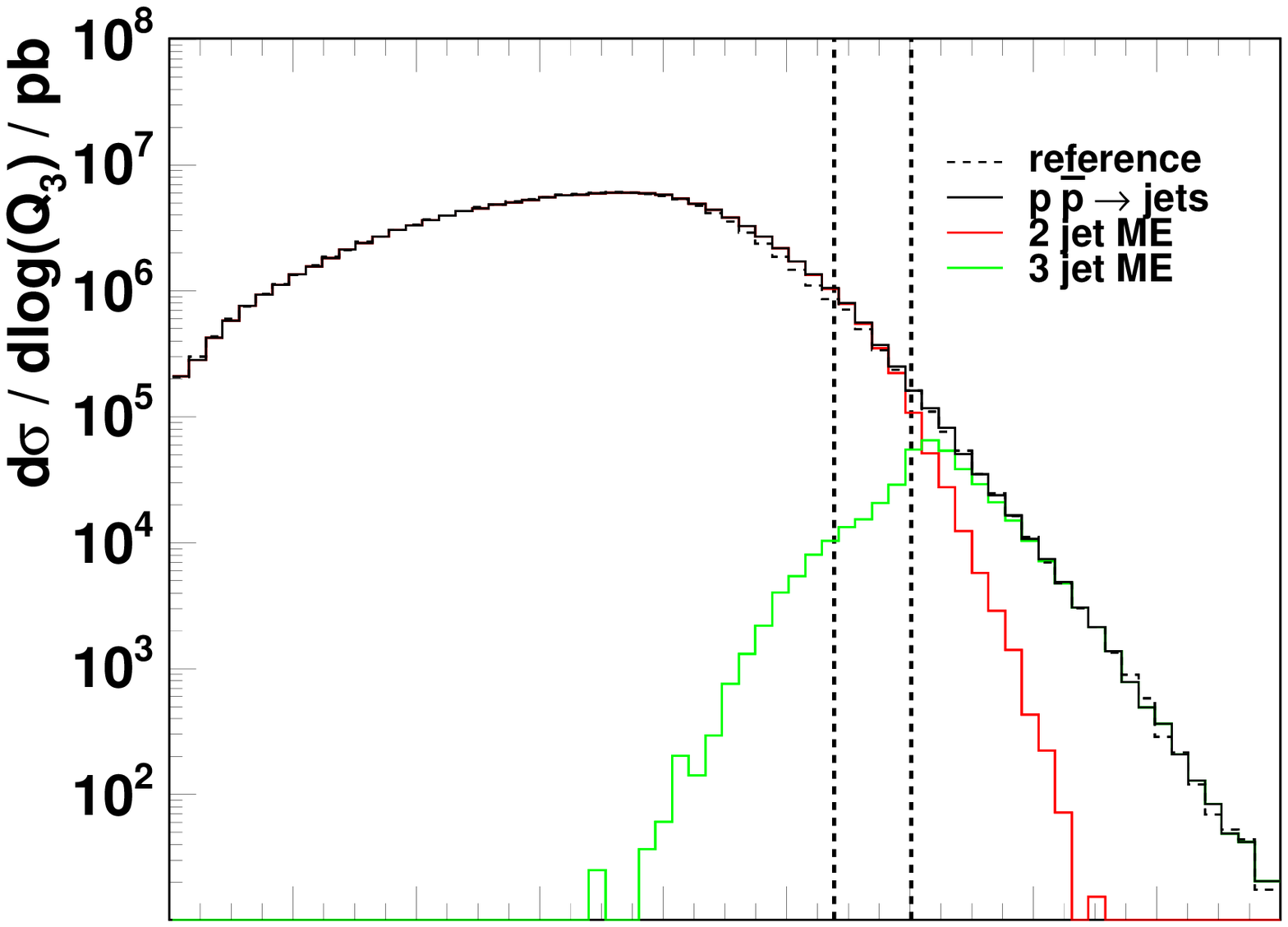}}
      \put(145,0){\includegraphics[width=5.5cm]
        {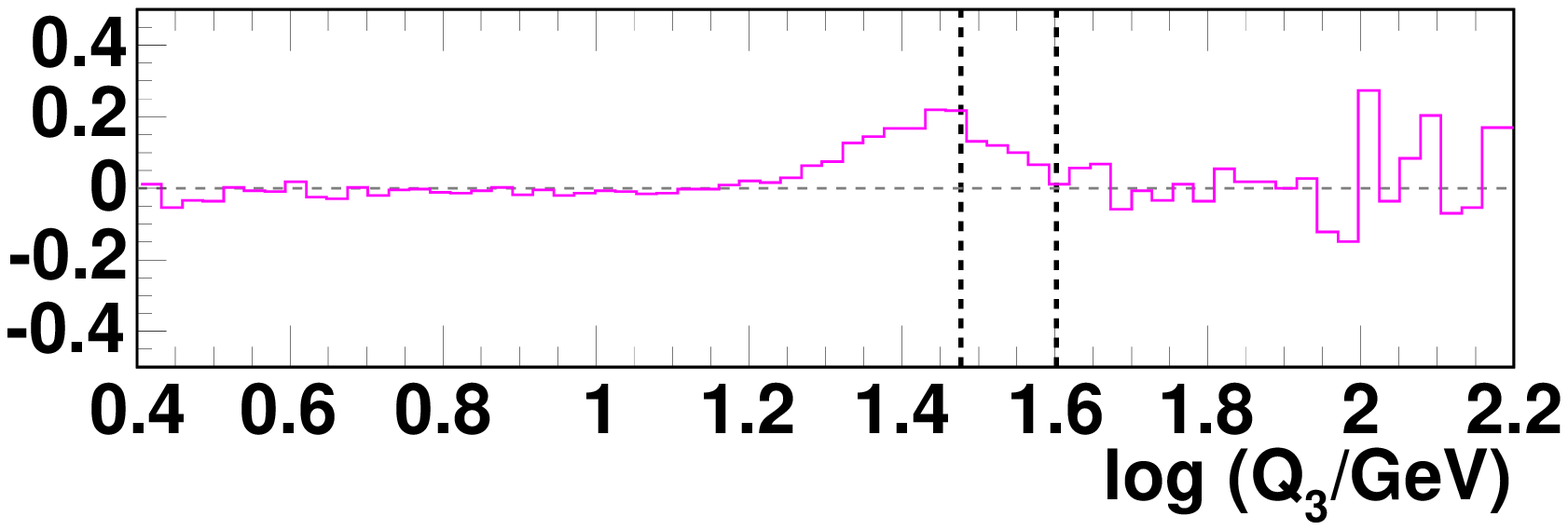}}
      \put(0,0){\includegraphics[width=5.5cm]
        {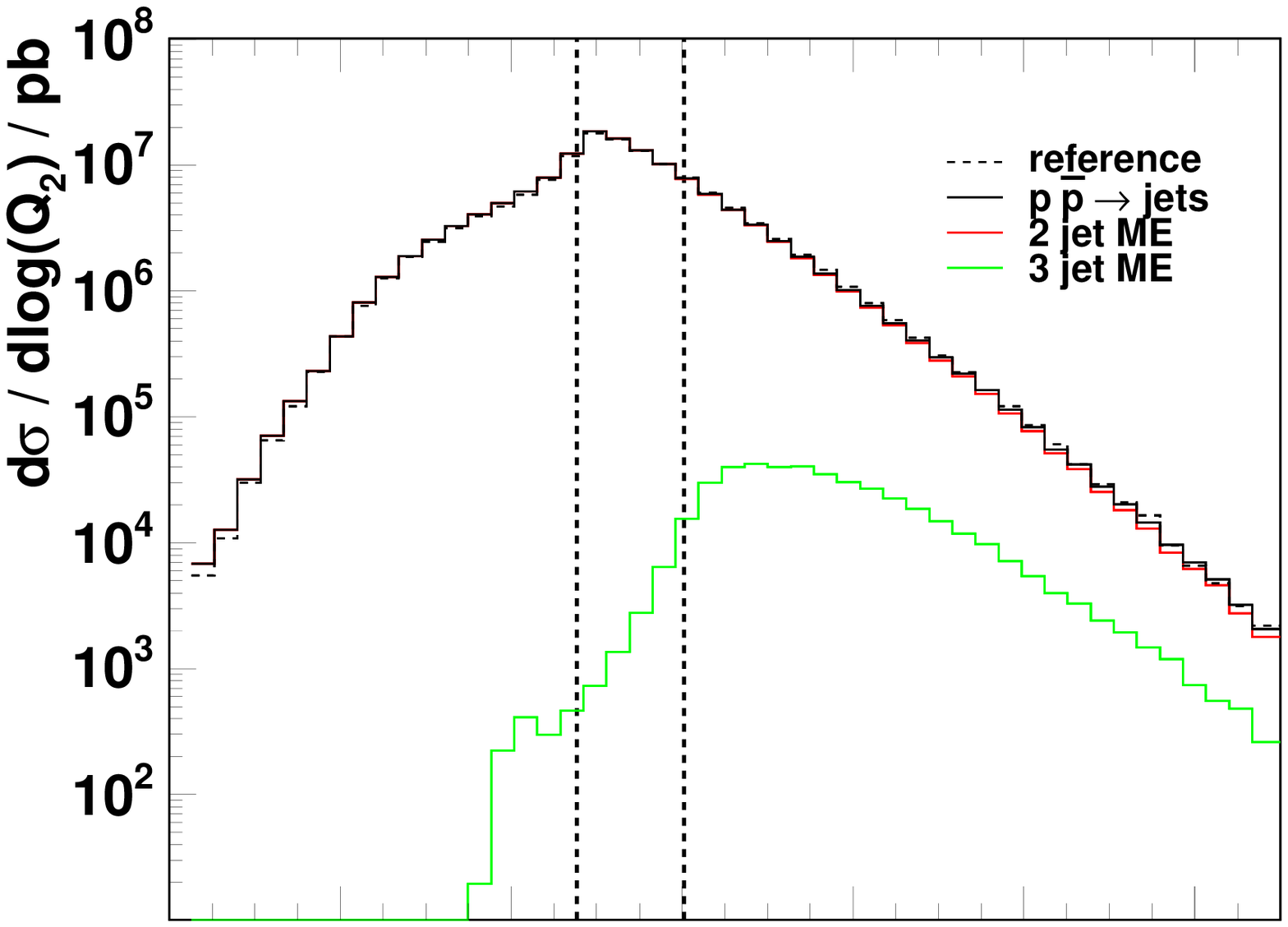}}
      \put(0,0){\includegraphics[width=5.5cm]
        {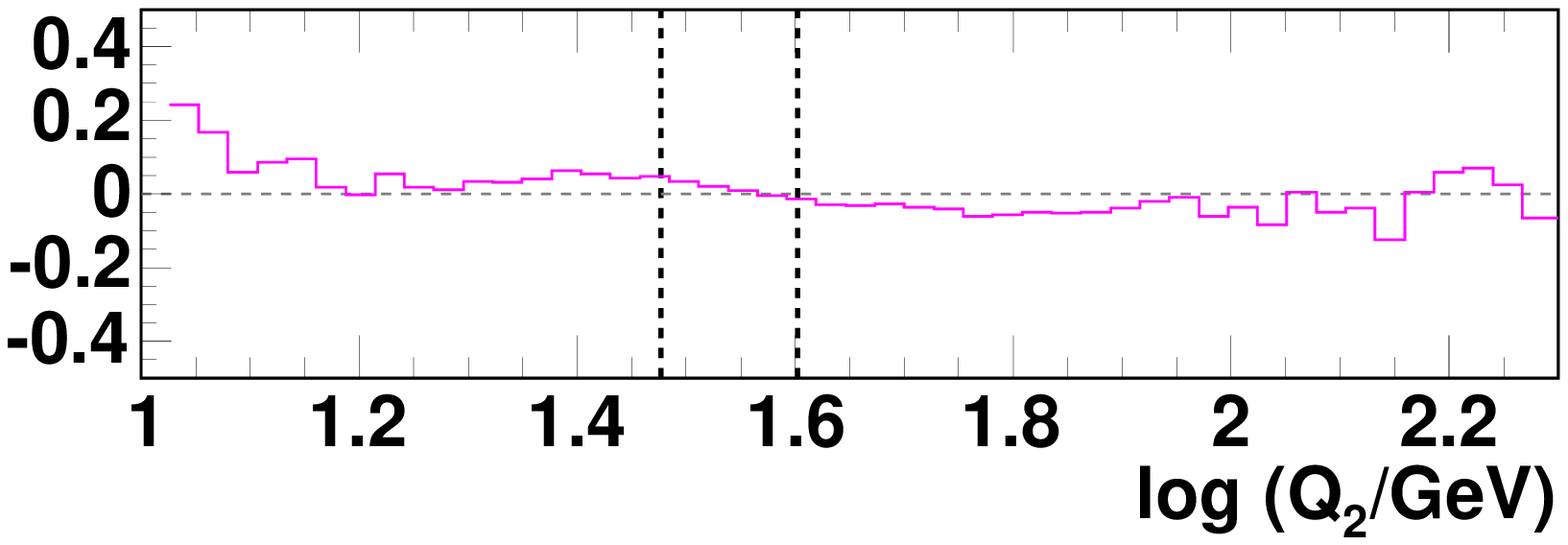}}
    \end{pspicture}
  \end{center}
  \caption{\label{fig:diff_jj_tev1_consistency}Differential $2\to 1$, 
    $3\to 2$, $4\to 3$ jet rate at the Tevatron, Run I. The mixed
    sample with two cuts ($Q_{\rm cut}^{(2)}=30$~GeV and 
    $Q_{\rm cut}^{(3)}=40$~GeV, as above), depicted with the solid
    black line is compared with a reference sample with one, common
    cut only ($Q_{\rm cut}^{(3)}=30$~GeV), displayed with the dashed
    black line. Deviations are maximally of the order of 20\%,
    indicating the success of the multi-cut treatment. Note also that
    in the $3\to 2$ jet rate, around 40 GeV the effect of merging the 
    2 with the three jet configuration becomes visible.    
  }
\end{figure*}

\begin{figure*}[h]
  \begin{center}
    \begin{pspicture}(440,150)
      \put(290,0){\includegraphics[width=5.5cm]
        {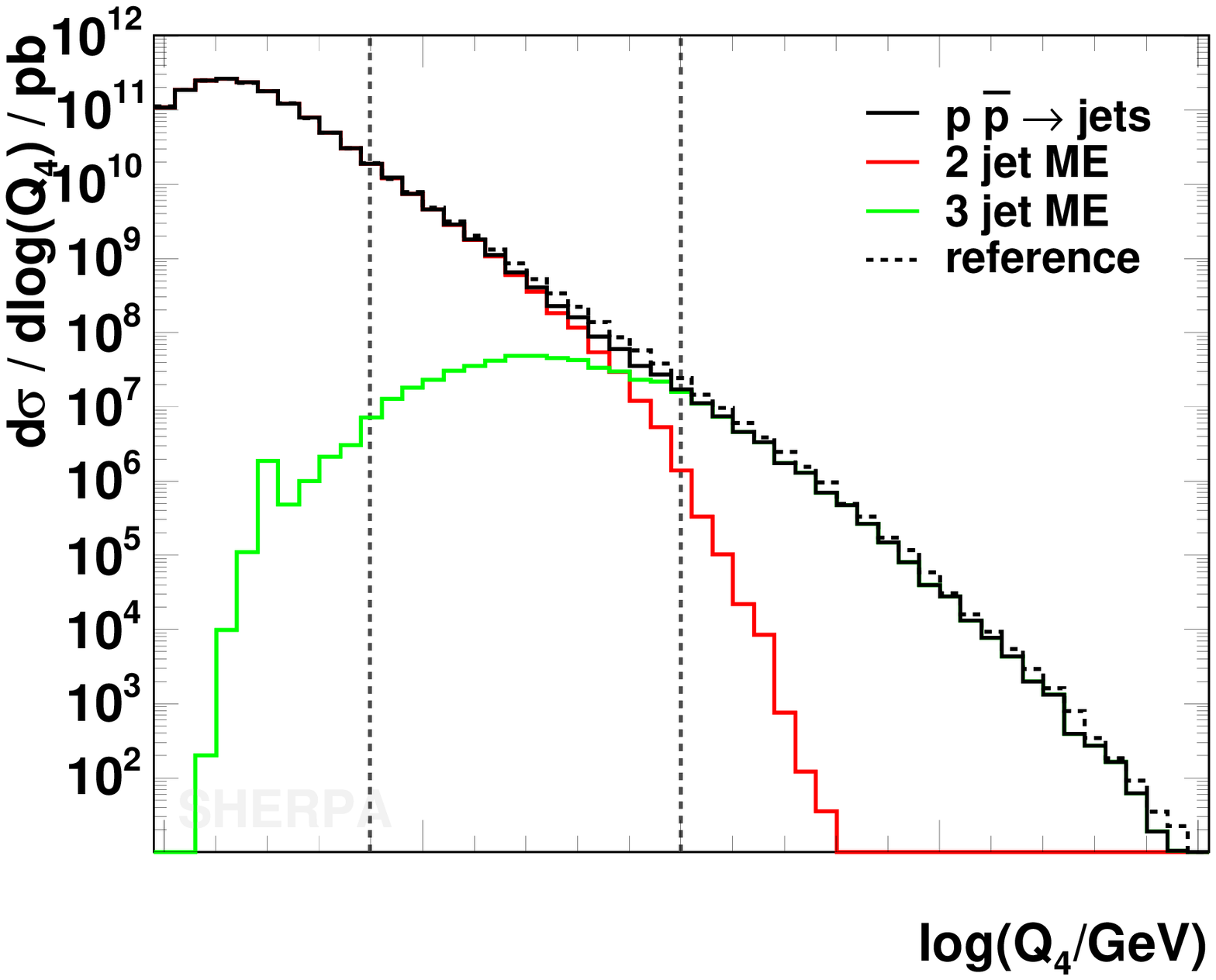}}
      \put(290,0){\includegraphics[width=5.5cm]
        {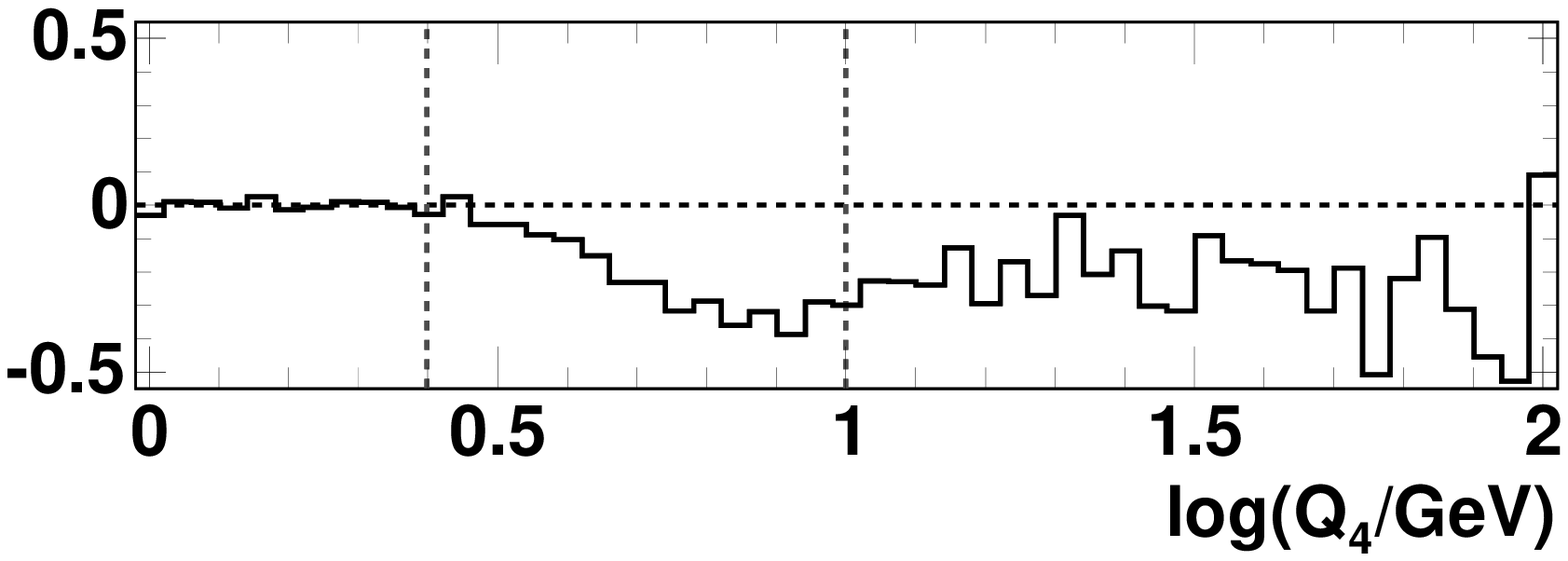}}
      \put(145,0){\includegraphics[width=5.5cm]
        {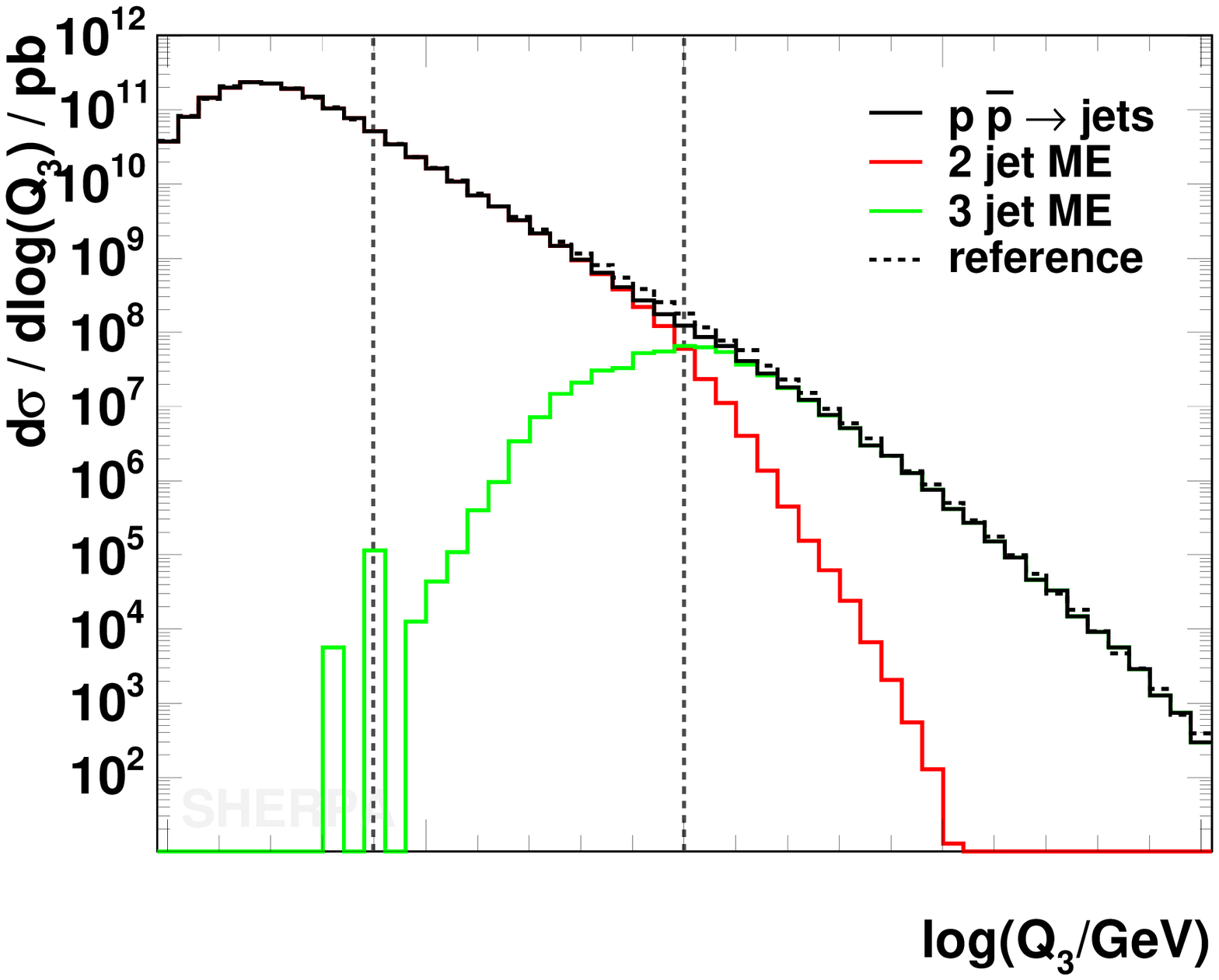}}
      \put(145,0){\includegraphics[width=5.5cm]
        {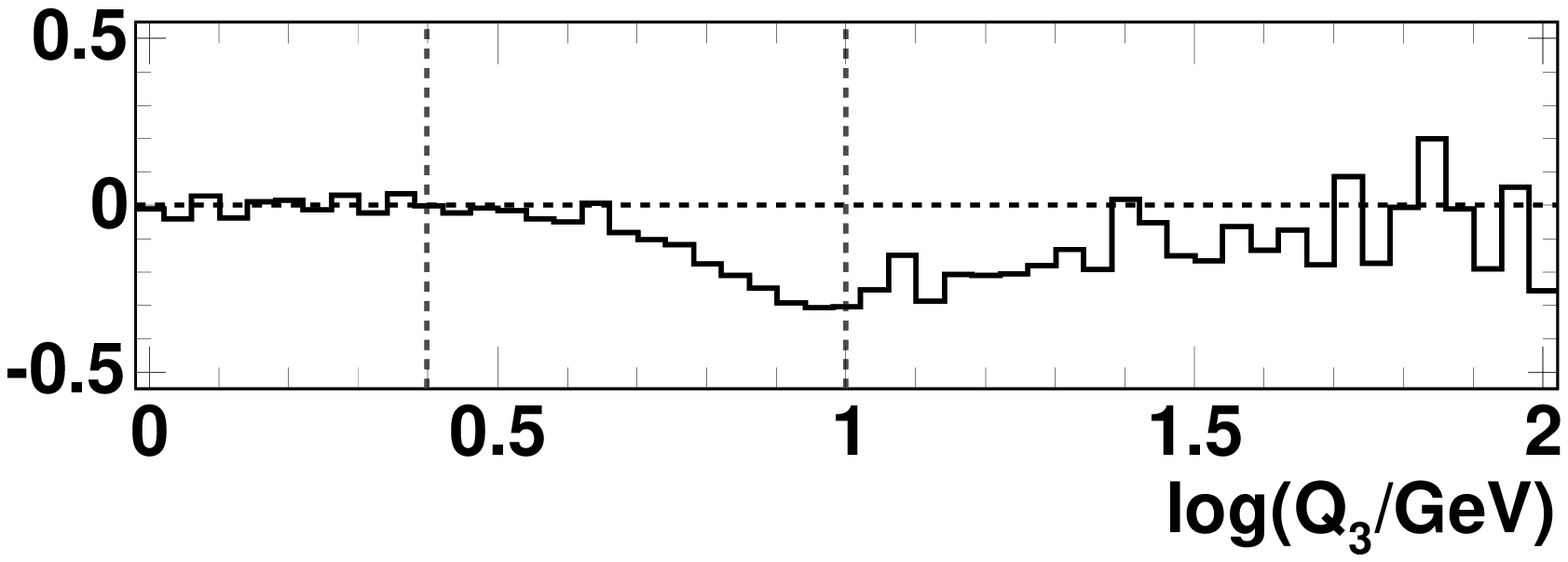}}
      \put(0,0){\includegraphics[width=5.5cm]
        {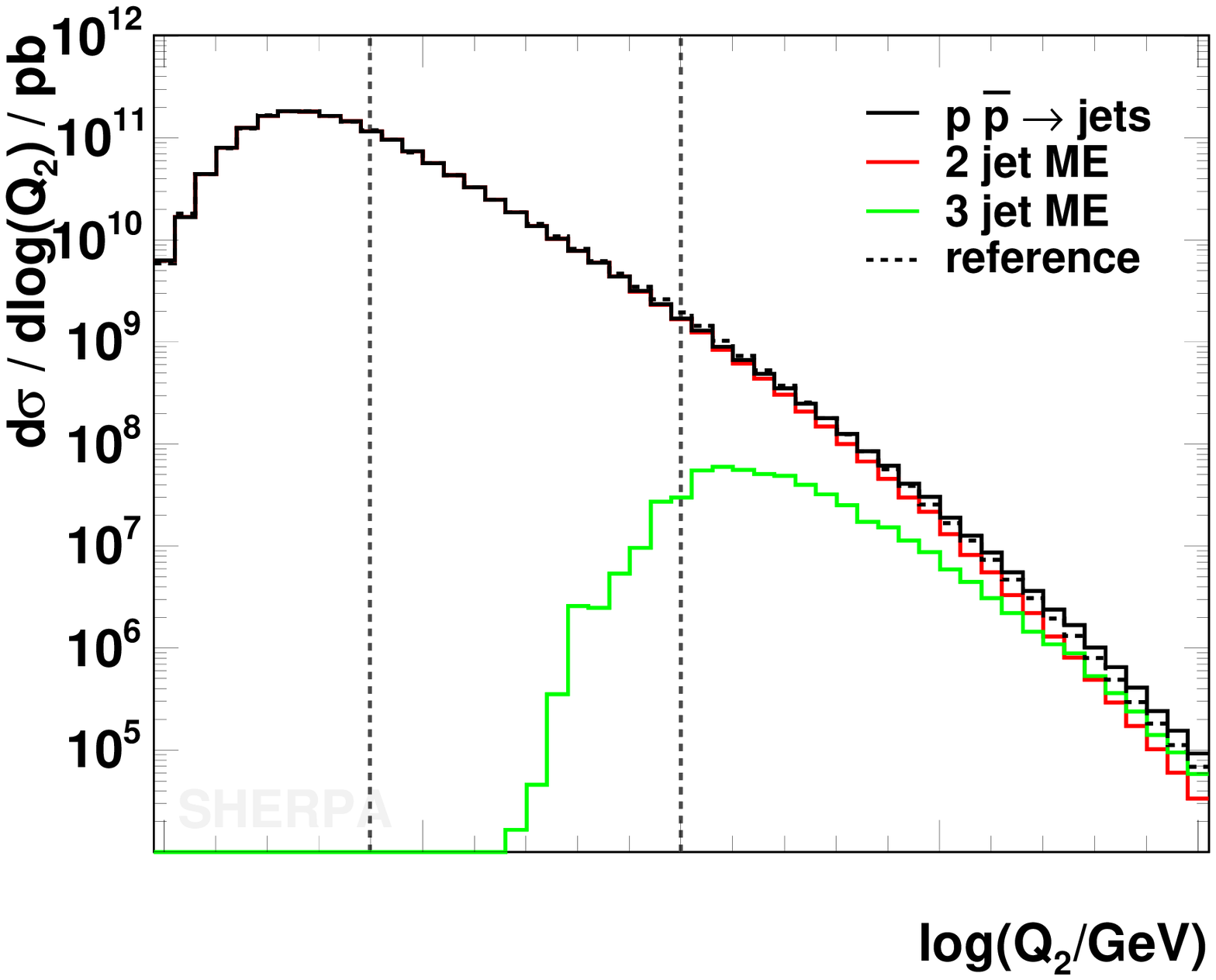}}
      \put(0,0){\includegraphics[width=5.5cm]
        {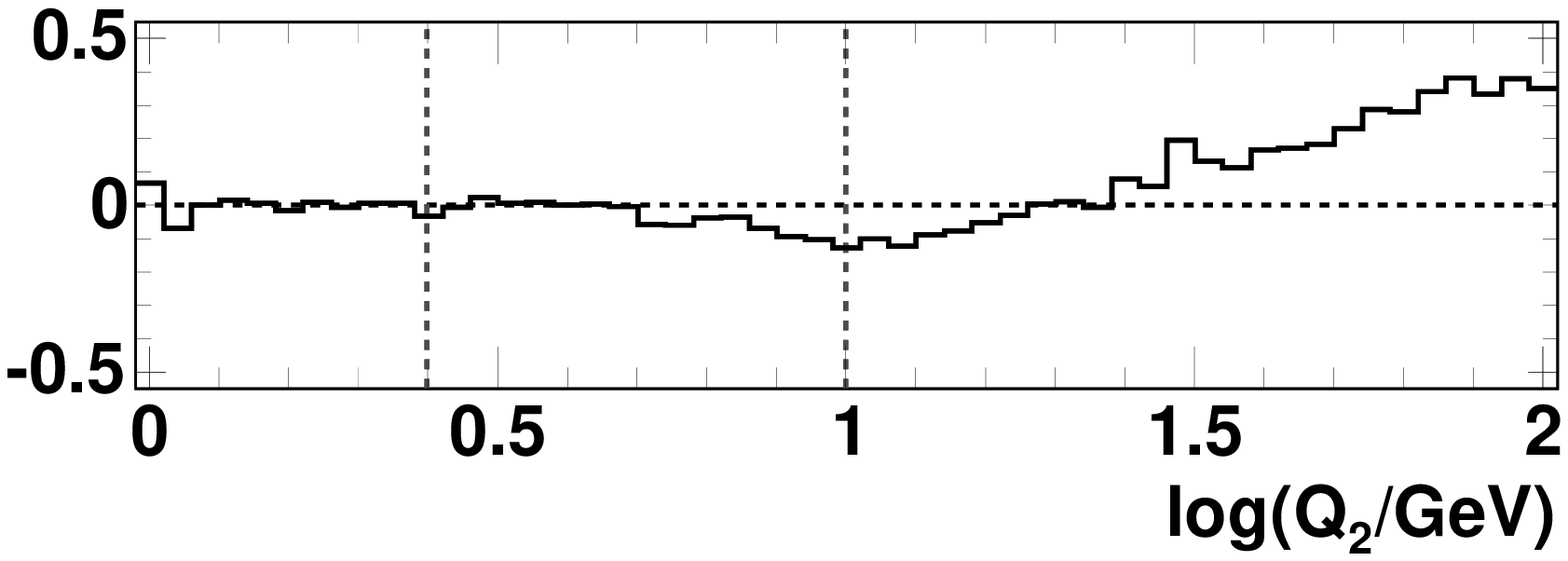}}
    \end{pspicture}
  \end{center}
  \caption{\label{fig:diff_jj_tev1} Differential jet rates at the
    Tevatron, Run I. From left to right, the rates for $2\to 1$, $3\to
    2$, and for $4\to 3$ transitions at the hadron level are
    exhibited; the samples are produced with $n_{\rm max}=3$ and  
    $Q_{\rm cut}^{(2)}=2.5$~GeV  \& $Q_{\rm cut}^{(3)}=10$~GeV.
    A reference curve is shown in black-dashed lines, contributions
    from different multiplicities are displayed in different colours.} 
\end{figure*}

\begin{figure*}[h]
  \begin{center}
    \begin{pspicture}(440,150)
      \put(290,0){\includegraphics[width=5.5cm]
        {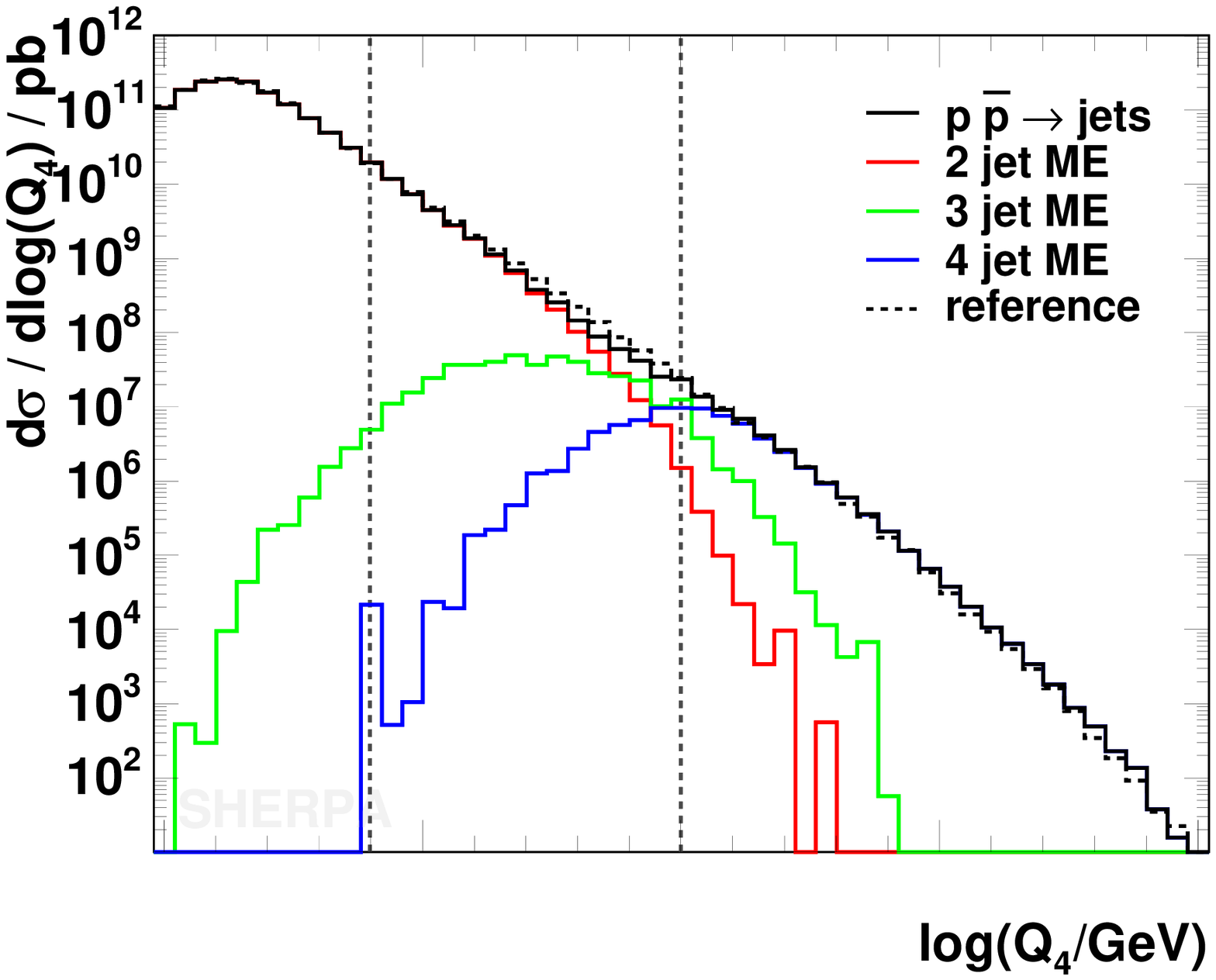}}
      \put(290,-8){\includegraphics[width=5.5cm]
        {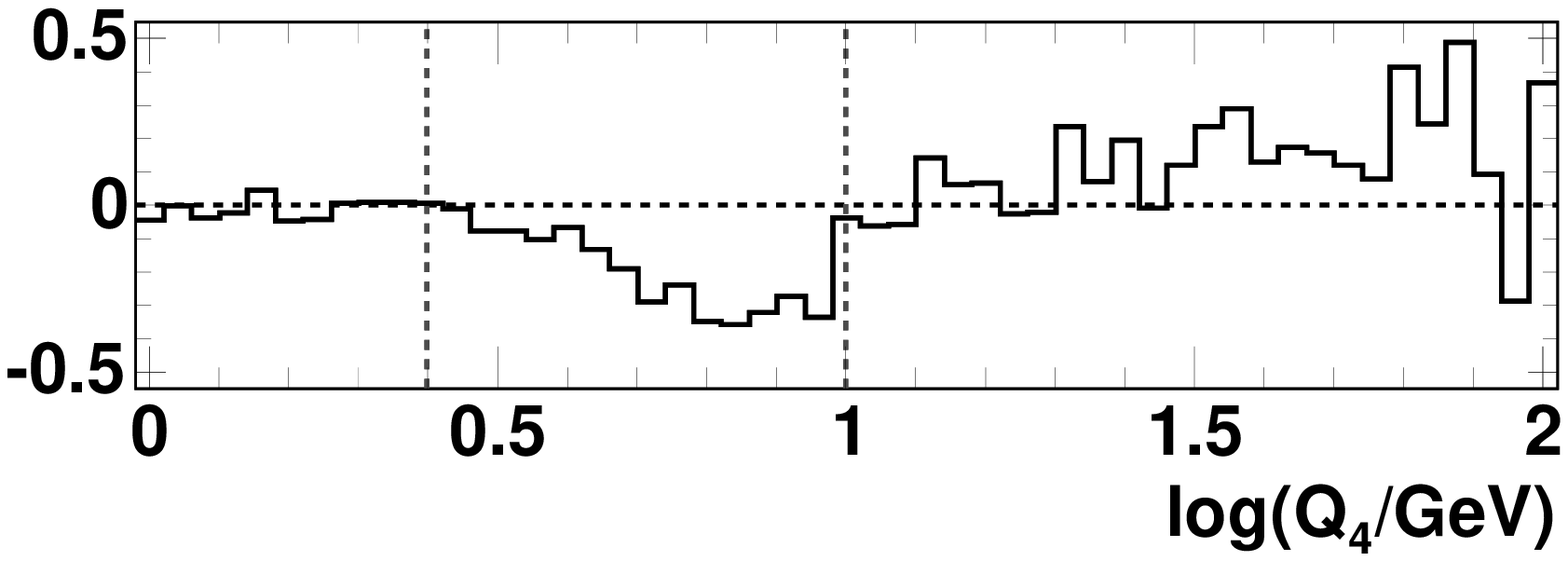}}
      \put(145,0){\includegraphics[width=5.5cm]
        {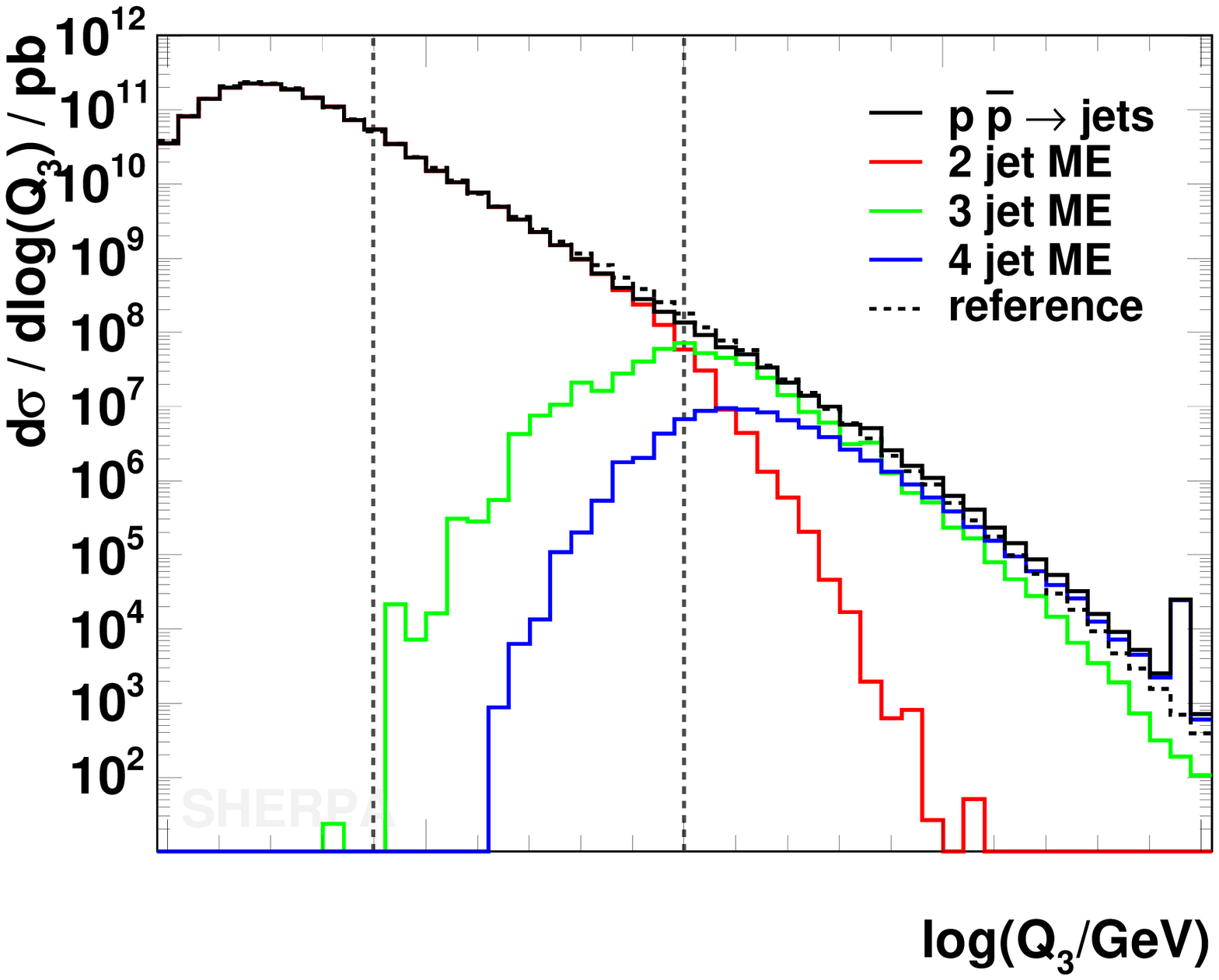}}
      \put(145,-8){\includegraphics[width=5.5cm]
        {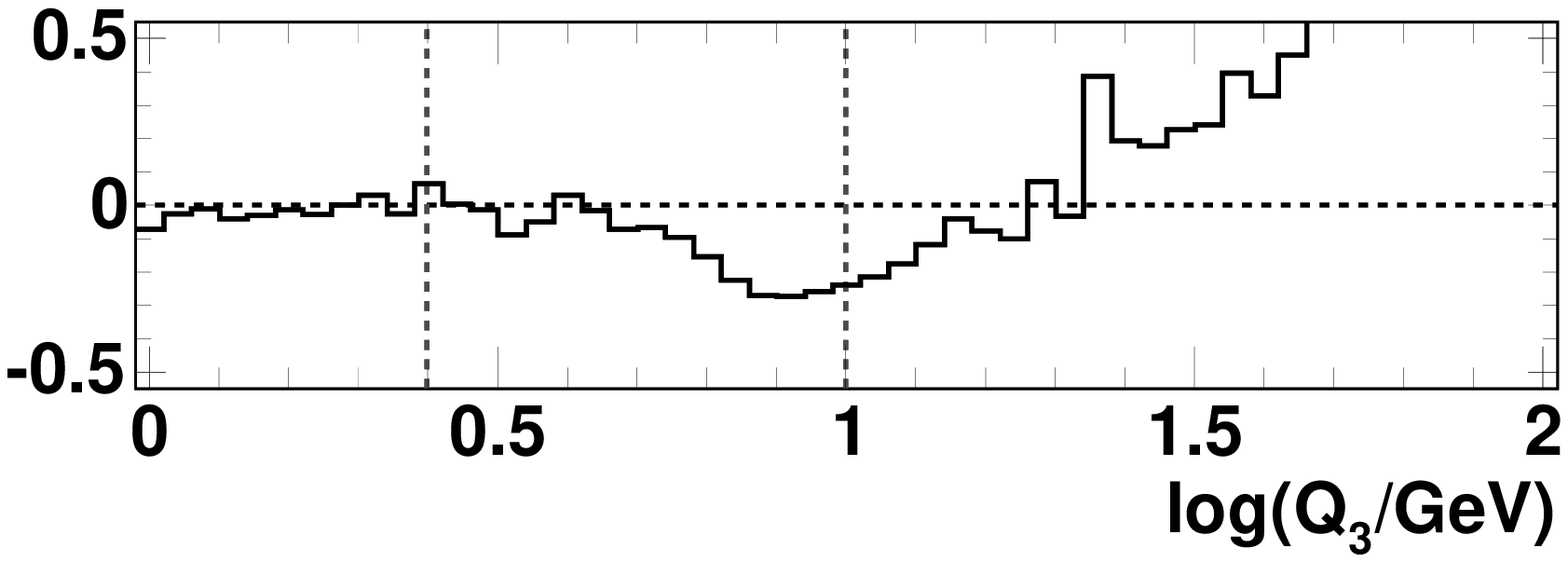}}
      \put(0,0){\includegraphics[width=5.5cm]
        {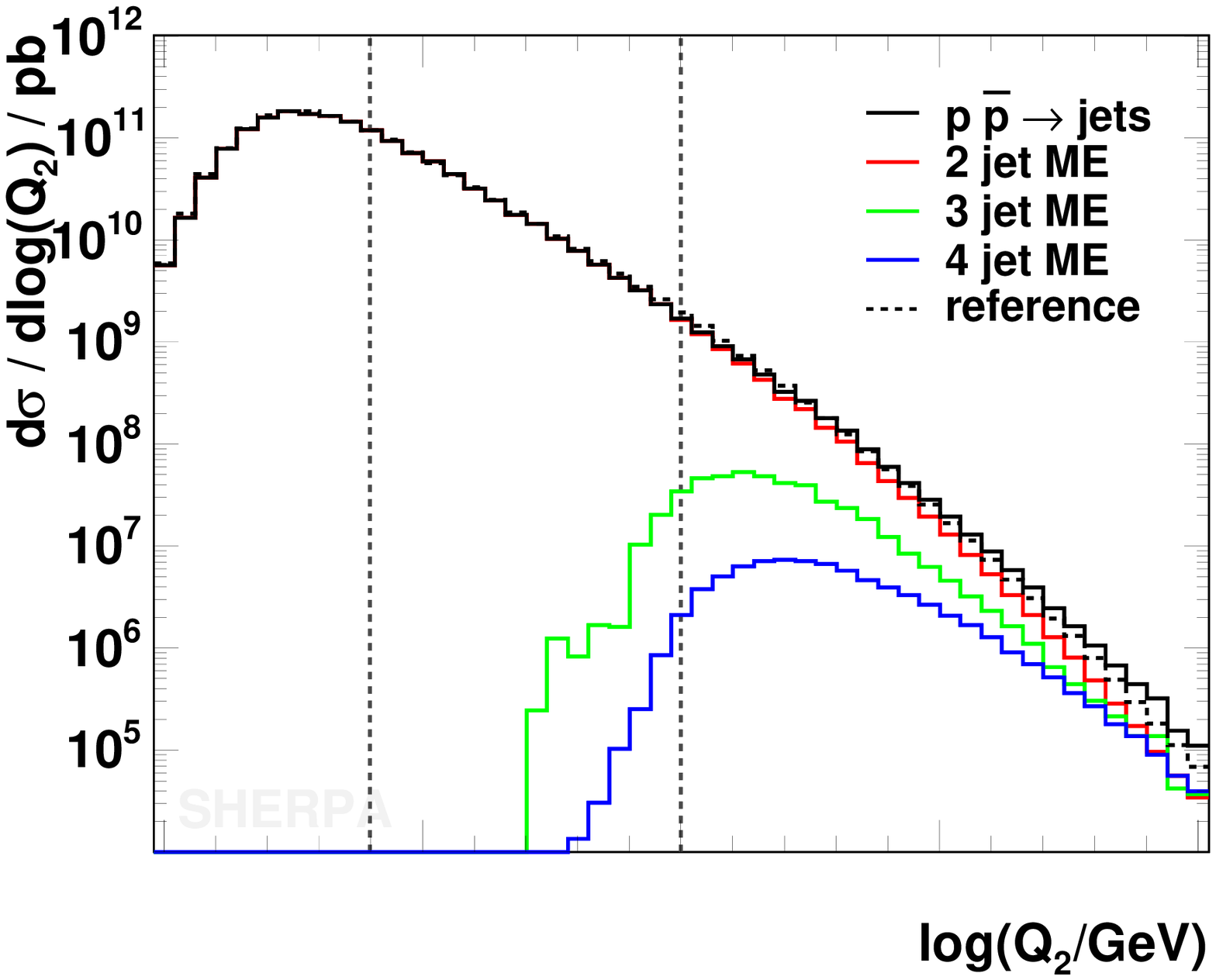}}
      \put(0,-8){\includegraphics[width=5.5cm]
        {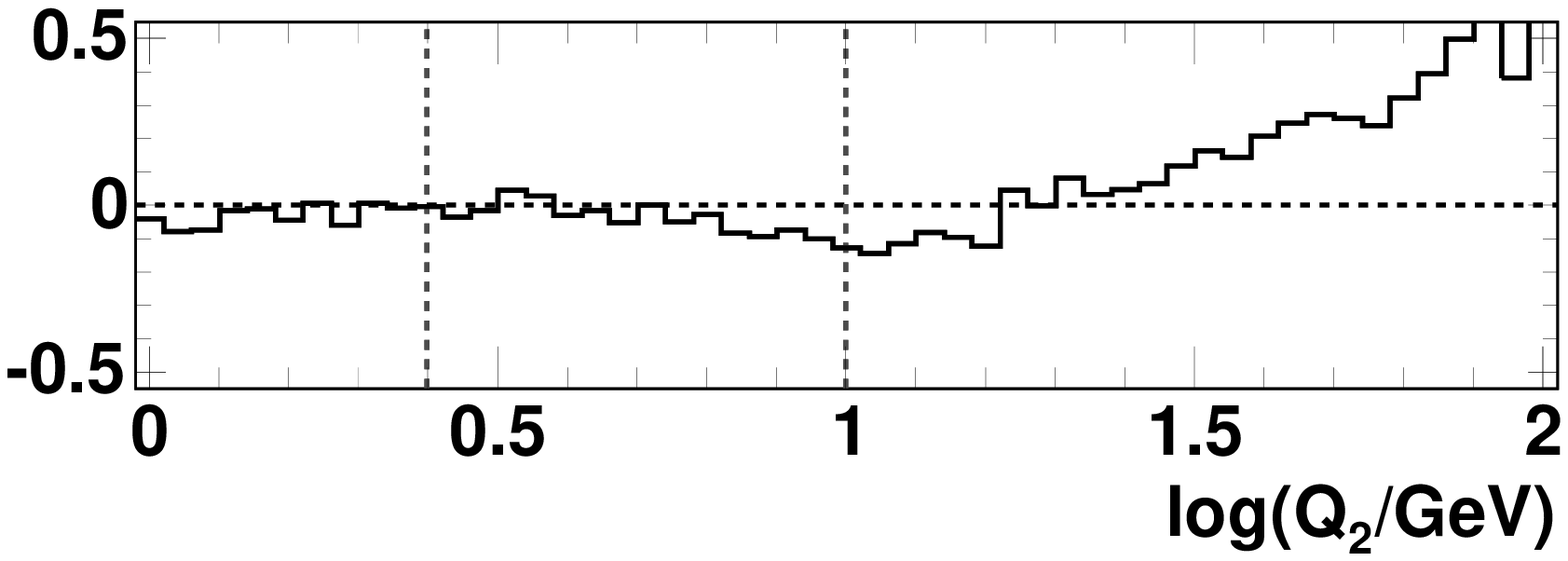}}
    \end{pspicture}
  \end{center}
  \caption{\label{fig:diff_jjj_tev1} Differential jet rates at the
    Tevatron, Run I. From left to right, the rates for $2\to 1$, $3\to
    2$, and for $4\to 3$ transitions at the hadron level are
    exhibited; the samples are produced with $n_{\rm max}=4$ and  
    $Q_{\rm cut}^{(2)}=2.5$~GeV  \& $Q_{\rm cut}^{(3,4)}=10$~GeV.
    A reference curve is shown in black-dashed lines, contributions
    from different multiplicities are displayed in different colours.}
\end{figure*}

\noindent
In Fig.\ \ref{fig:diff_jj_tev1_consistency}, the mixed sample from
above is further investigated. There, in addition to the summed
result, also the contributions from different jet multiplicities are
displayed. Clearly, the two samples fill quite separate regions of
phase space, i.e.\ above and below the jet resolution cut. Of course,
as before, there is some residual migration of the samples over the
respective jet resolution cut. The sum, however, is remarkably smooth 
over the cut. This allows to efficiently generate an inclusive QCD
sample with jets resolved at $40$ GeV, for example, where higher jet
configurations are accounted for by corresponding matrix elements and
the phase space below the matrix element cuts for them is properly
filled by the lowest multiplicity contribution. The quality of this
approach is further highlighted in Figs.\ \ref{fig:diff_jj_tev1} and
\ref{fig:diff_jjj_tev1}. There, again, differential jet rates are
depicted, this time the cuts have been chosen as 
$Q_{\rm cut}^{(2)}=2.5$~GeV  and  $Q_{\rm cut}^{(\ge 3)}=10$~GeV.
The plots cover up to ten orders of magnitude with an extremely smooth
prediction. 

\subsection{Results for $\epemto d \bar{d} u \bar{u} (g)$ at LEP II}

\noindent
In this section, the quality of the alternative algorithm will be
validated. To this end, $\epemto$jets at LEP~II are chosen as the
reference process. 

\subsubsection{QCD}

\begin{figure*}[h]
  \begin{center}
    \begin{pspicture}(450,200)
      \put(280,40){\includegraphics[width=5.5cm]
        {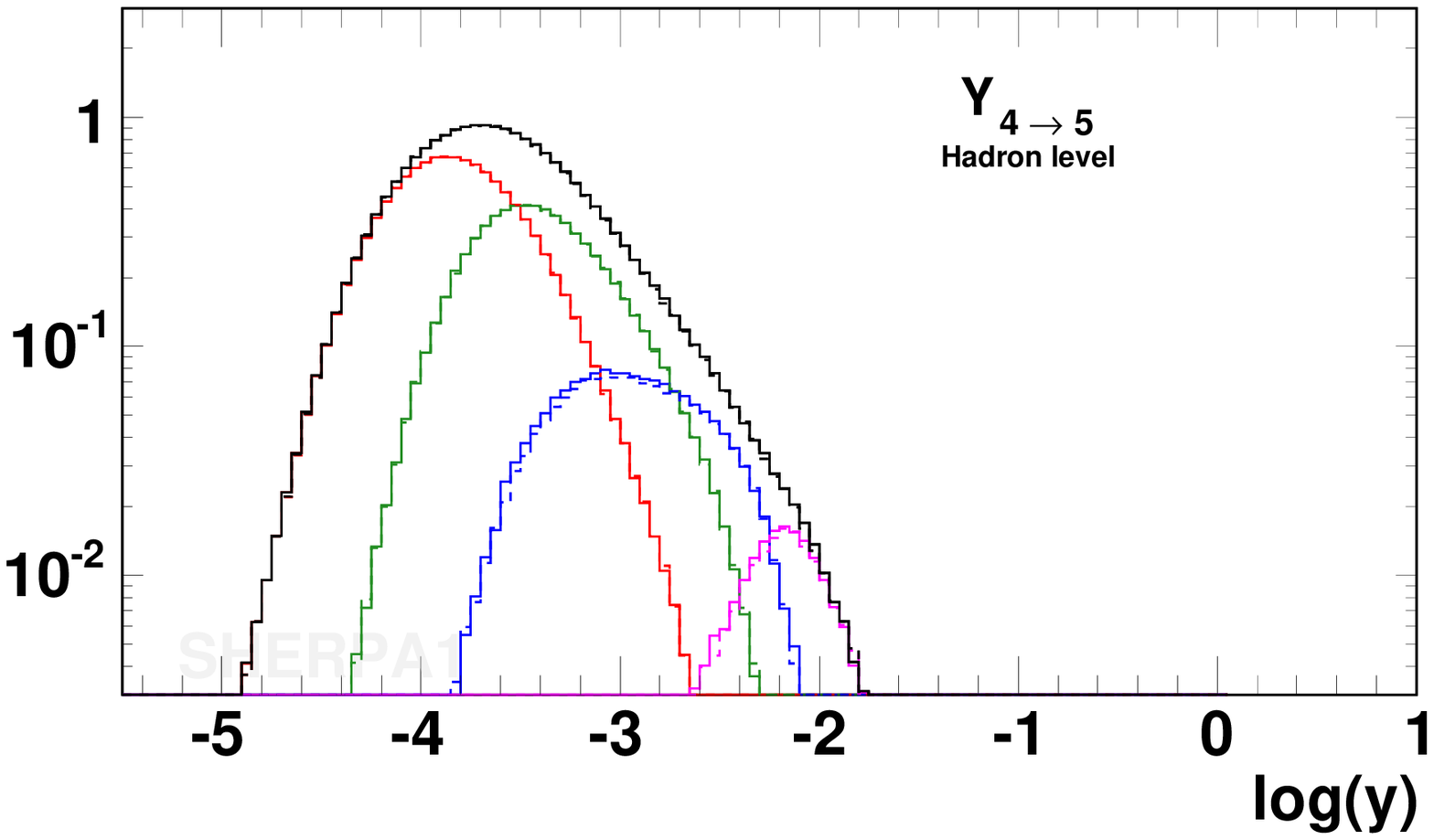}} 
      \put(140,40){\includegraphics[width=5.5cm]
        {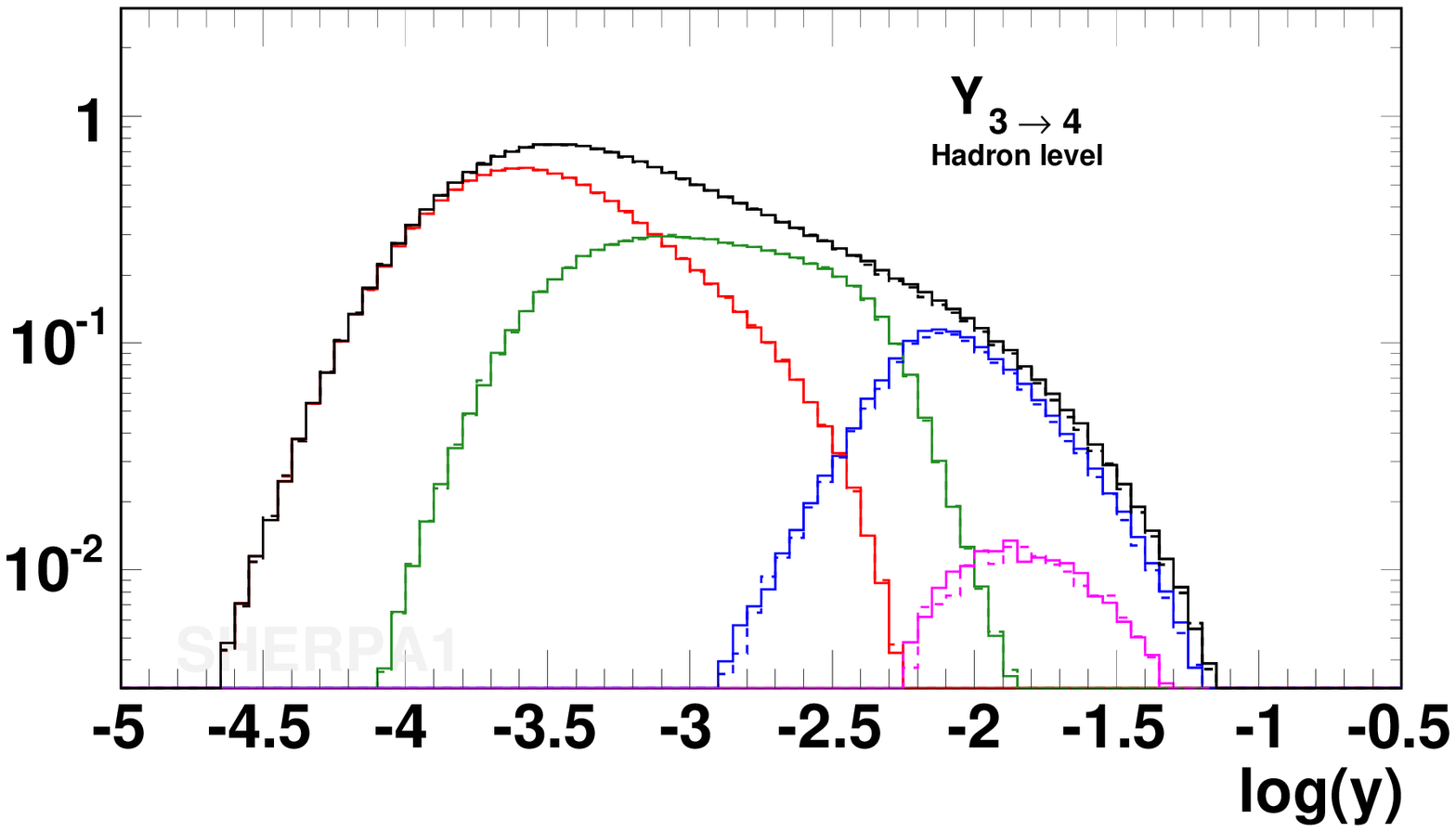}}
      \put(0,40){\includegraphics[width=5.5cm]
        {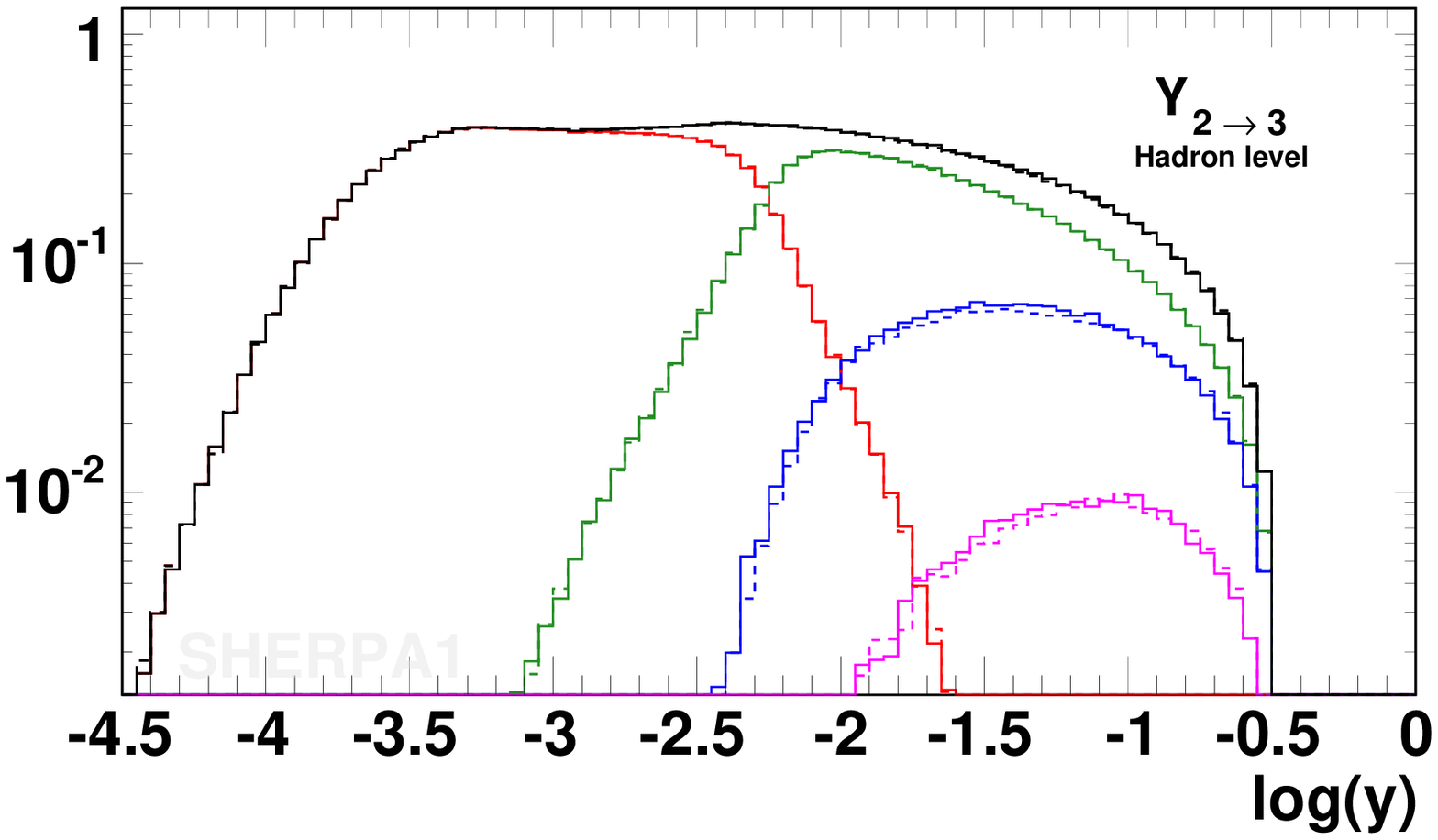}}
      \put(280,0){\includegraphics[width=5.5cm]
        {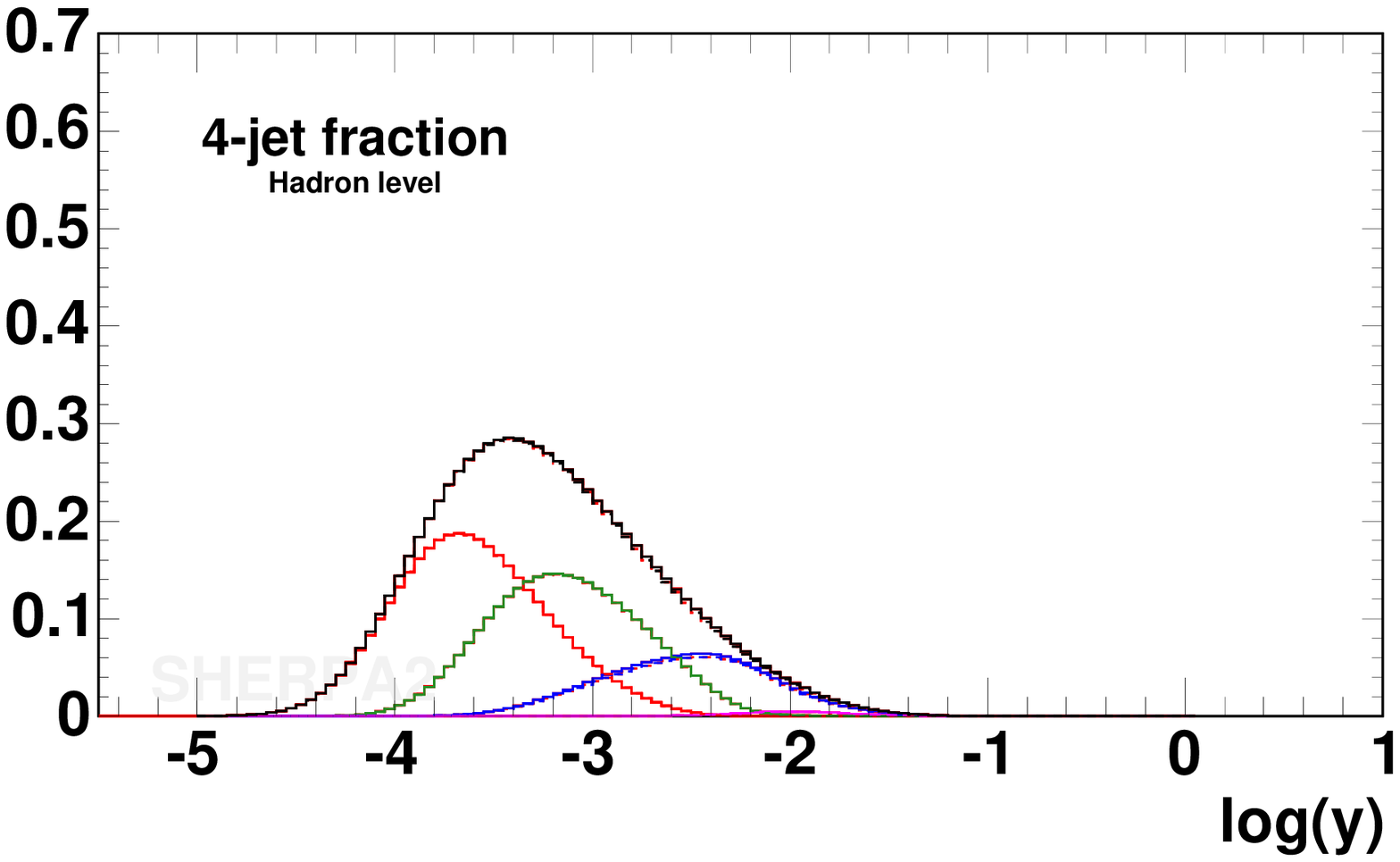}}
      \put(140,0){\includegraphics[width=5.5cm]
        {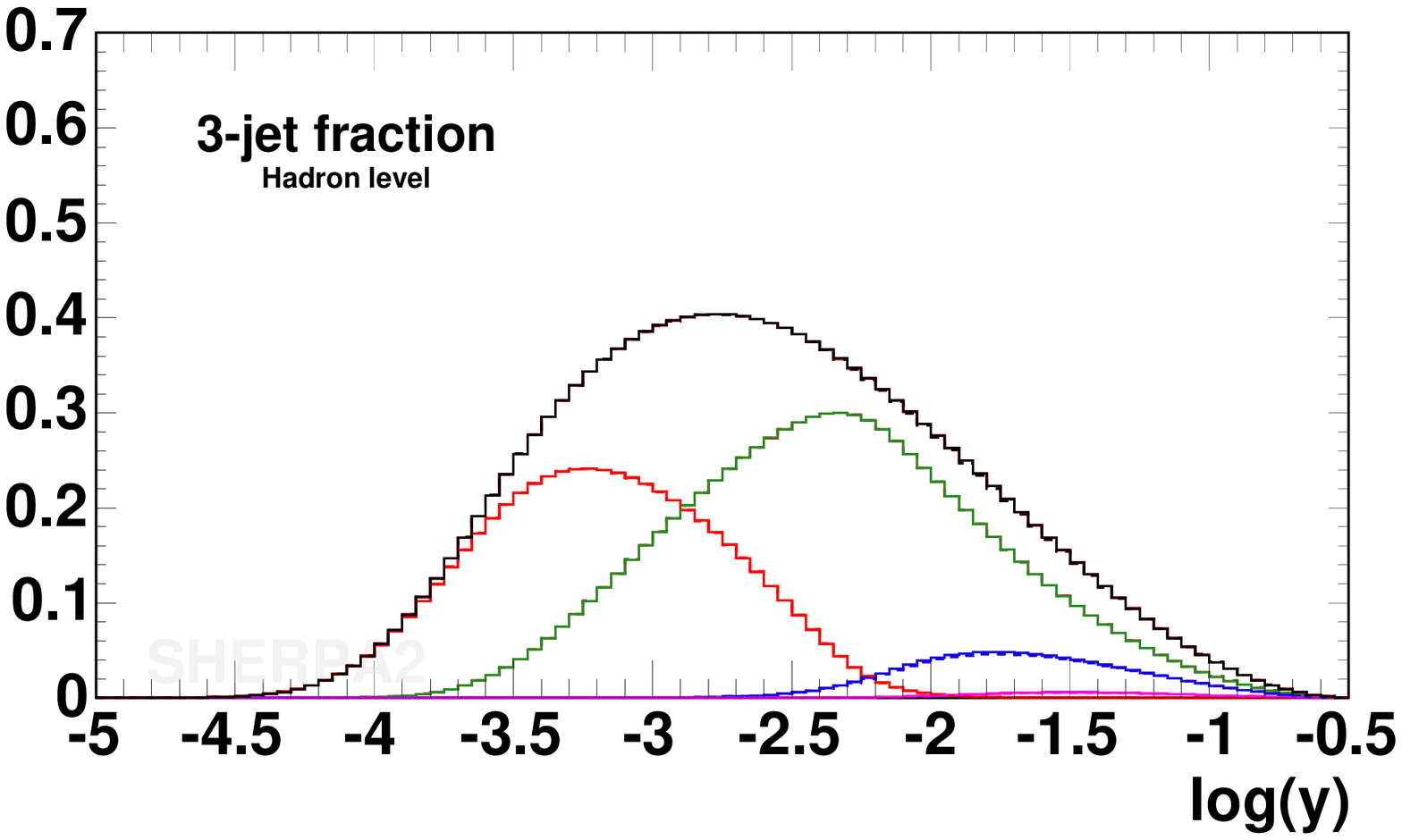}}
      \put(0,0){\includegraphics[width=5.5cm]
        {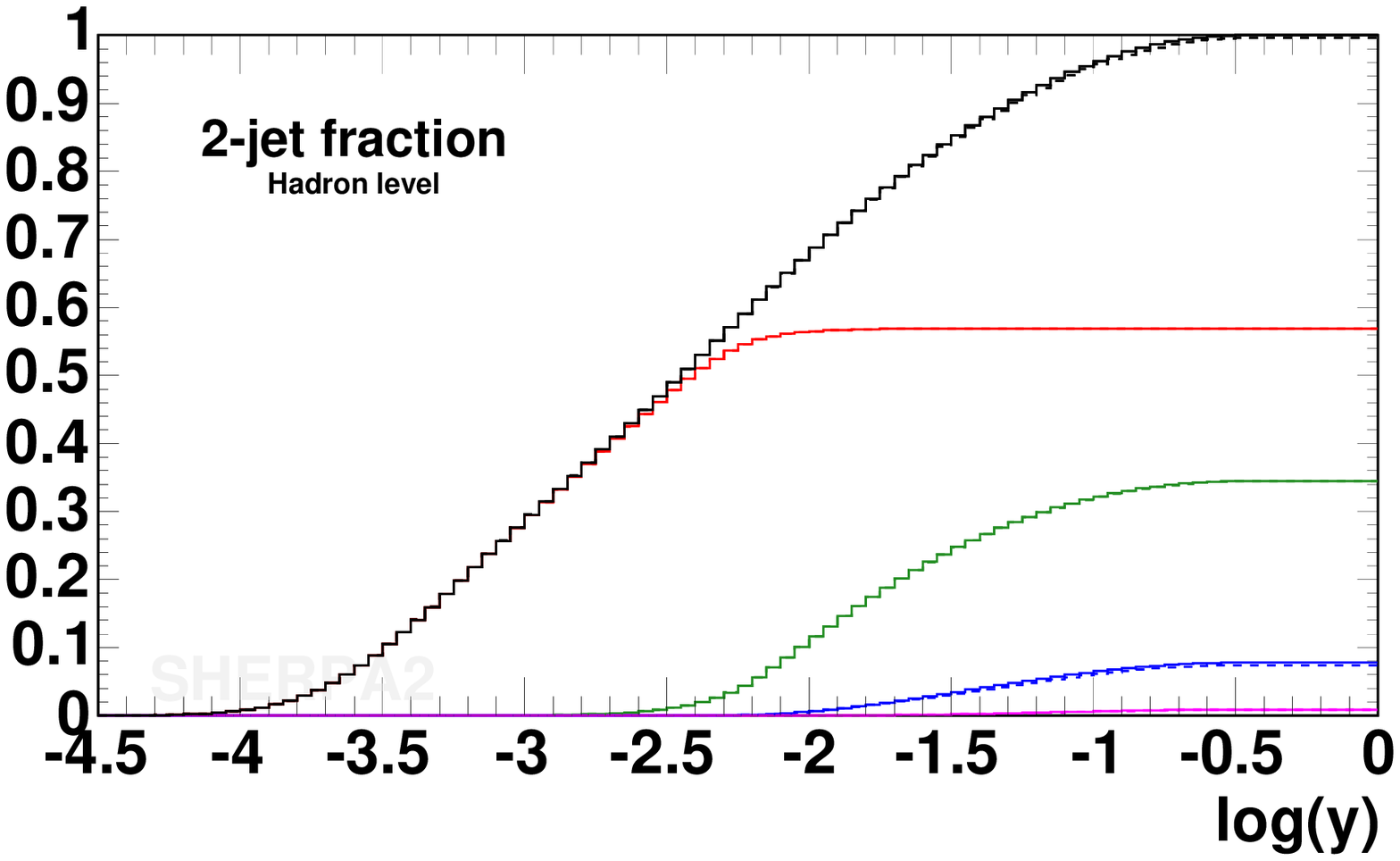}}
    \end{pspicture}
  \end{center}
  \caption{\label{fig:lep_qcd_diffjets} Differential (upper row) and
    total (lower row) jet rates in the Durham scheme at LEP~II for QCD events. 
    The result of the original and the alternative merging procedure
    are compared, (original=solid black, alternative=dashed red),
    differences are hardly visible.} 
\end{figure*}

\begin{figure*}[h]
  \begin{center}
    \begin{pspicture}(460,150)
      \put(280,0){\includegraphics[width=5.2cm]{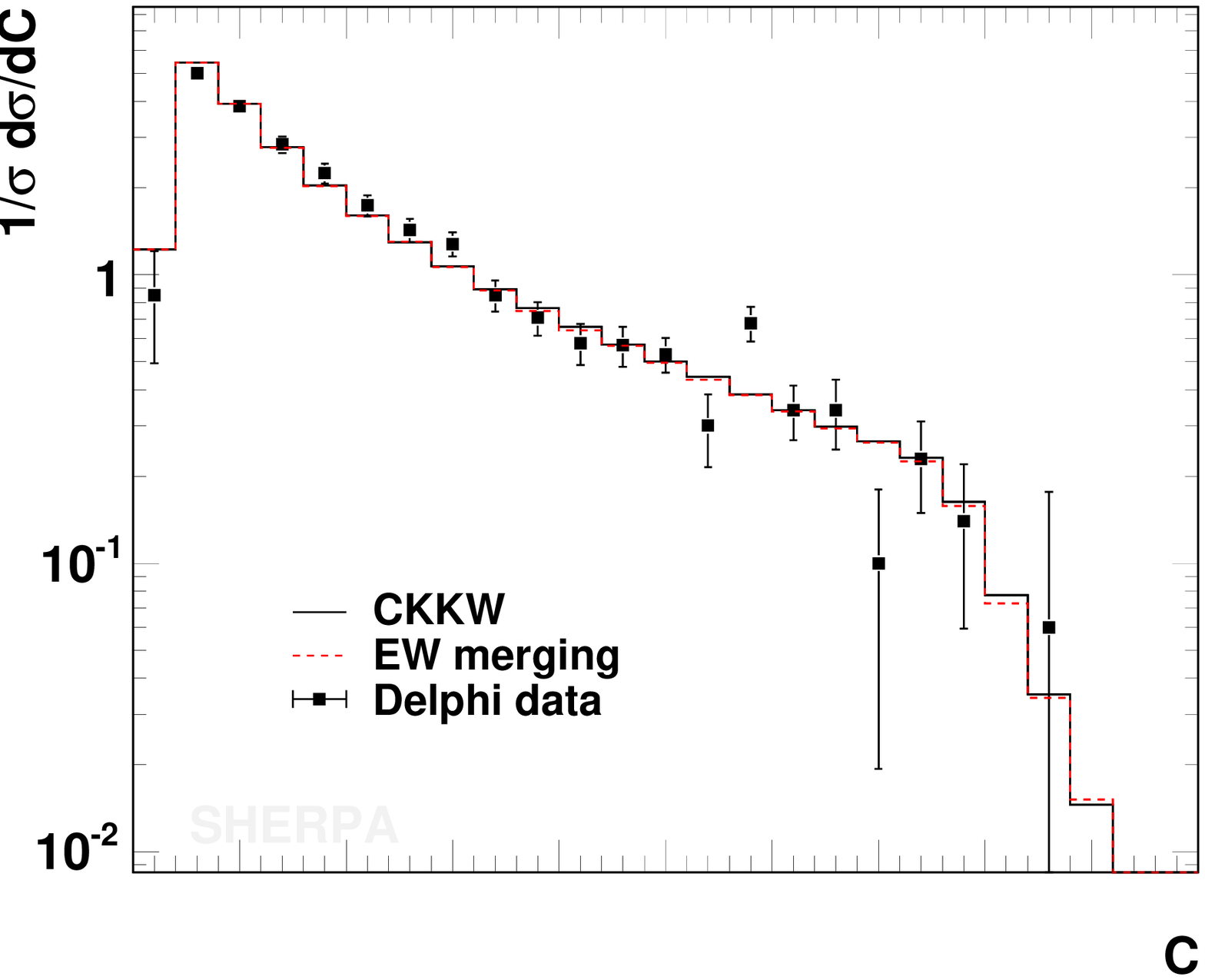}}
      \put(280,0){\includegraphics[width=5.2cm]{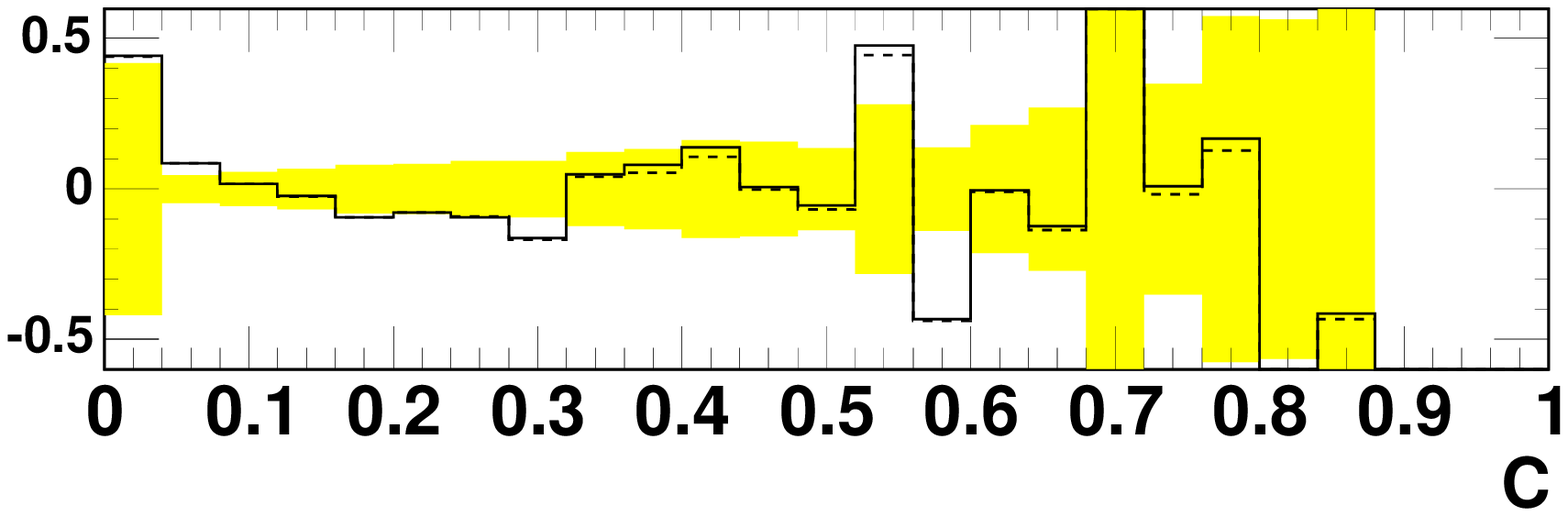}}
      \put(140,0){\includegraphics[width=5.2cm]{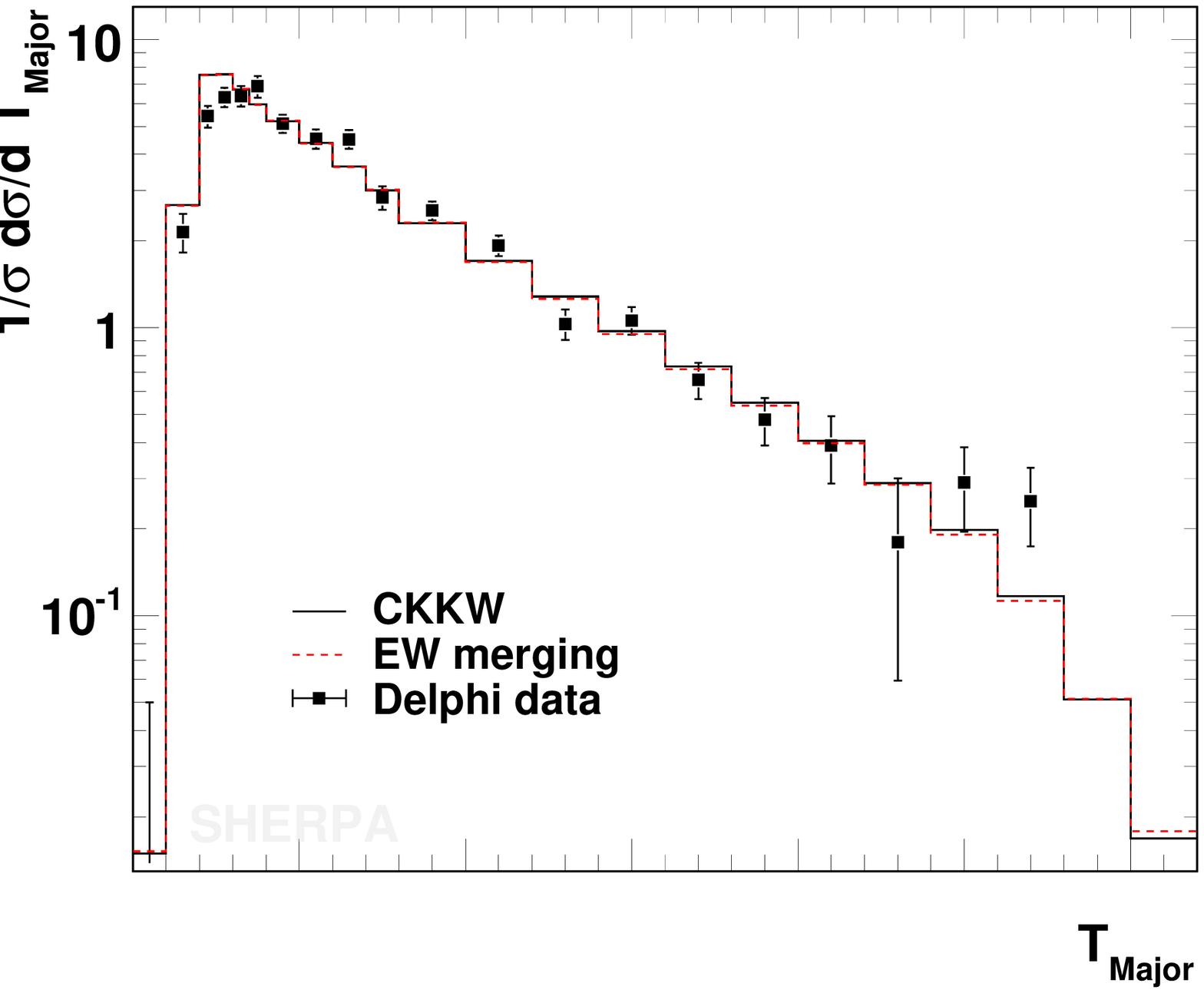}}
      \put(140,0){\includegraphics[width=5.2cm]{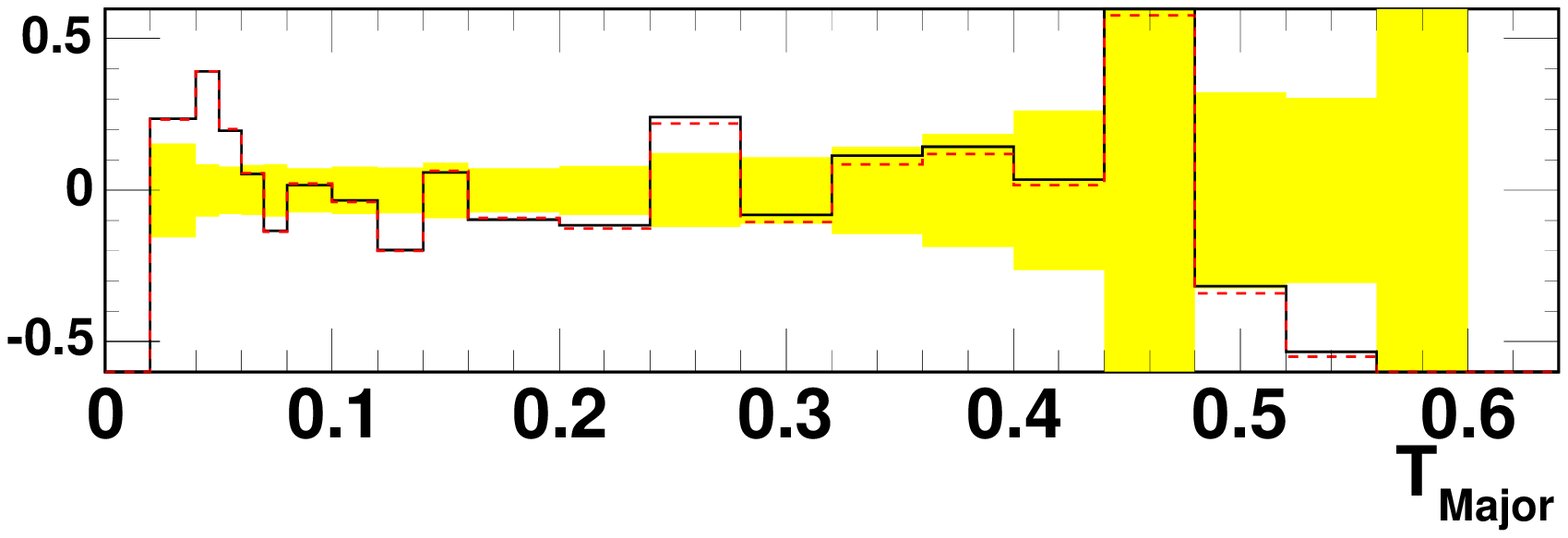}}
      \put(0,0){\includegraphics[width=5.2cm]{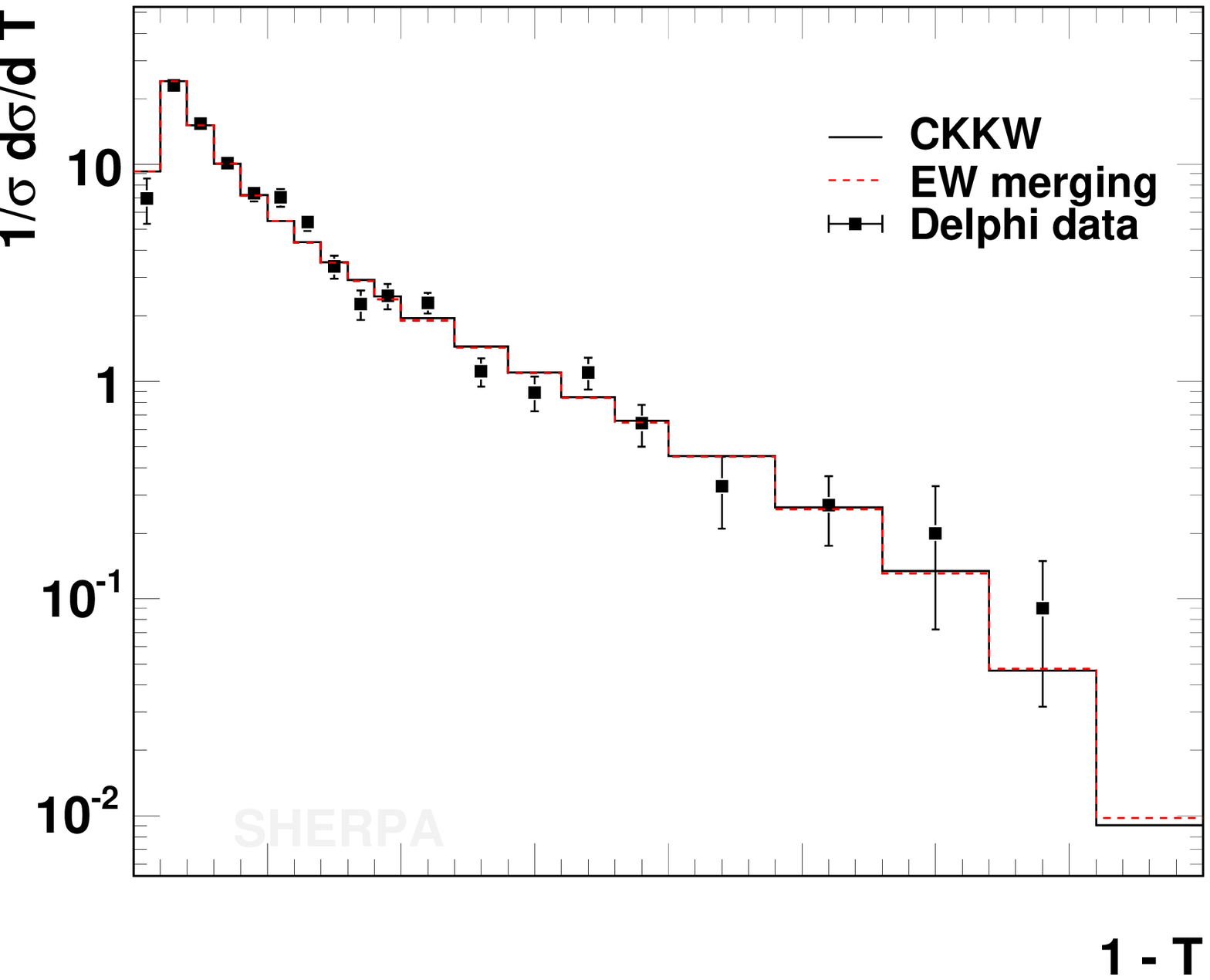}}
      \put(0,0){\includegraphics[width=5.2cm]{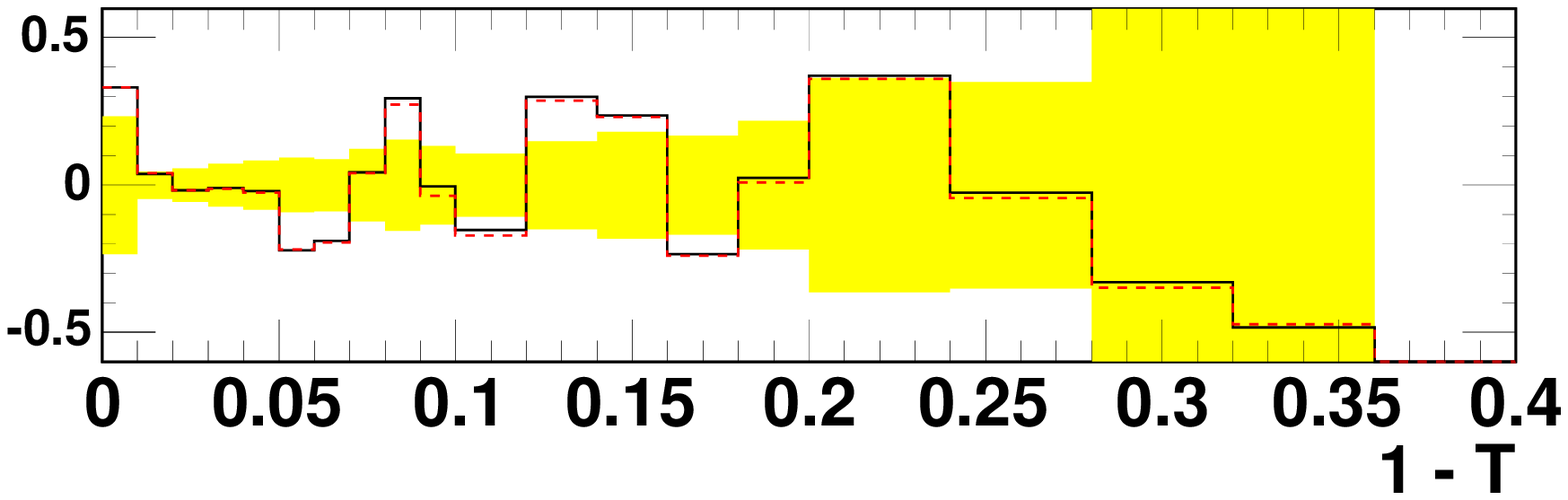}}
    \end{pspicture}
  \end{center}
  \caption{\label{fig:lep_qcd_event} Thrust, thrust major and
    $C$-parameter. Delphi data \cite{Abdallah:2003xz} taken at LEP~II events
    ($E_{\rm cms}=189$ GeV) are compared to simulation using the
    original and the alternative way of constructing the pseudo parton
    shower history (original=solid black, alternative=dashed red).}
\end{figure*}

\noindent
Concentrating first on the case where QCD alone contributes to the
production of extra jets, in Fig.\ \ref{fig:lep_qcd_diffjets}
differential and total jet rates in the Durham scheme at LEP~II as
described with the original and with the alternative approach are
compared. Clearly, the results are nearly indistinguishable. This
implies that in this case the ordering of the hardness of emissions
according to the $k_\perp$ measure is nearly identical with an
ordering according to the virtual masses occurring in the propagator
terms.  
\noindent
The same holds also true for event shape observables, depicted in
Fig.\ \ref{fig:lep_qcd_event}. There, measurements of thrust, thrust-major and the
$C$-parameter \cite{Abdallah:2003xz} are exhibited and compared to the simulation of
SHERPA. Again, the alternative and the original algorithm perform
equally well and both reproduce nicely the data. 

\subsubsection{Electroweak interactions}

\begin{figure*}[h]
  \begin{center}
    \begin{pspicture}(460,150)
      \put(280,0){\includegraphics[width=5.2cm]
        {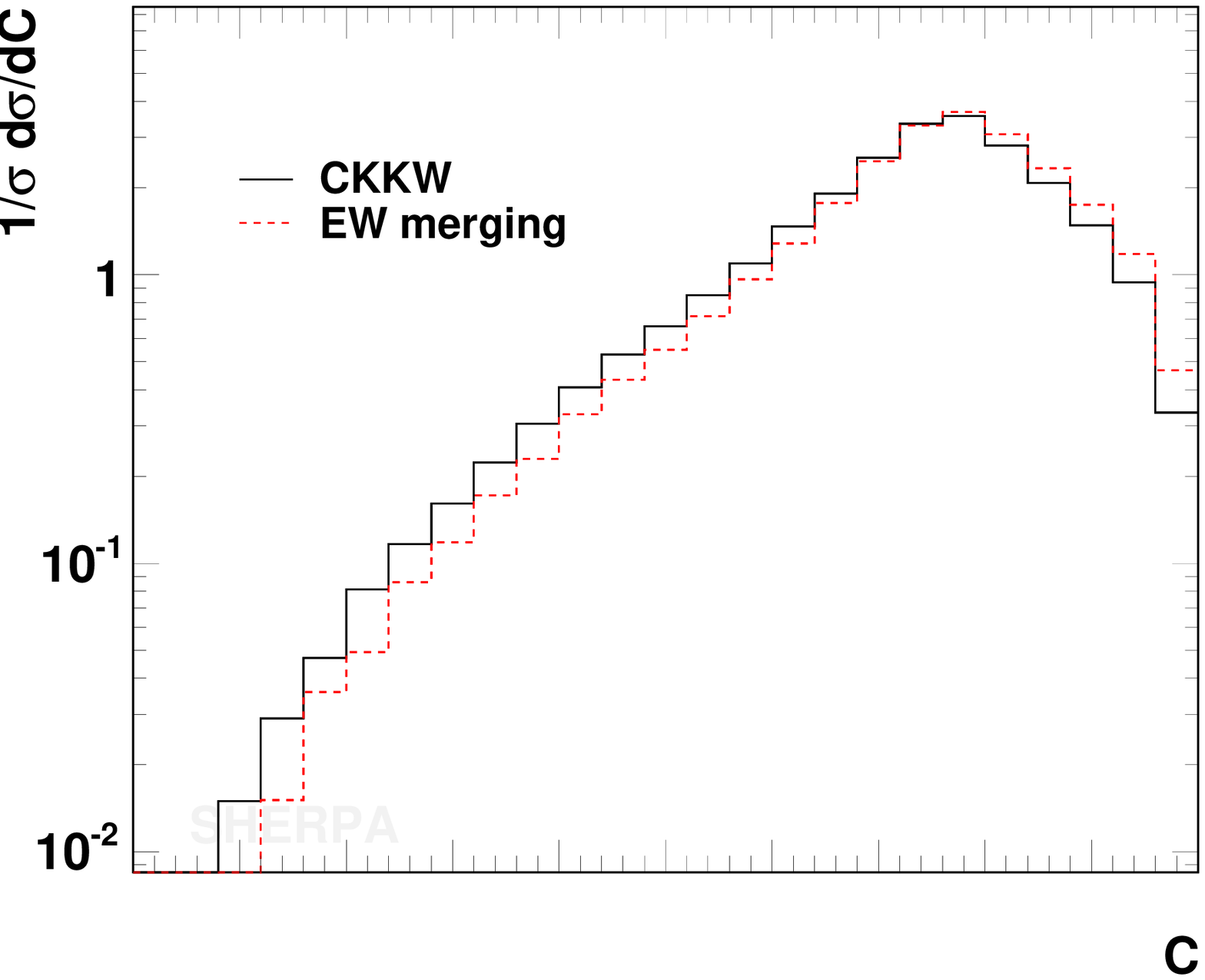}}
      \put(280,0){\includegraphics[width=5.2cm]
        {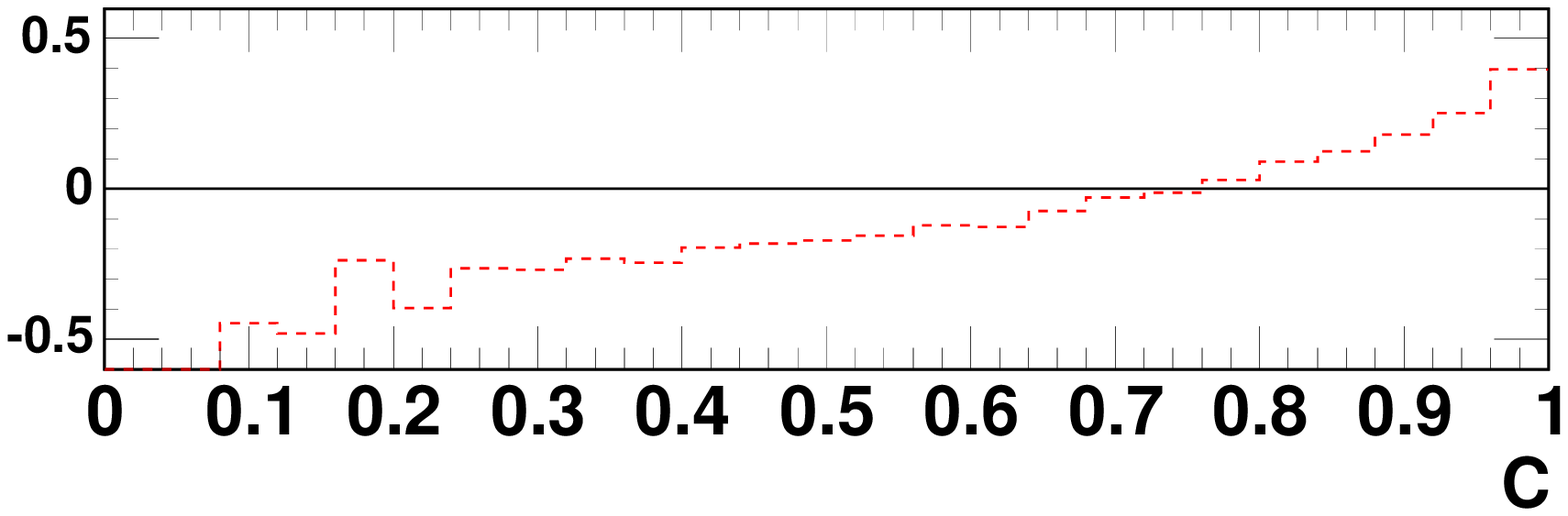}}
      \put(140,0){\includegraphics[width=5.2cm]
        {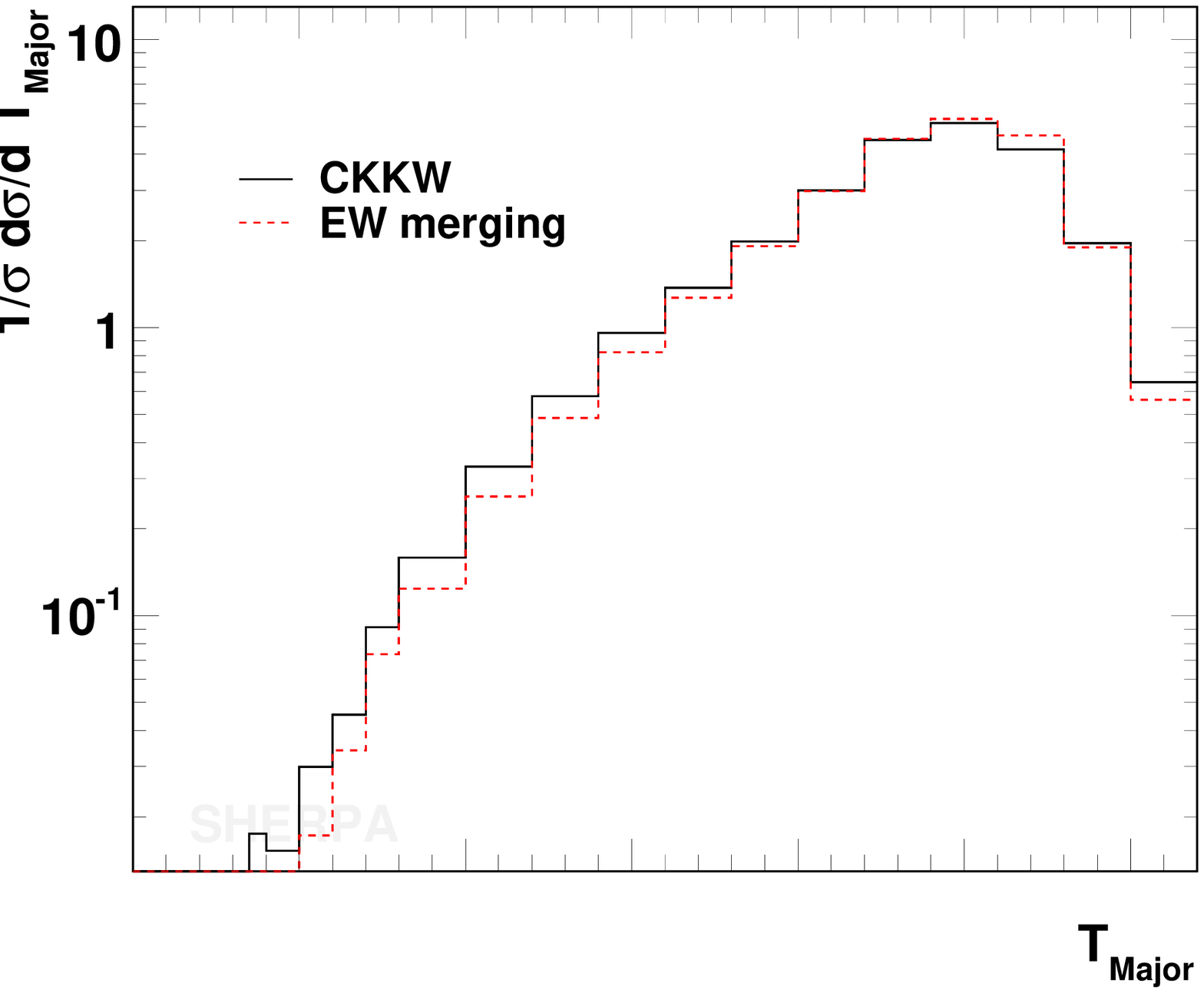}}
      \put(140,0){\includegraphics[width=5.2cm]
        {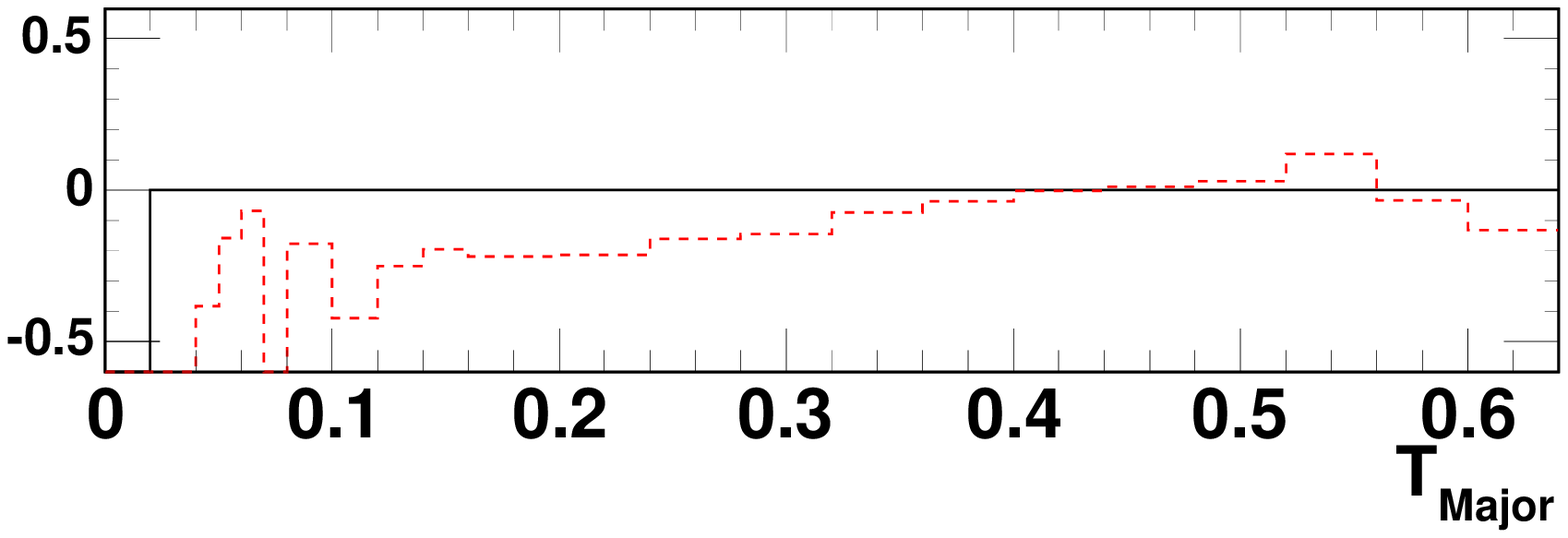}}
      \put(0,0){\includegraphics[width=5.2cm]
        {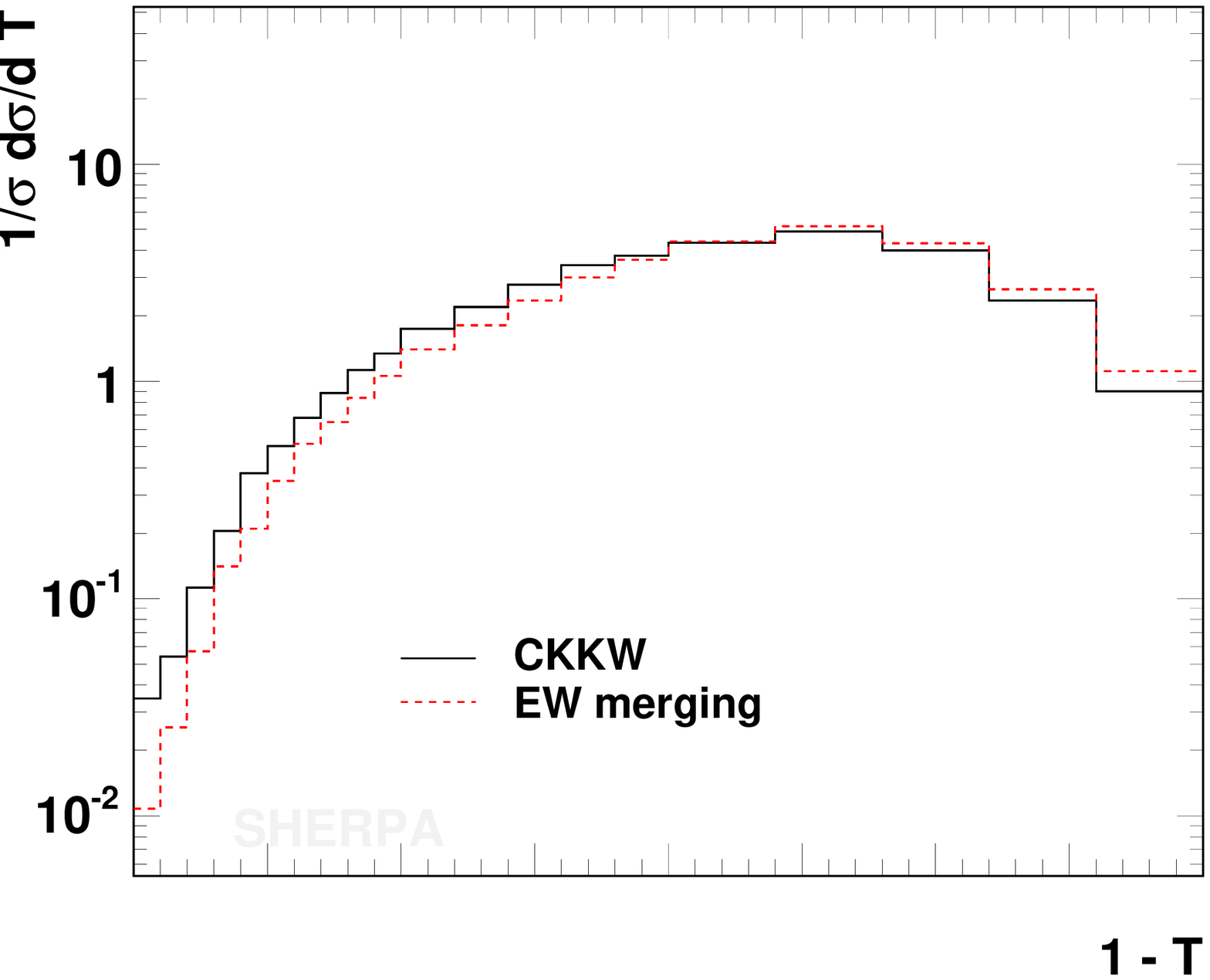}}
      \put(0,0){\includegraphics[width=5.2cm]
        {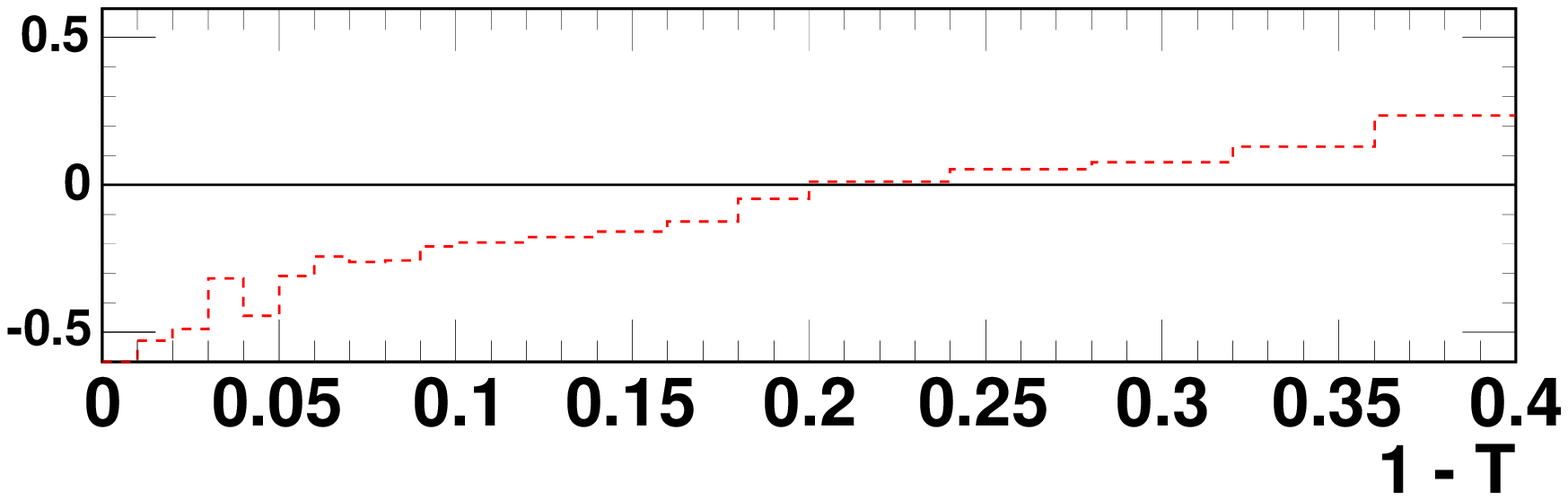}}
    \end{pspicture}
  \end{center}
  \caption{\label{fig:lep_ew1_event} Thrust, thrust-major and
    $C$-parameter in electroweak four jet events at LEP~II. This time,
    the results of the two merging prescriptions (original=solid black,
    alternative=dashed red) differ significantly, by up to 50\%.}
\end{figure*}

\begin{figure*}[h]
  \begin{center}
    \begin{pspicture}(450,190)
      \put(280,40){\includegraphics[width=5.5cm]
        {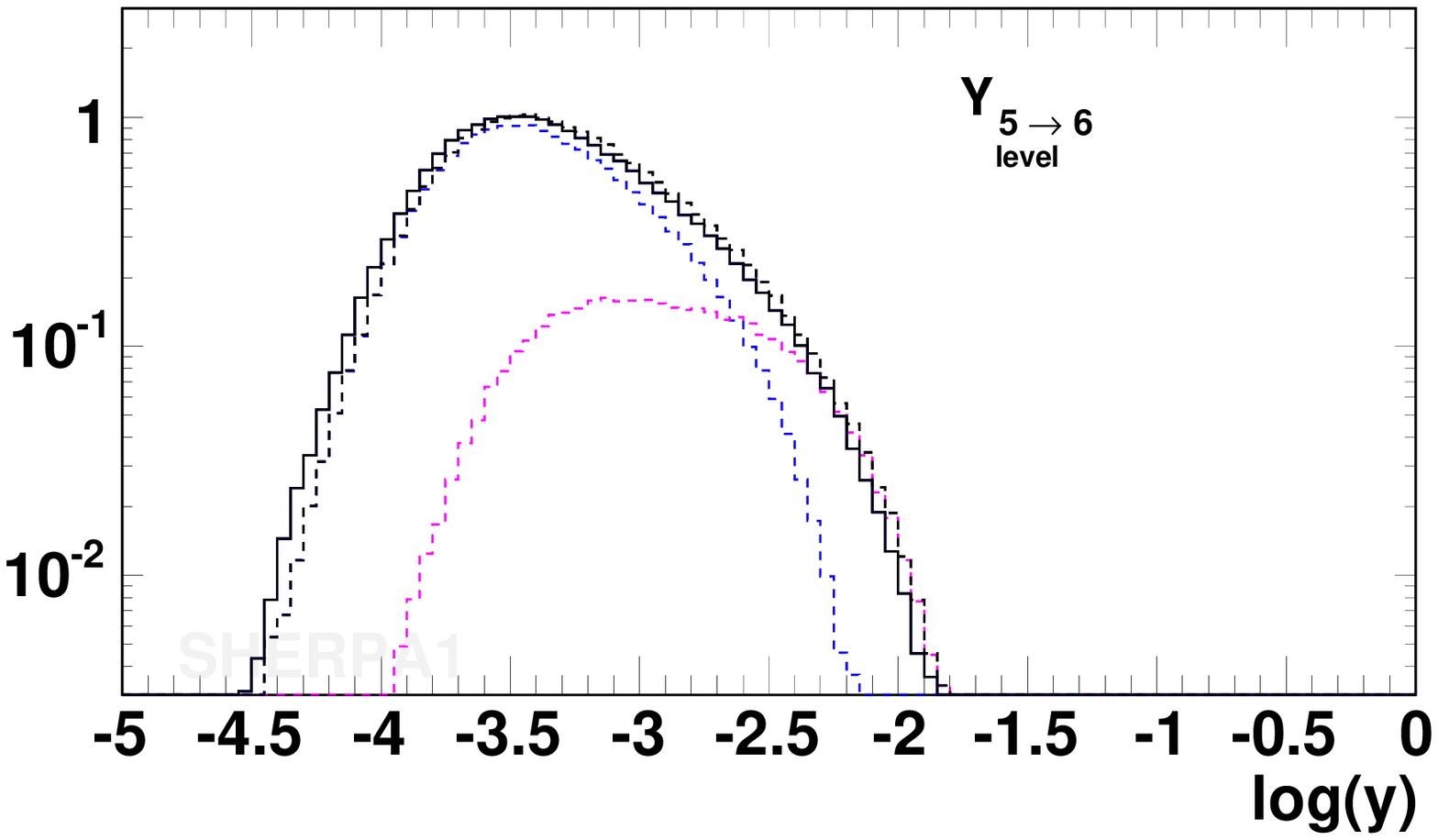}} 
      \put(140,40){\includegraphics[width=5.5cm]
        {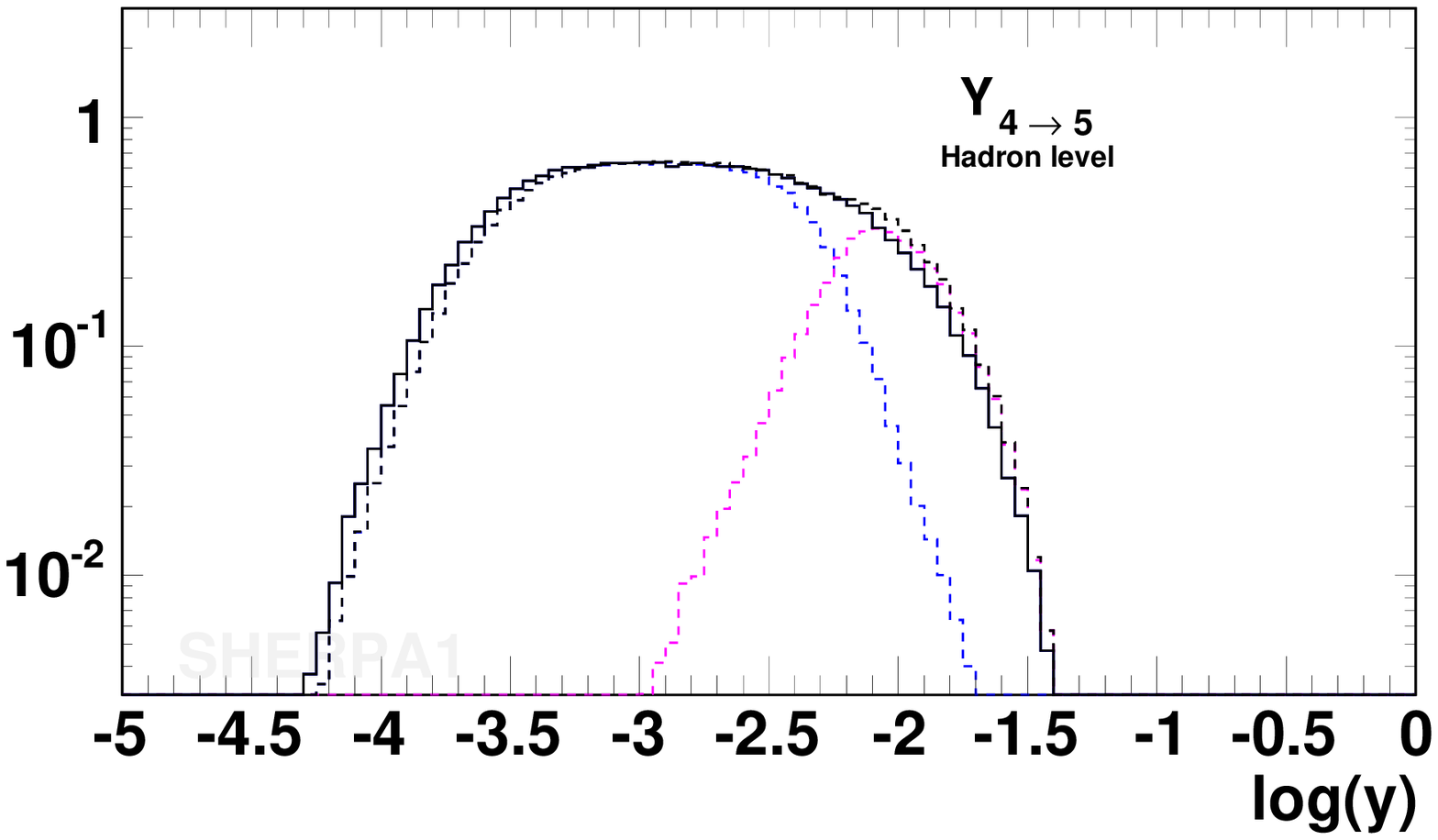}}
      \put(0,40){\includegraphics[width=5.5cm]
        {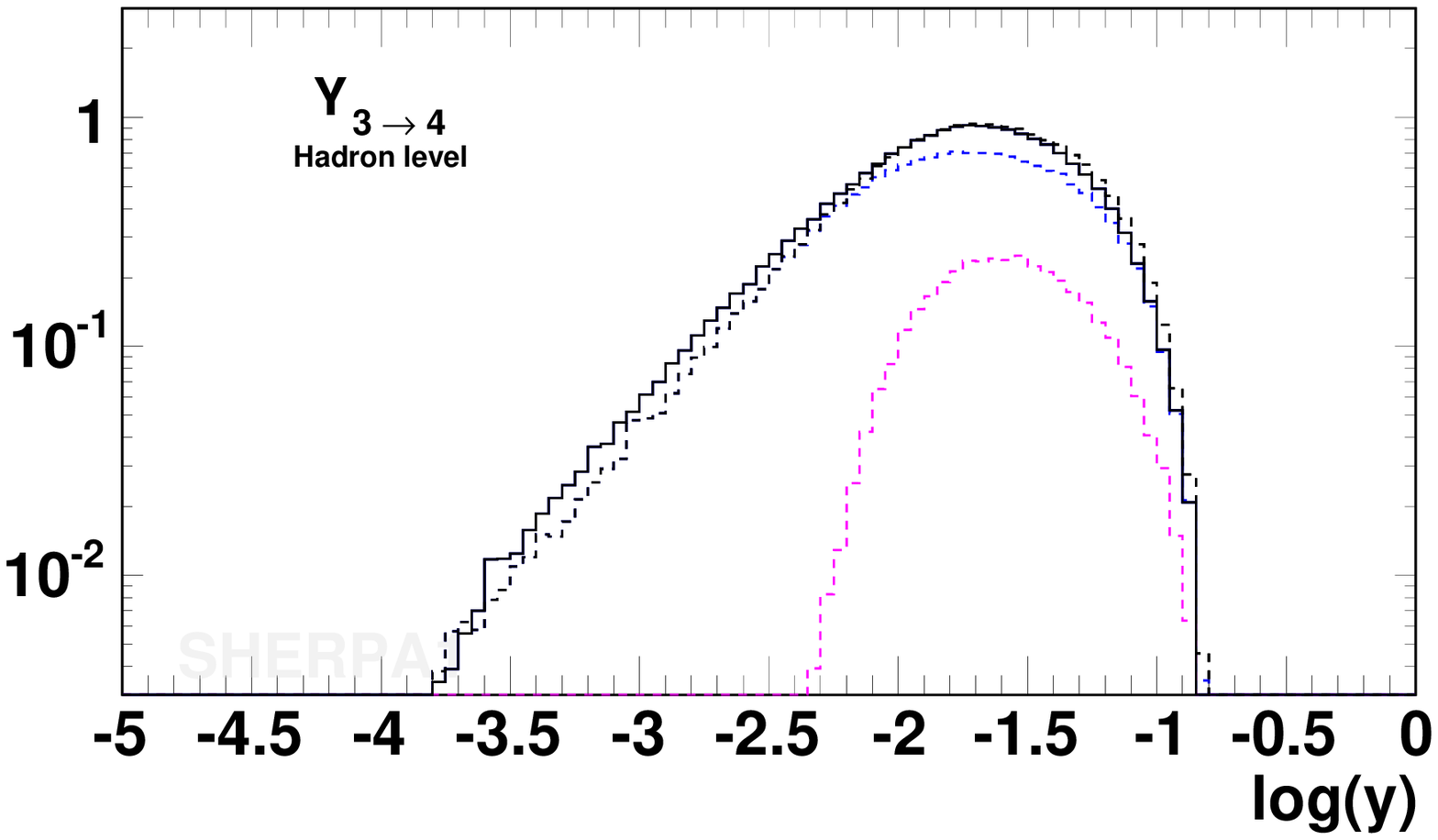}}
      \put(280,0){\includegraphics[width=5.5cm]
        {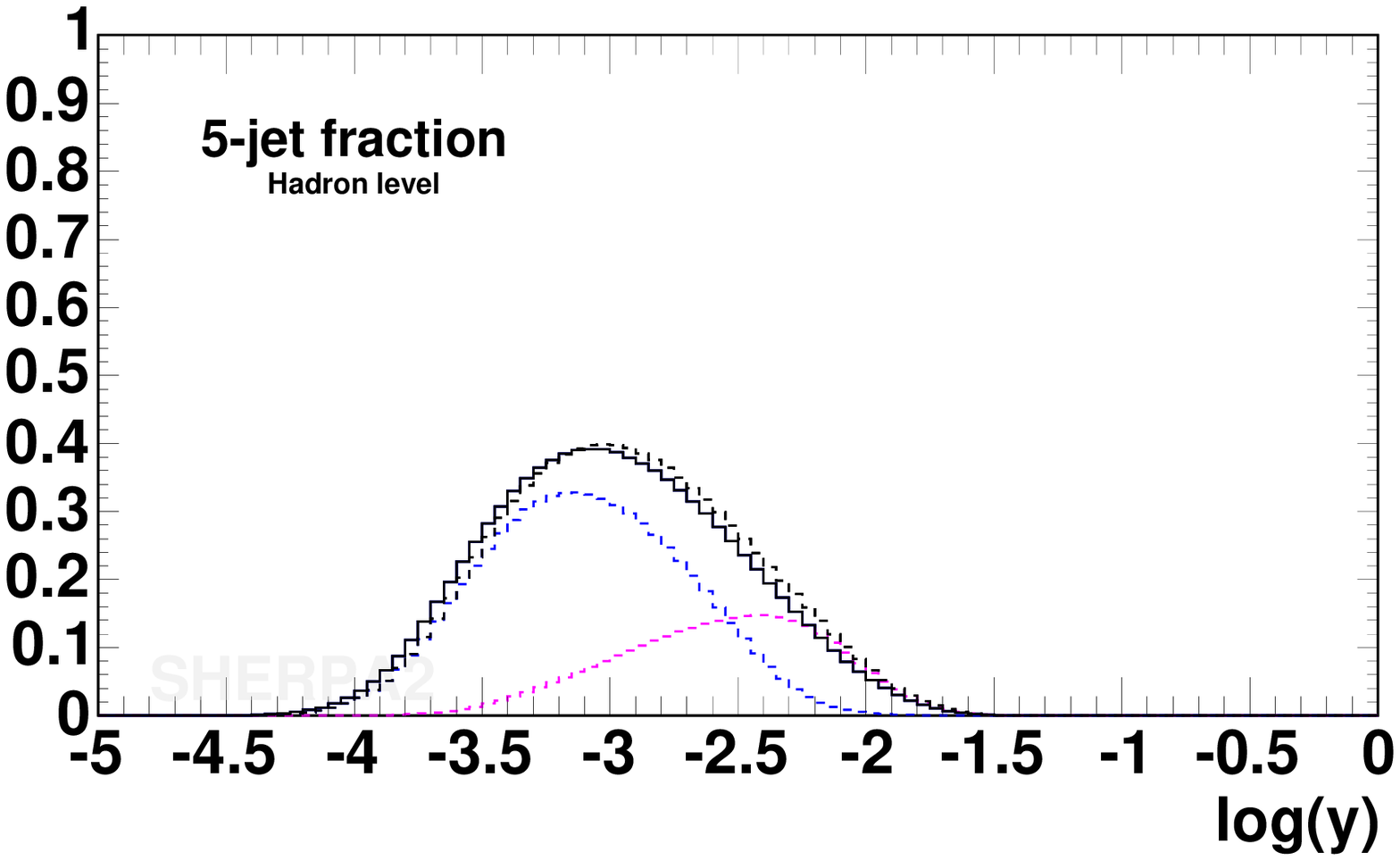}}
      \put(140,0){\includegraphics[width=5.5cm]
        {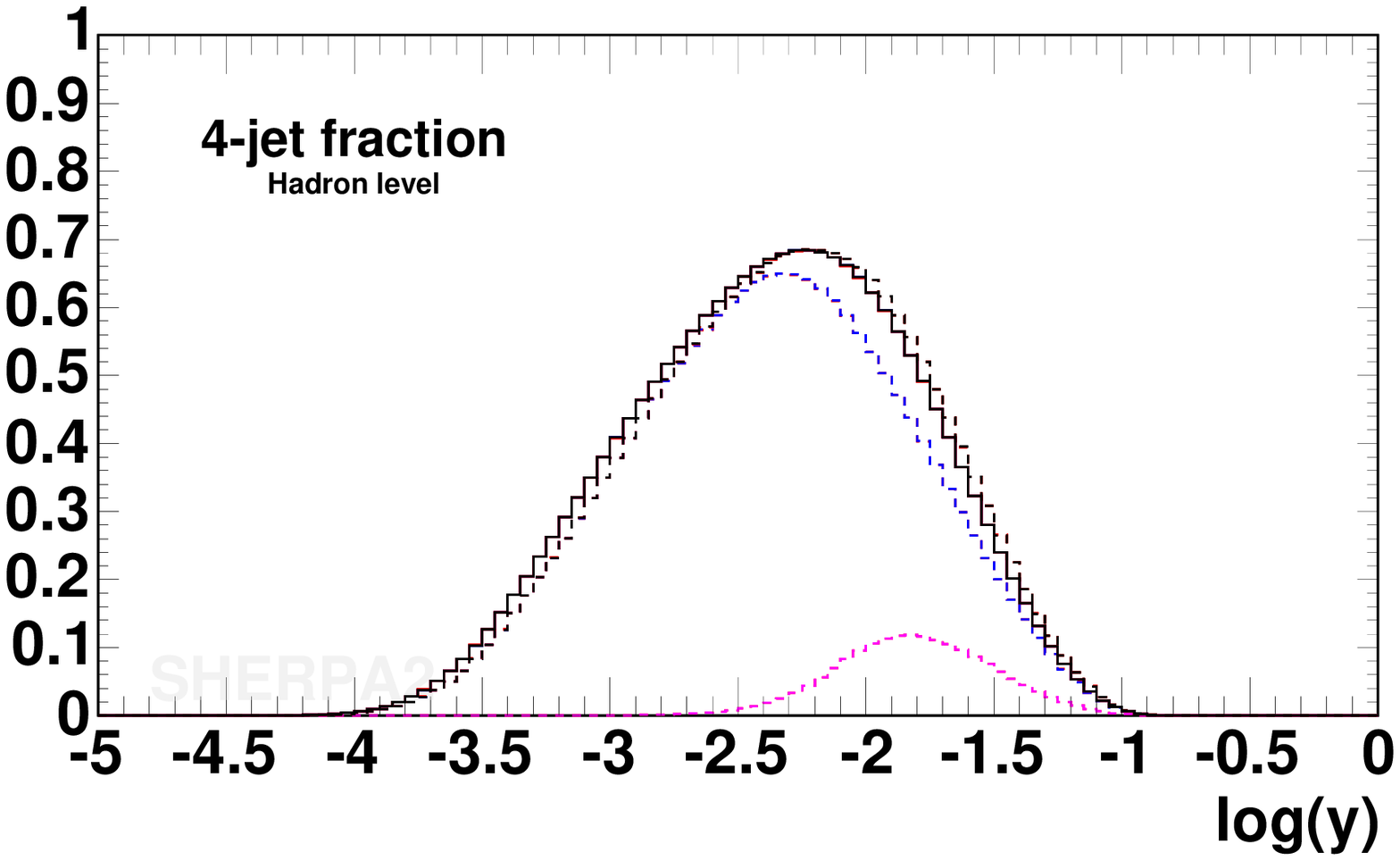}}
      \put(0,0){\includegraphics[width=5.5cm]
        {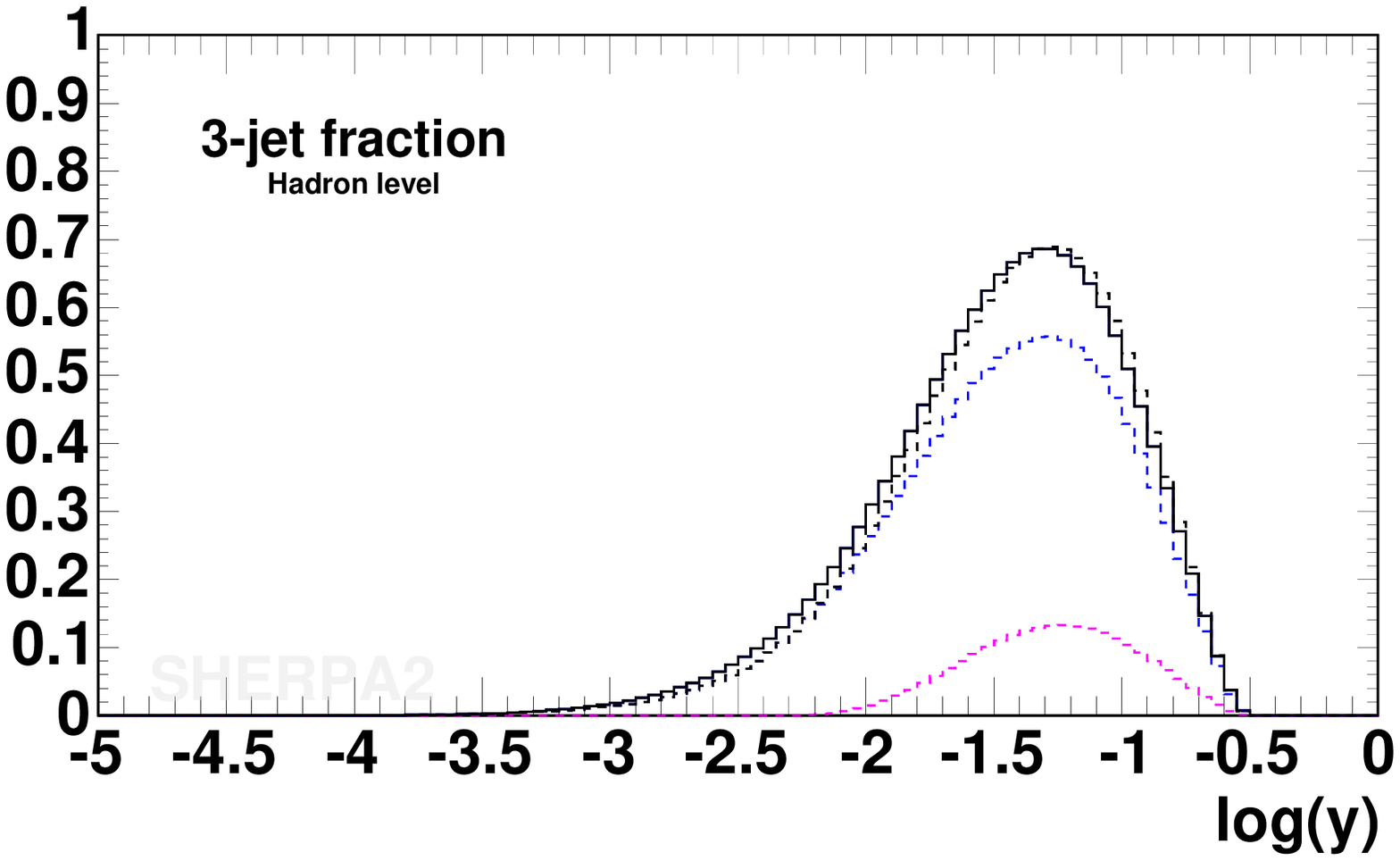}}
    \end{pspicture}
  \end{center}
  \caption{\label{fig:lep_ew1_diffjets} Differential (upper row) and
    total (lower row) jet rates in electroweak four jet events at LEP~II. 
    The results of the two merging prescriptions (original=solid black,
    alternative=dashed red) are compared with each other, differences
    are visible.}
\end{figure*}

\begin{figure*}[h]
  \begin{center}
    \begin{pspicture}(360,170)
      \put(200,0){\includegraphics[width=6cm]
        {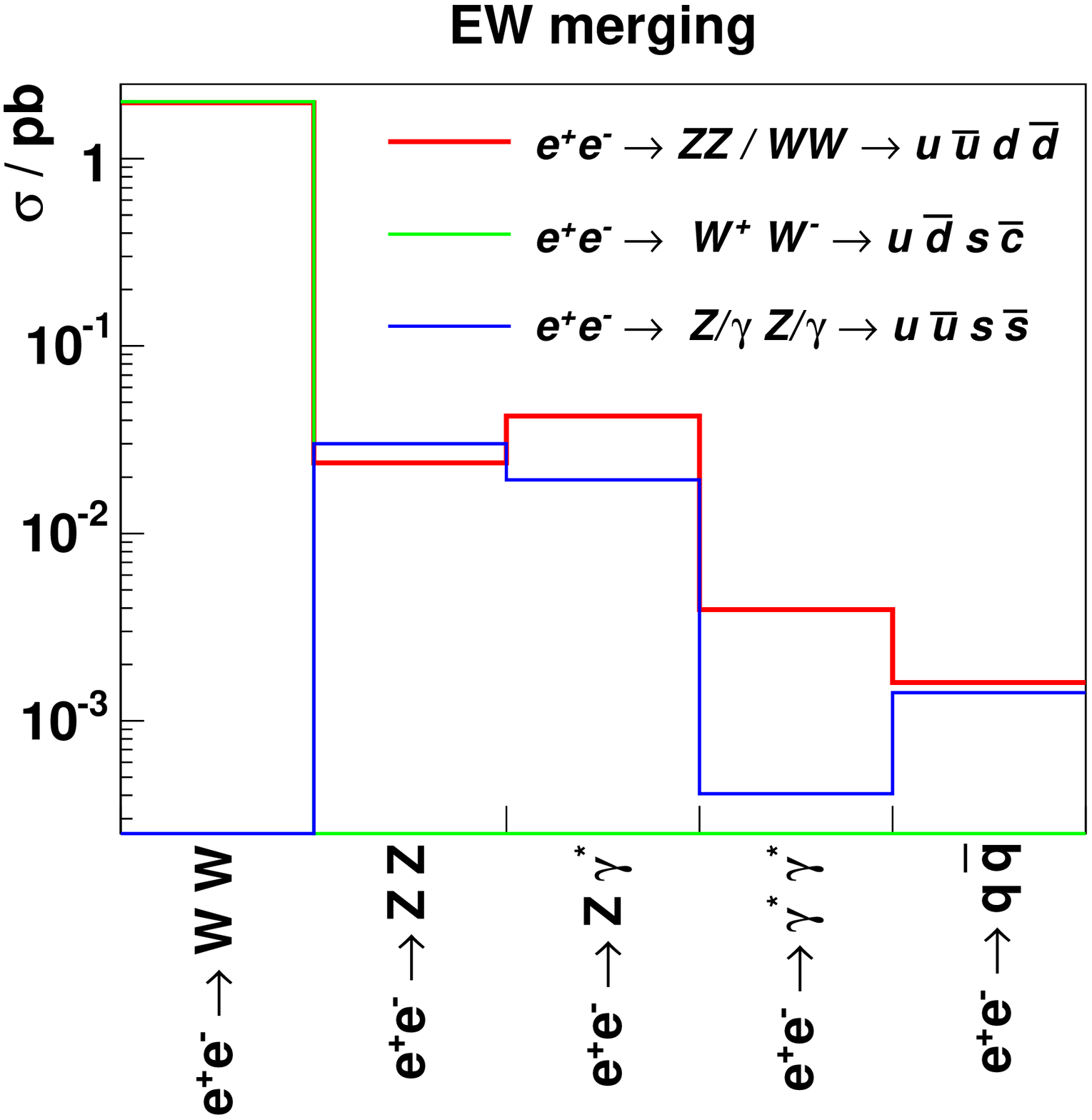}}
      \put(0,0){\includegraphics[width=6cm]
        {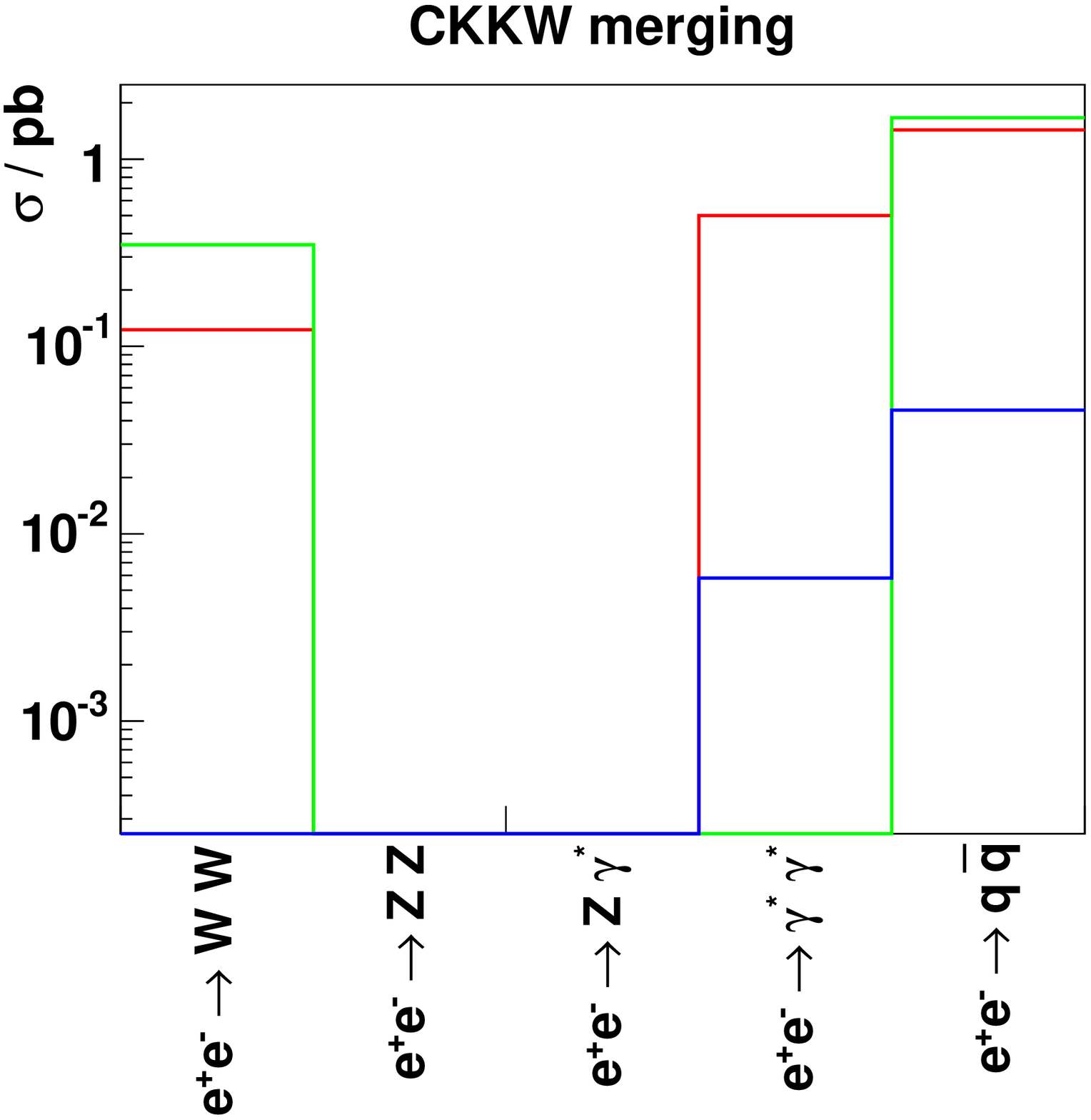}}
    \end{pspicture}
  \end{center}
  \caption{\label{fig:lep_ew1_coreprocesses}
    Statistics of selected ``core'' processes for different electroweak
    4-jet channels. The result of the original (left plot) with the
    alternative algorithm (right plot) is compared.}
\end{figure*}

\noindent
It is expected that differences in the two prescriptions to reconstruct the
pseudo parton shower history appear when the electroweak production of
four quarks is investigated. Both for event shape observables
displayed in Fig.\ \ref{fig:lep_ew1_event} and for total or
differential jet rates depicted in Fig.\ \ref{fig:lep_ew1_diffjets}
the differences are sizable, reaching up to 50\%. 
This can be easily understood. In Fig.\
\ref{fig:lep_ew1_coreprocesses} the cross sections for three typical
core $2\to 2$ processes are considered, namely $W$ pair production,
$Z/\gamma$ pair production, and the QED/electroweak analogue to the QCD
processes. Apparently, the alternative algorithm correctly reproduces the
expected cross sections for the $WW$ and $ZZ$ channel (2 pb and 0.02
pb). Hence, the relative contributions of the three considered
processes are consistent with the matrix element. In contrast, the
original algorithm fails to reproduce the correct rate for the $WW$
channel, because it triggers an unphysical migration into the QCD-like
configurations. Consequently, both samples differ in their colour
structure, in their Sudakov weights and, ultimately, in the starting
scales for their parton shower. 

\begin{figure*}[h]
\begin{center}
\begin{pspicture}(460,150)
\put(280,0){\includegraphics[width=5.2cm]
  {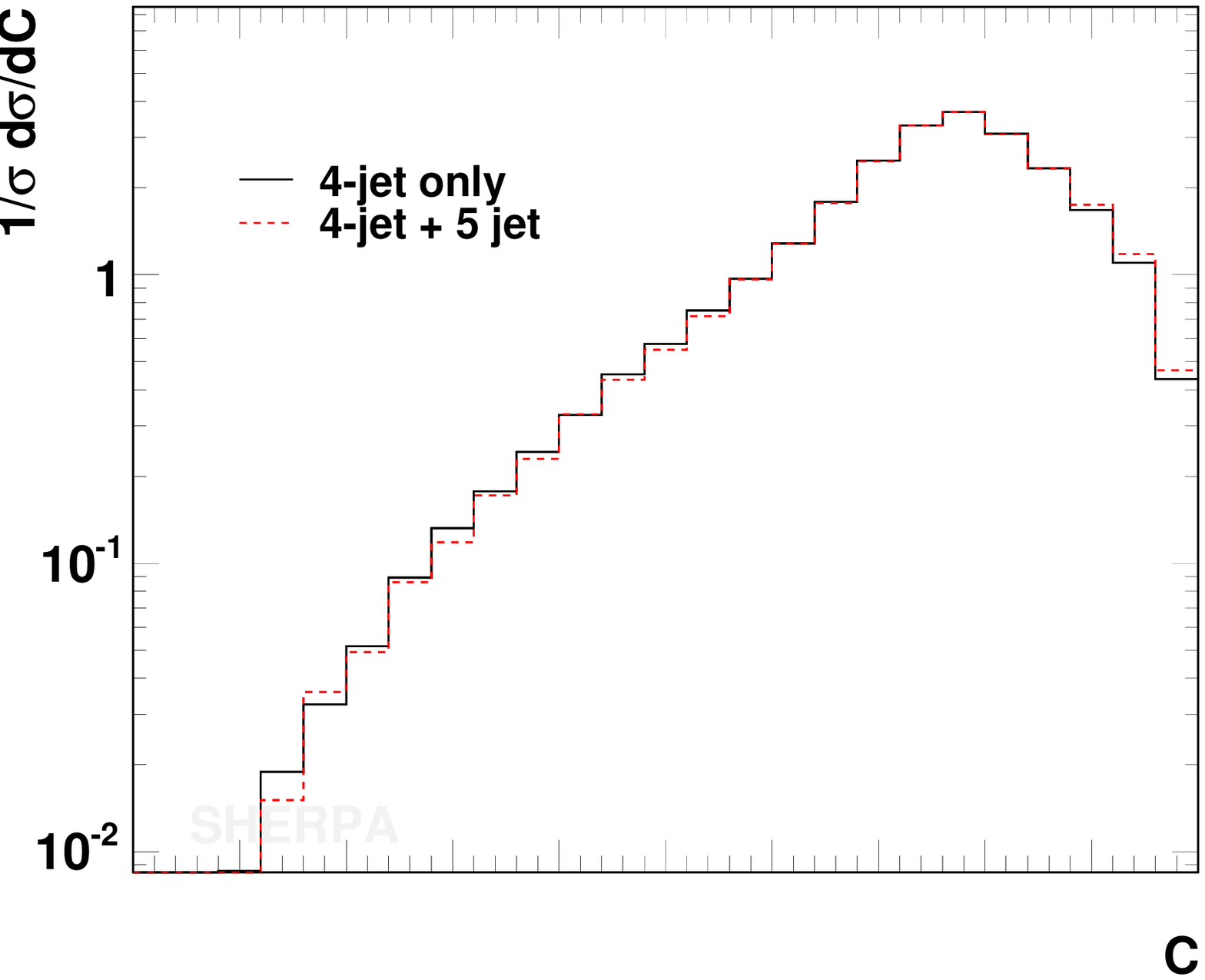}}
\put(280,0){\includegraphics[width=5.2cm]
  {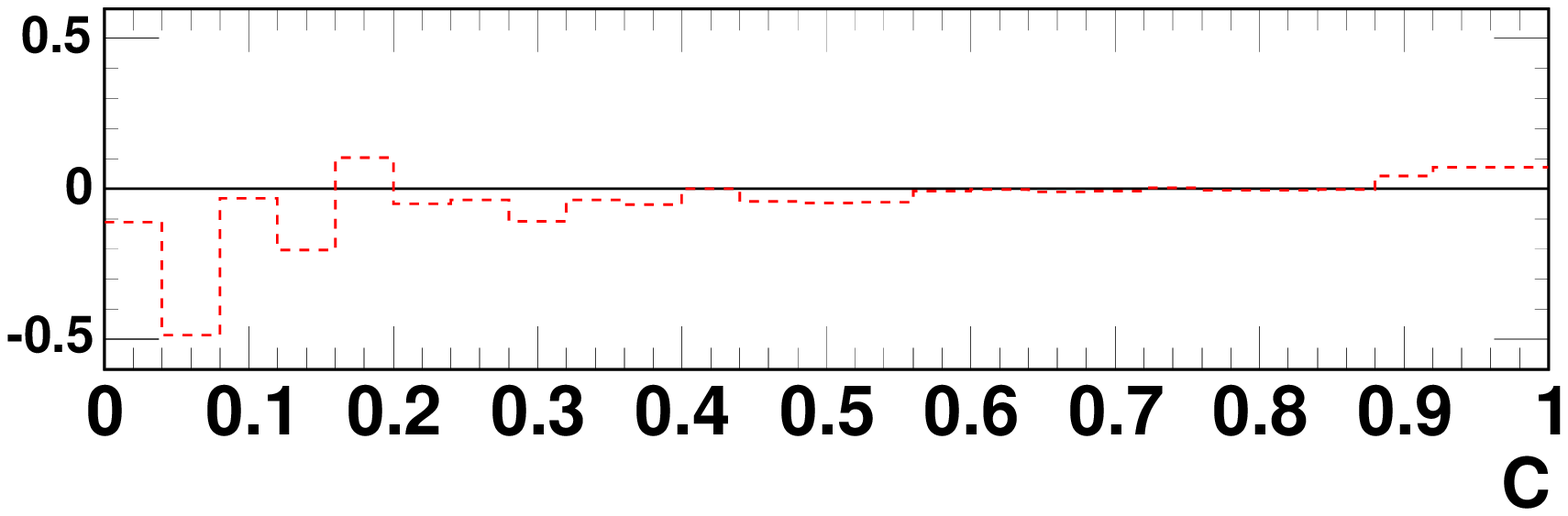}}
\put(140,0){\includegraphics[width=5.2cm]
  {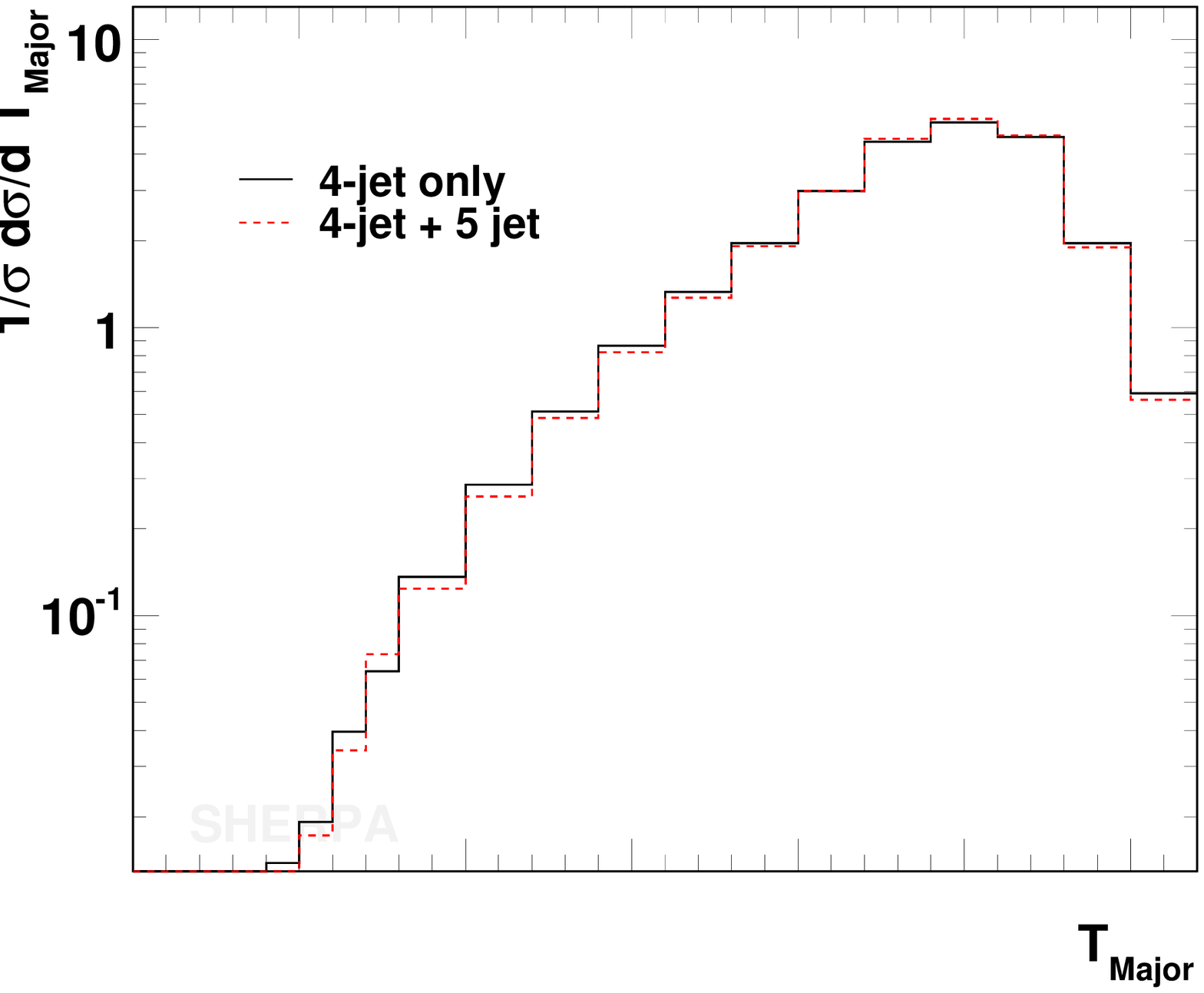}}
\put(140,0){\includegraphics[width=5.2cm]
  {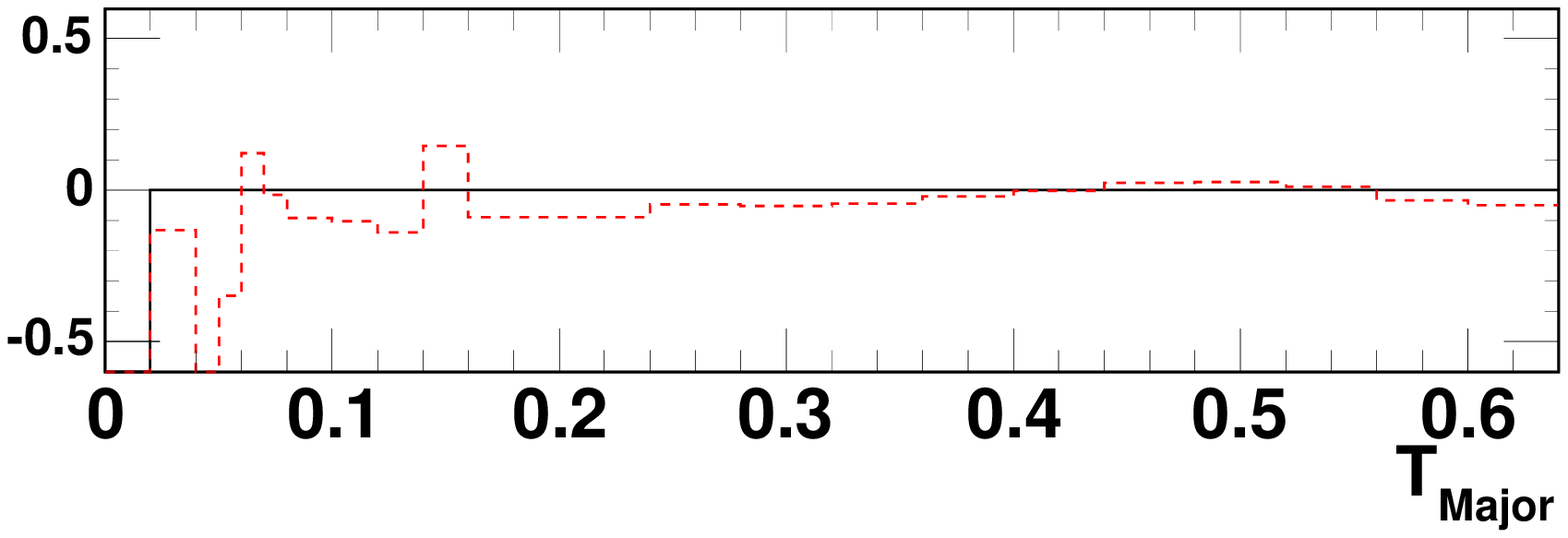}}
\put(0,0){\includegraphics[width=5.2cm]
  {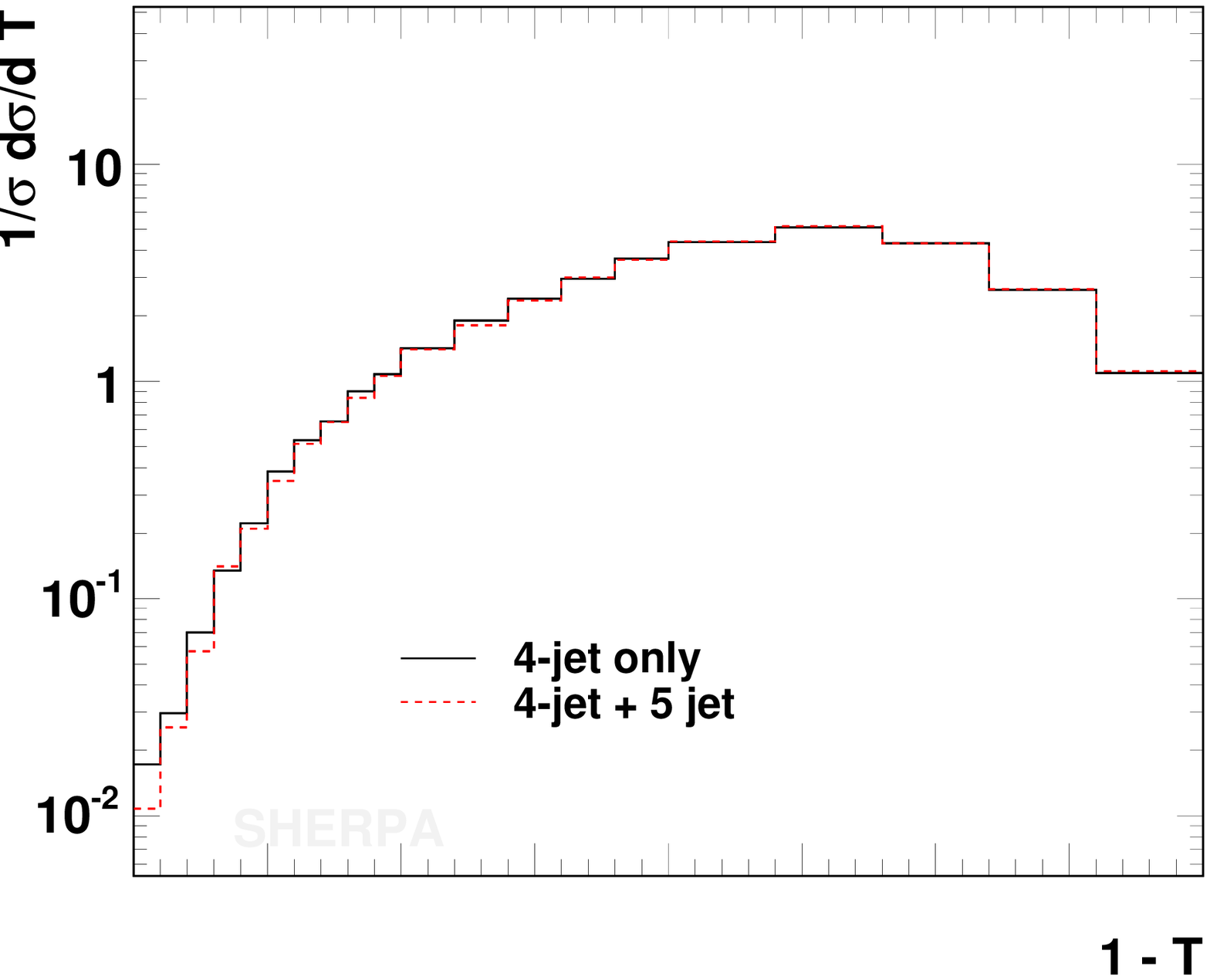}}
\put(0,0){\includegraphics[width=5.2cm]
  {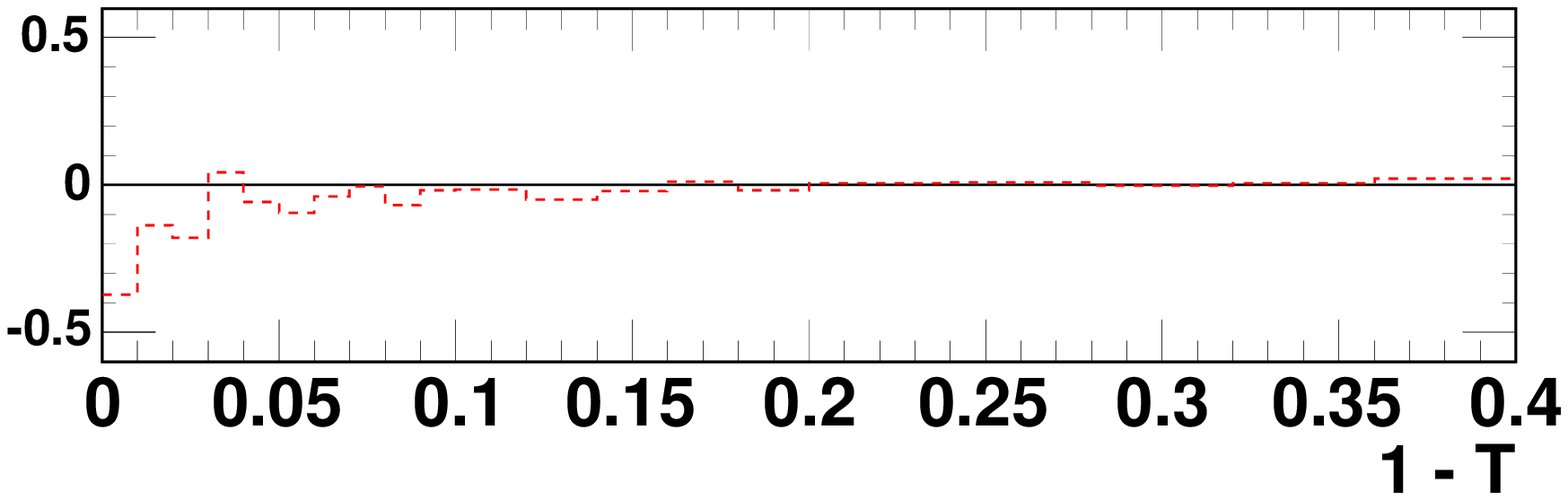}}
\end{pspicture}
\end{center}
\caption{\label{fig:lep_ew2_event} Thrust, thrust-major and
  $C$-parameter in electroweak four jet events at LEP2. Results of 
  SHERPA for a merged (4+5)-jet sample ($y_{\rm cut}=10^{-2.2}$)
  are contrasted with those of a pure 4-jet sample where the parton
  shower was running freely.}
\end{figure*}

\begin{figure*}[h]
\begin{center}
\begin{pspicture}(450,190)
\put(280,40){\includegraphics[width=5.5cm]
  {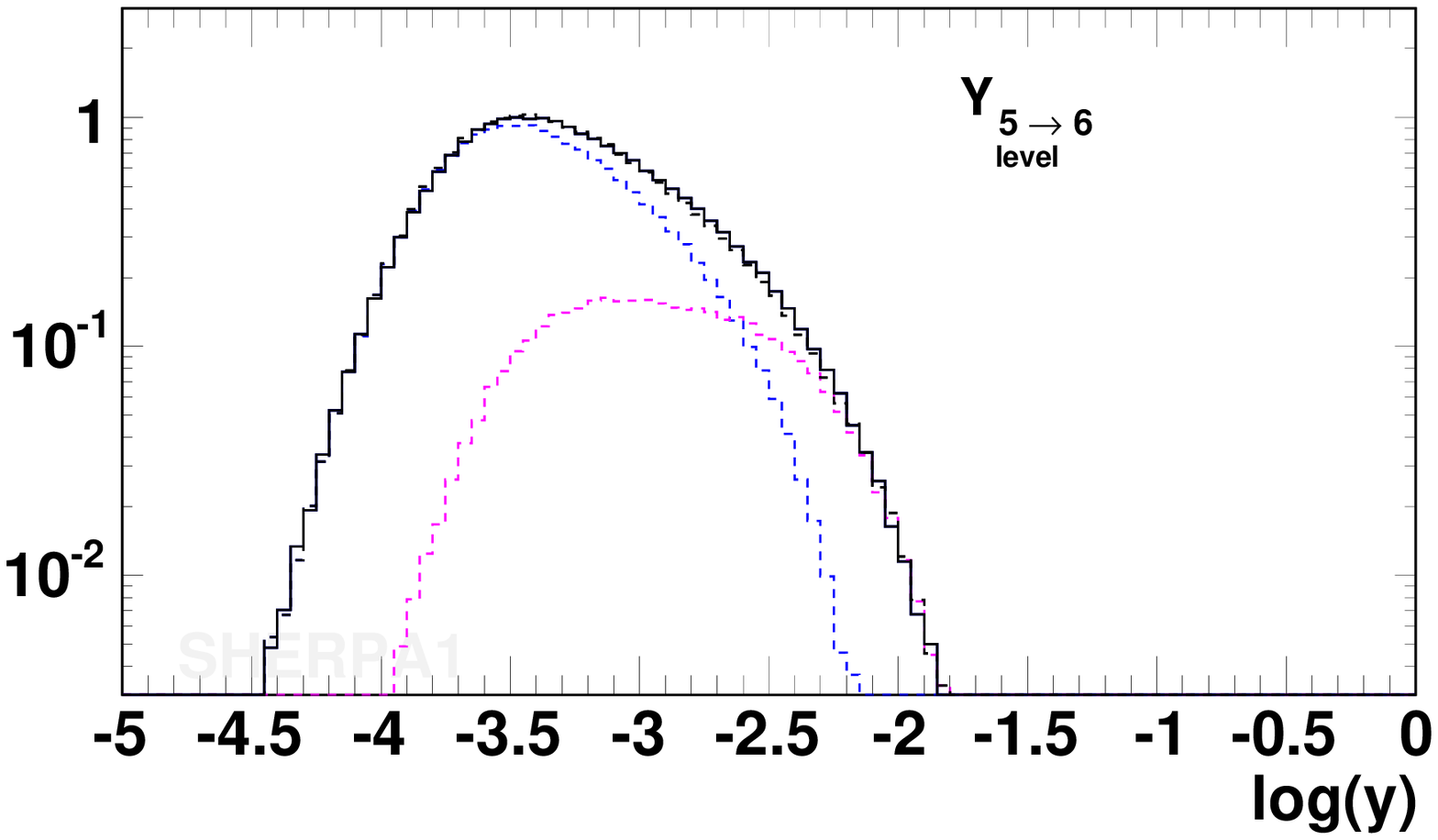}} 
\put(140,40){\includegraphics[width=5.5cm]
  {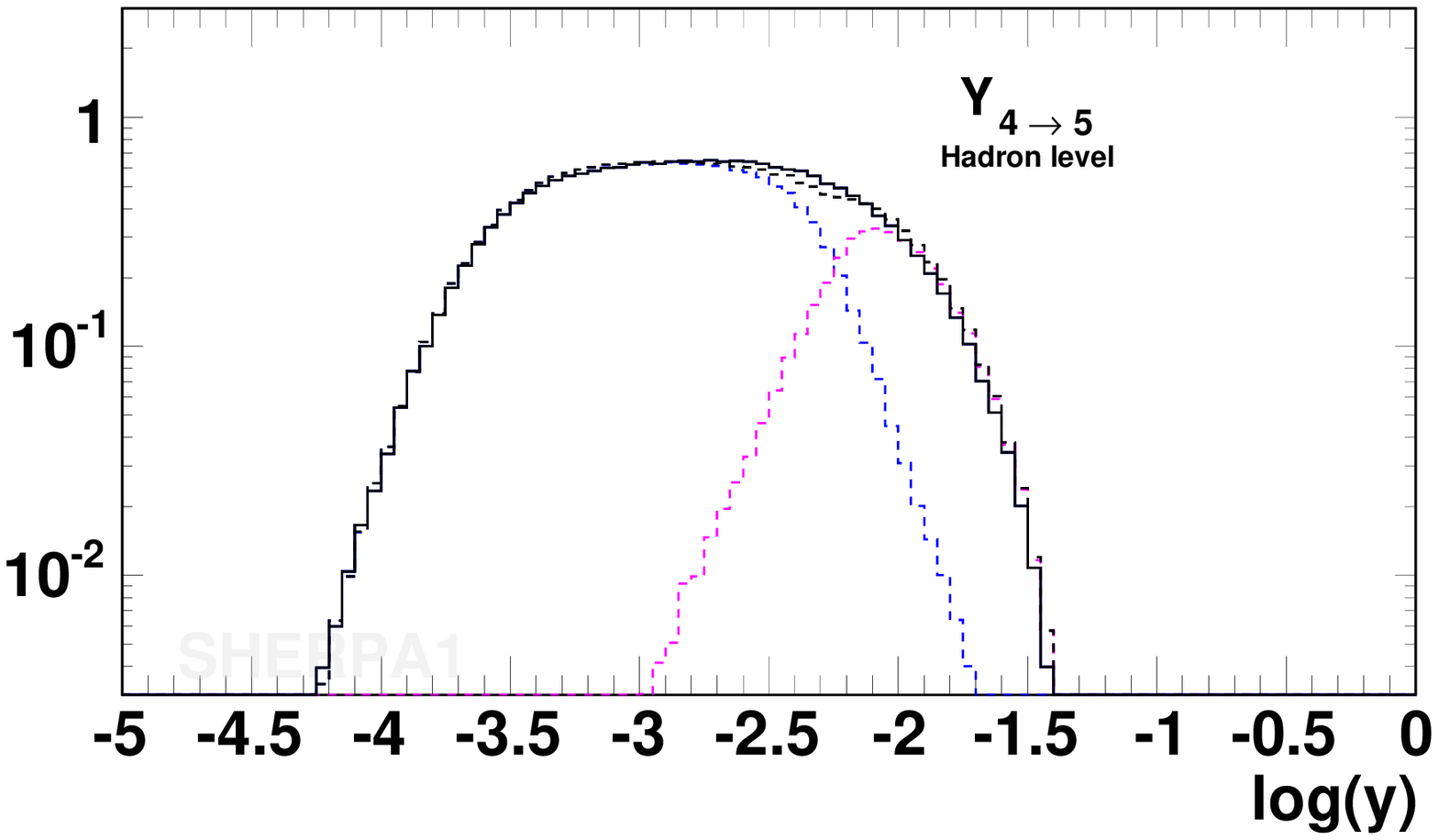}}
\put(0,40){\includegraphics[width=5.5cm]
  {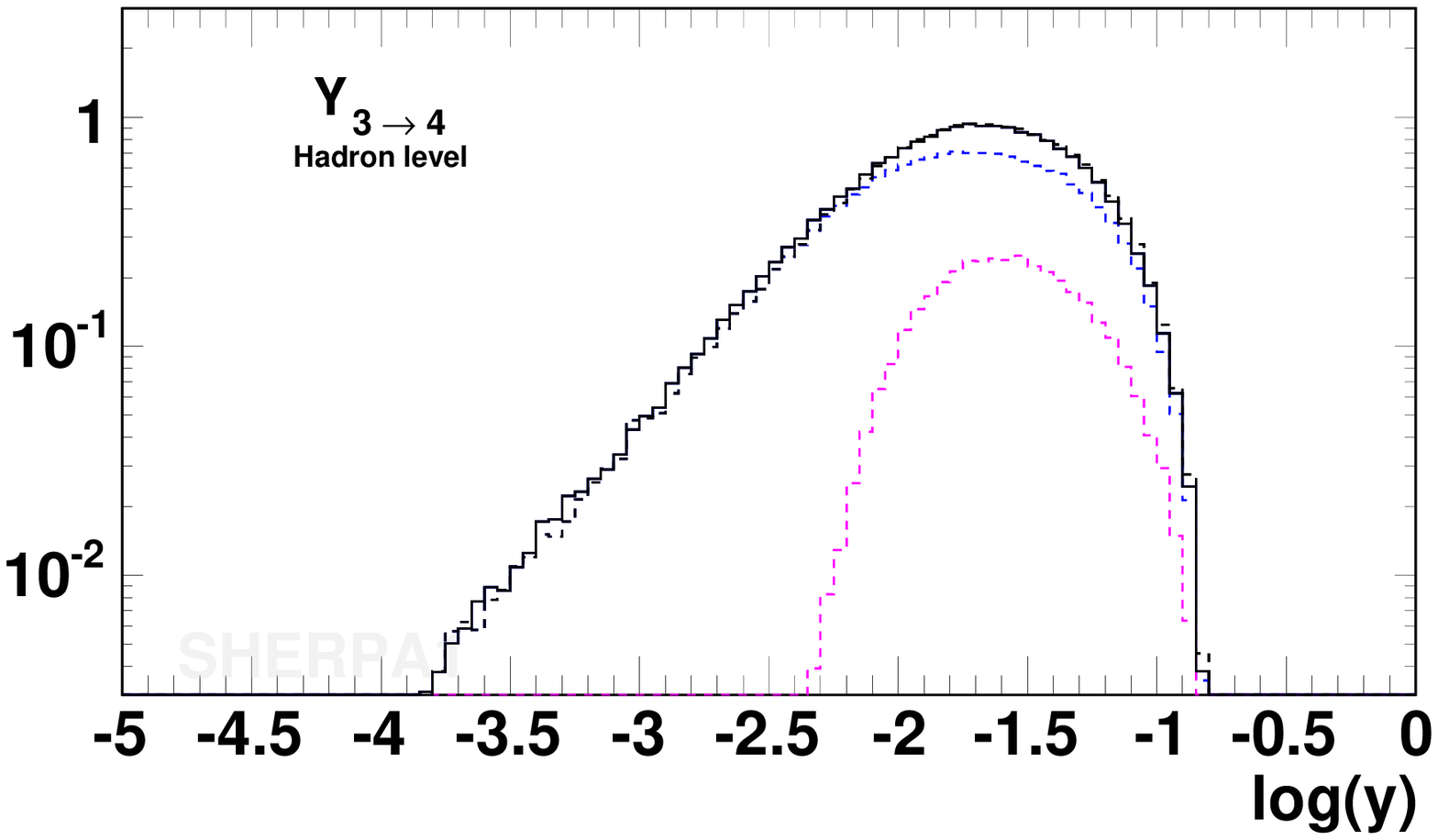}}
\put(280,0){\includegraphics[width=5.5cm]
  {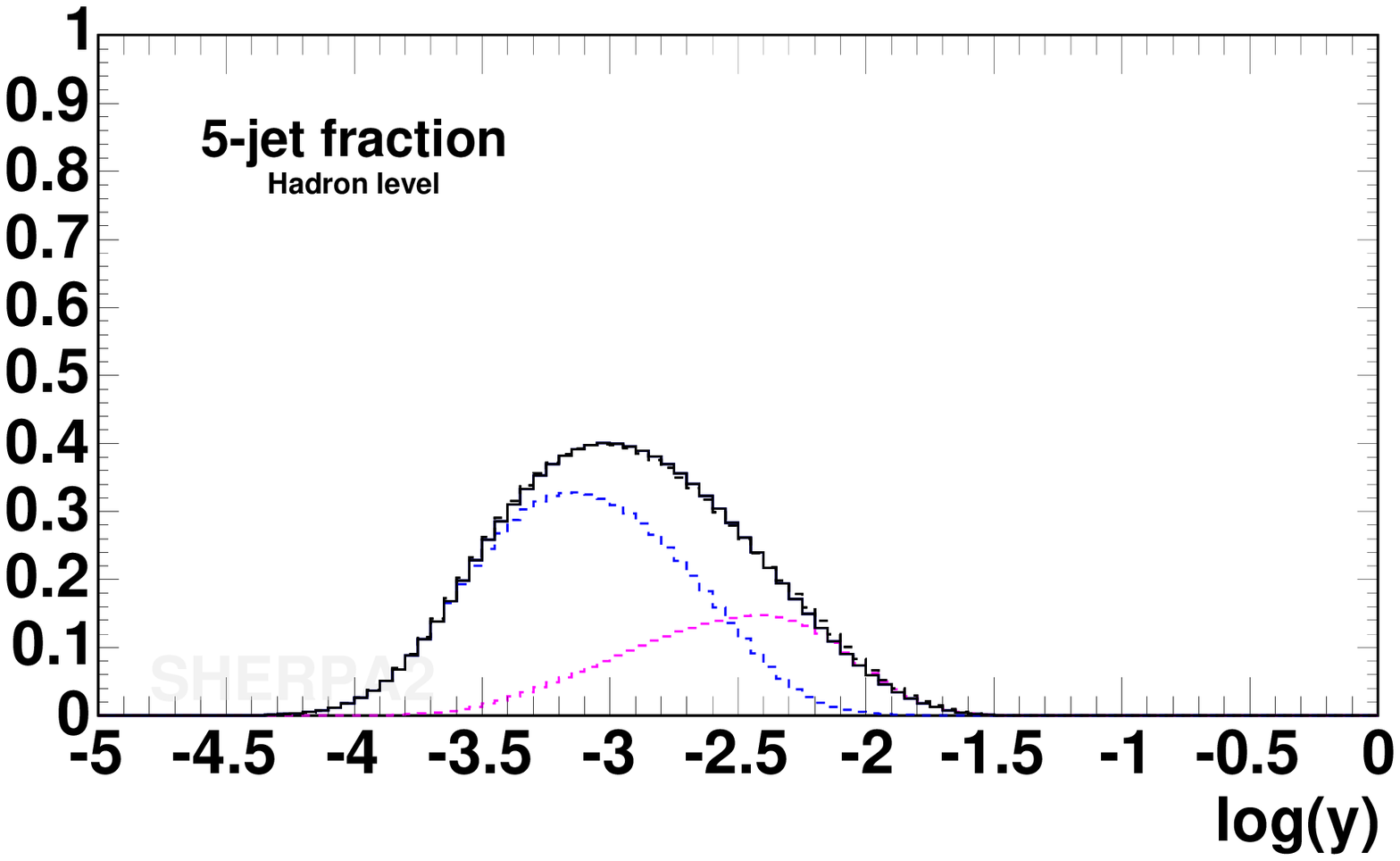}}
\put(140,0){\includegraphics[width=5.5cm]
  {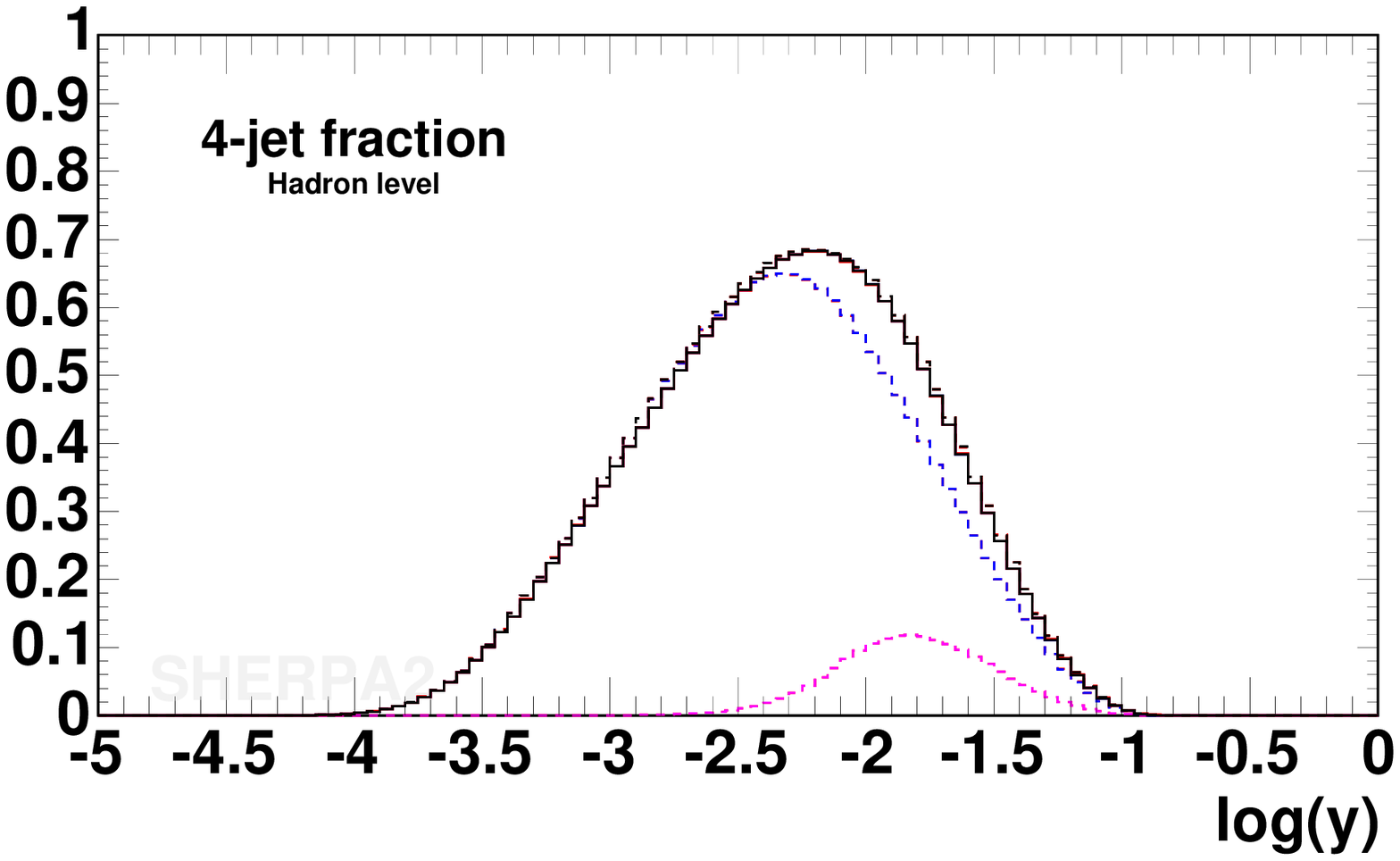}}
\put(0,0){\includegraphics[width=5.5cm]
  {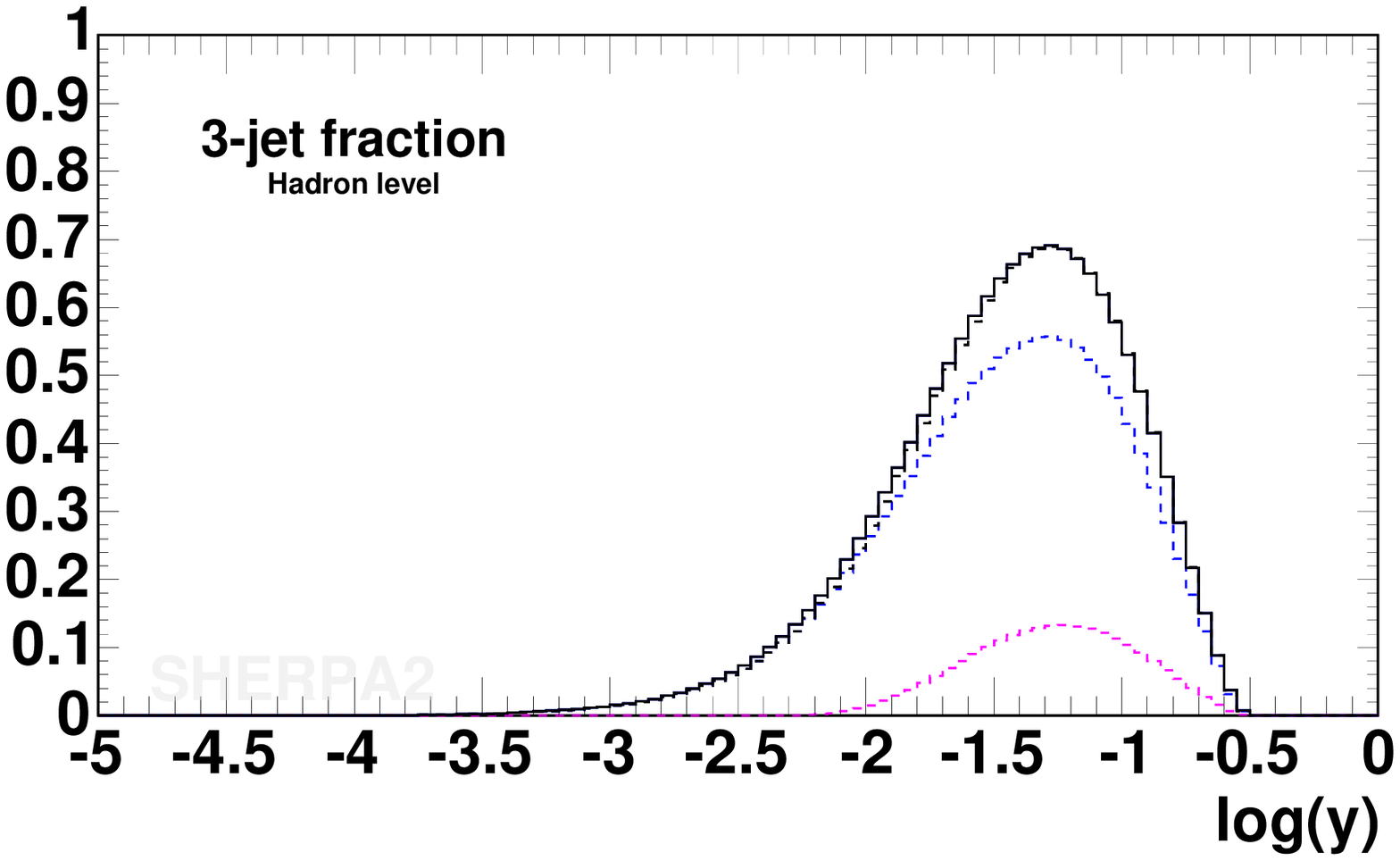}}
\end{pspicture}
\end{center}
\caption{\label{fig:lep_ew2_diffjets} Differential (upper row) and
  total (lower row) jet-rates in electroweak four jet events at
  LEP2. Results of a merged (4+5)-jet sample ($y_{\rm cut}=10^{-2.2}$) are
  contrasted with those of a pure 4-jet sample.}
\end{figure*}

\noindent
This finding gives a clear hint that the pole structure of
propagators has to be taken into proper account when merging such
matrix elements with the parton shower. Thus, in the following, the
focus will be on the self-consistency of such an approach. To
investigate this, again differential and total jet rates are
considered. In Fig.\ \ref{fig:lep_ew2_diffjets} corresponding results
for a $4$ jet and for a combined $(4+5)$-jet sample produced according
to the alternative algorithm are contrasted with each other. They are
in nice agreement, hinting that the combination of exclusive samples
into an inclusive one was successfully achieved. In Fig.\
\ref{fig:lep_ew2_event} the corresponding event shape observables are
shown. There, the differences between both samples are marginal; they
differ only in the low-statistics bins. Note that in all Figs.\
\ref{fig:lep_ew1_event}-\ref{fig:lep_ew2_diffjets} the multi-cut
treatment has been employed. The 4-jet matrix element cut has been
chosen to $y_{\rm cut}^{(4)}=10^{-4}$, while the 5-jet matrix
element was separated by $y_{\rm cut}^{(5)}=10^{-2.2}$. This is
important, since there exists no 3-jet matrix element, which could
compensate for the phase space cut in the 4-jet matrix element.

\clearpage

\section{Summary and Conclusion}

\noindent
In this publication, the procedure for a consistent merging of matrix
elements for the production of multi-particle final states at
tree-level and the parton shower has been discussed in great detail,
going beyond the scope of previous publications on that subject. In
particular, some improvements of the method have been presented which
consistently treat situations, where the parton shower must fill the
phase space for the production of jets which is not covered by
corresponding matrix elements. In addition, some ideas of how to
extent the original algorithm to cases were electroweak and strong
interactions compete have been set forth. A large number of examples
highlights how the algorithm works in various cases, results clearly
demonstrate its ability to yield reliable and predictive results in 
$e^+e^-$ annihilations and in hadronic collisions. 

\noindent
Apart from the presentation of the method, its implementation into the
new event generator SHERPA has been discussed. The relevant classes
are described in sufficient detail to allow users of SHERPA to
implement own ideas or to cross-check systematically the behaviour of
the algorithm in cases not covered here.

\end{fmffile}

\begin{appendix}

\section{\label{Det_Sec} Brief program documentation}

\noindent
The module, in which the merging algorithm is implemented, is an
integral part of the SHERPA framework. It is situated inside the main 
module SHERPA, and it employs SHERPAs basic physics tools, e.g., 
four-momenta, parton distribution functions, and jet algorithms. Of
course, in its present form it has many connections to specific
features of SHERPAs matrix element generator AMEGIC and its parton
shower module APACIC. An extension to other matrix element generators
or parton showers, however, is straightforward.

\noindent
This section gives a brief overview over the classes responsible for
the merging and their specific tasks within the algorithm. Where
needed, details on specific implementation issues are presented that
should, in principle, enable the interested user to implement and test
some of his or her own ideas. 

\subsection{Implementation}

\begin{figure*}[t]
\begin{center}
\includegraphics[width=12.5cm]{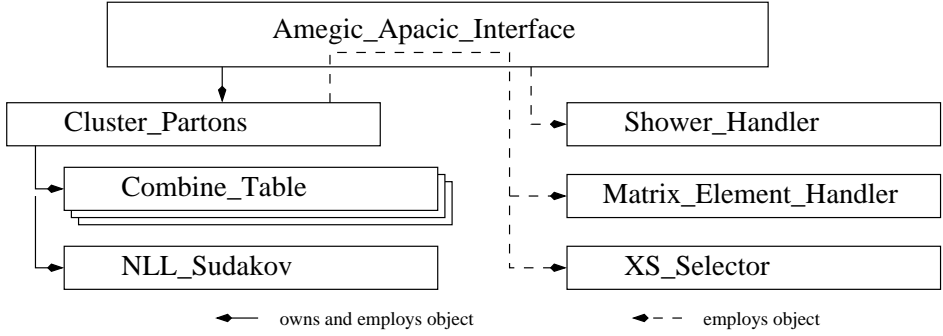}
\end{center}
\caption{\label{fig:merging_classes} Ownership of the main classes
  responsible for the merging of ME and PS. }
\end{figure*}

\noindent
The basic algorithmic steps underlying the realisation of the merging
algorithm in SHERPA can be summarised in the following way:
\begin{enumerate}
\item First of all, the pseudo parton shower history is
  reconstructed. To simplify the presentation, the focus here is on
  the implementation of the original approach only. Modifications to
  the extension described above can be found in the detailed
  description of the individual classes.
  \begin{itemize}
  \item Take all Feynman diagrams with a binary tree structure, i.e.\
    those that contain vertices with three legs only. For a given
    $2\to n$ process the resulting structure will have $n+2$ external particles.  
    In AMEGIC, this doubly linked binary tree structure is represented
    through the class {\tt Point}, each {\tt Point} contains pointers to
    its predecessor and offsprings. 
    In the merging procedure the {\tt Point}s of each Feynman diagram
    that correspond to an external particle are translated into a {\tt Leg}.
    The merging is
    performed in terms of the {\tt Leg}s, which ensures that the
    underlying Feynman diagram structure is not modified through the
    algorithms\footnote{
      Of course, any other matrix element generator with an
      internal representation of Feynman diagrams through doubly
      linked binary trees can easily be treated in the same way. If
      such a binary tree structure does not exist, it must be
      provided.}.  
  \item Test all pairs of external particles, i.e.\ {\tt Leg}s. In the
    original version of the merging algorithm, for each allowed pair the
    relative transverse momentum according to the $k_\perp$ algorithm is
    calculated. Pairings which do not correspond to a junction in the
    Feynman diagrams, are discarded. Each allowed pairing is stored in a
    table, conveniently represented as a class {\tt Combine\_\-Table},
    together with the list of diagrams where it occurs and with the
    $k_\perp$ value. Each {\tt Combine\_\-Table} has pointers to the
    previous one and its successor, i.e.\ to a {\tt Combine\_\-Table}
    with one {\tt Leg} more, and to another one, with one {\tt Leg}
    less.  
  \item In this {\tt Combine\_\-Table} the pairing with the smallest
    $k_\perp$ is selected. Their common predecessor is obtained from the
    first Feynman diagram(s) - in the original approach, the flavour of
    it is an unique choice anyhow. Its four-\-momentum is the sum of the
    two momenta of its offsprings, taken together this fully defines the
    intermediate particle, i.e.\ the corresponding {\tt Leg}. It
    replaces the two offsprings and it is used for the next round of
    clustering, operating on a duly reduced number of {\tt Leg}s. All
    diagrams that did not contain the selected splitting are discarded
    in the further procedure. 
  \item The procedure terminates as soon as a splitting results in a
    structure with four external legs, i.e.\ a $2\to 2$ process. 
  \end{itemize}
\item The Sudakov weight for the selected configuration is evaluated.
  \begin{itemize}
    \item The starting point is the core $2\to 2$ process. Its
      hard scale $Q_h$ is defined though the colour structure; in case
      there are different competing colour structures the winner is
      selected according to the relative weights. The details of this
      are implemented in an extra module of SHERPA, basically a
      library of $2\to 2$ processes called EXTRA\_XS. It incorporates
      the processes as realisations of an abstract base class, 
      {\tt XS\_\-Base}, the relevant one is chosen through an
      {\tt XS\_\-Selector}. 

      \noindent
      In addition, the $2\to 2$ process determines the scale for
      the coupling weight, $Q_{\rm QCD}$. 
      Then, however, this core process may result in a factor of
      \begin{align}
        \left[\alpha_S(Q_{\rm QCD})/
          \alpha_S(Q_{\rm cut})\right]^m\,,
      \end{align}
      where $m$ is the number of strong interactions in the core process.
      In most cases, these
      two scales are identical, exceptions are, for instance, the
      process $e^+e^-\to q\bar q$, which for sure has no strong 
      interaction, and therefore no scale $Q_{\rm QCD}$, cf.\ Sec.\
      \ref{Algo_Sec}. 

At that point, each {\tt Leg} is associated with a
      value $Q_1=Q_h$
    \item The previous {\tt Combine\_\-Table} ``unclusters'' one of
      the particles and yields the corresponding nodal $k_\perp$
      measure, $Q_{\rm 2}$. If the decay of this particle
      proceeds through the strong interaction, the weight is
      multiplied by 
      \begin{align}
        \left[\alpha_S(Q_2)/\alpha_S(Q_{\rm cut})\right]\,.
      \end{align}
      If the decaying particle $a$ is strongly interacting, a Sudakov
      weight is attached, namely
      \begin{align}
        \left[\Delta_a(Q_1,Q_{\rm cut})/
          \Delta_a(Q_2,Q_{\rm cut})\right]\,,
      \end{align}
      where $Q_1$ is the nodal value of the previous iteration step
      for this particle, i.e.\ the $k_\perp$ measure associated to the
      vertex, where it stems from. Then, for the two offsprings
      produced in the decay, their production scale is identified as
      $Q_1=Q_2$.  
    \item If no previous {\tt Combine\_\-Table} exists, there is no
      decaying particle left, and all {\tt Leg}s are external. Then,
      each {\tt Leg} with strong quantum numbers results in a factor
      \begin{align}
        \Delta_a(Q_1,Q_{\rm cut})
      \end{align}
      attached to the Sudakov weight. 
  \end{itemize}
  \item If the event is accepted after the Sudakov weight, the parton
    shower has to be attached. For this, the binary tree structure of
    the {\tt Point}s is translated to the {\tt Tree} structure of
    APACIC. APACIC, however, does not order its shower in terms of
    transverse momenta. Instead it employs an ordering by
    virtuality. Therefore, for each particle, the virtual mass of its 
    production vertex is identified and used as the starting scale of
    the parton shower evolution. Again, this is easily accomplished by
    just following the {\tt Combine\_\-Table}, starting from the core
    process\footnote{ 
      Any other parton shower algorithm can be used in a similar
      fashion, even when it is operating in terms of dipoles.}.    
\end{enumerate} 

\subsection{Steering}

\subsubsection*{The class {\tt Amegic\_Apacic\_Interface}}

\noindent
is the central interface class, steering the various steps of the
merging procedure. It is derived from the abstract class 
{\tt Perturbative\_\-Interface}. Each {\tt Perturbative\_\-Interface}
owns pointers to a {\tt Shower\_\-Handler} and to a 
{\tt Matrix\_\-Element\_\-Handler}, which provide access to the
internal structure of the parton shower and to the matrix elements,
respectively. In principle, other shower algorithms or another matrix
element treatment can easily be connected to the SHERPA framework -
from SHERPAs point of view merely the two handler classes (the 
{\tt Shower\_\-Handler} and the {\tt Matrix\_\-Element\_\-Handler})
would have to be suitably extended, and a corresponding interface of
the type {\tt ME\_PS\_Interface} would have to be constructed. 

\noindent
The {\tt Amegic\_Apacic\_Interface} is used through consecutive calls
to the following methods.
\begin{itemize}
\item {\tt DefineInitialConditions()}\\
  performs all steps necessary in order to construct a
  pseudo parton shower history and its corresponding weight, to 
  accept or reject this configuration, and, eventually, to initialise
  the parton shower. In particular for the first task, it heavily
  relies on the helper class {\tt Cluster\_\-Partons}, presented
  below. Depending on the success of the procedure, the integer return
  value of this method is ``0'', ``1'' or ``3'' indicating a rejected
  event, an accepted event, or a rejected event after the 
  {\em lose-jet-veto}, respectively. 

  \noindent
  {\tt DefineInitialConditions()} executes the following steps: 
  \begin{enumerate}
  \item
    Cluster the matrix element configuration to a $2\to 2 $ core process
    by calling {\tt Cluster\-Configuration()}.
  \item
    Determine the starting scale and colour connections of this core
    process with the help of an {\tt XS\_Base} from the EXTRA\_XS
    library. The corresponding process is selected through a call of
    {\tt Cluster\_\-Partons::\-GetXS()}.
  \item
    Evaluate the NLL Sudakov weight used for reweighting the ME
    kinematics through {\tt Cluster\_\-Par\-tons::\-CalculateWeight()}.
    Accept or reject the configuration accordingly.
  \item
    If accepted initialise the parton shower evolution by employing 
    {\tt Cluster\_\-Par\-tons::\-Fill\-Trees()}.
  \end{enumerate}
\item {\tt ClusterConfiguration()}\\
  is used to obtain a pseudo parton shower history. The clustering
  actually is achieved in the helper class, through the method 
  {\tt Cluster\_\-Partons::\-Cluster\-Confi\-guration()}. This method
  merely forms an intelligent wrapper around it, and it prepares
  merging {\tt Blob}s of the type "ME PS Interface". These {\tt Blob}s
  are used to translate the on-shell partons from the matrix element
  into the off-shell partons experiencing the parton shower. Therefore
  they are filled after the parton shower evolution. The latter is
  triggered by
\item {\tt PerformShowers()},\\
  which calls the appropriate routine in the {\tt Shower\_\-Hander}. 
  The jet-veto scale and the renormalisation scale, which have
  been determined in the merging procedure before\footnote{
    Remember, they may change because of, e.g., the highest
    multiplicity treatment described in Sec.\ \ref{Algo_Sec}.}, are
  handed over to the parton shower, also through the 
  {\tt Shower\_\-Hander}. If the shower evolution was successful,
\item {\tt FillBlobs()}\\
  inserts the prepared and filled ``ME PS Interface'' and the shower
  {\tt Blob}s into the event record.
\end{itemize}
Apart from {\tt ClusterConfiguration()}, which is obsolete for
instance for $2\to 2$ processes, these general methods have to be
provided by any realisation of a {\tt Perturbative\_\-Interface}.

\subsubsection*{The class {\tt Cluster\_Partons}}

\noindent
is the class central to the implementation of the merging
algorithm. It has three main routines, and a number of helper methods,
which will be discussed in the following:
\begin{itemize}
\item {\tt ClusterConfiguration()}\\
  is the method that clusters a given $2 \to n$ process until a $2
  \to 2$ core process remains. In so doing, it creates a history
  of successive emissions, each of which is associated with a
  specific emission scale, the nodal value of the respective
  clustering. The algorithm for the clustering implemented here
  proceeds as follows:
  \begin{enumerate}
  \item All possible Feynman graphs are iterated over. In AMEGIC, a
    diagram consists of a doubly linked tree of {\tt Point}s. They
    represent vertices, whereas the links are the propagating
    particles. Also, the external particles of each diagram are
    represented as {\tt Point}s, but with all but one of the links
    empty. The number of diagrams and these {\tt Point} structures
    themselves are accessible through the methods 
    {\tt Matrix\_\-Element\_\-Handler::Number\-Of\-Diagrams()} and 
    {\tt Matrix\_\-Element\_\-Handler::GetDiagram()}, respectively.
    However, the external particles of each diagram, both incoming and
    outgoing are translated into {\tt Leg}s, on which the actual
    clustering is performed without disturbing the {\tt Point}s
    underneath. 
  \item These first {\tt Leg}s and their four-momenta are stored in a 
    {\tt Combine\_\-Table}. Ultimately, it is this class, which, step
    by step, clusters two particles, i.e.\ {\tt Leg}s into an
    intermediate particle, i.e.\ {\tt Leg}. Its four-momentum in due
    course will be given by the appropriate combination of the two
    incident particles. As a result of this particular step, a new 
    {\tt Combine\_\-Table} emerges with the number of {\tt Leg}s 
    diminished by one, which is linked to the previous one. 
    
    \noindent
    Having thus filled the first {\tt Combine\_\-Table} through
    its method {\tt Fill\-Table}, the list of all emerging 
    {\tt Combine\_\-Table}s is constructed by calling {\tt CalcJet()}
    of the fist one. 
  \end{enumerate}
\item {\tt GetXS()}\\
  identifies the hard $2 \to 2$ core process. In particular, it
  determines the colour structure of it, and the relevant hard
  scale(s). The preferred way to carry out this task is to employ an
  internal library of analytical $2\to 2$ processes provided by the
  module EXTRA\_XS. An implemented cross section calculator can be
  obtained by {\tt XS\_Selector::GetXS()}, selecting the process in
  question through the flavours of its external particles. The cross
  section calculator is realised as an {\tt XS\_\-Base}, and it has
  suitable routines available for selecting colour connections
  ({\tt XS\_Base::SetColours()}) and for retrieving a renormalisation
  scale ({\tt XS\_Base::Scale()}).    

  \noindent
  An alternative solution exists for those processes which are not
  implemented yet but for which the colour connections are
  unambiguously defined. This is actually always the case if the
  number of strongly interacting particle involved is smaller than
  four. Then, the colour connections are explicitely constructed,
  using the routine {\tt Cluster\_\-Partons::\-Set\-Colours()}.
  In this case the hard scale reads
  \begin{align}
    Q_{\rm hard}^2 &=
    \left\{   
      \begin{array}[c]{lll}
        p_\perp^2 + p_3^2 + p_4^2 & \mbox{if initial and final state
          are colour connected, and} \\[2mm]
        (p_1+p_2)^2 & \parbox{8cm}{if there is no colour connection
          between \\[-1mm] initial and final state,}
      \end{array}
    \right.
  \end{align}
  with $p_1$/$p_2$ and $p_3$/$p_4$ denoting the four-momenta of the
  incoming and outgoing particles, respectively. In case there are 3
  coloured particles involved in the hard process, a scale for the
  evaluation of the strong coupling has to be determined too, which is
  identified with the transverse momentum $p_\perp$ of the outgoing
  coloured particle. 
\item {\tt CalculateWeight()}\\
  follows the previously obtained history and calculates from it the
  corresponding Sudakov weight according to the merging
  prescription described above. For this, the nodal values $Q_i$ 
  determined before in {\tt Cluster\-Configuration()} are employed.
  The start scale for the Sudakov weight $Q_h$ and the scale
  $Q_{\rm QCD}$ for possible $\alpha_S$ factors have been prepared by 
  {\tt GetXS()}/{\tt SetColours()}, see above.

  The construction of the weight starts with the core $2\to 2$ process.
  Hence, the first part of the weight is given by a factor
  $({\alpha_S(Q_{\rm QCD})}/{\alpha_S(Q_{\rm cut})})^m$, with $m$ specifying the
  number of strong couplings involved in the hard process. The
  clustering is followed backward through the sequence of 
  {\tt Combine\_\-Table}s, adding a factor   
  \begin{align}
    w &= \frac{\Delta(Q_i,Q_{\rm cut})}{\Delta(Q_j,Q_{\rm cut})} 
    \frac{\alpha_S(Q_j)}{\alpha_S(Q_{\rm cut})} \,,
  \end{align}
  for each internal line constructed during the backward
  clustering. The Sudakov form factors $\Delta(Q_i,Q_{\rm cut})$ are
  provided by  the method {\tt Delta()} of the class {\tt NLL\_Sudakov}.
  The algorithm ends with a factor as $\Delta(Q_i,Q_{\rm cut})$ for any
  dangling coloured particle, with possible another coupling weight in
  case of the extended merging algorithm (see
  Sec.\ \ref{Algo_Sec})\footnote{
    Note that the treatment of  matrix element events with a maximal
    number of outgoing particles is slightly modified, however, the
    general algorithm remains the same.}.
\item {\tt FillTrees()}\\
  translates the pseudo parton shower history into the {\tt Tree}
  structures, which APACIC uses to represent the parton shower. This
  history includes the starting scales of the shower evolution of each
  parton and possible constraints on shower emissions such as, e.g.,
  opening angles determined from colour connections of the partons.
  The {\tt Knot}s forming the {\tt Tree} are taken from the 
  pseudo parton shower through {\tt Point2Knot()}, the mutual
  relations are constructed through {\tt Establish\-Relations()}.
\item {\tt EstablishRelations()}\\
  builds a tree by creating mutual links between a given set of three
  {\tt Knot}s. At each step, the actual {\tt Tree} represents a
  partially performed shower. For the mutual relations, three cases
  are distinguished  
  \begin{enumerate}
  \item two incoming partons from the hard $2 \to 2$ process:\\
    the energy fractions $x_1$ and $x_2$ are filled from the information 
    in the {\tt Combine\_\-Table}.
  \item two outgoing partons from a common mother:\\ 
    The two final state particles are initialised using
    {\tt Final\_\-State\_\-Shower::Esta\-blish\-Relations()}. There,
    the more energetic parton is initialised with the angle and
    virtuality of the mother, the less energetic parton is initialised
    with angle and virtuality of the current branch. 
  \item one incoming parton, its mother and its sister:\\
    The incoming parton, its mother and its sister are initialised
    using {\tt Initial\_\-State\_\-Shower::Set\-Colours()}. Note,
    angle conditions inside the shower are fixed only during shower
    evolution. The starting scale of the shower is given by the
    virtual mass of the mother due to APACICs shower evolution in
    terms of virtualities.  
  \end{enumerate}
\item {\tt DetermineColourAngles()}\\ 
  determines the maximum angle between colour connected partons of
  a hard $2 \to 2$ process. These angles are used in the explicit
  angular vetoes of the parton shower.

  \noindent
  For initial state particles the colour angle is determined  in the
  lab frame after a boost along the $z$-axis, whereas starting angles for
  the final state system are determined in its c.m. frame.
  The starting angles are stored in the variable {\em ``thcrit''} of
  each knot. 
\end{itemize}

A number of simple access methods make the result of the clustering
process available to the interface class {\tt Amegic\_Apacic\_Interface}.
\begin{itemize}
\item {\tt Weight()} 
  returns the weight calculated in {\tt Calculate\-Weight()}.
\item {\tt Scale()}
  returns the hardest scale (of the core process) as determined in
  {\tt SetColours()}. 
\item {\tt AsScale()} 
  returns the scale associated with the strong coupling in the core
  process as determined in {\tt SetColours()}. 
\item {\tt Flav()} 
  provides the flavours of the core $2 \to
  2$ process. 
\item {\tt Momentum()} 
  returns the momenta of the core $2 \to 2$
  process.  
\end{itemize}

\subsection{Clustering}

\subsubsection*{The class {\tt Combine\_\-Table}}

\noindent
provides the structure for storing histories of successive
clusterings. The structure fills itself recursively, with the only
input being the Feynman diagrams of the process under consideration
and the four-momenta of the current event. 

\noindent
Each {\tt Combine\_\-Table}
consists of a list of possible clusterings (particles $i$, $j$ and the
flavour of the resulting intermediate particle) and the $k_\perp$
values and Feynman diagrams associated with them. These informations
are realised through the classes {\tt Combine\_\-Key} and 
{\tt Combine\_\-Data}, see below. In addition, a number of methods 
allows a {\tt Combine\_\-Table} to create these data and to construct
the sequence of {\tt Combine\_\-Table}s representing the clustering
history:
\begin{itemize}
\item {\tt FillTable()}\\  
  has two tasks to fulfil. First of all, a set of given {\tt Leg}s,
  i.e.\ particles, are filled into the table. 
  Then, all pairs of them are checked whether they can be clustered.
  A clustering is possible only, if it occurs in a corresponding
  Feynman diagram, which disables unphysical parton histories. This 
  check is performed through the method {\tt Combinable()}, see below.
\item {\tt CalcJet()}\\
  evaluates the $k_\perp$ distance of all allowed parton pairs ($i$,$j$) 
  created by {\tt FillTable()} with the {\tt Jet\_Finder}. After that,
  a pair to be clustered is selected according to the merging
  prescription, and a new {\tt Combine\_Table} is constructed, where
  the number of {\tt Leg}s is reduced by one. Consequently, after each
  clustering step, the set of Feynman diagrams is pruned, to include only
  those where the selected combination is possible. The four-momentum of
  the new (joined) {\tt Leg} is given by the corresponding combination
  of the two individual four momenta. The algorithm continues
  recursively with corresponding calls to {\tt FillTable()} and 
  {\tt CalcJet()} until only a $2\to 2$ process remains.
\item {\tt CalcPropagator()}\\
  performs all basic calculations for the determination of cluster
  probability for a given pair ($i$,$j$). This usually includes the
  evaluation of the $k_\perp$ measure, and the invariant mass
  $s_{ij}$. In case of the extended clustering algorithm, an estimate
  for the branching probability is also computed, which includes the
  couplings of that branching process, as well as
  the corresponding propagator. The couplings are available in the
  Feynman diagrams provided by {\tt AMEGIC}.
\item {\tt Combinable}\\
  determines whether two particles, i.e.\ {\tt Leg}s can be
  clustered. To this end, the two {\tt Points} related to the 
  {\tt Leg}s are checked whether they have a common third {\tt Point},
  i.e.\ vertex, linked to them. 
\end{itemize}

To exemplify the description above, consider the representation of a 
{\tt Combine\_\-Table} below. 
\begin{center}
  \begin{tabular}{r@{\&}llll}
    $i$&$j$ & $y_{ij}$     & $graphs$   &  $down$ link to the next table \\
    \hline
    0&2     & 0.0810366    & 2,3,4,5,8  &     \\
    0&4     & 0.0691623    & 6,7        &   $\to$ {\tt Combine\_Table} 2  \\
    1&3     & 0.0844243    & 0,1,6,7,8  &     \\
    1&4     & 0.293399     & 4,5        &     \\
    2&4     & 0.111385     & 0,1        &     \\
    3&4     & 0.215127     & 2,3        &   
  \end{tabular}
\end{center}
For each combination ($i$,$j$) a $k_\perp$ measure $y_{ij}$ and a list
of contributing Feynman $graphs$ is stored. For the winner combination a
$down$ link to a subsequent {\tt Combine\_\-Table} is provided. In
addition the {\tt Combine\_\-Table} contains a list of four-momenta of
the current configuration, and a matrix of all dangling legs (one row
for each graph), as well as a reference to the winner combination
($i$,$j$), and an $up$-link to the table with one combination less
performed. 

\noindent
In case the extended merging algorithm is applied, the 
{\tt Combine\_\-Table} also  has to include further information needed
for the winner determination: the virtuality of the resulting
propagator $s_{ij}$, the estimate of the propagator $prop$, and the
$coupling$ of the corresponding vertices. 

\subsubsection*{The class {\tt Leg}}

\noindent                                %
represents a particle dangling from a Feynman diagram. It stores
all information of a {\tt AMEGIC::\-Point} and an extra
"anti"-flag.

\noindent                                %
Since {\tt AMEGIC::\-Point} is the basic component of a Feynman
graph representation in AMEGIC, it can be conveniently used in the
clustering process to determine possible combinations and resulting
propagators. The additional $anti$-flag helps to keep track 
of charge conjugations during the clustering process. 
In order to access all information of a {\tt Point} more easily the
{\tt operator->} is overloaded. 
\footnote{The overloaded {\tt operator->} can sometimes lead to
  confusion, especially the {\em anti} flag can not be accessed via
  this operator in case a pointer to a leg is used. In this case the
  {\tt operator*} together with the dot has to be used. 
  So always think {\tt Leg} as a synonym for {\tt Point*}.}

\subsubsection*{The class {\tt Combine\_Key}}
is one of the basic elements for the creation of a
{\tt Combine\_\-Table}.  
It includes the numbers of combinable legs ($i$ and $j$) and the
flavour of the resulting propagator. It is used as a key in a fast
access map in {\tt Combine\_\-Table} in order to access the
information placed in a {\tt Combine\_\-Data} object.

\subsubsection*{The class {\tt Combine\_Data}}
is the basic element for the determination of clustering
when using a {\tt Combine\_\-Table}. 
It includes the distance of two legs $i$ and $j$ according to a
$k_\perp$ measure ($y_{ij}$), a list of numbers of
graphs where this combination is possible ($graphs$) and a link to the
new table where those legs have been combined ($down$).

\noindent
In case the extended merging algorithm is active, additional information is
included, namely: the virtuality of the resulting propagator $s_{ij}$,
the estimate of the propagator $prop$, and the $coupling$ of the
corresponding vertices.

\subsection{Weighting}

\begin{figure*}[t]
  \begin{center}
    \includegraphics[width=12.5cm]{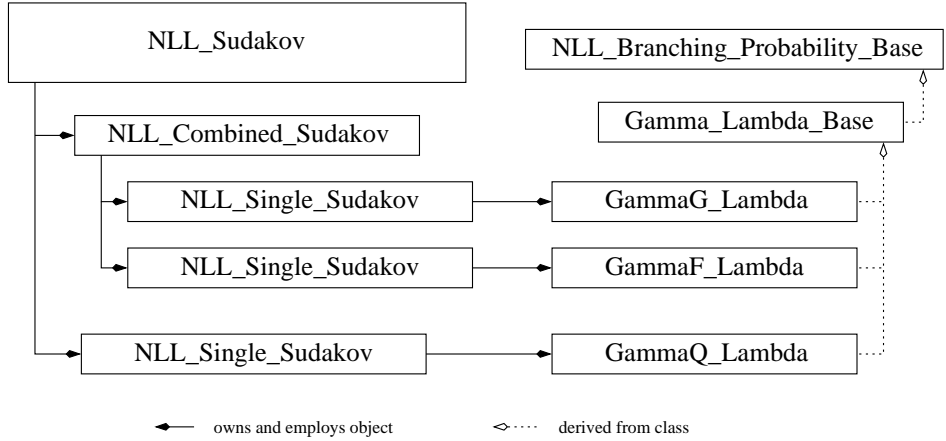}
  \end{center}
  \caption{\label{fig:nll_sudakov_classes} Ownership and inheritance diagram
    of the main classes related to the numerical evaluation of NLL
    Sudakov form factors. The sketch corresponds to the status after the
    initialisation via {\tt NLL\_\-Sudakov::\-Prepare\-Map()}.
  }
\end{figure*}

\subsubsection*{The class {\tt NLL\_\-Sudakov}}

\noindent
provides the numerical values for the Sudakov form factors used in the
merging procedure of parton shower and matrix elements. Consequently,
the main routine is {\tt Delta(const ATOOLS::Flavour \&)}, which
returns the Sudakov form factor for a given flavour. A corresponding
table of Sudakov form factor objects for all possible flavours is
created by respective calls to {\tt Prepare\-Map()} or 
{\tt Prepare\-Massive\-Map()}. 

\noindent
In the following a short description of the individual methods of this
class is given. 
\begin{itemize}
\item {\tt Delta (const ATOOLS::Flavour  \&)} \\
  is the main access method to NLL Sudakov form factors. It
  returns the appropriate NLL Sudakov form factor (in form of a
  {\tt NLL\_\-Sudakov\_\-Base} object) for any given flavour. For not 
  strongly interacting particles a reference to a 
  {\tt NLL\_\-Dummy\_\-Sudakov} object is provided.

  \noindent
  For instance, a typical call to determine the numerical value of the
  gluon Sudakov form factor at a scale $Q$ with a jet resolution scale
  $Q_0$ would look like
\begin{verbatim}
      double  dg = sud.Delta(Flavour(kf::gluon))(Q,Q0);
\end{verbatim}
\item {\tt PrepareMap()} \\
  initialises a map with all massless Sudakov form factors
  needed in the Standard Model. In so doing, a Sudakov form factor
  (cf. {\tt NLL\_\-Single\_\-Sudakov} and {\tt NLL\_\-Combined\_\-Sudakov}) 
  is initialised for each strongly interacting flavour (d-, u-, s-,
  c-, b-quark or anti-quark, and gluon) and put into a map for fast
  access. For the sake of completeness, a {\tt NLL\_Dummy\_Sudakov}
  (always one) is added to the map, which will be returned for any
  flavour without a dedicated Sudakov form factor\footnote{
    In order to keep track of all Sudakov objects inserted into the
    map, a list of unique {\tt NLL\_Sudakov\_Base} objects is
    maintained. It is used for proper destruction at the end of a
    run. This double book-keeping allows the usage of the same Sudakov
    object for all quark flavours, since (massless) QCD is flavour blind.}.
  The default massless integrated splitting functions used for the
  evaluation of the Sudakov form factors are {\tt GammaQ\_\-Lambda}, 
  {\tt GammaG\_\-Lambda}, and {\tt GammaF\_\-Lambda}.
\item {\tt PrepareMassiveMap() } \\
  initialises a map with all massive Sudakov form factors. This method
  is very similar to {\tt Prepare\-Map()}. However, using the massive
  version of Sudakov form factors necessitates the initialisation of a
  {\tt NLL\_Single\_Sudakov} for each flavour (d-, u-, s-, c-,
  b-quark) individually being now distinguishable by their mass (cf.\ 
  \cite{Krauss:2003cr,Rodrigo:1999qg}).
  The default massive integrated splitting functions used are  
  {\tt GammaQ\_Lambda\_Massive}, {\tt GammaG\_Lambda\_Massive}, and
  {\tt GammaF\_Lambda\_Massive}. An overview of the implemented
  branching probabilities is given in
  Tab.\ \ref{tab:branching_probabilities}.
\end{itemize}

\subsubsection*{The class {\tt NLL\_\-Sudakov\_\-Base}}

\noindent
is a pure virtual base class, providing an interface to any Sudakov
form factor like object.
 
\subsubsection*{The class {\tt NLL\_\-Single\_\-Sudakov}}

\noindent
provides the Sudakov form factor for a single given integrated
splitting function.
\subsubsection*{The class {\tt NLL\_\-Combined\_\-Sudakov}}

\noindent
provides the Sudakov form factor for a sum of integrated splitting
functions.
\subsubsection*{The class {\tt NLL\_\-Dummy\_\-Sudakov}}

\noindent
is a simple example of an Sudakov returning always one. It can be used
for only weakly interacting flavours. 

\subsubsection*{The class {\tt NLL\_\-Branching\_\-Probability\_\-Base}}

\noindent
represents a prototype for a branching probability (integrated
splitting function), which can be used in the evaluation of Sudakov
form factors (cf.\ class {\tt NLL\_\-Sudakov}). All realisations are
derived from this class. A list of available branching probabilities
can be found in Tab. \ref{tab:branching_probabilities}.
In general single integrated splitting functions have the form 
\[ 
\Gamma(Q,q) = \int_{z-(q/Q)}^{z+(q/Q)} dz \frac{\alpha_S(q)}{\pi} P(z)\,,
\]
where $\alpha_S$ is the (running) strong coupling and  $P(z)$ is the
splitting kernel.

The class provides methods to access the branching probability
$\Gamma(Q,q)$ through {\tt Gamma(}$q$,$Q${\tt )} as well as the value of the
integrated branching probability 
\[
-\log(\Delta(Q,Q_0)) = \int_{Q_0}^{Q} \,dq \, \Gamma(Q,q)\,.
\]
The latter is accessible through {\tt IntGamma(}$Q_0$,$Q${\tt)}, which
is used as the basis of Sudakov form factors. 
  
\begin{table}[ht!]
\begin{pspicture}(460,520)
\put(0,473){\fbox{%
\begin{minipage}{435pt}
\begin{tabular}{ll}
 \multicolumn{2}{l}{ {\tt Gamma\_AlphaS} and {\tt Gamma\_Lambda}}   \\
& $\displaystyle\Gamma_q(Q,q) = \frac{2C_F}{\pi}\frac{\as(q)}{q}
\left(\log\frac Q q -\frac 3 4\right)$ \\
& $\displaystyle\Gamma_g(Q,q) = \frac{2C_A}{\pi}\frac{\as(q)}{q}
\left(\log\frac Q q -\frac{11}{12}\right)$ \\
& $\displaystyle\Gamma_f^{n_f}(q) = n_f \frac{2
  T_R}{3\pi}\frac{\as(q)}{q} $ \\[3mm]
\end{tabular}
\end{minipage}}}

\put(30,350){\fbox{%
\begin{minipage}{405pt}
\begin{tabular}{ll}
  \multicolumn{2}{l}{ {\tt Gamma\-Q\_\-Lambda}}   \\
& $\displaystyle \int_{Q_0}^Q dq \Gamma_q(Q,q) = 
   \frac{2 C_F}{\beta_0} \left\{\log\frac{Q_0}{Q} + \left( \xi_1 -\frac
       3 4\right) \log\left|\frac{\xi_1}{\xi_0}\right|\right\} $ \\
  \multicolumn{2}{l}{ {\tt Gamma\-G\_\-Lambda}}   \\
& $\displaystyle \int_{Q_0}^Q dq \Gamma_q(Q,q) = 
   \frac{2 C_A}{\beta_0} \left\{\log\frac{Q_0}{Q} + \left( \xi_1 -\frac
       {11}{12}\right) \log\left|\frac{\xi_1}{\xi_0}\right|\right\} $ \\
  \multicolumn{2}{l}{ {\tt Gamma\-F\_\-Lambda}}   \\
& $\displaystyle \int_{Q_0}^Q dq \Gamma_f(q) = 
   n_f \frac{T_R}{3 \beta_0} \log\left|\frac{\xi_1}{\xi_0}\right|$ \\
\end{tabular}
\end{minipage}}}

\put(30,210){\fbox{%
\begin{minipage}{405pt}
\begin{tabular}{ll}
  \multicolumn{2}{l}{ {\tt Gamma\-Q\_\-Alpha\-S}}   \\
&  $ \displaystyle
       \int_{Q_0}^Q dq \Gamma_q(Q,q) = \frac{2 C_F}{\beta_0} \left\{ \log\frac{Q_0}{Q} 
            + 2 \left[  \log\frac{Q}{\mu} - \frac{3}{4}  + 
              \frac{2 \pi}{\beta_0 \as(\mu)} \right] 
            \log\frac{1+\eta_1}{1-\eta_0}
          \right\}
   $ \\
  
  \multicolumn{2}{l}{ {\tt Gamma\-G\_\-Alpha\-S}}   \\
&  $ \displaystyle
       \int_{Q_0}^Q dq \Gamma_g(Q,q) = \frac{2 C_A}{\beta_0} \left\{ \log\frac{Q_0}{Q} 
            + 2 \left[  \log\frac{Q}{\mu} - \frac{11}{12}  + 
              \frac{2 \pi}{\beta_0 \as(\mu)} \right] 
            \log\frac{1+\eta_1}{1-\eta_0}
          \right\}
   $ \\
  \multicolumn{2}{l}{ {\tt Gamma\-F\_\-Alpha\-S}}   \\
    & $\displaystyle \int_{Q_0}^Q dq \Gamma_f(q) =  \frac{2 n_f}{3 \beta_0}
    \log\frac{1+\eta_1}{1+\eta_0} $ \\
\end{tabular}
\end{minipage}}}

\put(0,87){\fbox{%
\begin{minipage}{435pt}
\begin{tabular}{ll}
  \multicolumn{2}{l}{ {\tt Gamma\_Lambda\_Massive}}  \\
& $\displaystyle
\Gamma_Q(Q,q,m) = \Gamma_q(Q,q) + \frac{C_F}{\pi}\frac{\as(q)}{q}
                    \left[\frac12 - \frac{q}{m}
                             \arctan \left(\frac{m}{q}\right)
                              -\frac{2m^2-q^2}{2m^2}
                             \log \left(\frac{m^2+q^2}{q^2}\right)
                         \right] $ \\
& $\displaystyle
\Gamma_F(q,m) =   \frac{T_R}{\pi}\frac{\as(q)}{q}\,\frac{q^2}{q^2+m^2}\,
                    \left[1 - \frac13\frac{q^2}{q^2+m^2}\right] $ \\   
\end{tabular}
\end{minipage}}}

\put(30,20){\fbox{%
\begin{minipage}{405pt}
\begin{tabular}{ll}
  \multicolumn{2}{l}{ {\tt Gamma\-Q\_\-Lambda\_\-Massive},}  \\
  \multicolumn{2}{l}{ {\tt Gamma\-G\_\-Lambda\_\-Massive}, and }\\
  \multicolumn{2}{l}{ {\tt Gamma\-F\_\-Lambda\_\-Massive}   
  use numerical integration.}
\end{tabular}
\end{minipage}}}
\end{pspicture}

\begin{align}
  \as^\mu(Q) &=
  \frac{\as(\mu)}{1 - \displaystyle\frac{\beta_0}{4 \pi} 
     \frac{\as(\mu)}{\log(\mu^2/Q^2)}} &
 \eta_0 &= \frac{\beta_0 \as(\mu)}{4 \pi}
 \log\left(\frac{Q_0^2}{\mu^2}\right)   &
 \eta_1&= \frac{\beta_0 \as(\mu)}{4 \pi}
 \log\left(\frac{Q^2}{\mu^2}\right)     \nnb \\
   \as^\Lambda(Q) &= \frac{2 \pi}{\beta_0 \log(Q/\Lambda)} &
   \xi_0 &= \log\frac{Q_0}{\Lambda} &
   \xi_1 &= \log\frac{Q}{\Lambda}   \nnb
\end{align}
\caption{\label{tab:branching_probabilities} Available implementations
  of NLL branching probabilities.}
\end{table}

\section{\label{app:observables} Observables}
The global properties of hadronic events may be characterised by a set
of observables, usually called event shapes. In section
\ref{Results_Sec} the following shape observables have been
considered. 
\begin{itemize}
 \item Thrust $T$ : \\
   The thrust axis $\vec{n}_T$ maximises the
   following quantity
  \begin{equation}
    T = \max_{\vec{n}_T} \left( \frac{\sum_i | \vec{p}_i \cdot
        \vec{n}_T |}{\sum_i | \vec{p}_i |} \right) \,,
  \end{equation}
   where the sum extends over all particles in the event.
   The thrust $T$ tends to 1 for events that has two thin back-to-back 
   jets (``pencil-like'' event), and it tends towards $1/2$ for
   perfectly isotropic events. 
 \item Thrust Major $T_{\rm Major}$ :  \\
   The thrust major vector $\vec{n}_{\rm Major}$ is defined in the same way as
   the thrust vector, but with the additional condition that
   $\vec{n}_{\rm Major}$ must lie in the plane perpendicular to $\vec{n}_T$:
  \begin{equation}
    T_{\rm Major} = \max_{\vec{n}_{\rm Major} 
      \perp \vec{n}_T } \left( \frac{\sum_i | \vec{p}_i \cdot
        \vec{n}_{\rm Major} |}{\sum_i | \vec{p}_i |} \right) \,.
  \end{equation}
\item Thrust Minor $T_{\rm Minor}$ : \\
  The minor axis is perpendicular to both the thrust axis and
  the major axis, 
  $\vec{n}_{\rm Minor} = \vec{n}_T \times \vec{n}_{\rm Major}$. The 
  value of thrust minor is then given by
  \begin{equation}
    T_{\rm Minor} = \frac{\sum_i | \vec{p}_i \cdot
        \vec{n}_{\rm Minor} |}{\sum_i | \vec{p}_i |} \,.
  \end{equation}
\item Oblateness $O$ : \\
  The oblateness is defined as the difference between thrust major
  $T_{\rm Major}$ and  thrust minor $T_{\rm Minor}$ :
  \begin{equation}
    O = T_{\rm Major} - T_{\rm Minor}
  \end{equation}

\item C-parameter $C$ : \\
  The C-parameter is derived from the eigenvalues of the linearised
  momentum tensor $\Theta^{\alpha\beta}$ , defined by
  \begin{equation}
    \Theta^{\alpha\beta} = \frac{1}{\sum_i |\vec{p}_i|} \sum_i
    \frac{p_i^\alpha p_i^\beta}{|\vec{p}_i|} \,, 
    \; \alpha,\beta = \{x,y,z\} \;.
  \end{equation}
  The three eigenvalues $\lambda_i$ of this tensor define $C$ with
  \begin{equation}
    C = 3\, (\lambda_1 \lambda_2 + \lambda_2 \lambda_3 + \lambda_1
    \lambda_3) \,.
  \end{equation}
\end{itemize}

\end{appendix}


\end{document}